\pdfoutput=1
\documentclass[letterpaper,12pt,leqno]{article}
\usepackage{paper}
\usepackage{cleveref}
\hypersetup{pdftitle={Productivity Shocks and Input Misallocation: A Decomposition}}

\begin{document}
\title{Productivity Shocks and Input Misallocation: A Decomposition}
\author{Davide Luparello
\thanks{Davide Luparello: dluparello@fsu.edu. I would like to thank Mark Roberts, Paul Grieco, and Jim Tybout for their invaluable comments and suggestions. I am also grateful to Carlo Altomonte, Gianmarco Ottaviano, Italo Colantone, Jonathan Eaton, Marc Henry, Karl Schurter, Ewout Verriest, Conor Ryan, Bradley Setzler, Fernando Parro, Jingting Fan, Kai-Jie Wu, Stephen Yeaple, Kala Krishna, and Benedicta Marzinotto for their helpful feedback. My appreciation extends to the organizers and participants of the Penn State Trade and Development and Applied Microeconomics Brownbag seminars, Bocconi Trade Tea seminar, MICROPROD Final Event Conference, and the 12th CompNet Annual Conference. Lastly, I would like to acknowledge the research assistance provided by Anthropic's Claude LLM within the guidelines specified in \cite{korinek2023generative}. This paper previously circulated under the names "Productivity, Inputs Misallocation, and the Financial Crisis" and "Do Productivity Shocks Cause Input Misallocation?". The data used in this study were collected during the development of the EU Horizon 2020 funded MICROPROD project (Grant Agreement No. 822390), which concluded in 2022. Additional project information is available at \url{https://cordis.europa.eu/project/id/822390/results}.}}
\affil{\it Florida State University and BAFFI}
\date{September 21, 2026\\
 \textcolor{blue}{{\href{https://drive.google.com/file/d/1mi7LMpdVqPVgNn1Ru0whGg2qpDYy4G7r/view?usp=sharing}{\small[click here for the most updated version]}}}}                      
\begin{titlepage}\maketitle

This paper asks how much input misallocation productivity uncertainty generates and at which stage of input decisions it arises. I separate revenue productivity by when each component is revealed and trace each into the gap between an input's marginal revenue product and its price. In six European countries, shocks revealed after an input is committed account for 20 percent of capital gap dispersion and 5 percent of labor gap dispersion. An unanticipated one percent rise in productivity raises the capital gap by 0.92 percent and the labor gap by 0.19 percent, because most of the shock passes into the wage.

\vspace{1em}\noindent \textbf{JEL codes:} D22, D24, L25, L60, O47.

\end{titlepage}

\doublespacing
\section{Introduction}\label{s:introduction}
A firm's forecast productivity informs key executive decisions about operations and investments. When leadership anticipates high productivity, they may increase orders for manufacturing inputs, expand staff and facilities, or allocate more funding to research and development for new products. Conversely, if lower productivity is projected, executives may reduce input orders to contain costs. However, firms face dynamic uncertainty in forecasting future productivity. As productivity shocks occur unpredictably over time, initial forecasts may diverge from realized productivity. When wedges emerge between anticipated and realized productivity, input decisions set ex ante can substantially diverge from optimal ex post quantities, generating ex post misallocation even when choices are ex ante optimal. 

How much input misallocation does productivity uncertainty generate, and at which stage of the firm's input decisions does it arise? This paper quantifies and decomposes that misallocation for every predetermined input of the production function. The mechanism is one of timing. Revenue total factor productivity (TFPR) is revealed in stages as the firm makes its input decisions, and I separate it into components by the stage at which each is revealed. An input committed before a component is revealed cannot respond to it within the period, so that component passes into the input gap, the ratio of the input's marginal revenue product (MRP) to its price. Following \cite{restuccia2008policy} and \cite{hsieh2009misallocation}, I measure misallocation by the dispersion of that gap across the firms of an industry: under some assumptions, the gap should be zero for every firm, so its dispersion indicates frictions that prevent resources from flowing from less to more efficient uses.

To infer the relationship between the input gap and TFPR and its components, I estimate a partial equilibrium production function model with the \cite{gandhi2020identification} estimator on 2000--2017 firm-level panel data from six European countries. The productivity process has period-specific coefficients, which capture regime shifts such as the Great Recession and the European debt crisis. The model separates TFPR into an expected productivity level, known when every input is chosen; an ex-ante innovation, revealed after capital and labor are committed and before materials is allocated; and an ex-post shock, unresolved until production is complete. Capital and labor are thus the predetermined inputs and materials the flexible one. Input prices differ in their timing as well: the rental rate of capital is known when capital is committed, whereas the wage is settled after the ex-post shock and may respond to it. The model retains productivity uncertainty as its sole explicit source of misallocation. The estimator recovers the three components, and their pass-through into the gap of each predetermined input gives, to my knowledge, the first decomposition of input gap dispersion that traces misallocation to the stage at which each component is revealed.

I obtain two results. First, productivity accounts for a substantial share of input gap dispersion, and the timing of the productivity behind that share differs by input: for capital, most of it is attributable to shocks revealed after the input is committed; for labor, most of it is attributable to productivity known when the input was chosen. Within a country-industry-year cell, the three components of TFPR together account for 28.5 percent of the dispersion of the capital gap and 18.4 percent of that of the labor gap. By component, the ex-post shock accounts for 16.5 percent for capital, the ex-ante innovation for 3.3, and expected productivity, which the firm knew when it chose its inputs, for 8.6; for labor the ex-post shock accounts for 1.9 percent, the ex-ante innovation for 3.1, and expected productivity for 13.5. The misallocation that productivity uncertainty generates is therefore about a fifth of capital gap dispersion and one twentieth of labor gap dispersion. For capital that share rises over the sample period, from 15 percent before the Great Recession to 22 percent from the European debt crisis on, through the ex-post shock; for labor it stays near one twentieth throughout.

Second, the model predicts that the ex-post shock passes one for one into the gap of any input committed before it realizes, unless the input's price responds to the shock. In the data, an unanticipated one percent rise in TFPR raises the marginal revenue product of capital by 0.92 percent relative to its rental rate, close to the predicted one and indistinguishable from it in Germany, Poland and Italy. For labor the same shock raises the marginal revenue product of labor by only 0.19 percent relative to the wage, because the wage responds to the shock: the wage bill rises by 0.54 percent, 0.41 through pay per head and 0.13 through headcount.

These results hinge on the \cite{gandhi2020identification} estimator. In Monte Carlo simulations, a factor shares estimator, by imposing constant returns to scale, overstates capital elasticities and attenuates the persistence of productivity, and a proxy-variable estimator is unbiased on average but too imprecise for inference on elasticity dispersion. In the firm-level panel, factor shares with sectoral expenditure shares delivers more capital gap dispersion than the baseline estimator and about the same labor gap dispersion. \cite{ruzic2023returns} document the same sensitivity in U.S. Census microdata, where industry-specific markups and returns to scale turn a 29 percent rise in measured misallocation over 1982 to 2007 into a 13 percent decline. Re-estimation under imperfect competition, with period-specific markups and EUROSTAT producer price deflators, yields dispersion series for TFPR and the input gaps and pass-through coefficients that closely track those of the competitive baseline.

This paper contributes to two literatures. The first measures misallocation by the dispersion of marginal revenue products across firms \citep{restuccia2008policy,hsieh2009misallocation}.\footnote{Determinants of misallocation include regulations and institutions \citep{bartelsman2009measuring,bartelsman2013cross}, employment protection and labor regulations \citep{restuccia2017causes,amirapu2020labor}, credit supply shocks and financial frictions \citep{gopinath2017capital,ben2023tfp}, markup dispersion \citep{meier2024monetary}, suboptimal firm selection \citep{yang2021micro}, and factor adjustment costs and imperfect information about productivity \citep{asker2014dynamic,david2019sources}. Measured dispersion may also reflect measurement error and data imputation \citep{gollin2021heterogeneity,bils2021misallocation,white2018imputation} and heterogeneity in technology and demand \citep{restuccia2013misallocation,bento2017misallocation,blackwood2021macro,haltiwanger2018misallocation}.} Part of this literature establishes that adjustment costs and productivity uncertainty generate dispersion in marginal revenue products even when every firm chooses its inputs optimally given its information at the time of the choice. \cite{asker2014dynamic} show this for capital modeled as a dynamic input with adjustment costs; \cite{david2016information} for inputs chosen on noisy signals about productivity, with the information structure inferred from production and stock market data; and \cite{david2019sources} decompose capital misallocation into adjustment costs, uncertainty, firm-specific factors and other distortions. \cite{feng2025firm} and \cite{berman2019demand} add learning as a source of that dispersion: inputs are chosen on beliefs about own productivity or own demand that later signals revise. In Chinese firm-level data, the dispersion of the marginal revenue product of capital falls with firm age as beliefs converge. French exporters set export quantities on beliefs about demand that noisy sales later revise; younger firms revise more strongly, and firms in more uncertain markets revise less. In this paper, productivity is revealed in stages and its components are recovered from the production function estimator, so the misallocation each unresolved component generates in each predetermined input's gap is quantified directly.

The second literature estimates production functions from firm-level panel data. \cite{gandhi2020identification} show that the proxy-variable methods of \cite{olley1996dynamics}, \cite{levinsohn2003estimating} and \cite{ackerberg2015identification} do not identify a gross output production function, and identify it instead from the first-order condition of the flexible input: its revenue share determines its elasticity, and the residual of that share equation is the ex-post shock. This paper turns the timing structure of that estimator into a measurement of misallocation. Each component of productivity is dated relative to the input decisions, so the pass-through of the components into the input gaps measures misallocation by the stage at which productivity is revealed, and the timing restriction on the ex-post shock, which the estimator maintains, becomes a testable prediction for the input gaps.

The remainder of the paper is organized as follows. Section \ref{s:theory} presents the model. Section \ref{s:data} describes the data. Section \ref{s:empirics} presents the estimation procedure. Section \ref{s:results} reports the elasticity and returns to scale estimates, the evolution of input gap and TFPR dispersion, the pass-through of productivity and its components into the input gaps, and the test of the ex-post restriction. Section \ref{s:robustness} examines the forecastable part of the recovered ex-post shock, the association between institutional quality and the dispersion of the productivity components, and the sensitivity of the results to imperfect competition and to the estimation method. Section \ref{s:conclusion} concludes.

\section{Theoretical Framework}\label{s:theory}
The theoretical framework follows \cite{gandhi2020identification}, augmented to allow time-varying productivity processes. The framework incorporates firms' imperfect information about productivity as the only modeled source of misallocation. Specifically, the model assumes multiple productivity shocks occurring between and after input allocations. The model remains intentionally agnostic on how capital and labor are allocated: the only restriction on that choice is that both are committed before either of the period's productivity shocks realizes. This flexible approach allows the framework to generate dynamics consistent with a wide range of frictions that could affect these inputs in both static and intertemporal settings\footnote{Examples are resale losses due to transaction costs, the market for lemons phenomenon \citep{akerlof1970market}, the physical cost of resale and refit costs for capital, hiring, firing, and training losses for workers \citep{bloom2009impact}, working capital and borrowing constraints, government regulations, transportation costs, subsidies, and taxes \citep{hsieh2009misallocation}.}.

Let $Y_{jt}$ denote the revenue and $K_{jt}$, $L_{jt}$, and $M_{jt}$ the capital, labor, and materials allocations of firm $j$ at time $t$. 
Lowercase variables indicate natural logarithms. 
The log revenue function is non-parametrically specified as
\begin{equation}
\label{eq:revenue}
    y_{jt}=f(k_{jt}, l_{jt}, m_{jt})+\nu_{jt}
\end{equation}
With a slight abuse of notation, let $F(k_{jt},l_{jt},m_{jt}) \equiv e^{f(k_{jt},l_{jt},m_{jt})}$ denote the corresponding level revenue function written as a function of log inputs, so that the level revenue equation reads $Y_{jt} = F(k_{jt},l_{jt},m_{jt})\, e^{\nu_{jt}}$. Here $\nu_{jt}$ is revenue total factor productivity, specified as the sum of a persistent and an unexpected component \citep{olley1996dynamics}
\begin{equation}
\label{eq:tfpr}
\nu_{jt}=\omega_{jt}+\varepsilon_{jt}
\end{equation} 
$\omega_{jt}$ is a persistent productivity factor that firm $j$ observes at the beginning of period $t$ but does not know at $t-1$. 
By contrast, $\varepsilon_{jt}$ represents a residual, short-term idiosyncratic revenue fluctuation observed only as the period ends.\footnote{
$\varepsilon_{jt}$ admits several observationally equivalent alternative interpretations. First, $\varepsilon_{jt}$ may reflect ex-post measurement error in firm revenue: \cite{bils2021misallocation} show that revenue noise alone can account for a nontrivial share of measured input misallocation. Second, because the estimation targets revenue rather than physical output, $\varepsilon_{jt}$ also absorbs firm-specific demand shocks \citep{foster2008reallocation}. Following \cite{gandhi2020identification}, I adopt the \textit{ex post productivity shock} interpretation: a shock to revenue productivity observed only at the end of period $t$. Under the model's timing assumptions, the supply-side versus demand-side origin of $\varepsilon_{jt}$ does not affect the misallocation argument, since any unforecastable revenue component revealed after inputs are committed generates dispersion in input MRPs that firms cannot eliminate ex post. Following the convention in the production-function estimation literature, I assume no measurement error in the input variables; \cite{collard2016production} and \cite{il2016estimating} study measurement error in capital.
}

Let $\mathcal{I}_{jt}$ denote the information set that firm $j$ holds at time $t$ before the realization of $\varepsilon_{jt}$, and $\tilde{\mathcal{I}}_{jt}$ the information set that firm $j$ holds at the end of period $t$.
The following assumptions characterize the pricing behavior, the input allocation, the persistent productivity process, and the realization of the productivity shocks. 
\begin{assumption}[Pricing Behavior]\label{assp}
Firms are price takers in the output and input markets. 
The firm faces the nominal input prices $w_{jt}$, $r_{jt}$, and $\rho_{jt}$ as the wage rate, the rental rate of capital, and the unit cost of materials. 
The rental rate and the unit cost of materials are known to the firm when it allocates materials. The wage rate is determined after the ex-post shock realizes and may respond to it. Assume the log-linear specification $w_{jt} = \bar{w}_{jt}\,e^{\theta \varepsilon_{jt}}$, where $\bar{w}_{jt} \in \mathcal{I}_{jt}$ is the part of the wage rate fixed before the ex-post shock realizes and the parameter $\theta \geq 0$ is the elasticity of the wage rate with respect to the ex-post shock. The case $\theta = 0$ is a wage rate fixed before the ex-post shock realizes.
\end{assumption}
\begin{assumption}[Input Allocations]\label{assa}
The firm chooses $K_{jt}$ and $L_{jt}$ as a function of $\tilde{\mathcal{I}}_{jt-1}$, so capital and labor are predetermined with respect to the period's productivity shocks. The model places no further restriction on how either input is chosen. The firm chooses $M_{jt}$ as a function of $\mathcal{I}_{jt}$, by solving a static value-added maximization problem taking $K_{jt}$ and $L_{jt}$ as given, so materials is the flexible production input.
\end{assumption}

Together, Assumptions~\ref{assp} and~\ref{assa} fix the quantity of capital and labor before either shock realizes, and the nominal outlay on capital before the ex-post shock realizes; the outlay on labor is fixed up to the wage response $\theta$.
\begin{assumption}[Persistent Productivity Process]\label{asspr}
$\omega_{jt}$ follows a time-varying Markov process with a mean-zero innovation that realizes in period $t$. Specifically,
\begin{equation}
\begin{aligned}
\label{eq:omega}
 &\omega_{jt}=g_{t}(\omega_{jt-1})+\eta_{jt}\\
&\omega_{jt-1}\in\tilde{\mathcal{I}}_{jt-1},\quad\eta_{jt}\notin\tilde{\mathcal{I}}_{jt-1}\quad\forall j,t\\
&E(\eta_{jt}\mid \tilde{\mathcal{I}}_{jt-1})=E(\eta_{jt})=0\\
    \end{aligned}
\end{equation} 
where $g_t(\cdot)$ is a continuous, strictly monotonic function, and 
$\eta_{jt}$ is the productivity innovation that realizes in period $t$.
\end{assumption}

The innovation $\eta_{jt}$ is not necessarily identically distributed each period since the only restriction pertains to its first moment.
Let its cross-sectional variance be time-varying, $\sigma^2_t$.
\begin{assumption}[Timing of the Productivity Shocks]\label{asst}
$\eta_{jt}$ realizes at the beginning of period $t$.
The persistent productivity component, $\omega_{jt}$, is then known at the beginning of period $t$. 
The unexpected component, $\varepsilon_{jt}$, is mean-independent of $\mathcal{I}_{jt}$ and realizes towards the end of period $t$. Specifically,
\begin{equation}
\begin{aligned}
    &\omega_{jt}\in\mathcal{I}_{jt},\quad
    \varepsilon_{jt}\notin \mathcal{I}_{jt},\quad
    \varepsilon_{jt}\in \tilde{\mathcal{I}}_{jt}\quad\forall j,t\\
    &E(\varepsilon_{jt}\mid \mathcal{I}_{jt})=E(\varepsilon_{jt})=0\\    
    &E(e^{\varepsilon_{jt}}\mid \mathcal{I}_{jt})=E(e^{\varepsilon_{jt}})=\mathcal{E}\\
    \end{aligned}
\end{equation}
where $\mathcal{E}$ is a scalar constant.
\end{assumption}

Similarly, the short-term fluctuation $\varepsilon_{jt}$ is not necessarily identically distributed.
Let its cross-sectional variance be time-varying, $\tau^2_t$.

The following timeline summarizes the assumptions above.
\begin{enumerate}
\item Firm $j$ ends period $t-1$ having observed the revenue $Y_{jt-1}$, and therefore knowing the period's TFPR $\nu_{jt-1}$ and both of its components, $\omega_{jt-1}$ and $\varepsilon_{jt-1}$. 
That is, the firm has information set $\tilde{\mathcal{I}}_{jt-1}$.
\item The firm now chooses $K_{jt}$ and $L_{jt}$ as a function of $\tilde{\mathcal{I}}_{jt-1}$.
\item Firm $j$ enters period $t$. $\eta_{jt}$ realizes and the rental rate $r_{jt}$, the unit cost of materials $\rho_{jt}$, and the pre-shock wage $\bar{w}_{jt}$ are known. 
The firm now knows $\omega_{jt}$ and has information set $\mathcal{I}_{jt}$.
\item The firm chooses the period $t$ materials allocation, $M_{jt}$, as a function of $\mathcal{I}_{jt}$, solving the following value-added maximization problem given the materials unit price $\rho_{jt}$
\begin{equation}
\label{eq:MAX}
    \max_{M_{jt}}\left[{E(F(k_{jt},l_{jt},m_{jt})e^{\nu_{jt}}\mid \mathcal{I}_{jt})-\rho_{jt} M_{jt}}\right]
\end{equation}
\item Finally, right before period $t$ ends, the firm observes the revenue $Y_{jt}$ and therefore the period's TFPR $\nu_{jt}$. Since $\omega_{jt}$ was already known, the ex-post shock $\varepsilon_{jt}$ follows as the residual. The wage $w_{jt}$ is settled at the same time.
The firm now has the information set $\tilde{\mathcal{I}}_{jt}$.
\end{enumerate}

Throughout this paper, I refer to $\eta_{jt}$ as the \textit{ex-ante productivity shock} and $\varepsilon_{jt}$ as the \textit{ex-post productivity shock}, relative to the materials allocation.

\subsection{How Productivity Shocks Map into Input Gap Dispersion}\label{mechanism}
Under the model's assumptions, I derive how each productivity component maps into the marginal revenue product of each input at the firm level, so that dispersion in firm-level productivities translates into dispersion of input marginal revenue products across firms at the industry level.

The marginal revenue product of input $X$ is the derivative of revenue $Y_{jt}$ with respect to the input allocation $X_{jt}$. By the chain rule, the MRP of any input $X$ can be written as a function of final revenue $Y_{jt}$, input allocation $X_{jt}$, and the revenue elasticity of that input, $\mathrm{elas}^X_{jt} \equiv \partial y_{jt}/\partial x_{jt}$.
\begin{equation}
\begin{aligned}
\label{eq:MP}
    \mathrm{MRP}^X_{jt}=\frac{\partial Y_{jt}}{\partial X_{jt}}=\frac{\partial Y_{jt}}{\partial y_{jt}}\frac{\partial y_{jt}}{\partial x_{jt}}\frac{\partial x_{jt}}{\partial X_{jt}}=\frac{Y_{jt}}{X_{jt}}\mathrm{elas}^X_{jt}\quad\forall X \in \{K,L,M\}
    \end{aligned}
\end{equation}
Taking the natural logarithm and totally differentiating with respect to TFPR leads to\footnote{Indeed,
\begin{equation*}
    \frac{\partial{\mathrm{mrp}}^X_{jt}}{\partial y_{jt}}=-\frac{\partial{\mathrm{mrp}}^X_{jt}}{\partial x_{jt}}=\frac{\partial {\mathrm{mrp}}^X_{jt}}{\partial\log \mathrm{elas}^X_{jt}}=1.
\end{equation*}}
\begin{equation}
\label{eq:tfp_growth}
\frac{d{\mathrm{mrp}}^X_{jt}}{d\nu_{jt}}=\frac{d{y_{jt}}}{d\nu_{jt}}-\frac{d{x_{jt}}}{d\nu_{jt}}+\frac{d{\log \mathrm{elas}^X_{jt}}}{d\nu_{jt}}
\end{equation}

Equation (\ref{eq:tfp_growth}) shows that three factors determine the elasticity of an input's marginal revenue product with respect to a hypothetical change in the firm's TFPR. The first and second terms are the elasticities of revenue and input allocation with respect to TFPR. The third term captures TFPR's effect on the input's revenue elasticity.

By combining equations (\ref{eq:tfpr}) and (\ref{eq:omega}), TFPR can be further decomposed into three components based on the firm's productivity information.
\begin{equation}
\label{eq:TFP_dec}
\nu_{jt} = g_t(\omega_{jt-1})+\eta_{jt}+\varepsilon_{jt}
\end{equation}
By the chain rule, the total effect of a change in TFPR on an input's MRP combines the effects of the changes in each of its three components.\footnote{$\mathrm{mrp}^X_{jt}$ depends on $\nu_{jt}$ only through its three components $(g_t(\omega_{jt-1}), \eta_{jt}, \varepsilon_{jt})$. Each path-specific response, defined as $d\mathrm{mrp}^X_{jt}/d\nu_{jt}$ when $\nu_{jt}$ varies through a single component while the other two are held fixed, is given by the chain rule: $(\partial \mathrm{mrp}^X_{jt}/\partial \omega_{jt-1})\,(\partial g_t(\omega_{jt-1})/\partial \omega_{jt-1})^{-1}$ along the $\omega_{jt-1}$-path, $\partial \mathrm{mrp}^X_{jt}/\partial \eta_{jt}$ along the $\eta_{jt}$-path, and $\partial \mathrm{mrp}^X_{jt}/\partial \varepsilon_{jt}$ along the $\varepsilon_{jt}$-path. The total response $d\mathrm{mrp}^X_{jt}/d\nu_{jt}$ depends on the direction in $(\omega_{jt-1}, \eta_{jt}, \varepsilon_{jt})$-space along which $\nu_{jt}$ is taken to vary. The estimator's identification argument relies only on the mean-independence conditions in Assumptions~\ref{asspr}--\ref{asst}.}
More formally,
\begin{equation}
\label{eq:combination}
\frac{d{\mathrm{mrp}}_{jt}^X}{d{ \nu_{jt}}}=c\!\left(\frac{d {\mathrm{mrp}}_{jt}^X}{d \omega_{jt-1}}\!\left(\frac{\partial g_t(\omega_{jt-1})}{\partial \omega_{jt-1}}\right)^{\!-1}\!,\frac{d{\mathrm{mrp}}_{jt}^X}{d\eta_{jt}},\frac{d{\mathrm{mrp}}_{jt}^X}{d\varepsilon_{jt}} \right)
\end{equation}
where $c(\cdot)$ aggregates the three path-specific responses.
In turn, the effect of each productivity component can be decomposed as
\begin{equation}
\label{eq:channels}
    \frac{d{\mathrm{mrp}}_{jt}^X}{d\theta}=\frac{d y_{jt}}{d \theta}-\frac{d{x_{jt}}}{d\theta}+\frac{d{\log \mathrm{elas}^X_{jt}}}{d\theta}\quad\forall \theta \in \{\omega_{jt-1},\eta_{jt}, \varepsilon_{jt}\}
\end{equation}

In summary, the framework implies that expected productivity, the ex-ante productivity shock, and the ex-post productivity shock affect three model objects: the firm's final revenue, the input allocation, and the input's revenue elasticity. The combined response of these three pathways yields the model-implied derivative of the MRP of a given input with respect to TFPR variation. Figure \ref{fig:mechanism} summarizes this mechanism. At the industry level, dispersion in firm-level productivity components is therefore predicted to generate dispersion in input MRPs.

The derivation uses no property of an input beyond its being predetermined, so it extends to any number of predetermined inputs, each adding a margin on which the effect of productivity uncertainty on input gap dispersion can be assessed. Materials is the one input for which this prediction cannot be tested: under the model, its allocation satisfies the first-order condition of the static value-added maximization problem of Assumption~\ref{assa}, so its gap equals the ex-post shock by construction.

\begin{figure}[ht]
    \centering
    \caption{How Productivity Shocks Map into Input Gap Dispersion: Mechanism}\label{fig:mechanism}
\begin{tikzpicture}[node distance=1.5cm and 2.2cm, auto, >=Latex]  %

\tikzstyle{mynode} = [rectangle, rounded corners, fill=white, draw=black, minimum width=2.5cm, minimum height=1cm, text centered]

\node[mynode] (output) {Realized revenue};
\node[mynode, below=of output] (allocation) {Input allocation};
\node[mynode, below=of allocation] (elasticity) {Revenue elasticity};

\node[mynode, left=of allocation] (ex-ante) {Ex-ante productivity shock};
\node[mynode, above=of ex-ante] (hetero) {Expected productivity};
\node[mynode, below=of ex-ante] (ex-post) {Ex-post productivity shock};

\node[mynode, right=of allocation] (mp)  {MRP};

\draw[->, thick, shorten >=1mm, shorten <=1mm] (hetero) -- (output);
\draw[->, thick, shorten >=1mm, shorten <=1mm] (hetero) -- (allocation);
\draw[->, thick, shorten >=1mm, shorten <=1mm] (hetero) -- (elasticity);

\draw[->, thick, shorten >=1mm, shorten <=1mm] (ex-ante) -- (output);
\draw[->, thick, shorten >=1mm, shorten <=1mm] (ex-ante) -- (allocation);
\draw[->, thick, shorten >=1mm, shorten <=1mm] (ex-ante) -- (elasticity);

\draw[->, thick, shorten >=1mm, shorten <=1mm] (ex-post) -- (output);
\draw[->, thick, shorten >=1mm, shorten <=1mm] (ex-post) -- (allocation);
\draw[->, thick, shorten >=1mm, shorten <=1mm] (ex-post) -- (elasticity);

\draw[->, thick, shorten >=1mm, shorten <=1mm] (output) -- (mp);
\draw[->, thick, shorten >=1mm, shorten <=1mm] (allocation) -- (mp);
\draw[->, thick, shorten >=1mm, shorten <=1mm] (elasticity) -- (mp);
\end{tikzpicture}
\note{\scriptsize Each productivity component (left) affects the MRP of an input (right) through three pathways: realized revenue, input allocation, and revenue elasticity. Arrows represent the partial effects formalized in equations~(\ref{eq:tfp_growth})--(\ref{eq:channels}).}
\end{figure}

\subsection{Materials: The Marginal Revenue Product Gap Is the Ex-Post Shock}\label{materials_identity}
The mechanism just described applies to all three inputs, but for materials the mapping degenerates into an identity, because materials is the only input the model allocates through a first-order condition. Taking the derivative of the value-added problem in equation~(\ref{eq:MAX}) with respect to $M_{jt}$, and applying the distributional restrictions in Assumptions~\ref{asspr} and~\ref{asst}, gives $\frac{\partial F(k_{jt},l_{jt},m_{jt})}{\partial M_{jt}}e^{\omega_{jt}}\mathcal{E}=\rho_{jt}$. Let $s_{jt}\equiv\log\left(\rho_{jt}M_{jt}/Y_{jt}\right)$ denote the log nominal materials share of revenue. Rearranging the first-order condition in terms of the revenue elasticity of materials yields\footnote{Taking logs and substituting $\omega_{jt}=y_{jt}-f(k_{jt},l_{jt},m_{jt})-\varepsilon_{jt}$ gives $\log\rho_{jt}=y_{jt}-f(k_{jt},l_{jt},m_{jt})-\varepsilon_{jt}+\log\mathcal{E}+\log\frac{\partial F(k_{jt},l_{jt},m_{jt})}{\partial M_{jt}}$. Since $\log\frac{\partial F(k_{jt},l_{jt},m_{jt})}{\partial M_{jt}}-f(k_{jt},l_{jt},m_{jt})=\log\mathrm{elas}^M_{jt}-m_{jt}$, adding $m_{jt}-y_{jt}$ to both sides delivers equation~(\ref{eq:share_theory}).}
\begin{equation}
\label{eq:share_theory}
    s_{jt}=\log \mathrm{elas}^M_{jt}+\log\mathcal{E}-\varepsilon_{jt}
\end{equation}
From equation~(\ref{eq:MP}), $\mathrm{MRP}^M_{jt}=(Y_{jt}/M_{jt})\,\mathrm{elas}^M_{jt}$, so the materials gap, the log marginal revenue product of materials less the log of its unit cost $\rho_{jt}$, is
\begin{equation}
\label{eq:mrpm_identity}
    \mathrm{mrp}^M_{jt}-\log\rho_{jt}=\log \mathrm{elas}^M_{jt}-s_{jt}=\varepsilon_{jt}-\log\mathcal{E}
\end{equation}
where the first equality uses the definition of $s_{jt}$ and the second uses equation~(\ref{eq:share_theory}). Hence the materials gap and the ex-post shock differ by the constant $\log\mathcal{E}$, firm by firm and period by period.

Two consequences follow. First, $\mathrm{Var}(\mathrm{mrp}^M-\log\rho)=\mathrm{Var}(\varepsilon)$ exactly, within any group over which $\mathcal{E}$ is constant. Second, a regression of the materials gap on $\varepsilon$ returns a slope of one and an $R^2$ of one, with no residual variance left to explain. The relation in equation~(\ref{eq:mrpm_identity}) holds whatever process generated the data. The testable content of the mechanism therefore lies in capital and labor, whose marginal revenue products are not tied to $\varepsilon_{jt}$ by construction since equation~(\ref{eq:share_theory}) restricts only the materials margin.

\section{Data and Descriptive Statistics}\label{s:data}
The annual firm-level harmonized balance sheet data used in this paper are obtained from the MICROPROD project's Micro Data Infrastructure for European manufacturing firms (Statistical Classification of Economic Activities, NACE, Rev2 code C), hereafter MP.\footnote{\url{https://cordis.europa.eu/project/id/822390}} Funded by the European Union (EU) and built on Bureau van Dijk's Orbis database, MICROPROD integrates international European microdata to inform policymakers on growth and reform policies. 

As of 2020, MP contained approximately 500,000 unique manufacturing firms for Italy, France, and Spain operating between 2000 and 2017. Additional German, Polish, and Romanian manufacturing firms operating between 2004 and 2018 have since been included, and I use them through 2017. MP's careful examination of firm operating status\footnote{i.e., active or inactive.}, conservative imputation of missing values, and selection of a \textit{productivity sample} minimally subject to imputation produce a representative sample that closely replicates Eurostat's aggregate Structural Business Statistics.\footnote{\url{https://ec.europa.eu/eurostat/web/structural-business-statistics}} \cite{altomonte2020employment} extensively detail MP's data collection and cleansing. Recent applications of MP include \cite{abele2021one} and \cite{altomonte2022intangible}; the underlying Orbis data have been used more widely (e.g., \citealp{asker2014dynamic,gopinath2017capital,kalemliconstruct,altomonte2025liquidity}).

I use the variable \textit{Operating Revenue (Turnover)} to approximate the firm's revenue $Y$, and the \textit{Number of Employees} to approximate the workforce $L$. MP does not have separate information on prices and quantities for intermediates and capital. I deflate the variables \textit{Material Costs} for intermediates $M$, and \textit{Total Fixed Assets} (tangible and intangible) for capital $K$. I source industry-level intermediate inputs (for materials) and gross output (for capital) deflators from EU-KLEMS (Capital, Labor, Energy, Materials, Services), which reports manufacturing at thirteen industrial groups of one to three NACE Rev2 two-digit codes.

Table \ref{tab:Descriptive} reports summary statistics for the main variables used in the analysis. From the three quartiles (Q1, median, and Q3), it is apparent that all variables' distributions are heavily right-skewed, with many small firms and few big firms. The minima and maxima indicate that I capture both very small and very large firms. However, the comparison of medians and means across countries indicates that the German and Polish samples underrepresent smaller firms, which may attenuate the estimated variance of productivity shocks for those countries. This problem is less pronounced for the other countries.\footnote{Coverage also varies across countries along sector composition and time-series length, so cross-country comparisons should be read as conditional on the observed sample.} The sample is an unbalanced panel spanning 2000--2017 (2004--2017 for Germany, Poland, and Romania): firms may enter or exit, but those with gaps in consecutive-year coverage are dropped to ensure valid lag construction.

\begin{table}[!h]
    \centering
    \caption{Descriptive Statistics for Main Variables}
    \label{tab:Descriptive}
\begin{adjustbox}{width=\textwidth}
        \begin{tabular}{llrrrrrr}
  \toprule
Variable & Statistic & Germany & Spain & France & Italy & Poland & Romania \\
  \midrule
Operating & Mean & 82,820 & 5,305 & 14,335 & 7,209 & 18,848 & 3,246 \\
  Revenue, & SD & 682,296 & 98,994 & 263,397 & 90,308 & 96,147 & 35,602 \\
   Th. Euros & Median & 19,132 & 681 & 1,201 & 1,673 & 4,137 & 160 \\
   & Q1 & 5,390 & 261 & 399 & 679 & 1,523 & 49 \\
   & Q3 & 53,200 & 2,063 & 4,654 & 4,562 & 11,196 & 650 \\
   & Min & 1 & 1 & 1 & 1 & 1 & 1 \\
   & Max & 65,336,438 & 25,537,175 & 51,905,000 & 29,382,602 & 4,818,801 & 4,990,944 \\
  \midrule
Cost & Mean & 51,540 & 3,459 & 7,336 & 3,945 & 12,132 & 1,936 \\
 of Materials, & SD & 1,201,597 & 85,857 & 181,084 & 68,914 & 72,588 & 27,217 \\
  Th. Euros, & Median & 7,946 & 305 & 331 & 644 & 2,051 & 62 \\
   Undeflated & Q1 & 1,816 & 99 & 94 & 198 & 632 & 16 \\
   & Q3 & 25,716 & 1,073 & 1,704 & 2,077 & 6,406 & 281 \\
   & Min & 1 & 1 & 1 & 1 & 1 & 1 \\
   & Max & 298,700,068 & 23,475,881 & 36,833,000 & 24,205,783 & 4,239,812 & 4,192,198 \\
  \midrule
Fixed & Mean & 23,210 & 1,980 & 3,644 & 2,284 & 7,112 & 1,516 \\
  Assets, & SD & 177,734 & 29,474 & 98,443 & 36,201 & 42,499 & 13,156 \\
  Th. Euros, & Median & 2,589 & 180 & 167 & 314 & 1,034 & 41 \\
Undeflated & Q1 & 483 & 50 & 53 & 81 & 260 & 10 \\
   & Q3 & 9,852 & 644 & 589 & 1,164 & 3,403 & 195 \\
   & Min & 1 & 1 & 1 & 1 & 1 & 1 \\
   & Max & 20,739,803 & 5,867,352 & 17,469,000 & 12,144,733 & 2,420,068 & 956,784 \\
  \midrule
Number & Mean & 219 & 23 & 48 & 28 & 156 & 56 \\
  of Employees & SD & 641 & 120 & 306 & 154 & 367 & 264 \\
   & Median & 87 & 8 & 10 & 11 & 70 & 8 \\
   & Q1 & 32 & 4 & 4 & 5 & 26 & 3 \\
   & Q3 & 209 & 19 & 31 & 24 & 159 & 28 \\
   & Min & 1 & 1 & 1 & 1 & 1 & 1 \\
   & Max & 38,383 & 17,284 & 49,425 & 33,636 & 14,600 & 18,456 \\
  \midrule
  Observations &  & 76,403 & 1,000,134 & 606,514 & 1,261,767 & 55,581 & 168,393 \\
   \bottomrule
\end{tabular}
\end{adjustbox}
\note{This table presents summary statistics for the key variables used in the analysis by country. The variables shown are \textit{Operating Revenue}, a proxy for output ($Y$); \textit{Cost of Materials}, a proxy for intermediates ($M$), reported before deflation; \textit{Fixed Assets}, a proxy for capital ($K$), reported before deflation; and \textit{Number of Employees} (headcount), a proxy for labor ($L$). Materials and capital are deflated using industry-level intermediate inputs and gross output deflators from EU-KLEMS, respectively. The table displays the mean, standard deviation (SD), median, first and third quartiles (Q1 and Q3), minimum and maximum for each variable and country. The sample period is 2000--2017 for France, Spain, and Italy, and 2004--2017 for Germany, Poland, and Romania. Observations denote firm-year observations.}
\end{table}

\section{Empirical Framework}\label{s:empirics}
Building on the nonparametric approach of \cite{gandhi2020identification} (hereafter GNR), I adapt the estimation procedure to allow for time-varying productivity dynamics. Additional details can be found in Appendix \ref{Gandhi}.

Stage one estimates the share equation~(\ref{eq:share_theory}), approximating the materials revenue elasticity by a polynomial. Applying least squares delivers two key outputs: materials revenue elasticities and firm-level ex-post productivity shocks. Integrating the estimated materials elasticity over materials and subtracting both the integral and the estimated shocks from revenue uncovers the sum of two unobserved components: the portion of revenue unrelated to materials and the persistent component of TFPR.

In stage two, polynomials approximate the Markov process for the persistent component of TFPR and the remaining part of the revenue function. The polynomial parameters are estimated via generalized method of moments (GMM), exploiting the orthogonality of the ex-ante productivity shock to predetermined input allocations and past persistent productivity.

The procedure restricts the number of flexible inputs to one, not the number of predetermined inputs: further predetermined inputs enlarge the first-stage polynomial and the portion of revenue unrelated to materials, and supply the moment conditions that identify it. Validity of the nonparametric bootstrap standard errors is discussed in Appendix \ref{app:asym}. The estimated model recovers firm-level input elasticities, marginal revenue products, and productivity components.

\section{Results}\label{s:results}
I let the productivity Markov process $g_t(\cdot)$ vary across five time periods: \textit{Beginning of the millennium} (2001--2002); \textit{Pre-crisis} (2003--2006); \textit{Great Recession} (2007--2009); \textit{European debt crisis} (2010--2013); and \textit{Post-crisis} (2014--2017). Estimation for Germany, Poland, and Romania starts in 2005. I estimate the production function separately for each of the six countries and eight industry groups.\footnote{The eight groups, with their NACE Rev.~2 codes, are
\label{industry_list}\begin{enumerate}
    \item Food, beverages and tobacco (10, 11, and 12);
    \item Textiles, apparel and leather (13, 14, and 15);
    \item Wood, paper, and printing (16, 17, and 18);
    \item Coke, chemicals, and pharmaceuticals (19, 20, 21);
    \item Rubber, plastics, metallic and non-metallic mineral products, fabricated metal products (22, 23, 24, and 25);
    \item Electronic, optical products and electrical equipment (26 and 27);
    \item Machinery, motor vehicles and other transport equipment (28, 29, and 30);
    \item Furniture and other manufacturing (31, 32, and 33).
\end{enumerate}} I compute standard errors by a non-parametric bootstrap that resamples firms, with 100 replications per estimation. Appendix~\ref{app:est_res} reports the estimates.\footnote{The log marginal revenue product of an input contains the log of its revenue elasticity, which is undefined for a non-positive value. I therefore exclude the firm-year observations in which any of the three estimated elasticities is non-positive, 5.46 percent of the sample, from the dispersion series and the regressions that follow.}

\begin{figure}[p]
    \caption{Revenue Elasticities and Returns to Scale Distributions}
    \label{fig:elas}
    \centering
    \begin{subfigure}[h]{0.48\textwidth}
        \centering
        \includegraphics[width=0.95\textwidth]{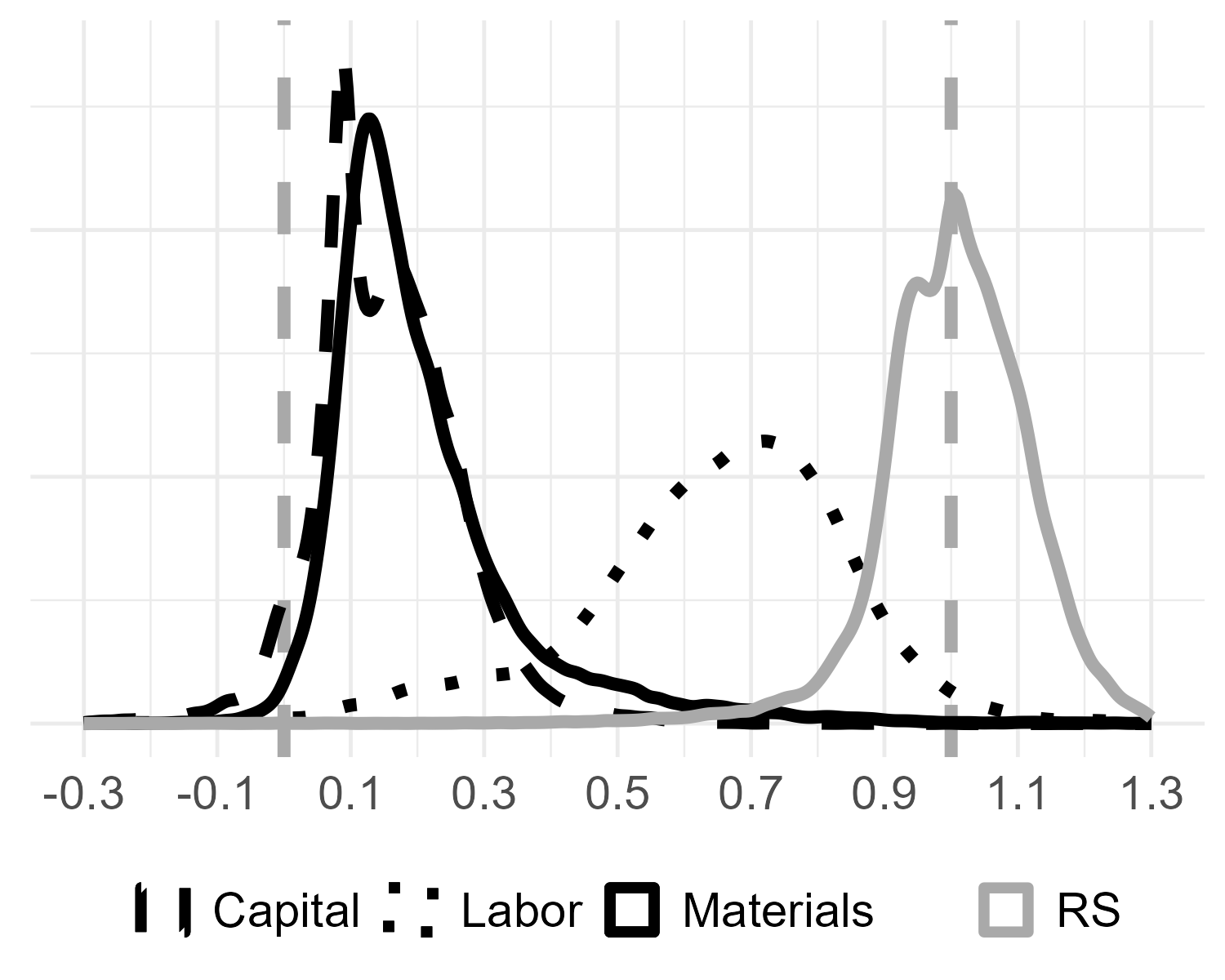}
        \caption{Germany}
        \label{fig:DEU}
    \end{subfigure}
    \hfill
    \begin{subfigure}[h]{0.48\textwidth}
        \centering
        \includegraphics[width=0.95\textwidth]{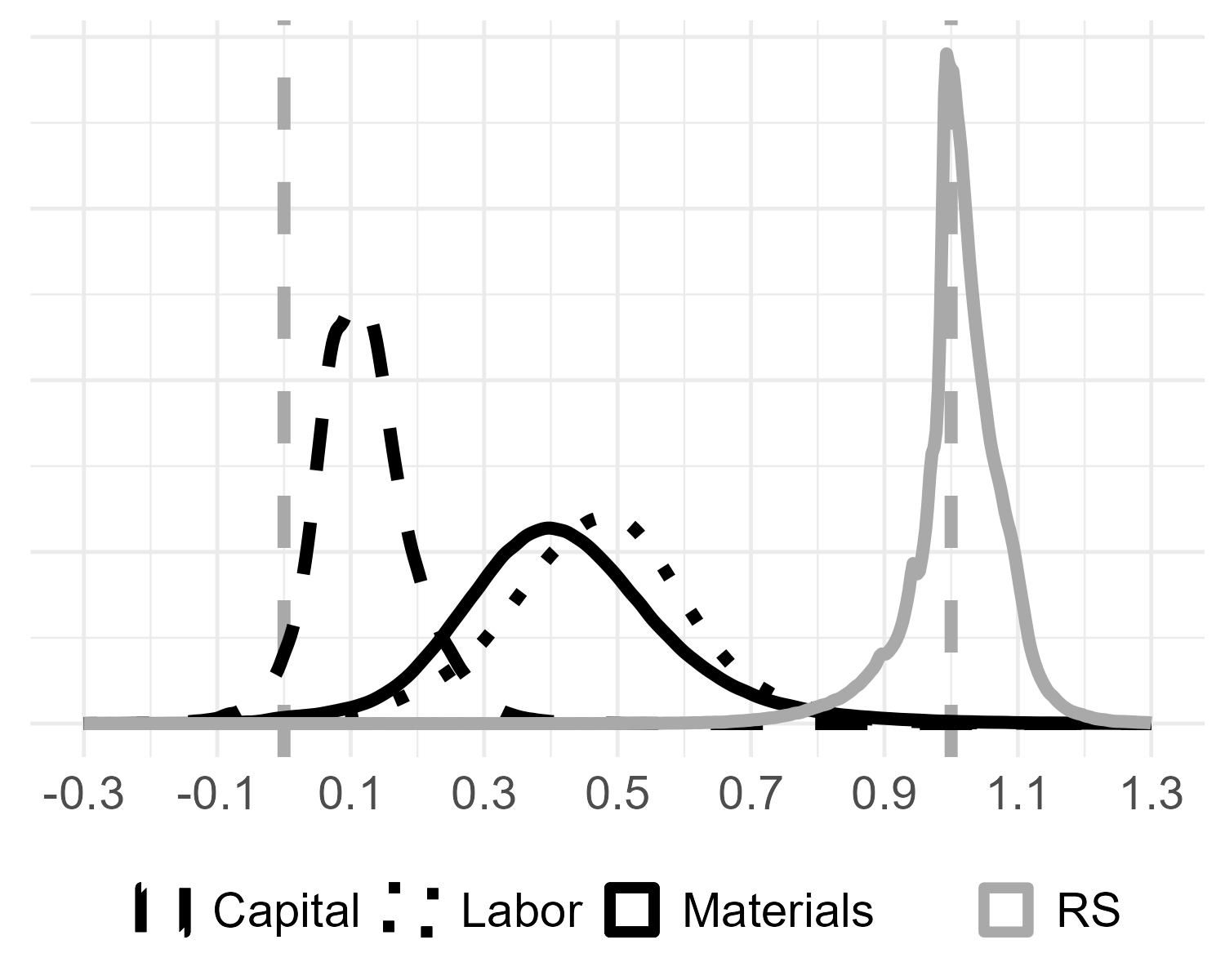}
        \caption{Spain}
        \label{fig:ESP}
    \end{subfigure}

    \begin{subfigure}[h]{0.48\textwidth}
        \centering
        \includegraphics[width=0.95\textwidth]{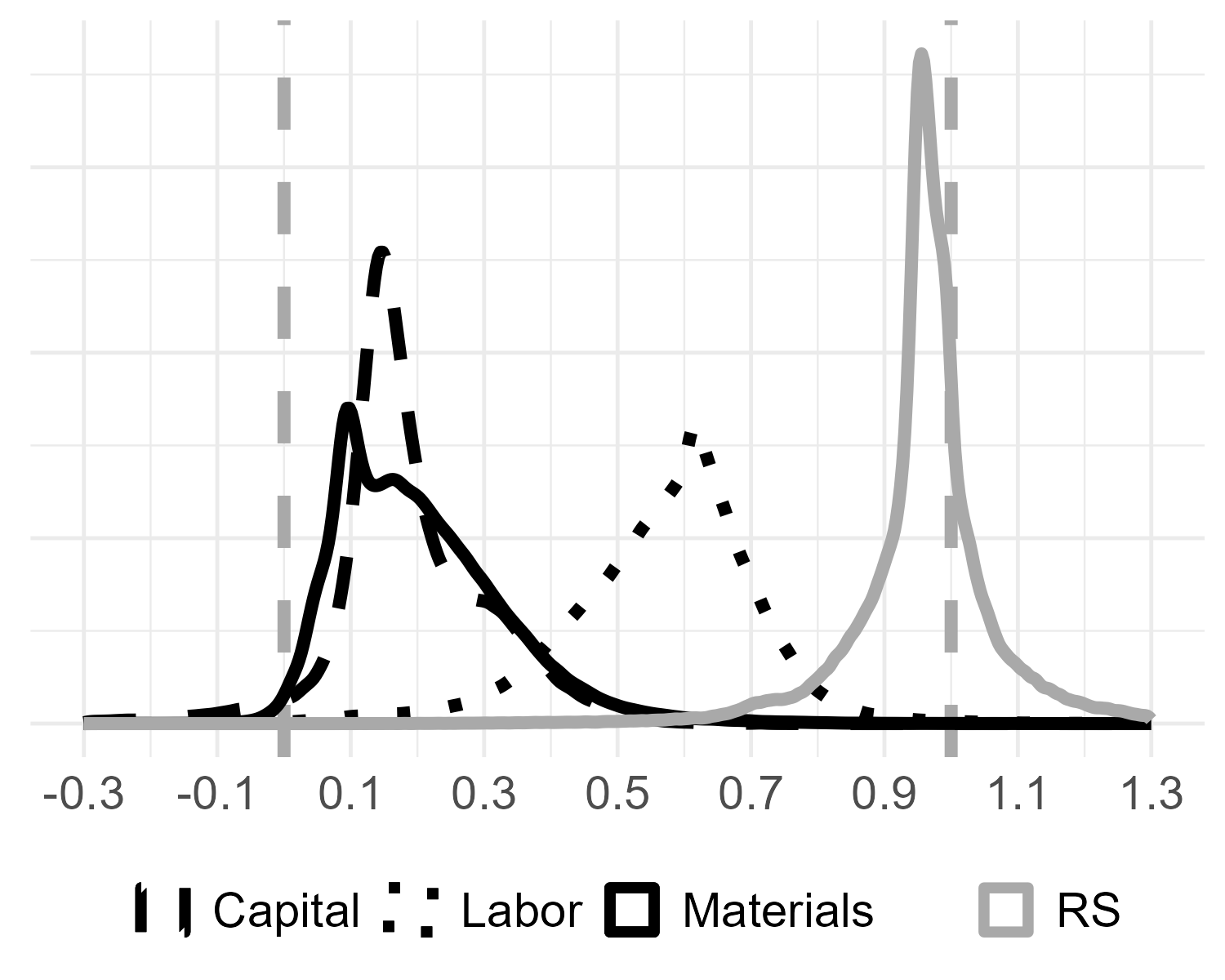}
        \caption{France}
        \label{fig:FRA}
    \end{subfigure}
        \hfill
    \begin{subfigure}[h]{0.48\textwidth}
        \centering
        \includegraphics[width=0.95\textwidth]{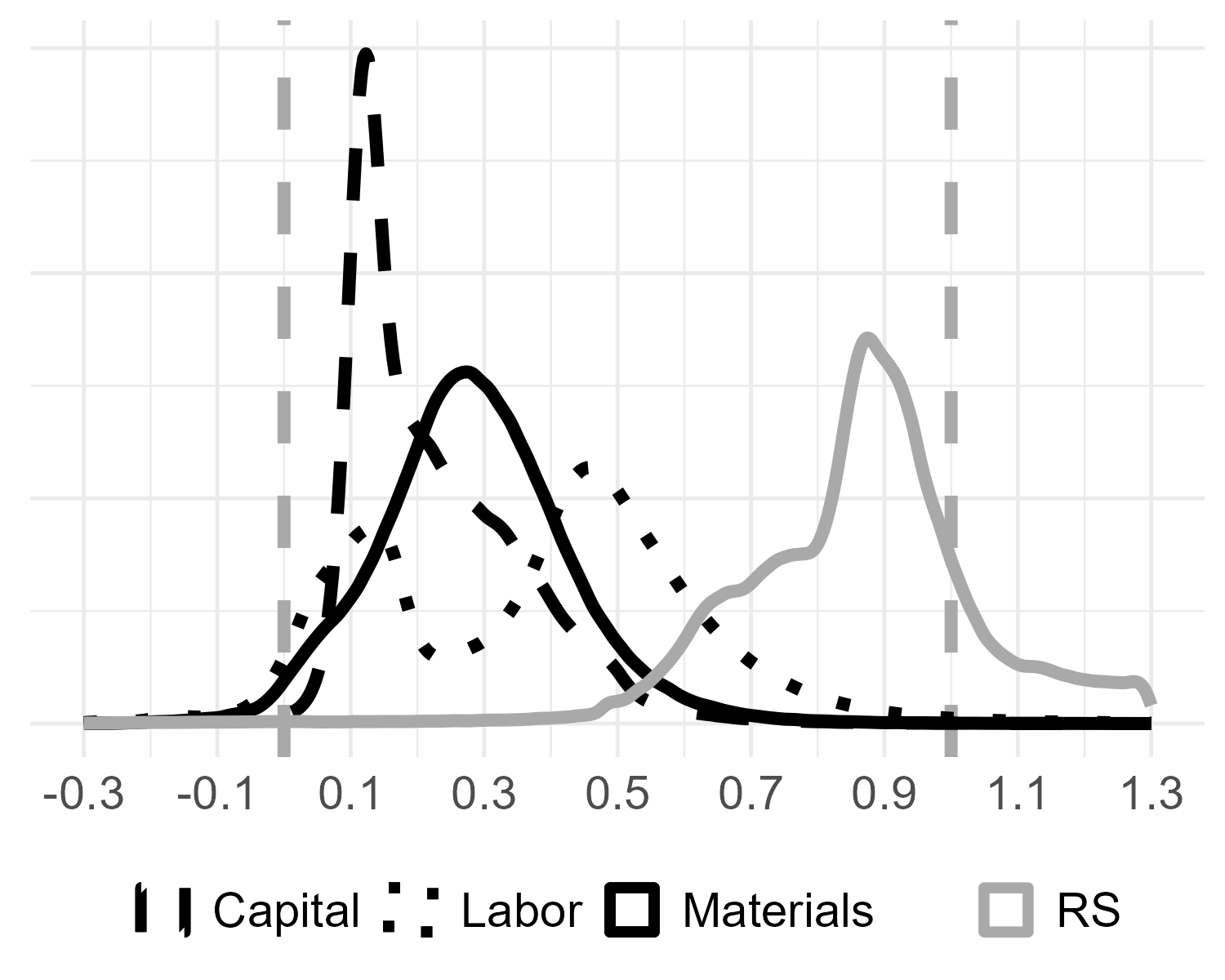}
        \caption{Italy}
        \label{fig:ITA}
    \end{subfigure}
    
    \begin{subfigure}[h]{0.48\textwidth}
        \centering
        \includegraphics[width=0.95\textwidth]{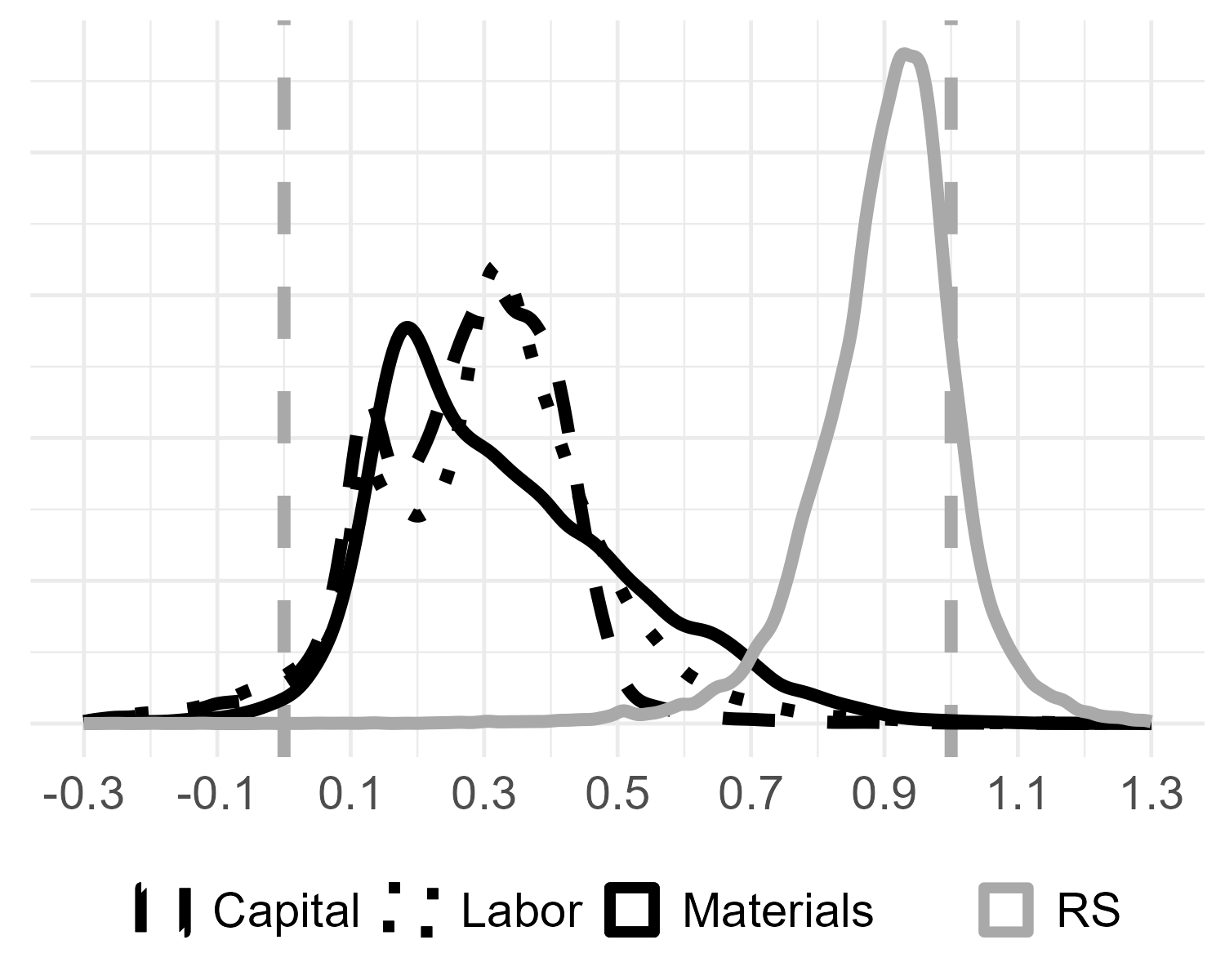}
        \caption{Poland}
        \label{fig:POL}
    \end{subfigure}
    \hfill
    \begin{subfigure}[h]{0.48\textwidth}
        \centering
        \includegraphics[width=0.95\textwidth]{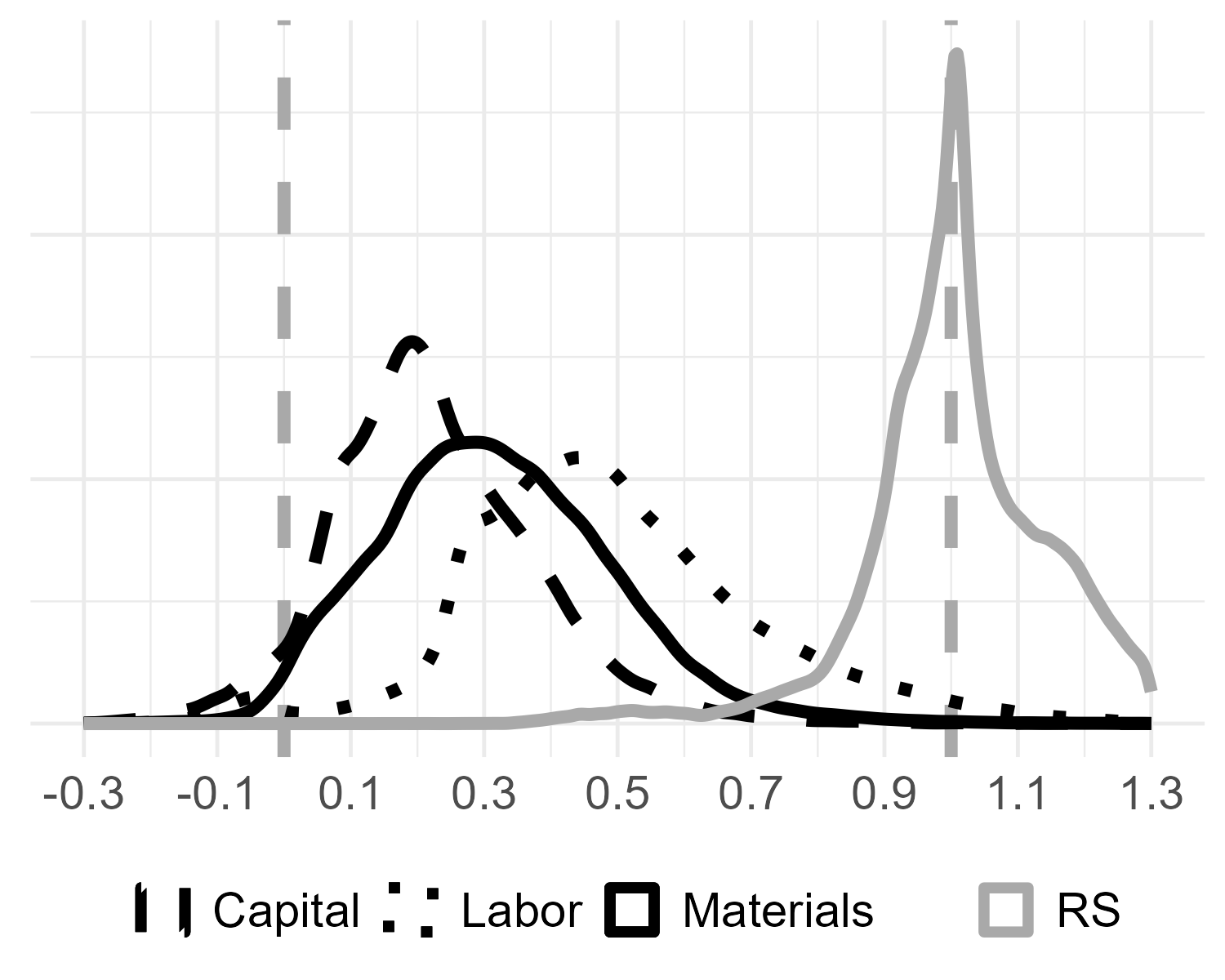}
        \caption{Romania}
        \label{fig:ROM}
    \end{subfigure}
    
    \note{\scriptsize This figure displays the industry and time pooled distributions of estimated input revenue elasticities and returns to scale (RS) by country: Germany (Fig. \ref{fig:DEU}), Spain (Fig. \ref{fig:ESP}), France (Fig. \ref{fig:FRA}), Italy (Fig. \ref{fig:ITA}), Poland (Fig. \ref{fig:POL}), and Romania (Fig. \ref{fig:ROM}). The inputs shown are materials (black solid line), capital (black dashed line), and labor (black dotted line). A solid grey line depicts the distribution of RS. The two dashed grey vertical lines delimit the interval $(0,1)$. Sample period: 2000--2017 (2004--2017 for Germany, Poland, and Romania).}
\end{figure}

Across historical periods, the parameter estimates and standard errors for the Markov process $\omega_{jt}$ indicate it can be closely approximated by a first-order autoregressive (AR(1)) process with drift. The estimated input revenue elasticities and returns to scale distributions, which are functions of the model parameters\footnote{For derivations of the firm-time specific elasticities, see the system of equations (\ref{eq:elas}). I compute firm-time specific returns to scale as the sum of the input elasticities.}, are informative about the underlying production technology. Figure~\ref{fig:elas} plots the distribution densities of elasticities and returns to scale for each country, pooled across years and industries.

Most revenue elasticity distributions lie within the unit interval. France and Germany are the most labor-intensive, with median\footnote{Because the distributions are highly skewed, I use the median as the measure of central tendency.} labor elasticities of 0.58 in France and 0.67 in Germany, against median materials elasticities of 0.18 and 0.17. Spain has the highest median materials elasticity at 0.41, followed by Romania at 0.31, Poland at 0.30, and Italy at 0.28. Poland has the highest median capital elasticity at 0.28 and the lowest median labor elasticity at 0.31. Italy's labor elasticity distribution is double-peaked, with modes near 0.1 and 0.45. Median returns to scale are close to one in Germany, Spain, and Romania (1.01), slightly below one in France (0.96), and lower in Poland and Italy (0.92 and 0.88). Input intensity therefore varies considerably across the six countries.

Throughout this section I measure misallocation by the input gap: the gap between an input's marginal revenue product and the price the firm paid for that input. Marginal revenue product has the units of a price and equals the input price at an undistorted optimum, so its dispersion mixes dispersion in the input gap with dispersion in input prices themselves. Dividing equation~(\ref{eq:MP}) by the input price and taking logs gives
\begin{equation}
\label{eq:gap}
    \underbrace{\mathrm{mrp}^{X}_{jt} - \log P^X_{jt}}_{\text{Input Gap}} = \log \mathrm{elas}^X_{jt} + \log\!\left(\frac{Y_{jt}}{P^X_{jt}X_{jt}}\right), \qquad X\in\{K,L,M\},
\end{equation}
where $P^X_{jt}$ is the nominal input price of Assumption~\ref{assp} ($r_{jt}$, $w_{jt}$, and $\rho_{jt}$ for capital, labor, and materials) and $P^X_{jt}X_{jt}$ is the firm's nominal outlay on that input. I measure it from the firm's accounts, as the wage bill for labor and materials costs for materials.\footnote{The labor gap is the log of the labor elasticity times revenue over the wage bill, so it prices a euro of labor. Differences in pay per worker across firms that reflect worker quality raise the marginal revenue product per worker and the wage bill together and cancel in the gap, so they do not enter its dispersion. In the production function, labor is headcount.} For capital I use the book value of fixed assets.\footnote{For capital the outlay is therefore the nominal stock of fixed assets at book value, so the capital gap omits the log rental rate $\log r_{jt}$ that the log rental payment $\log r_{jt} + \log K_{jt}$ would add. A rental rate common to the firms of a country-industry-year cell shifts every capital gap in the cell by the same amount, so the within-cell dispersion and the regressions with cell fixed effects are unaffected. In the production function, capital is the deflated stock of Section~\ref{s:data}.}

\subsection{The (Co)Evolution of the Dispersions of Input Gaps and TFPR}
\label{sec:disp_evolution}

This subsection documents how the dispersion of TFPR and of the input gaps of equation~(\ref{eq:gap}) evolved over the sample period, pooled across the six countries. Under the timing assumptions of Section~\ref{s:theory}, when an input is committed before a productivity shock realizes its input gap absorbs that shock, so the framework implies that the dispersion of the capital and labor gaps should co-move with the dispersion of TFPR. The aggregate series are a first, descriptive look at that implication.

\begin{figure}[!h]
    \caption{Pooled Evolution of TFPR and Input Gap Dispersion}
    \label{fig:pooled_disp}
    \centering
    \begin{subfigure}[h]{0.48\textwidth}
        \centering
        \includegraphics[width=\textwidth]{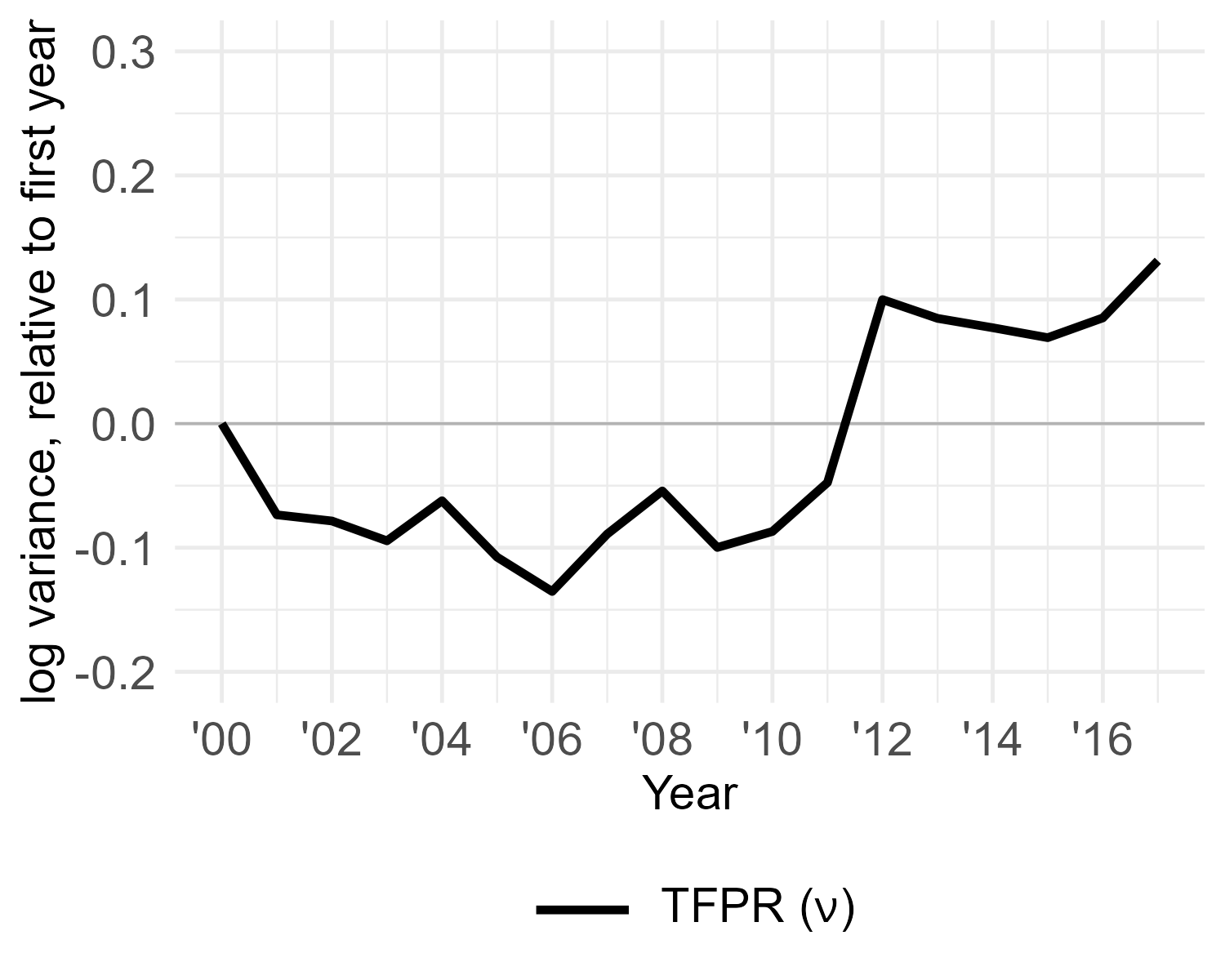}
        \caption{TFPR}
        \label{fig:pooled_prod}
    \end{subfigure}
    \hfill
    \begin{subfigure}[h]{0.48\textwidth}
        \centering
        \includegraphics[width=\textwidth]{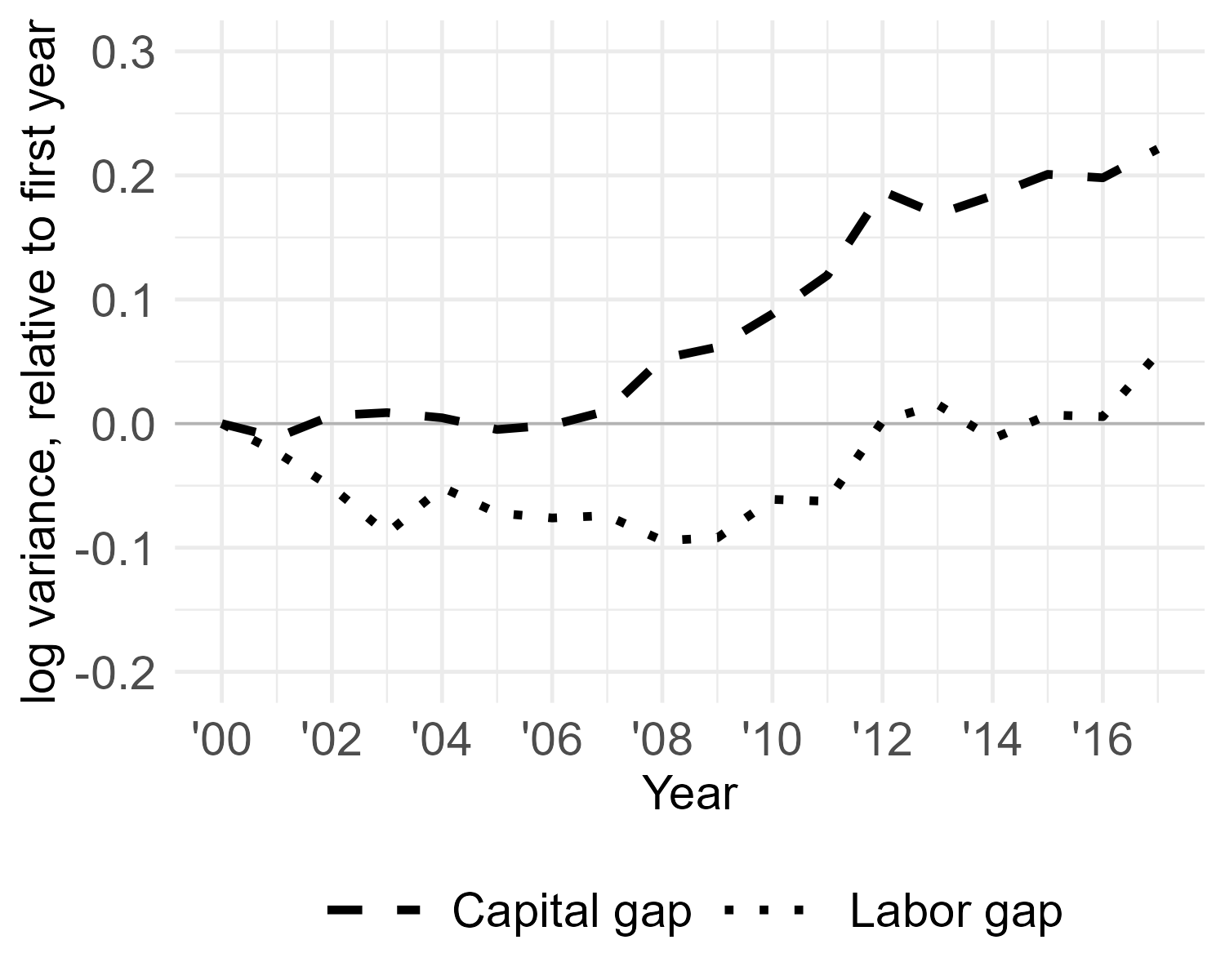}
        \caption{Capital and labor gaps}
        \label{fig:pooled_gap}
    \end{subfigure}
    \note{\scriptsize For each country and year, the aggregate variance is the weighted average across industries of the within-industry variance, with time-invariant weights equal to each industry's average share of the country's manufacturing revenue. Each line is the equal-weight average across countries of the log aggregate variance of the variable named, expressed relative to the country's first sample year (2000 for Spain, France, and Italy; 2004 for Germany, Poland, and Romania), so that every series starts at zero and the entry of the three later countries does not shift the level. Figure~\ref{fig:pooled_prod}: TFPR, $\nu_{jt} = \omega_{jt} + \varepsilon_{jt}$. Figure~\ref{fig:pooled_gap}: the capital gap (dashed) and the labor gap (dotted), each $\mathrm{mrp}^X_{jt} - \log P^X_{jt}$. Three countries are observed before 2004 and six from 2004; the small rise at 2004 in the TFPR and labor series reflects the three entrants starting at zero. Appendix~\ref{app:country_disp} reports the series country by country.}
\end{figure}

Figure~\ref{fig:pooled_disp} plots the pooled dispersion of TFPR (Figure~\ref{fig:pooled_prod}) and of the capital and labor gaps (Figure~\ref{fig:pooled_gap}) over time, constructed in three steps. For each country and year, the aggregate variance is the weighted average across industries of the within-industry variance, with time-invariant weights equal to each industry's average share of the country's manufacturing revenue.\footnote{This approach reflects solely within-industry variation over time, as suggested in \cite{gopinath2017capital}.} Each country's log aggregate variance is then expressed relative to its own first sample year.\footnote{Germany, Poland, and Romania enter the sample in 2004, so pooling the levels would reflect the change in country composition in that year rather than the dynamics of dispersion.} Finally, the countries observed in a year are averaged with equal weights. Appendix~\ref{app:country_disp} reports the series country by country, together with the mean of TFPR.

TFPR dispersion and capital gap dispersion both end the sample above their starting values, the capital gap rising first. In Figure~\ref{fig:pooled_prod}, the aggregate variance of TFPR ends the sample 14 percent above its starting value, and the whole of the rise comes after 2011. In Figure~\ref{fig:pooled_gap}, the aggregate variance of the capital gap is flat until 2007 and then rises by 24 percent to 2017, with most of the increase between 2007 and 2012, the years of the Great Recession and the European debt crisis. The aggregate variance of the labor gap instead falls by 9 percent to 2008 and ends the sample only 6 percent above its starting value. Under Assumption~\ref{assp} the wage is settled after the ex-post shock and may respond to it, so a rise in the marginal revenue product of labor is matched in part by a rise in its price and the labor gap absorbs only the remainder, whereas the rental rate of capital is set before the shock and the capital gap absorbs all of it.

Under the maintained model, a shock that arrives after an input is committed is absorbed into the input gap. Co-movement between TFPR dispersion and capital gap dispersion is therefore what Section~\ref{mechanism} leads one to expect. Aggregate co-movement does not quantify the pass-through, however. How much of a shock enters the input gap, and how much of each input gap's dispersion it accounts for, are questions about firms rather than countries, and the next subsection turns to them.

\subsection{Pass-Through of Productivity into Input Gaps}
\label{sec:productivity_passthrough}

The firm-level counterpart of Figure~\ref{fig:pooled_disp} regresses the input gap on productivity within a country-industry-year cell,
\begin{equation}
\label{eq:gap_reg}
    \mathrm{mrp}^{X}_{jt} - \log P^X_{jt} = \iota_{cst} + \beta\, x_{jt} + \xi_{jt},
\end{equation}
where $x_{jt}$ is a measure of productivity and $\iota_{cst}$ is a country-industry-year fixed effect, so $\beta$ is identified by comparing a firm with others producing in the same market in the same year. This subsection reports $\beta$ for total TFPR and for its three components.

Panel~A of Table~\ref{tab:components} reports the results setting $x_{jt}$ equal to TFPR, $\nu_{jt} = \omega_{jt} + \varepsilon_{jt}$. TFPR passes into the capital gap at 1.218 in the pooled specification and between 1.113 (Romania) and 1.273 (Spain) by country, and into the labor gap at 0.569 pooled and between 0.438 (Poland) and 0.711 (Romania). Panel~A of Table~\ref{tab:gapshares} reports the share of the within-cell variance of the input gap that TFPR accounts for, the within-cell $R^2$ of equation~(\ref{eq:gap_reg}): 28.4 percent for the capital gap and 13.5 percent for the labor gap.\footnote{Under the lognormal approximation of \cite{hsieh2009misallocation}, the loss in industry TFP from the misallocation of an input is proportional to the within-cell variance of the input's log gap, with a proportionality factor that depends on the elasticity of demand and on the input's revenue elasticity. The factor is common to every component of that variance, so a component's share of gap dispersion equals its share of the loss. The level of the loss requires a demand elasticity and a market structure that the model does not impose.}

\begin{table}[!h]
    \centering
    \caption{Pass-Through of Total Factor Productivity and Its Components into Input Gaps}
    \label{tab:components}
{\setlength{\tabcolsep}{4pt}%
\begin{adjustbox}{max width=\textwidth}
\begin{tabular}{lccccccc}
\toprule
 & Germany & Spain & France & Italy & Poland & Romania & Pooled \\
\midrule
\multicolumn{8}{l}{\textit{Panel A: total TFPR, $\omega_{jt} + \varepsilon_{jt}$}} \\
  \textbf{Capital} & 1.194 & 1.273 & 1.148 & 1.239 & 1.189 & 1.113 & 1.218 \\
   & (0.057) & (0.014) & (0.028) & (0.014) & (0.037) & (0.014) & (0.009) \\
  \textbf{Labor} & 0.707 & 0.480 & 0.601 & 0.572 & 0.438 & 0.711 & 0.569 \\
   & (0.030) & (0.014) & (0.038) & (0.021) & (0.074) & (0.019) & (0.014) \\
\midrule
\multicolumn{8}{l}{\textit{Panel B: the three components, entered jointly}} \\
  \textbf{Capital} &  &  &  &  &  &  &  \\
  \hspace{1em}$g(\omega_{-1})$ & 1.167 & 1.453 & 1.226 & 1.105 & 1.323 & 1.279 & 1.216 \\
   & (0.107) & (0.025) & (0.069) & (0.022) & (0.118) & (0.041) & (0.019) \\
  \hspace{1em}$\eta_{jt}$ & 1.163 & 1.411 & 1.206 & 1.324 & 1.277 & 1.185 & 1.307 \\
   & (0.049) & (0.018) & (0.029) & (0.017) & (0.088) & (0.026) & (0.011) \\
  \hspace{1em}$\varepsilon_{jt}$ & 1.234 & 1.179 & 1.064 & 1.283 & 1.092 & 1.060 & 1.204 \\
   & (0.062) & (0.015) & (0.017) & (0.016) & (0.027) & (0.012) & (0.009) \\
\midrule
  \textbf{Labor} &  &  &  &  &  &  &  \\
  \hspace{1em}$g(\omega_{-1})$ & 0.898 & 0.767 & 0.889 & 1.055 & 0.786 & 1.052 & 0.947 \\
   & (0.043) & (0.018) & (0.021) & (0.033) & (0.108) & (0.033) & (0.016) \\
  \hspace{1em}$\eta_{jt}$ & 0.735 & 0.723 & 0.552 & 0.788 & 0.720 & 0.857 & 0.739 \\
   & (0.040) & (0.016) & (0.014) & (0.026) & (0.074) & (0.022) & (0.014) \\
  \hspace{1em}$\varepsilon_{jt}$ & 0.461 & 0.328 & 0.335 & 0.332 & 0.177 & 0.601 & 0.352 \\
   & (0.041) & (0.010) & (0.011) & (0.026) & (0.031) & (0.013) & (0.014) \\
\midrule
  Firm-years & 58,881 & 873,704 & 529,529 & 1,077,031 & 44,005 & 127,626 & 2,710,776 \\
  Cells & 104 & 136 & 136 & 136 & 104 & 104 & 720 \\
\bottomrule
\end{tabular}
\end{adjustbox}}
\note{\scriptsize Each cell reports $\beta$ from a firm-level regression of the input gap $\mathrm{mrp}^X_{jt} - \log P^X_{jt}$ on productivity, absorbing country-industry-year fixed effects. Panel A uses total TFPR. Panel B enters its three components jointly: the Markov conditional mean $g(\omega_{-1})$, the ex-ante shock $\eta_{jt}$, and the ex-post shock $\varepsilon_{jt}$. Standard errors in parentheses are non-parametric bootstrap standard errors, the standard deviation of $\beta$ across 100 replications of the \cite{gandhi2020identification} estimation. Materials is omitted because $\mathrm{mrp}^M - \log \rho = \varepsilon - \log \mathcal{E}$ (Section~\ref{materials_identity}), so its coefficients hold by construction.}
\end{table}

\begin{table}[!h]
    \centering
    \caption{Share of Input Gap Dispersion Accounted For by Productivity and by Its Components}
    \label{tab:gapshares}
{\setlength{\tabcolsep}{4pt}%
\begin{adjustbox}{max width=\textwidth}
\begin{tabular}{lccccccc}
\toprule
 & Germany & Spain & France & Italy & Poland & Romania & Pooled \\
\midrule
\multicolumn{8}{l}{\textit{Panel A: total TFPR, $\omega_{jt} + \varepsilon_{jt}$}} \\
  \textbf{Capital} & 39.3 & 21.9 & 39.6 & 27.5 & 47.9 & 50.1 & 28.4 \\
   & (5.8) & (0.8) & (2.3) & (0.9) & (5.8) & (2.6) & (0.6) \\
  \textbf{Labor} & 42.0 & 13.4 & 27.2 & 10.1 & 7.8 & 27.1 & 13.5 \\
   & (4.3) & (0.8) & (5.2) & (1.0) & (2.6) & (2.3) & (1.0) \\
\midrule
\multicolumn{8}{l}{\textit{Panel B: the three components, entered jointly}} \\
  \textbf{Capital} &  &  &  &  &  &  &  \\
  \hspace{1em}$g(\omega_{-1})$ & 18.8 & 8.1 & 20.9 & 5.5 & 20.0 & 12.0 & 8.6 \\
   & (3.9) & (0.5) & (2.3) & (0.3) & (4.2) & (1.3) & (0.4) \\
  \hspace{1em}$\eta_{jt}$ & 3.3 & 2.6 & 3.8 & 3.3 & 6.0 & 7.4 & 3.3 \\
   & (0.6) & (0.1) & (0.2) & (0.2) & (0.8) & (0.8) & (0.1) \\
  \hspace{1em}$\varepsilon_{jt}$ & 17.2 & 11.5 & 15.3 & 18.9 & 22.4 & 31.2 & 16.5 \\
   & (2.4) & (0.5) & (0.8) & (0.7) & (2.0) & (1.4) & (0.4) \\
  \hspace{1em}All three & 39.3 & 22.2 & 40.0 & 27.7 & 48.4 & 50.5 & 28.5 \\
   & (5.7) & (0.8) & (2.2) & (0.9) & (5.8) & (2.6) & (0.6) \\
\midrule
  \textbf{Labor} &  &  &  &  &  &  &  \\
  \hspace{1em}$g(\omega_{-1})$ & 35.7 & 10.9 & 30.2 & 11.5 & 8.5 & 11.4 & 13.5 \\
   & (3.7) & (0.7) & (5.9) & (1.1) & (3.1) & (1.6) & (1.0) \\
  \hspace{1em}$\eta_{jt}$ & 5.0 & 3.6 & 2.5 & 2.8 & 2.7 & 5.8 & 3.1 \\
   & (0.5) & (0.2) & (0.2) & (0.3) & (0.6) & (0.7) & (0.2) \\
  \hspace{1em}$\varepsilon_{jt}$ & 6.0 & 2.8 & 2.2 & 1.0 & 0.6 & 12.4 & 1.9 \\
   & (1.6) & (0.2) & (0.2) & (0.3) & (0.3) & (0.7) & (0.2) \\
  \hspace{1em}All three & 46.7 & 17.4 & 35.0 & 15.4 & 11.8 & 29.5 & 18.4 \\
   & (4.0) & (0.9) & (5.9) & (1.4) & (3.7) & (2.6) & (1.2) \\
\midrule
  Firm-years & 58,881 & 873,704 & 529,529 & 1,077,031 & 44,005 & 127,626 & 2,710,776 \\
  Cells & 104 & 136 & 136 & 136 & 104 & 104 & 720 \\
\bottomrule
\end{tabular}
\end{adjustbox}}
\note{\scriptsize Each cell reports the share of within-cell dispersion of the input gap $\mathrm{mrp}^X_{jt} - \log P^X_{jt}$ accounted for by productivity, in percent, with non-parametric bootstrap standard errors in parentheses: the standard deviation across 100 replications of the \cite{gandhi2020identification} estimation. Panel A reports the within-cell $R^2$ of a firm-level regression of the input gap on $\omega_{jt} + \varepsilon_{jt}$ absorbing country-industry-year fixed effects. Panel B splits that share by component from the joint regression in Table~\ref{tab:components}: the share of component $\kappa$ is $\hat{\beta}_\kappa \mathrm{Cov}(\kappa, \mathrm{gap}) / \mathrm{Var}(\mathrm{gap})$, which sums to the regression $R^2$ reported in the row labelled All three. The standard error of All three is the standard deviation of the summed shares across the replications. Materials is omitted because its input gap equals $\varepsilon_{jt}$ up to a within-cell constant (Section~\ref{materials_identity}).}
\end{table}

Panel~B of Table~\ref{tab:components} enters the three components of TFPR jointly, on the same firm-years and with the same fixed effects as Panel~A. The joint specification serves two purposes. First, it estimates the coefficient on $\varepsilon_{jt}$ with the other two components, $g(\omega_{-1})$ and $\eta_{jt}$, held constant. Second, it allocates the within-cell variance of the input gap to the three components: a component's share is the product of its coefficient and its within-cell covariance with the gap, divided by the within-cell variance of the gap, so the three shares sum to the within-cell $R^2$ of the joint specification.

For capital the three coefficients of the pooled specification differ by no more than 0.1, at 1.216 on $g(\omega_{-1})$, 1.307 on $\eta_{jt}$, and 1.204 on $\varepsilon_{jt}$. By country, the coefficient on $g(\omega_{-1})$ exceeds that on $\varepsilon_{jt}$ by 0.16 to 0.28 in Spain, Romania, France, and Poland, falls short of it by 0.18 in Italy, and lies within one bootstrap standard error of it in Germany, so the capital gap responds to known productivity more strongly than to the ex-post shock in four countries, less strongly in one, and about equally in the pooled specification. For labor the coefficients fall from 0.947 on $g(\omega_{-1})$ through 0.739 on $\eta_{jt}$ to 0.352 on $\varepsilon_{jt}$, an ordering that holds in every country.

Panel~B of Table~\ref{tab:gapshares} allocates the share of Panel~A to the three components.\footnote{The row labelled All three exceeds the share in Panel~A because Panel~A imposes a common coefficient on the three components, the coefficient on total TFPR, whereas Panel~B estimates one for each. The two shares coincide only when the three coefficients are equal; the difference is 0.1 percentage points for capital, whose coefficients differ by no more than 0.1, and 4.9 for labor, whose coefficients differ by up to 0.6.} For capital the ex-post shock is the largest contributor in the pooled specification, at 16.5 percent against 8.6 for $g(\omega_{-1})$ and 3.3 for $\eta_{jt}$; it is the largest in Spain, Italy, Poland, and Romania, and $g(\omega_{-1})$ is in Germany and France, although in Germany and Poland the two shares lie within one bootstrap standard error of each other. For labor $g(\omega_{-1})$ is the largest contributor in every country except Romania, where the ex-post shock exceeds it by 1.0 percentage point; the pooled shares are 13.5, 3.1, and 1.9 percent for $g(\omega_{-1})$, $\eta_{jt}$, and $\varepsilon_{jt}$, and sum to the 18.4 of the row All three. The two shocks, both unknown when the input was committed, together account for 19.8 percent of capital gap dispersion and 5.0 percent of labor gap dispersion.

The shares change across the periods of the productivity process (Appendix~\ref{app:shares_period}). For capital, the share of the ex-post shock rises from 11.6 percent in 2001--2002 to 18.8 in 2010--2013, the level it keeps in 2014--2017, while the share of expected productivity falls from 9.4 to 7.5 percent; the two shocks together account for 15 to 16 percent of capital gap dispersion before the Great Recession and 22 percent from the European debt crisis on. For labor the share of the two shocks lies between 4.4 and 5.9 percent in every period. The model restricts only the coefficient on $\varepsilon_{jt}$; the next subsection tests that restriction.

\subsection{The Ex-Post Shock Pass-Through}
\label{sec:passthrough}
\label{sec:expost_shock}

Assumptions~\ref{assp}, \ref{assa}, and~\ref{asst} deliver a point prediction for the response of the input gap in equation~(\ref{eq:gap}) to the ex-post shock. Capital and labor are committed before $\varepsilon_{jt}$ realizes. The capital outlay is fixed when the shock arrives, so a firm drawing a favorable shock earns more revenue on a capital stock it cannot adjust within the period, and the shock passes into the capital gap one for one. Labor is fixed in headcount but not in pay, so the shock passes into the labor gap at $1 - \theta$, one for one less the wage response of Assumption~\ref{assp}. Both are point restrictions, and both are testable firm by firm.

I regress the input gap on the ex-post shock within a country-industry-year cell:
\begin{equation}
\label{model:passthrough}
    \mathrm{mrp}^{X}_{jt} - \log P^X_{jt} = \iota_{cst} + \beta\,\varepsilon_{jt} + \xi_{jt}.
\end{equation}
Table~\ref{tab:passthrough} reports estimates of equation~(\ref{model:passthrough}) together with three variations of it: specification (2) adds firm fixed effects, which absorb any time-invariant firm-level input gap; specification (3) replaces $\varepsilon_{jt}$ with its within-firm AR(1) innovation, the component unforecastable from the firm's own history; and specification (4) keeps the firm fixed effects and replaces the estimation cell with the finer cell at which the deflators are defined, country by two-digit industry by year, so that the relative price of materials to output, which is constant within that cell and enters both $\varepsilon_{jt}$ and the input gap through the nominal revenue and materials outlay, cannot generate co-movement between them.\footnote{Materials is not reported. Section~\ref{materials_identity} shows that the materials gap equals $\varepsilon_{jt}$ up to a constant by construction. Specification (3) is the one case where the identity does not bind, since replacing the shock with its innovation changes the regressor but not the dependent variable. The materials coefficient in that specification is nonetheless 0.997, so replacing the shock with its innovation alters the coefficient only marginally.}

\begin{table}[!h]
    \centering
    \caption{Pass-Through of the Ex-Post Productivity Shock into Input Gaps}
    \label{tab:passthrough}
{\setlength{\tabcolsep}{4pt}%
\begin{adjustbox}{max width=\textwidth}
\begin{tabular}{llccccccc}
\toprule
 & & Germany & Spain & France & Italy & Poland & Romania & Pooled \\
\midrule
  \textbf{Capital} & & & & & & & &  \\
  \hspace{1em}Cell FE & (1) & 1.000 & 0.912 & 0.756 & 0.973 & 1.023 & 0.886 & 0.919 \\
   & & (0.057) & (0.019) & (0.017) & (0.017) & (0.030) & (0.013) & (0.010) \\
  \hspace{1em}+ Firm FE & (2) & 0.926 & 0.909 & 0.842 & 0.919 & 0.998 & 0.785 & 0.893 \\
   & & (0.043) & (0.011) & (0.017) & (0.012) & (0.037) & (0.010) & (0.007) \\
  \hspace{1em}+ Firm FE, innovation & (3) & 0.894 & 0.861 & 0.858 & 0.939 & 0.915 & 0.670 & 0.873 \\
   & & (0.050) & (0.010) & (0.015) & (0.014) & (0.036) & (0.008) & (0.007) \\
  \hspace{1em}Deflator-cell FE & (4) & 0.927 & 0.914 & 0.843 & 0.918 & 0.997 & 0.786 & 0.894 \\
   & & (0.043) & (0.011) & (0.017) & (0.012) & (0.037) & (0.010) & (0.007) \\
  \midrule
  \textbf{Labor} & & & & & & & &  \\
  \hspace{1em}Cell FE & (1) & 0.330 & 0.204 & 0.149 & 0.145 & 0.186 & 0.501 & 0.191 \\
   & & (0.045) & (0.009) & (0.010) & (0.026) & (0.030) & (0.012) & (0.013) \\
  \hspace{1em}+ Firm FE & (2) & 0.515 & 0.374 & 0.266 & 0.354 & 0.400 & 0.544 & 0.365 \\
   & & (0.060) & (0.007) & (0.013) & (0.018) & (0.034) & (0.009) & (0.009) \\
  \hspace{1em}+ Firm FE, innovation & (3) & 0.504 & 0.413 & 0.296 & 0.379 & 0.378 & 0.475 & 0.387 \\
   & & (0.057) & (0.006) & (0.010) & (0.019) & (0.034) & (0.008) & (0.009) \\
  \hspace{1em}Deflator-cell FE & (4) & 0.516 & 0.379 & 0.267 & 0.354 & 0.402 & 0.545 & 0.366 \\
   & & (0.060) & (0.007) & (0.013) & (0.018) & (0.034) & (0.009) & (0.009) \\
   \bottomrule
\end{tabular}
\end{adjustbox}%
}
\note{\scriptsize Each cell reports $\beta$ from $\mathrm{mrp}^X_{jt} - \log P^X_{jt} = \iota_{cst} + \beta\,\varepsilon_{jt} + \xi_{jt}$ (equation~\ref{model:passthrough}), estimated on firm-year observations. Specification (1) absorbs country-industry-year fixed effects. Specification (2) adds firm fixed effects, absorbing any time-invariant firm-level input gap. Specification (3) replaces $\varepsilon_{jt}$ with its within-firm AR(1) innovation. Specification (4) keeps the firm fixed effects of (2) and replaces the estimation cell with the deflator cell (country, 2-digit NACE, year). Standard errors in parentheses are non-parametric bootstrap standard errors, the standard deviation of $\beta$ across 100 replications of the \cite{gandhi2020identification} estimation. Materials is omitted because $\mathrm{mrp}^M - \log\rho = \varepsilon - \log\mathcal{E}$ holds by construction (Section~\ref{materials_identity}).}
\end{table}

For capital the pooled coefficient is 0.919, close to the prediction of one, and it is stable across the three variations: 0.893 with firm fixed effects, 0.873 on the innovation, and 0.894 at the deflator cell. By country, the capital coefficient is statistically indistinguishable from one in Germany (1.000), Poland (1.023), and Italy (0.973), modestly below it in Spain (0.912) and Romania (0.886), and lowest in France (0.756). The transmission into the capital gap is then close to one for one, as the timing model predicts. The univariate estimate of 0.919 lies below the 1.204 that Table~\ref{tab:components} reports when $g(\omega_{-1})$ and $\eta_{jt}$ enter jointly with $\varepsilon_{jt}$: both components are negatively correlated with $\varepsilon_{jt}$ within cell (Appendix~\ref{app:orthogonality}), so their omission biases the univariate coefficient downward.

For labor the pooled coefficient is 0.191 with cell fixed effects and 0.365 with firm fixed effects, well below one in every specification and every country, ranging from 0.145 in Italy to 0.501 in Romania. Under Assumption~\ref{assp}, the distance from one is attributable to the response of the wage to the shock. Quantifying that response requires decomposing the pass-through into its margins.

Because the input gap in equation~(\ref{eq:gap}) is a sum, the pass-through coefficient decomposes into three regressions with the same right-hand side as specification (2):
\begin{equation}
\label{eq:margins}
    z_{jt} = \iota_{cst} + \alpha_j + \beta_z\, \varepsilon_{jt} + \xi_{jt}, \qquad z \in \{\log \mathrm{elas}^X_{jt},\ y_{jt},\ \log(P^X_{jt} X_{jt})\},
\end{equation}
where $\alpha_j$ is a firm fixed effect and $\beta = \beta_{\mathrm{elas}} + \beta_y - \beta_{PX}$ exactly; for labor $\log(w_{jt} L_{jt}) = \log w_{jt} + \log L_{jt}$, so the wage-bill margin splits further into $\beta_{wL} = \beta_w + \beta_L$. Under Assumptions~\ref{assa} and~\ref{asst} every input, materials included, is chosen before $\varepsilon_{jt}$ realizes, so the elasticity margin is zero, because the revenue elasticities are functions of the input quantities alone, and the capital and headcount margins are zero; the revenue margin is one, and under Assumption~\ref{assp} the wage margin is $\theta$. Table~\ref{tab:margins} reports the margins for both inputs.

\begin{table}[!h]
    \centering
    \caption{Margins of the Pass-Through of the Ex-Post Shock}
    \label{tab:margins}
{\setlength{\tabcolsep}{4pt}%
\begin{adjustbox}{max width=\textwidth}
\begin{tabular}{lcccccccc}
\toprule
 & Model-Implied Value & Germany & Spain & France & Italy & Poland & Romania & Pooled \\
\midrule
  \textbf{Capital} & & & & & & & &  \\
  \hspace{1em}Elasticity margin, $\log \mathrm{elas}^K_{jt}$ & 0 & -0.017 & -0.091 & -0.012 & -0.092 & -0.096 & 0.003 & -0.069 \\
   & & (0.022) & (0.006) & (0.009) & (0.008) & (0.027) & (0.006) & (0.004) \\
  \hspace{1em}Revenue margin, $y_{jt}$ & 1 & 0.687 & 0.761 & 0.590 & 0.602 & 0.834 & 0.562 & 0.641 \\
   & & (0.040) & (0.009) & (0.014) & (0.008) & (0.030) & (0.012) & (0.005) \\
  \hspace{1em}Outlay margin, $\log(P^K_{jt} K_{jt})$ & 0 & -0.257 & -0.240 & -0.263 & -0.409 & -0.260 & -0.220 & -0.321 \\
   & & (0.045) & (0.008) & (0.011) & (0.011) & (0.034) & (0.009) & (0.006) \\
  \hspace{1em}Pass-through (sum of margins) & 1 & 0.926 & 0.909 & 0.842 & 0.919 & 0.998 & 0.785 & 0.893 \\
   & & (0.043) & (0.011) & (0.017) & (0.012) & (0.037) & (0.010) & (0.007) \\
  \midrule
  \textbf{Labor} & & & & & & & &  \\
  \hspace{1em}Elasticity margin, $\log \mathrm{elas}^L_{jt}$ & 0 & -0.005 & -0.042 & -0.054 & 0.043 & -0.031 & 0.059 & 0.005 \\
   & & (0.018) & (0.005) & (0.006) & (0.017) & (0.023) & (0.007) & (0.008) \\
  \hspace{1em}Revenue margin, $y_{jt}$ & 1 & 0.687 & 0.761 & 0.590 & 0.602 & 0.834 & 0.562 & 0.641 \\
   & & (0.040) & (0.009) & (0.014) & (0.008) & (0.030) & (0.012) & (0.005) \\
  \hspace{1em}Wage-bill margin, $\log(w_{jt} L_{jt})$ & $\theta$ & 0.167 & 0.344 & 0.270 & 0.291 & 0.403 & 0.077 & 0.281 \\
   & & (0.034) & (0.007) & (0.010) & (0.006) & (0.032) & (0.010) & (0.004) \\
  \hspace{2em}Pay per head, $\log w_{jt}$ & $\theta$ & 0.270 & 0.483 & 0.532 & 0.367 & 0.584 & 0.240 & 0.413 \\
   & & (0.043) & (0.008) & (0.012) & (0.008) & (0.033) & (0.007) & (0.005) \\
  \hspace{2em}Headcount, $\log L_{jt}$ & 0 & -0.103 & -0.139 & -0.262 & -0.076 & -0.182 & -0.163 & -0.132 \\
   & & (0.027) & (0.009) & (0.010) & (0.007) & (0.022) & (0.011) & (0.005) \\
  \hspace{1em}Pass-through (sum of margins) & $1 - \theta$ & 0.515 & 0.374 & 0.266 & 0.354 & 0.400 & 0.544 & 0.365 \\
   & & (0.060) & (0.007) & (0.013) & (0.018) & (0.034) & (0.009) & (0.009) \\
   \bottomrule
\end{tabular}
\end{adjustbox}%
}
\note{\scriptsize Each cell reports $\beta_z$ from equation~(\ref{eq:margins}), the regression of one component of the input gap on $\varepsilon_{jt}$ with country-industry-year and firm fixed effects, on the firm-year sample of specification (2) of Table~\ref{tab:passthrough}. The model-implied value follows from Assumptions~\ref{assp}, \ref{assa}, and~\ref{asst}. Standard errors in parentheses are non-parametric bootstrap standard errors, the standard deviation of $\beta_z$ across 100 replications of the \cite{gandhi2020identification} estimation.}
\end{table}

For labor the wage-bill margin is 0.281, the net of a rise in pay per head of 0.413 and a fall in headcount of 0.132.\footnote{The pay response is contemporaneous: with $\varepsilon_{jt}$ and $\varepsilon_{jt-1}$ entered jointly, the coefficients are 0.355 and 0.057. It is larger in firms with fewer than 50 employees, classified by their median headcount over the sample period, in four of the six countries.} Pay per head therefore responds to the shock about three times as strongly as headcount, and in the opposite direction: the wage-bill response is mostly price, and the pay-per-head margin, which equals $\theta$ under Assumption~\ref{assp}, yields an estimate of 0.41. The remaining difference, from $1 - \theta = 0.59$ to the estimated 0.365, arises from two margins: the revenue margin, which labor shares with capital, and the headcount margin, whose decline of 0.132 offsets part of that difference.

The revenue margin is 0.641, below the one the model implies. Log revenue contains the other two components of TFPR in addition to $\varepsilon_{jt}$, and both are negatively correlated with $\varepsilon_{jt}$ within cell (Appendix~\ref{app:orthogonality}): on average, a favorable ex-post shock coincides with lower expected productivity and a lower ex-ante shock, which dampens the response of revenue.

\section{Discussion and Robustness}\label{s:robustness}
\subsection{The Forecastable Part of the Ex-Post Shock}
\label{sec:eps_timeseries}\label{sec:information_content}

The estimated ex-post shock is serially correlated. Within a country-industry-year cell, the first-order autocorrelation of $\varepsilon_{jt}$ is 0.759, and it declines to 0.660 at a lag of two years and to 0.511 at five. Demeaning within firm leaves a first-order autocorrelation coefficient of 0.387, and the within-firm estimate rises with panel length, reaching 0.544 for firms observed for fifteen years or more. A richer approximation of the share equation does not remove the persistence: replacing the second-degree polynomial with a complete third-degree polynomial moves the within-firm autocorrelation of the refitted residual only from 0.378 at degree two to 0.367. Since $\varepsilon_{jt-1} \in \mathcal{I}_{jt}$ and Assumption~\ref{asst} sets $E(\varepsilon_{jt} \mid \mathcal{I}_{jt}) = 0$, the lagged shock should not predict the capital gap in period $t$. However, adding $\varepsilon_{jt-1}$ to the pass-through specification (2) of Table~\ref{tab:passthrough} for capital yields a coefficient of 0.132 on the lagged shock, with the coefficient on the contemporaneous shock at 0.877 against 0.893 without the lag.

The recovered residual is the distance between the materials share and the materials elasticity, $\log \mathrm{elas}^M_{jt} - s_{jt}$, and persistent variation reaches it by three routes. First, the materials quantity may deviate from the quantity the static first-order condition prescribes by an amount the firm knows: a supply contract fixes the quantity before the period's productivity is known, an inventory makes the choice intertemporal, and lumpy purchasing moves the quantity in steps. The share equation attributes the whole distance between the share and the elasticity to the shock, so a known deviation that persists is recorded as a persistent $\varepsilon_{jt}$. Second, the marginal revenue product of materials may exceed its price by a persistent firm-specific factor, the leading case being a firm-specific markup, which enters the share equation exactly where $\varepsilon_{jt}$ does. Third, either term of the residual may carry a persistent error. The elasticity is the cell's estimated technology evaluated at the firm's inputs, so a vertically integrated firm, which produces inputs the other firms purchase, or a multi-product firm, whose mix sets its own materials intensity, receives an elasticity that differs from its own by a firm-specific amount, and the share is understated for a firm whose accounts record part of its intermediate purchases outside the material costs item. Either error enters $\varepsilon_{jt}$ as a permanent component.

Table~\ref{tab:passthrough} points to the second route. Specification (2) adds firm fixed effects to specification (1), which removes the permanent component of $\varepsilon_{jt}$ from the regressor. If the gap responds to the permanent component less than to the time-varying one, removing it raises the coefficient; if more, removing it lowers the coefficient. For capital the coefficient falls slightly, from 0.919 to 0.893, so the capital gap responds to the permanent component at least as strongly as to the time-varying one; for labor it rises from 0.191 to 0.365, so the labor gap responds to the permanent component far less. A firm-specific markup may produce this pattern. A markup scales revenue, and with it the marginal revenue product of every input, by the same factor. The capital gap records it in full as long as the rental price of capital is independent of the firm's markup; the labor gap records it only partly if the wage rises with the markup, under rent sharing or under the sorting of higher-quality workers into high-markup firms.

Let $u_{jt} = \varepsilon_{jt} - \lambda\,\varepsilon_{jt-1}$ denote the innovation of a first-order autoregression of the recovered residual fitted within firm, with $\lambda = 0.43$ the within-firm regression slope. The innovation is uncorrelated with the lagged residual by construction, so it satisfies at one lag the zero-autocorrelation restriction that Assumption~\ref{asst} implies. Specification (3) of Table~\ref{tab:passthrough} replaces $\varepsilon_{jt}$ with $u_{jt}$ under the same firm fixed effects: the capital coefficient is 0.873 against 0.893 for $\varepsilon_{jt}$, and the labor coefficient 0.387 against 0.365. The pass-through coefficients move by 0.020 and 0.022, so the test of the point predictions of Section~\ref{sec:passthrough} does not depend on whether the forecastable part of the shock is removed from the regressor.\footnote{The choice of $\lambda$ matters more than the removal itself. The within-firm estimate is biased toward zero in short panels \citep{nickell1981biases}; the first-difference estimator of \cite{anderson1982formulation}, which instruments the lagged difference with the second lag of the residual, gives 0.28, and the value at which $u_{jt}$ is uncorrelated with $\varepsilon_{jt-2}$ is 0.57. Across $\lambda$ from 0.28 to 0.54 the capital coefficient of specification (3) runs from 0.94 to 0.79 and the labor coefficient from 0.40 to 0.36, while the shares move by less than one percentage point.}

However, the interpretation of the input gap does depend on it. Under Assumptions~\ref{assp}, \ref{assa}, and~\ref{asst}, the within-cell dispersion of the input gap of a predetermined input is ex post misallocation, because the input was committed before the shock was known, and that reading applies to the dispersion that $u_{jt}$ generates. The dispersion that the forecastable part of $\varepsilon_{jt}$ generates reflects a component the firm rationally knew when it committed the input. The distinction matters for capital and not for labor. I split $\varepsilon_{jt}$ within firm into the part known when the input was committed, the firm's sample mean of $\varepsilon_{jt}$ plus $\lambda$ times last period's deviation from it, and the innovation $u_{jt}$, and enter the two jointly with $g(\omega_{-1})$ and $\eta_{jt}$ in the decomposition of Table~\ref{tab:gapshares}.\footnote{Constructing $u_{jt}$ requires $\varepsilon_{jt-1}$, so the decomposition is estimated on the 2,416,277 firm-years of Table~\ref{tab:gapshares}'s sample for which the shock of the previous year is observed; on that subsample the two shocks account for 20.2 percent of capital gap dispersion and 4.9 percent of labor gap dispersion, against 19.8 and 5.0 percent on the full sample.}

For capital the innovation accounts for 4.5 percent of gap dispersion and the known part for 12.3, so the share of capital gap dispersion that productivity uncertainty generates falls from about a fifth to 7.8 percent. That share is of the order that \cite{david2019sources} attribute to uncertainty, modeled as a noisy signal of next period's productivity, in the dispersion of the average revenue product of capital: about 10 percent among Chinese manufacturing firms and 7 percent among publicly traded firms in the United States. For labor the innovation accounts for 2.3 percent of gap dispersion and the known part for 0.4, so the share of labor gap dispersion that productivity uncertainty generates moves from 4.9 to 5.6 percent and remains about one twentieth.

\subsection{Institutional Quality and Productivity Dispersion}\label{s:robustness_institutions}

This subsection asks whether the dispersion of expected productivity and of the two shocks varies with the institutional environment in which firms operate; Appendix~\ref{app:macro_reg} reports the exercise in full. For each of nine standardized indices of institutional quality (the six Worldwide Governance Indicators of \citealp{kaufmann2011worldwide}, the World Bank Ease of Doing Business score, the Organisation for Economic Co-operation and Development (OECD) Product Market Regulation indicator, and the World Economic Forum Global Competitiveness Index), I regress the log variance of $g(\omega_{-1})$, of $\eta_{jt}$ and of $\varepsilon_{jt}$ across the firms of a country-industry-year cell on the index. The indices vary only at the country-year level, so the regressions absorb industry and year fixed effects.

Across the indices, better institutions are associated with lower dispersion of both shocks and with higher dispersion of expected productivity. An improvement in the Ease of Doing Business score is associated with a lower variance of $\eta_{jt}$ and of $\varepsilon_{jt}$ and with a higher, less precisely estimated, variance of $g(\omega_{-1})$, and Government Effectiveness, Rule of Law, Control of Corruption and the Global Competitiveness Index show the same pattern. Product Market Regulation stands out as the only index whose association with the dispersion of expected productivity is negative, and Political Stability as the only one of the governance indicators whose association with shock dispersion is positive. The evidence is suggestive rather than causal. With that caveat, the pattern is consistent with institutions reducing the uncertainty firms face while letting productivity differences across firms persist.

\subsection{Sensitivity to Imperfect Competition}\label{s:robustness_ic}\label{s:robustness_markups}

The baseline framework assumes that firms are price takers in the output market (Assumption~\ref{assp}). Unmodeled market power could contaminate the productivity estimates and, consequently, the input gap dispersion patterns and the pass-through documented in Section~\ref{s:results}. If firms face downward-sloping demand curves, observed revenue reflects both physical output and an endogenous price response, so that the baseline revenue elasticities may conflate production technology with demand structure. Markups correlated with productivity could then generate spurious input gap dispersion that is attributed to productivity uncertainty rather than to market power.

To address this concern, I re-estimate the production function under imperfect competition (IC) using the extension in Appendix O6-4 of \cite{gandhi2020identification}. The approach allows firms to face time-varying, industry-specific demand elasticities by introducing period-specific markup parameters into the first-stage share equation. The second stage uses revenue deflated by EUROSTAT producer price indices (PPI) and an aggregate demand shifter as an additional instrument to separately identify a markup constant from the production function coefficients. The full derivation and estimation details are in Appendix~\ref{app:ic}.

I estimate the IC model for all six countries and eight industry groups on the samples of the competitive baseline; Appendix~\ref{app:ic} reports the estimates. The markups are common to all firms within an industry-period cell and indicate modest market power, with country means between 0.99 and 1.16. Allowing for market power leaves the results of Section~\ref{s:results} in place. On the same industry-year cells, the IC series of TFPR and input gap dispersion follow the competitive ones in level and over time, with France's capital gap as the one visible departure, and the capital gap remains more dispersed than the labor gap. Re-estimating the pass-through of TFPR and of its components on the IC gaps and components leaves the capital and labor coefficients close to their competitive values. Both the dispersion patterns and the pass-through of productivity are therefore robust to industry-period-level deviations from price-taking.

The extension addresses market power at the level of the cell; it does not address markup variation across firms within a cell. That variation cannot be separated from $\varepsilon_{jt}$ inside the share equation: a firm pricing at a firm-specific markup $\mu_{jt}$ and facing no ex-post shock delivers a share equation of the same form as equation~(\ref{eq:share_theory}), with $-\log \mu_{jt}$ in place of $-\varepsilon_{jt}$.\footnote{Frictions in the materials market enter the share equation where the markup does. \cite{shenoy2021estimating} models them as a wedge between the market price of materials and the shadow price the firm faces, and shows that the share equation then understates the materials elasticity. Its test rejects the assumption that the materials choice depends only on the other inputs and on productivity in Chilean and Colombian manufacturing industries.} In Section~\ref{sec:eps_timeseries} I point to a persistent firm-specific markup as the leading account of the permanent component of $\varepsilon_{jt}$, which the firm fixed effects of specification (2) in Table~\ref{tab:passthrough} absorb and the innovation of specification (3) removes. In the decomposition of that section the permanent component is part of the productivity known when the input was committed, so it does not enter the share of gap dispersion attributed to productivity uncertainty. A markup that varied unpredictably across firms from year to year would remain in the innovation, where the share equation cannot distinguish it from a productivity surprise.\footnote{Firm-level markups in the manner of \cite{de2012markups} are recovered from the same materials first-order condition, so using them here would be circular.} For the misallocation measured here the distinction is immaterial: the input was committed before the fluctuation was known, so the gap it generates is the same.

\subsection{Sensitivity to the Estimation Method}\label{s:robustness_method}

The results in this paper rely on the \cite{gandhi2020identification} estimator to recover firm-level input elasticities, marginal revenue products, and TFPR components. One question is whether the main findings are an artifact of this particular estimation method. Appendix~\ref{app:mc} addresses this question in two steps: Monte Carlo experiments that compare GNR with two widely used alternatives, the factor shares approach (see \citealp{de2021industrial} for a review) and the proxy-variable method of \cite{ackerberg2015identification} extended to gross output as in Section III of \cite{gandhi2020identification}, and an empirical comparison with the factor shares approach on the MICROPROD data.

The Monte Carlo experiments show that the factor shares approach introduces systematic biases directly relevant to the misallocation analysis. The constant returns to scale assumption overstates capital elasticities by 58--76\%, and the ex-post shock $\varepsilon_{jt}$ contaminates expenditure shares through output $y_{jt}$, creating spurious elasticity heterogeneity. These biases propagate into the input gaps. Factor shares sets the elasticity of a flexible input equal to its expenditure share, so the gap of that input is zero for every firm by construction. In the simulation this leaves the capital gap as the only object on which factor shares can be compared with the other estimators. That gap inherits both defects through its log elasticity. The overstatement raises its level, and the ex-post shock in each share raises its dispersion: the capital elasticity has a standard deviation across firms of 0.28 against a true 0.04. The proxy-variable estimator, while approximately unbiased on average, delivers estimates too imprecise for reliable inference on elasticity dispersion.

The experiments also provide benchmarks for the firm-level test of Section~\ref{sec:passthrough}, for capital and labor. On simulated data the capital gap responds to the true ex-post shock one for one, at 1.001, and the labor gap at the realized wage responds at 0.589, the model-implied $1 - \theta = 0.59$ (Appendix~\ref{app:mc_reg}). GNR reproduces both benchmarks, at 1.000 for capital and 0.590 for labor. The proxy-variable estimator reproduces the coefficients on $\varepsilon_{jt}$ and returns imprecise ones on the other two components, and factor shares, which recovers no ex-post shock, cannot run the test.

The TFPR variance decomposition also depends on the estimator. GNR recovers the true decomposition, in which the ex-ante shock $\eta_{jt}$ and the ex-post shock $\varepsilon_{jt}$ account for the majority of TFPR dispersion. Factor shares cannot separate $\varepsilon_{jt}$ from $\omega_{jt}$ and absorbs the shock into the lagged Solow residual, which inflates the variance share of the expected productivity level and attenuates the estimated persistence of productivity: the AR(1) coefficient falls from 0.80 to 0.68 under the translog process.

The empirical comparison in Appendix~\ref{app:mc_empirical} applies the factor shares estimator to the MICROPROD data on the same samples as the GNR estimates. The proxy-variable estimator is not taken to the data: the input gaps and the components $g(\omega_{-1})$ and $\eta_{jt}$ require the elasticities of its second stage, which returned estimates too imprecise in the Monte Carlo for the decomposition or for a comparison of gap dispersion to be informative. 

I use sectoral factor shares, the form in which the approach is normally applied. In each country-industry-year cell the materials and labor elasticities are the cell's materials expenditure and wage bill, each divided by the cell's revenue, and the capital elasticity is one minus the two, under constant returns to scale. GNR delivers less dispersion of the capital gap than the sectoral factor shares benchmark, a within-cell standard deviation of 1.12 against 1.31, and about the same dispersion of the labor gap, 0.76 against 0.73. The difference for capital is consistent with firm-level heterogeneity in technology accounting for part of the dispersion the sectoral benchmark reads as capital gaps. The two estimators deliver similar TFPR dispersion. The factor shares residual is less persistent than $\omega_{jt}$, at 0.79 against 0.82, a fifth of the Monte Carlo's attenuation: the sectoral share is the revenue-weighted mean of the firms' shares, so the ex-post shock, which enters each firm's own share in full, enters the cell's share only through its mean across the firms of the cell.

\section{Summary and Concluding Remarks}\label{s:conclusion}
In this paper, I asked how much input misallocation productivity uncertainty generates and at which stage of the firm's input decisions it arises. I decomposed firm-level TFPR into expected productivity $g(\omega_{-1})$, known when every input is chosen; the ex-ante shock $\eta_{jt}$, revealed after capital and labor are committed and before materials is allocated; and the ex-post shock $\varepsilon_{jt}$, unresolved until production is complete. An input committed before a component is revealed cannot adjust to it within the period. That component therefore enters the input's gap, the ratio of its marginal revenue product to its price. The share of the gap's dispersion that each component accounts for is then the misallocation that arises at the stage at which that component is revealed. 

Estimating the model with the \cite{gandhi2020identification} procedure on firm-level panel data from six European countries over 2000--2017, I found that the three components of TFPR together account for 28.5 percent of capital gap dispersion within a country-industry-year cell and for 18.4 percent of labor gap dispersion. The timing behind those shares differs by input. For capital the ex-post shock is the largest contributor, at 16.5 percent, and the two shocks together account for about a fifth of capital gap dispersion; for labor expected productivity leads, at 13.5 percent, and the two shocks together account for one twentieth.

The model predicts that the ex-post shock passes one for one into the gap of any input committed before it realizes, unless the input's price responds to the shock. In the data, an unanticipated one percent rise in TFPR raises the marginal revenue product of capital by 0.92 percent relative to its rental rate, close to the predicted one and indistinguishable from it in Germany, Poland and Italy. For labor the same shock raises the marginal revenue product of labor by 0.19 percent relative to the wage, because the wage responds to the shock: the wage bill rises by 0.54 percent, 0.41 through pay per head and 0.13 through headcount. 

The estimates survive checks of the estimation method. In a Monte Carlo with the model's timing, GNR recovers the input elasticities, the TFPR decomposition and the model-implied pass-through of the ex-post shock for both inputs; the factor shares approach overstates the capital elasticity under constant returns to scale and absorbs the ex-post shock into the Solow residual, and the proxy-variable estimator returns imprecise estimates of elasticity dispersion and of the expected and ex-ante components. Re-estimation under imperfect competition, with period-specific markups, yields dispersion series and pass-through coefficients that closely track those of the competitive baseline.

Three directions would extend the framework. Embedding the decomposition in a general equilibrium model with input-output linkages, along the lines of \cite{baqaee2020productivity}, would map input gap dispersion into aggregate efficiency losses by the stage at which productivity is revealed. Estimating productivity uncertainty jointly with adjustment costs and financial frictions would attribute the part of input gap dispersion that productivity does not account for. Finally, separating the ex-post shock from measurement error and from firm-specific demand shocks, with a capital measure less exposed to the error in book values \citep{collard2016production} and firm-level price deflators, would sharpen the productivity-uncertainty interpretation of the dispersion it generates.

\clearpage
\appendix
\section{Identification and Estimation of the Partial Equilibrium Model \citep{gandhi2020identification}}\label{Gandhi}
Taking the first-order condition (FOC) for $M_{jt}$ in equation (\ref{eq:MAX}) yields
\begin{equation}\label{eq:foc_exp}
     \frac{\partial E(F(k_{jt},l_{jt},m_{jt})e^{\nu_{jt}}\mid \mathcal{I}_{jt})}{\partial M_{jt}}-\rho_{jt}=0
\end{equation}
By the distributional assumptions on $\omega_{jt}$ and $\varepsilon_{jt}$ (see Assumptions \ref{asspr} and \ref{asst}), the FOC reads
    \begin{equation}\label{eq:foc_simp}
    \frac{\partial F(k_{jt},l_{jt},m_{jt})}{\partial M_{jt}}e^{\omega_{jt}}\mathcal{E}-\rho_{jt}=0
\end{equation}
Taking logs, and substituting $\omega_{jt}=y_{jt}-f(k_{jt},l_{jt},m_{jt})-\varepsilon_{jt}$ delivers
\begin{equation}
\label{eq:foc_log}
    \log(\rho_{jt})=y_{jt}-f(k_{jt},l_{jt},m_{jt})-\varepsilon_{jt}+\log\mathcal{E}+\log\frac{\partial F(k_{jt},l_{jt},m_{jt})}{\partial M_{jt}}
\end{equation}
Let $s_{jt}=\log\left(\frac{\rho_{jt}M_{jt}}{Y_{jt}}\right)$, the (log) intermediates cost share of revenue of firm $j$ at time $t$. 
Rearranging
\footnote{
Indeed,
\begin{equation*}
    \begin{aligned}
    &\log\frac{\partial F(k_{jt},l_{jt},m_{jt})}{\partial M_{jt}}-f(k_{jt},l_{jt},m_{jt})\\
    =&\log\frac{\partial F(k_{jt},l_{jt},m_{jt})}{\partial M_{jt}}-\log F(k_{jt},l_{jt},m_{jt})\\
    =&\log\left(\frac{\partial F(k_{jt},l_{jt},m_{jt})}{F(k_{jt},l_{jt},m_{jt})}/\frac{\partial M_{jt}}{M_{jt}}\right)-\log M_{jt}\\
    =&\log \left(\frac{\partial f(k_{jt},l_{jt},m_{jt})}{\partial m_{jt}}\right) -m_{jt}
    \end{aligned}
\end{equation*}
where the last equality follows from $\frac{\partial F(k_{jt},l_{jt},m_{jt})}{F(k_{jt},l_{jt},m_{jt})}/\frac{\partial M_{jt}}{M_{jt}}=\frac{\partial f(k_{jt},l_{jt},m_{jt})}{\partial m_{jt}}$, being the revenue elasticity of material inputs.
} equation (\ref{eq:foc_log}) yields an expression of the log intermediates share as a function of the log revenue elasticity of intermediate inputs, a scalar constant, and the ex-post productivity shock
\begin{equation}
\label{eq:share_reg}
    s_{jt}= \log \left(\frac{\partial f(k_{jt},l_{jt},m_{jt})}{\partial m_{jt}}\right)+\log\mathcal{E}-\varepsilon_{jt}
\end{equation}
A second-degree complete polynomial in capital, labor, and materials approximates the intermediate inputs elasticity $\frac{\partial f(k_{jt},l_{jt},m_{jt})}{\partial m_{jt}}$. Substituting it into equation (\ref{eq:share_reg}) yields
\begin{equation}
\label{eq:share_poly}
    s_{jt}= \log \left(\begin{aligned}
&\gamma_{0}+\gamma_{k} k_{jt}+\gamma_{l} l_{jt}+\gamma_{m} m_{jt}+\gamma_{k k} k_{jt}^{2}+\gamma_{ll} l_{jt}^{2} \\
&+\gamma_{m m} m_{jt}^{2}+\gamma_{k l} k_{jt} l_{jt}+\gamma_{k m} k_{jt} m_{jt}+\gamma_{l m} l_{jt} m_{jt}
\end{aligned}\right)+\log\mathcal{E}-\varepsilon_{jt}
\end{equation}

Nonlinear least squares consistently estimates the $\gamma$ parameters in equation (\ref{eq:share_poly}) up to the scalar constant $\mathcal{E}$. 
Denote this estimate $\hat\gamma'$.\footnote{From equation (\ref{eq:share_poly}), regressing just
\begin{equation*}
    s_{jt}=\log \left(\begin{aligned}
&\gamma_{0}+\gamma_{k} k_{jt}+\gamma_{l} l_{jt}+\gamma_{m} m_{jt}+\gamma_{k k} k_{jt}^{2}+\gamma_{ll} l_{jt}^{2} \\
&+\gamma_{m m} m_{jt}^{2}+\gamma_{k l} k_{jt} l_{jt}+\gamma_{k m} k_{jt} m_{jt}+\gamma_{l m} l_{jt} m_{jt}
\end{aligned}\right)-\varepsilon_{jt}
\end{equation*}
delivers the vector $\hat\gamma'$ which is an estimate for $\gamma\mathcal{E}$.} However, since the (negative) regression residuals are estimates for the ex-post productivity shock $\varepsilon_{jt}$, by Assumption \ref{asst}, the residuals yield an estimate $\hat{\mathcal{E}}=\frac{1}{N}\sum_{(j,t)} e^{\hat{\varepsilon}_{jt}}$, where $N=\sum_t J_t$ is the total number of firm-year observations, and $J_t$ denotes the number of firms observed at time $t$.

Having an estimate for $\mathcal{E}$, I recover an estimate\footnote{$\hat{\gamma} = \hat{\gamma}^{\prime} / \hat{\mathcal{E}}$.} $\hat\gamma$ for the parameter vector free of the constant $\mathcal{E}$ and an estimate for the intermediate input elasticity, $\frac{\partial f(k_{jt},l_{jt},m_{jt})}{\partial m_{jt}}$. Finally, I estimate the cross-sectional variance $\tau^2_t$ by taking the cross-sectional sample variance of the estimates $\hat{\varepsilon}_{jt}$.
\begin{equation}
    \hat{\tau}^2_t = \frac{1}{J_t-1} \sum_{j=1}^{J_t} \left(\hat{\varepsilon}_{jt} - \frac{1}{J_t}\sum_{j=1}^{J_t}\hat{\varepsilon}_{jt}\right)^2
\end{equation}

The Fundamental Theorem of Calculus\footnote{
For a more detailed discussion about the use of integration for identification, see Section IV in \cite{gandhi2020identification}.
} derives the deterministic part of the revenue function by integrating the revenue elasticity of material inputs
\begin{equation}
\label{eq:integral}
\int \frac{\partial f\left(k_{jt}, l_{jt}, m_{jt}\right)}{\partial m_{jt}} d m_{jt}=f\left(k_{jt}, l_{jt}, m_{jt}\right)+\mathcal{C}\left(k_{jt}, l_{jt}\right)
\end{equation}
where $\mathcal{C}\left(k_{jt}, l_{jt}\right)$ is a constant of integration.
The integral has a closed form solution since a complete, second-order polynomial approximates $\frac{\partial f\left(k_{jt}, l_{jt}, m_{jt}\right)}{\partial m_{jt}}$: 
\begin{equation}
\int \frac{\partial f\left(k_{jt}, l_{jt}, m_{jt}\right)}{\partial m_{jt}} d m_{jt}=\left(\begin{aligned}
&\gamma_{0}+\gamma_{k} k_{jt}+\gamma_{l} l_{jt}+\frac{\gamma_{m}}{2} m_{jt}\\
&+\gamma_{k k} k_{jt}^{2}+\gamma_{ll} l_{jt}^{2}+\frac{\gamma_{m m}}{3} m_{jt}^{2}\\
&+\gamma_{k l} k_{jt} l_{jt}+\frac{\gamma_{k m}}{2} k_{jt} m_{jt}+\frac{\gamma_{lm}}{2} l_{jt} m_{jt}
\end{aligned}\right) m_{jt}
\end{equation}
Substituting the $\gamma$ parameter vector with its estimate $\hat\gamma$ is enough to retrieve the sample analog for the integral.

Finally, consider constructing the following random variable $\mathcal{Y}_{j t}$
\begin{equation}
\label{eq:Y_def}
\begin{aligned}
\mathcal{Y}_{j t} &\equiv y_{j t}-\varepsilon_{j t}-\int \frac{\partial f\left(k_{jt}, l_{jt}, m_{jt}\right)}{\partial m_{jt}} d m_{jt}\\
&=y_{j t}-\varepsilon_{j t}- f\left(k_{jt}, l_{jt}, m_{jt}\right)-\mathcal{C}\left(k_{jt}, l_{jt}\right)\\ &=\omega_{j t}-\mathcal{C}\left(k_{jt}, l_{jt}\right)
\end{aligned}
\end{equation}
where the second line follows by equation (\ref{eq:integral}) and the third line by equations (\ref{eq:revenue}) and (\ref{eq:tfpr}). Revenue $y_{jt}$ is observed, and the previous steps provide estimates for the ex-post productivity shock $\varepsilon_{jt}$ and $\int \frac{\partial f\left(k_{jt}, l_{jt}, m_{jt}\right)}{\partial m_{jt}} d m_{jt}$. I can therefore compute $\mathcal{Y}_{j t}$; denote its values $\hat{\mathcal{Y}}_{j t}$.

From the last line of equation (\ref{eq:Y_def}), $\mathcal{Y}_{j t}$ is also a function of $\omega_{jt}$, the persistent component of TFPR, and $\mathcal{C}(\cdot)$, the residual function from integrating in equation (\ref{eq:integral}). The $\mathcal{C}(\cdot)$ function is approximated by a second-order complete polynomial. Formally\footnote{
The $\mathcal{C}(\cdot)$ function is normalized to contain no constant since this latter cannot be separately identified later from the mean productivity, $E(\omega_{jt})$.
}
\begin{equation}
\label{eq:C_poly}
\mathcal{C}\left(k_{jt}, l_{jt}\right)=\alpha_k k_{jt}+\alpha_l l_{jt}+\alpha_{kk}k_{jt}^2+\alpha_{ll}l_{jt}^2+\alpha_{kl}k_{jt}l_{jt}
\end{equation} 

The time-varying Markovian process for $\omega_{jt}$ is given by the following truncated third-order polynomial approximation with period-specific coefficients, indexed $1\leq p \leq P$, where $P$ is the number of time periods the country's sample spans: five for Spain, France, and Italy, and four for Germany, Poland, and Romania, whose data begin later
\begin{equation}
\label{eq:omega_markov}
\omega_{jt}=g_{t}(\omega_{jt-1})+\eta_{jt}=\sum_{0 \leq a \leq 3} \delta_{ap} \omega_{jt-1}^{a}+\eta_{jt},\qquad t\in p
\end{equation}
Substituting equations (\ref{eq:C_poly}) and (\ref{eq:omega_markov}) in (\ref{eq:Y_def}), and replacing for $\omega_{jt-1}$ using the third line of equation (\ref{eq:Y_def}), yields
\begin{equation}
\begin{aligned}
\label{eq:Y_subst}
&\hat{\mathcal{Y}}_{j t} =-(\alpha_k k_{jt}+\alpha_l l_{jt}+\alpha_{kk}k_{jt}^2+\alpha_{ll}l_{jt}^2+\alpha_{kl}k_{jt}l_{jt})+\\
&\sum_{0 \leq a \leq 3} \sum_{1 \leq p \leq P} I(t\in p) \delta_{ap} \left(\hat{\mathcal{Y}}_{j t-1}+\alpha_k k_{jt-1}+\alpha_l l_{jt-1}+\alpha_{kk}k_{jt-1}^2+\alpha_{ll}l_{jt-1}^2+\alpha_{kl}k_{jt-1}l_{jt-1}\right)^a+\eta_{j t}
\end{aligned}
\end{equation}
where $I(\cdot)$ is the indicator function.

By the allocation assumptions, $k_{jt}$ and $l_{jt}$ are predetermined inputs at period $t-1$. 
Given Assumptions \ref{asspr} and \ref{asst} regarding the information set available to firm $j$ at the end of period $t-1$, their allocation is made before realizing the productivity innovation $\eta_{jt}$. Thus, an exactly identified GMM system estimates the parameter vectors $\boldsymbol{\alpha}$ and $\boldsymbol{\delta}$ from (\ref{eq:Y_subst}) using the following unconditional moment conditions 
\begin{equation}\label{eq:gmm_moments}
\begin{aligned}
&E\left[\eta_{j t} k_{jt}^{\tau_{k}} l_{jt}^{\tau_{l}}\right]=0,\quad \forall \tau_k,\tau_l,~0<\tau_k+\tau_l\leq 2\\
&E\left[\eta_{j t} \hat{\mathcal{Y}}^a_{j t-1}\,\mathbf{1}(t\in p)\right]=0,\quad \forall a,p,~0\leq a\leq 3,~1\leq p\leq P\\
\end{aligned}
\end{equation}
The first set of conditions contains five moments identifying the five components of the $\boldsymbol{\alpha}$ parameter vector. 
The second set of moments contains $P\times 4$ conditions identifying the period-specific coefficients of the Markov process for $\omega_{jt}$.

The cross-sectional sample variance of the GMM residuals estimates the cross-sectional variance of the productivity innovation $\eta_{jt}$, $\sigma^2_t$:
\begin{equation}
    \hat{\sigma}^2_t = \frac{1}{J_t-1} \sum_{j=1}^{J_t} \left(\hat{\eta}_{jt} - \frac{1}{J_t}\sum_{j=1}^{J_t}\hat{\eta}_{jt}\right)^2
\end{equation}

Having the estimates $\boldsymbol{\hat\gamma}$ and $\boldsymbol{\hat\alpha}$ for the $\boldsymbol{\gamma}$ and $\boldsymbol{\alpha}$ parameter vectors allows computing the firm-time specific revenue elasticities for material inputs, capital, and labor. Respectively,
\begin{equation}
\resizebox{\textwidth}{!}{$
\begin{aligned}
        \widehat{\mathrm{elas}}^M_{jt}&=\hat\gamma_{0}+\hat\gamma_{k} k_{jt}+\hat\gamma_{l} l_{jt}+\hat\gamma_{m} m_{jt}+\hat\gamma_{k k} k_{jt}^{2}+\hat\gamma_{ll} l_{jt}^{2} +
\hat\gamma_{m m} m_{jt}^{2}+\hat\gamma_{k l} k_{jt} l_{jt}+\hat\gamma_{k m} k_{jt} m_{jt}+\hat\gamma_{l m} l_{jt} m_{jt}\\
    \widehat{\mathrm{elas}}^K_{jt}&=m_{jt}\left(\hat\gamma_k+2\hat\gamma_{kk}k_{jt}+\hat\gamma_{kl}l_{jt}+\frac{\hat\gamma_{km}}{2}m_{jt}\right)-\hat\alpha_k-2\hat\alpha_{kk}k_{jt}-\hat \alpha_{kl}l_{jt}\\
    \widehat{\mathrm{elas}}^L_{jt}&=m_{jt}\left(\hat\gamma_l+2\hat\gamma_{ll}l_{jt}+\hat\gamma_{kl}k_{jt}+\frac{\hat\gamma_{lm}}{2}m_{jt}\right)-\hat\alpha_l-2\hat\alpha_{ll}l_{jt}-\hat \alpha_{kl}k_{jt}
\end{aligned}
$}
\label{eq:elas}
\end{equation}
\subsection{Asymptotic Inference}\label{app:asym}
The empirical model is consistent with the just-identified case of the two-step sieve procedure in \cite{gandhi2020identification}, with the difference that the polynomial parameters of the Markovian process for TFPR are period-specific in this paper. Nonetheless, the setting in this paper can also be mapped to the setting in \cite{hahn2018nonparametric}, ensuring not only consistency and asymptotic normality of the non-parametric estimator but also the numerical equivalence of the asymptotic variance to its parametric counterpart.

Indeed the first-stage estimator is similar to \cite{gandhi2020identification} and takes the form 
\begin{equation}
    \hat\gamma_{n} = \arg\max_\gamma - \sum_{j,t} \left\{ s_{jt} - \ln \left( \sum_{r_k+r_l+r_m\leq r(n)} \gamma_{r_k,r_l,r_m}k_{jt}^{r_k}l_{jt}^{r_l}m_{jt}^{r_m}\right) \right\}^2
\end{equation}
which, using the notation in \cite{hahn2018nonparametric}, corresponds to 
\begin{equation}
    \hat{h}_n = \arg \max_{h \in \mathcal{H}_n} \sum_{i=1}^{n} \varphi (Z_{1,i}; h)
\end{equation}
where  $r(n)$ is a sequence growing to infinity with $n$, $\mathcal{H}_n$ is a sieve space with dimension increasing in $n$, $Z_{1,i}=\{s_{jt},k_{jt},l_{jt},m_{jt}\}$, and $h=\gamma$.

By contrast, the second-stage estimator is different from \cite{gandhi2020identification} since it accounts for period-specific Markov regime variations. It takes the form
\begin{equation}
\begin{aligned}
    &(\hat{\alpha}_n, \hat{\delta}_n) =\\ 
    &\arg\max_{\alpha, \delta} - \left\{ \sum_{0<\tau_k+\tau_l \leq \tau(n)} \left[\sum_{j,t} \eta_{jt}(\alpha, \delta) k_{jt}^{\tau_k} l_{jt}^{\tau_l}\right]^2 +\sum_{0\leq a\leq A(n)}  \sum_{1\leq p\leq P} \left[\sum_{j,t} I(t\in p)\, \eta_{jt}(\alpha, \delta) \hat{\mathcal{Y}}^a_{jt-1}(\hat\gamma_{n})\right]^2 \right\}
\end{aligned}
\end{equation}
where $\tau(n)$ and $A(n)$ are sequences growing to infinity with $n$, $I(\cdot)$ is the indicator function and
\begin{equation}
\begin{aligned}
  & {\eta}_{jt}(\alpha, \delta) = \hat{\mathcal{Y}}_{jt}(\hat\gamma_{n}) + \sum_{0<\tau_k +\tau_l\leq\tau(n) } \alpha_{\tau_k,\tau_l} k_{jt}^{\tau_k}l_{jt}^{\tau_l} \\
  & - \sum_{0\leq a \leq A(n)}\sum_{1\leq p \leq P} I(t\in p)\delta_{ap} \left( \hat{\mathcal{Y}}_{jt-1}(\hat\gamma_{n}) + \sum_{0<\tau_k +\tau_l\leq\tau(n)}  \alpha_{\tau_k,\tau_l} k_{jt-1}^{\tau_k}l_{jt-1}^{\tau_l} \right)^a 
\end{aligned}
\end{equation}

The second-stage estimator corresponds, using the notation in \cite{hahn2018nonparametric}, to 
\begin{equation}
    \hat{g}_n = \arg \max_{g \in \mathcal{G}_n} \sum_{i=1}^{n} \psi (Z_{2,i}; g, \hat{h}_n)
\end{equation}
where $\mathcal{G}_n$ is a sieve space with dimension increasing in $n$, $g$ collects the constant of integration and the Markov function, indexed by $\boldsymbol{\alpha}$ and $\boldsymbol{\delta}$ respectively, and $Z_{2,i}=\{y_{jt},k_{jt},l_{jt},m_{jt},y_{jt-1},k_{jt-1},l_{jt-1},m_{jt-1}\}$.

Similar to \cite{gandhi2020identification}, since I use linear polynomial sieves, under the standard assumptions in \cite{chen2007large} and \cite{hahn2018nonparametric}, the estimator is consistent and asymptotically normal. Hence, I use a non-parametric bootstrap procedure to compute confidence intervals for my estimates.
\newpage

\section{Model Estimates}\label{app:est_res}

This appendix reports the estimates of the production function for each of the six countries and eight industry groups, the groups listed in Footnote~\ref{industry_list}. The coefficients of the Markov process of productivity, $g_t(\cdot)$, are specific to each of five periods: \textit{Beginning of the millennium} (2001--2002); \textit{Pre-crisis} (2003--2006); \textit{Great Recession} (2007--2009); \textit{European debt crisis} (2010--2013); and \textit{Post-crisis} (2014--2017). I compute standard errors by a non-parametric bootstrap that resamples firms, with 100 replications per country-industry cell.\footnote{In cells with few firms the four coefficients of the cubic Markov process are imprecise, with bootstrap intervals that cover both signs.} \Crefrange{tab:Food}{tab:Furn} report the estimates, one table per industry group.

\newpage
    {\fontsize{9pt}{11pt}\selectfont
\setlength{\tabcolsep}{4pt}
\begin{longtable}{lcccccc}
    \caption{Food, Beverages and Tobacco}
    \label{tab:Food} \tabularnewline
    \caption*{Production Function Estimates} \tabularnewline
  \toprule
 & Germany & Spain & France & Italy & Poland & Romania \\ 
  \midrule
Observations & 5,153 & 156,587 & 143,753 & 111,367 & 6,959 & 26,761 \\ 
  \midrule
  $\gamma_0$ & 0.493 & 0.567 & 0.141 & 0.171 & 0.031 & 0.722 \\ 
   & (0.050) & (0.049) & (0.084) & (0.012) & (0.045) & (0.032) \\ 
  $\gamma_k$ & -0.033 & -0.079 & -0.053 & -0.082 & -0.277 & -0.099 \\ 
   & (0.012) & (0.005) & (0.031) & (0.006) & (0.013) & (0.008) \\ 
  $\gamma_l$ & -0.083 & -0.003 & 0.033 & 0.160 & 0.317 & 0.032 \\ 
   & (0.021) & (0.015) & (0.019) & (0.010) & (0.026) & (0.016) \\ 
  $\gamma_m$ & 0.148 & 0.135 & 0.041 & 0.043 & 0.075 & 0.164 \\ 
   & (0.012) & (0.011) & (0.025) & (0.004) & (0.014) & (0.011) \\ 
  $\gamma_{kk}$ & -0.004 & -0.001 & -0.000 & -0.008 & -0.016 & 0.002 \\ 
   & (0.001) & (0.000) & (0.000) & (0.001) & (0.001) & (0.001) \\ 
  $\gamma_{ll}$ & 0.003 & -0.019 & -0.012 & -0.046 & -0.053 & -0.026 \\ 
   & (0.003) & (0.002) & (0.007) & (0.003) & (0.003) & (0.003) \\ 
  $\gamma_{mm}$ & 0.015 & 0.013 & 0.003 & 0.005 & 0.008 & 0.010 \\ 
   & (0.001) & (0.001) & (0.002) & (0.000) & (0.002) & (0.001) \\ 
  $\gamma_{kl}$ & 0.003 & 0.007 & 0.015 & 0.012 & 0.059 & 0.010 \\ 
   & (0.003) & (0.001) & (0.009) & (0.001) & (0.003) & (0.002) \\ 
  $\gamma_{km}$ & -0.011 & -0.017 & -0.009 & -0.003 & -0.020 & -0.014 \\ 
   & (0.003) & (0.001) & (0.005) & (0.000) & (0.002) & (0.001) \\ 
  $\gamma_{lm}$ & -0.011 & 0.011 & 0.001 & 0.018 & 0.015 & 0.002 \\ 
   & (0.003) & (0.003) & (0.001) & (0.001) & (0.003) & (0.002) \\ 
  \midrule
  $\alpha_k$ & -0.088 & -0.138 & -0.551 & -0.302 & -0.283 & 0.006 \\ 
   & (0.091) & (0.026) & (0.070) & (0.035) & (0.280) & (0.048) \\ 
  $\alpha_l$ & -0.323 & -0.195 & 0.031 & 0.220 & 0.089 & -0.132 \\ 
   & (0.112) & (0.034) & (0.060) & (0.057) & (0.370) & (0.137) \\ 
  $\alpha_{kk}$ & -0.039 & -0.022 & -0.043 & -0.035 & -0.033 & -0.022 \\ 
   & (0.013) & (0.002) & (0.004) & (0.004) & (0.030) & (0.003) \\ 
  $\alpha_{ll}$ & -0.010 & -0.045 & -0.102 & -0.155 & -0.035 & -0.052 \\ 
   & (0.015) & (0.006) & (0.009) & (0.013) & (0.056) & (0.026) \\ 
  $\alpha_{kl}$ & -0.033 & -0.036 & 0.075 & -0.026 & -0.014 & -0.062 \\ 
   & (0.021) & (0.005) & (0.014) & (0.013) & (0.082) & (0.019) \\ 
  \bottomrule
  \newpage
    \caption*{Productivity Process Estimates} \tabularnewline
  \toprule
 & Germany & Spain & France & Italy & Poland & Romania \\ 
  \midrule
  \textbf{2001 - 2002} & & & & & & \\
  $\delta_0$ & NA & -0.184 & 0.026 & 0.058 & NA & NA \\ 
   & (NA) & (0.030) & (0.014) & (0.017) & (NA) & (NA) \\ 
  $\delta_1$ & NA & 0.540 & 0.927 & 0.802 & NA & NA \\ 
   & (NA) & (0.077) & (0.032) & (0.028) & (NA) & (NA) \\ 
  $\delta_2$ & NA & -0.029 & -0.032 & -0.016 & NA & NA \\ 
   & (NA) & (0.041) & (0.029) & (0.016) & (NA) & (NA) \\ 
  $\delta_3$ & NA & -0.002 & -0.008 & -0.009 & NA & NA \\ 
   & (NA) & (0.005) & (0.006) & (0.003) & (NA) & (NA) \\ 
  \textbf{2003 - 2006} & & & & & & \\
  $\delta_0$ & 0.074 & -0.106 & 0.010 & 0.068 & 0.071 & -0.070 \\ 
   & (0.082) & (0.019) & (0.016) & (0.012) & (0.306) & (0.093) \\ 
  $\delta_1$ & 0.758 & 0.636 & 0.920 & 0.860 & 0.826 & 0.973 \\ 
   & (0.221) & (0.046) & (0.023) & (0.021) & (0.425) & (0.247) \\ 
  $\delta_2$ & -0.075 & -0.042 & -0.005 & -0.030 & 0.069 & 0.013 \\ 
   & (0.489) & (0.030) & (0.030) & (0.011) & (0.316) & (0.265) \\ 
  $\delta_3$ & -0.081 & -0.006 & -0.001 & -0.008 & -0.067 & -0.043 \\ 
   & (0.312) & (0.006) & (0.005) & (0.004) & (0.100) & (0.103) \\ 
  \textbf{2007 - 2009} & & & & & & \\
  $\delta_0$ & 0.071 & -0.070 & 0.022 & 0.065 & 0.152 & -0.316 \\ 
   & (0.063) & (0.018) & (0.015) & (0.014) & (0.457) & (0.087) \\ 
  $\delta_1$ & 0.704 & 0.648 & 0.924 & 0.817 & 0.769 & 0.564 \\ 
   & (0.240) & (0.055) & (0.019) & (0.022) & (0.562) & (0.169) \\ 
  $\delta_2$ & 0.122 & -0.062 & -0.000 & 0.026 & -0.001 & -0.081 \\ 
   & (0.348) & (0.033) & (0.031) & (0.015) & (0.263) & (0.092) \\ 
  $\delta_3$ & 0.002 & -0.006 & 0.001 & -0.001 & -0.005 & -0.034 \\ 
   & (0.235) & (0.005) & (0.004) & (0.002) & (0.044) & (0.053) \\ 
  \textbf{2010 - 2013} & & & & & & \\
  $\delta_0$ & 0.063 & 0.005 & 0.037 & 0.077 & 0.244 & -0.036 \\ 
   & (0.038) & (0.010) & (0.014) & (0.011) & (0.206) & (0.036) \\ 
  $\delta_1$ & 0.854 & 0.768 & 0.940 & 0.845 & 0.679 & 0.926 \\ 
   & (0.132) & (0.031) & (0.011) & (0.014) & (0.149) & (0.099) \\ 
  $\delta_2$ & 0.223 & -0.073 & 0.007 & 0.013 & 0.009 & 0.078 \\ 
   & (0.281) & (0.025) & (0.027) & (0.011) & (0.097) & (0.104) \\ 
  $\delta_3$ & -0.192 & -0.019 & 0.000 & -0.001 & -0.007 & 0.001 \\ 
   & (0.175) & (0.005) & (0.005) & (0.001) & (0.027) & (0.040) \\ 
  \textbf{2014 - 2017} & & & & & & \\
  $\delta_0$ & 0.006 & -0.095 & 0.016 & 0.051 & 0.220 & -0.209 \\ 
   & (0.065) & (0.013) & (0.004) & (0.011) & (0.186) & (0.049) \\ 
  $\delta_1$ & 1.049 & 0.701 & 0.989 & 0.861 & 0.657 & 0.757 \\ 
   & (0.258) & (0.028) & (0.012) & (0.010) & (0.097) & (0.122) \\ 
  $\delta_2$ & -0.156 & -0.066 & -0.042 & -0.013 & -0.012 & 0.097 \\ 
   & (0.283) & (0.025) & (0.017) & (0.007) & (0.078) & (0.127) \\ 
  $\delta_3$ & -0.059 & -0.006 & -0.007 & -0.003 & 0.006 & -0.004 \\ 
   & (0.105) & (0.008) & (0.003) & (0.001) & (0.028) & (0.040) \\ 
  \bottomrule
\end{longtable}
\note{\scriptsize This table reports estimates from the two-stage GNR procedure (Appendix~\ref{Gandhi}). The upper panel presents production function parameters: $\gamma$ coefficients from the first-stage share regression (equation~\ref{eq:share_poly}) and $\alpha$ coefficients from the second-stage GMM (equation~\ref{eq:Y_subst}). The lower panel presents Markov process coefficients $\delta$ for each time period (equation~\ref{eq:omega_markov}), where $\delta_0$ is the drift, $\delta_1$ the AR(1) persistence, and $\delta_2$, $\delta_3$ higher-order terms. Bootstrap standard errors (in parentheses) from 100 non-parametric bootstrap replications. Six countries (Germany, Spain, France, Italy, Poland, Romania), 2001--2017 for Spain, France and Italy and 2005--2017 for Germany, Poland and Romania, whose samples begin in 2004. NA indicates insufficient data for that country-period cell.}

    \newpage
    \fontsize{9pt}{11pt}\selectfont
\setlength{\tabcolsep}{4pt}
\begin{longtable}{lcccccc}
    \caption{Textiles, Apparel and Leather}
    \label{tab:Textile} \tabularnewline
    \caption*{Production Function Estimates} \tabularnewline
  \toprule
 & Germany & Spain & France & Italy & Poland & Romania \\ 
  \midrule
Observations & 2,544 & 98,052 & 35,389 & 163,069 & 3,219 & 26,155 \\ 
  \midrule
  $\gamma_0$ & 0.320 & 0.891 & 0.536 & 0.127 & 0.312 & 0.803 \\ 
   & (0.093) & (0.015) & (0.029) & (0.011) & (0.032) & (0.019) \\ 
  $\gamma_k$ & -0.087 & -0.049 & -0.030 & -0.053 & -0.171 & -0.064 \\ 
   & (0.029) & (0.002) & (0.002) & (0.002) & (0.014) & (0.001) \\ 
  $\gamma_l$ & 0.082 & -0.176 & -0.133 & 0.044 & 0.053 & -0.173 \\ 
   & (0.047) & (0.004) & (0.007) & (0.003) & (0.012) & (0.004) \\ 
  $\gamma_m$ & 0.092 & 0.209 & 0.160 & 0.075 & 0.204 & 0.184 \\ 
   & (0.026) & (0.004) & (0.009) & (0.004) & (0.014) & (0.005) \\ 
  $\gamma_{kk}$ & -0.003 & -0.001 & -0.001 & 0.002 & -0.009 & -0.007 \\ 
   & (0.003) & (0.000) & (0.000) & (0.000) & (0.001) & (0.001) \\ 
  $\gamma_{ll}$ & -0.022 & 0.009 & 0.011 & -0.006 & -0.011 & 0.012 \\ 
   & (0.006) & (0.000) & (0.001) & (0.000) & (0.002) & (0.000) \\ 
  $\gamma_{mm}$ & 0.005 & 0.012 & 0.010 & 0.011 & 0.011 & 0.009 \\ 
   & (0.003) & (0.000) & (0.001) & (0.000) & (0.001) & (0.000) \\ 
  $\gamma_{kl}$ & 0.022 & 0.005 & 0.004 & -0.001 & 0.025 & 0.006 \\ 
   & (0.006) & (0.000) & (0.000) & (0.000) & (0.003) & (0.000) \\ 
  $\gamma_{km}$ & -0.005 & -0.006 & -0.001 & -0.010 & 0.000 & 0.003 \\ 
   & (0.003) & (0.000) & (0.000) & (0.000) & (0.002) & (0.001) \\ 
  $\gamma_{lm}$ & -0.002 & -0.021 & -0.016 & 0.008 & -0.017 & -0.016 \\ 
   & (0.006) & (0.001) & (0.001) & (0.001) & (0.002) & (0.001) \\ 
  \midrule
  $\alpha_k$ & -0.384 & -0.033 & -0.192 & -0.316 & -0.532 & -0.351 \\ 
   & (0.099) & (0.008) & (0.026) & (0.026) & (0.074) & (0.042) \\ 
  $\alpha_l$ & 0.664 & -0.116 & -0.374 & -0.124 & 0.194 & -0.065 \\ 
   & (0.262) & (0.023) & (0.033) & (0.042) & (0.154) & (0.037) \\ 
  $\alpha_{kk}$ & -0.019 & -0.009 & -0.003 & -0.014 & -0.047 & -0.029 \\ 
   & (0.010) & (0.001) & (0.003) & (0.003) & (0.006) & (0.003) \\ 
  $\alpha_{ll}$ & -0.166 & -0.054 & -0.009 & -0.004 & -0.044 & -0.033 \\ 
   & (0.035) & (0.005) & (0.005) & (0.006) & (0.018) & (0.006) \\ 
  $\alpha_{kl}$ & 0.092 & -0.038 & 0.006 & -0.013 & 0.037 & 0.005 \\ 
   & (0.023) & (0.003) & (0.006) & (0.007) & (0.014) & (0.009) \\ 
  \bottomrule
  \newpage
    \caption*{Productivity Process Estimates} \tabularnewline
  \toprule
 & Germany & Spain & France & Italy & Poland & Romania \\ 
  \midrule
  \textbf{2001 - 2002} & & & & & & \\
  $\delta_0$ & NA & -0.032 & -0.012 & 0.127 & NA & NA \\ 
   & (NA) & (0.008) & (0.016) & (0.019) & (NA) & (NA) \\ 
  $\delta_1$ & NA & 0.747 & 0.698 & 0.899 & NA & NA \\ 
   & (NA) & (0.033) & (0.040) & (0.032) & (NA) & (NA) \\ 
  $\delta_2$ & NA & -0.102 & -0.127 & -0.017 & NA & NA \\ 
   & (NA) & (0.049) & (0.028) & (0.016) & (NA) & (NA) \\ 
  $\delta_3$ & NA & -0.054 & -0.026 & -0.027 & NA & NA \\ 
   & (NA) & (0.020) & (0.008) & (0.006) & (NA) & (NA) \\ 
  \textbf{2003 - 2006} & & & & & & \\
  $\delta_0$ & 3.276 & -0.035 & -0.012 & 0.096 & 0.342 & -0.067 \\ 
   & (3.786) & (0.009) & (0.016) & (0.012) & (0.203) & (0.030) \\ 
  $\delta_1$ & -5.866 & 0.733 & 0.705 & 0.929 & 0.381 & 0.771 \\ 
   & (5.264) & (0.032) & (0.038) & (0.017) & (0.325) & (0.059) \\ 
  $\delta_2$ & 4.677 & -0.094 & -0.097 & -0.013 & 0.223 & -0.005 \\ 
   & (2.618) & (0.030) & (0.020) & (0.012) & (0.271) & (0.097) \\ 
  $\delta_3$ & -1.037 & -0.034 & -0.016 & -0.022 & 0.026 & -0.033 \\ 
   & (0.483) & (0.012) & (0.007) & (0.004) & (0.133) & (0.040) \\ 
  \textbf{2007 - 2009} & & & & & & \\
  $\delta_0$ & 1.008 & -0.027 & -0.017 & 0.098 & 0.132 & -0.220 \\ 
   & (1.439) & (0.008) & (0.012) & (0.018) & (0.266) & (0.053) \\ 
  $\delta_1$ & -1.130 & 0.716 & 0.774 & 0.890 & 0.863 & 0.391 \\ 
   & (2.464) & (0.026) & (0.040) & (0.019) & (0.637) & (0.057) \\ 
  $\delta_2$ & 1.512 & -0.158 & -0.099 & 0.000 & -0.039 & -0.082 \\ 
   & (1.531) & (0.045) & (0.026) & (0.015) & (0.572) & (0.023) \\ 
  $\delta_3$ & -0.353 & -0.045 & -0.015 & -0.021 & 0.043 & -0.013 \\ 
   & (0.360) & (0.016) & (0.013) & (0.006) & (0.201) & (0.007) \\ 
  \textbf{2010 - 2013} & & & & & & \\
  $\delta_0$ & 0.823 & -0.022 & 0.034 & 0.132 & 0.257 & -0.087 \\ 
   & (0.949) & (0.006) & (0.009) & (0.012) & (0.274) & (0.025) \\ 
  $\delta_1$ & -0.517 & 0.792 & 0.876 & 0.914 & 0.553 & 0.684 \\ 
   & (1.185) & (0.021) & (0.036) & (0.014) & (0.674) & (0.043) \\ 
  $\delta_2$ & 0.981 & -0.064 & -0.078 & -0.009 & 0.240 & -0.104 \\ 
   & (0.560) & (0.041) & (0.035) & (0.012) & (0.723) & (0.029) \\ 
  $\delta_3$ & -0.214 & -0.062 & -0.035 & -0.024 & -0.047 & -0.026 \\ 
   & (0.134) & (0.017) & (0.009) & (0.004) & (0.300) & (0.010) \\ 
  \textbf{2014 - 2017} & & & & & & \\
  $\delta_0$ & 0.281 & -0.053 & 0.006 & 0.109 & 0.324 & -0.072 \\ 
   & (0.877) & (0.010) & (0.007) & (0.014) & (0.180) & (0.030) \\ 
  $\delta_1$ & 0.434 & 0.633 & 0.896 & 0.910 & 0.749 & 0.661 \\ 
   & (1.311) & (0.046) & (0.040) & (0.017) & (0.264) & (0.036) \\ 
  $\delta_2$ & 0.361 & -0.181 & -0.032 & -0.013 & -0.048 & -0.123 \\ 
   & (0.693) & (0.036) & (0.035) & (0.012) & (0.234) & (0.022) \\ 
  $\delta_3$ & -0.075 & -0.025 & -0.016 & -0.019 & 0.006 & -0.026 \\ 
   & (0.135) & (0.019) & (0.011) & (0.004) & (0.103) & (0.006) \\ 
  \bottomrule
\end{longtable}
\note{\scriptsize This table reports estimates from the two-stage GNR procedure (Appendix~\ref{Gandhi}). The upper panel presents production function parameters: $\gamma$ coefficients from the first-stage share regression (equation~\ref{eq:share_poly}) and $\alpha$ coefficients from the second-stage GMM (equation~\ref{eq:Y_subst}). The lower panel presents Markov process coefficients $\delta$ for each time period (equation~\ref{eq:omega_markov}), where $\delta_0$ is the drift, $\delta_1$ the AR(1) persistence, and $\delta_2$, $\delta_3$ higher-order terms. Bootstrap standard errors (in parentheses) from 100 non-parametric bootstrap replications. Six countries (Germany, Spain, France, Italy, Poland, Romania), 2001--2017 for Spain, France and Italy and 2005--2017 for Germany, Poland and Romania, whose samples begin in 2004. NA indicates insufficient data for that country-period cell.}

    \newpage
    \fontsize{9pt}{11pt}\selectfont
\setlength{\tabcolsep}{4pt}
\begin{longtable}{lcccccc}
    \caption{Wood, Paper, and Printing}
    \label{tab:Wood} \tabularnewline
    \caption*{Production Function Estimates} \tabularnewline
  \toprule
 & Germany & Spain & France & Italy & Poland & Romania \\ 
  \midrule
Observations & 5,997 & 155,263 & 82,221 & 109,908 & 6,075 & 29,486 \\ 
  \midrule
  $\gamma_0$ & 0.123 & 0.572 & 0.456 & 0.124 & 0.416 & 0.628 \\ 
   & (0.097) & (0.050) & (0.079) & (0.006) & (0.115) & (0.034) \\ 
  $\gamma_k$ & -0.040 & -0.105 & -0.029 & -0.142 & -0.033 & -0.110 \\ 
   & (0.026) & (0.004) & (0.005) & (0.007) & (0.020) & (0.004) \\ 
  $\gamma_l$ & 0.010 & 0.005 & -0.094 & 0.068 & -0.037 & -0.049 \\ 
   & (0.016) & (0.022) & (0.016) & (0.003) & (0.027) & (0.012) \\ 
  $\gamma_m$ & 0.071 & 0.141 & 0.130 & 0.089 & 0.086 & 0.143 \\ 
   & (0.050) & (0.015) & (0.023) & (0.004) & (0.035) & (0.009) \\ 
  $\gamma_{kk}$ & -0.002 & -0.006 & 0.001 & -0.006 & 0.004 & -0.008 \\ 
   & (0.002) & (0.000) & (0.000) & (0.000) & (0.002) & (0.001) \\ 
  $\gamma_{ll}$ & -0.006 & -0.017 & 0.004 & -0.006 & -0.001 & -0.001 \\ 
   & (0.004) & (0.003) & (0.001) & (0.000) & (0.003) & (0.001) \\ 
  $\gamma_{mm}$ & 0.006 & 0.008 & 0.011 & 0.010 & 0.007 & 0.003 \\ 
   & (0.004) & (0.001) & (0.002) & (0.000) & (0.003) & (0.001) \\ 
  $\gamma_{kl}$ & 0.006 & 0.023 & 0.002 & 0.020 & -0.001 & 0.011 \\ 
   & (0.004) & (0.001) & (0.001) & (0.001) & (0.003) & (0.001) \\ 
  $\gamma_{km}$ & -0.003 & -0.008 & -0.008 & -0.016 & -0.013 & 0.001 \\ 
   & (0.002) & (0.001) & (0.001) & (0.001) & (0.004) & (0.001) \\ 
  $\gamma_{lm}$ & -0.006 & -0.004 & -0.011 & 0.007 & 0.003 & -0.004 \\ 
   & (0.005) & (0.003) & (0.002) & (0.000) & (0.005) & (0.002) \\ 
  \midrule
  $\alpha_k$ & -0.437 & -0.148 & -0.175 & -0.580 & -0.121 & -0.217 \\ 
   & (0.107) & (0.021) & (0.050) & (0.026) & (0.082) & (0.034) \\ 
  $\alpha_l$ & 0.106 & -0.084 & -0.394 & -0.088 & -0.641 & -0.030 \\ 
   & (0.200) & (0.032) & (0.039) & (0.041) & (0.240) & (0.033) \\ 
  $\alpha_{kk}$ & -0.033 & -0.019 & -0.017 & -0.055 & -0.023 & -0.021 \\ 
   & (0.007) & (0.002) & (0.003) & (0.003) & (0.006) & (0.003) \\ 
  $\alpha_{ll}$ & -0.109 & -0.065 & -0.031 & 0.012 & 0.034 & -0.044 \\ 
   & (0.023) & (0.007) & (0.006) & (0.007) & (0.029) & (0.006) \\ 
  $\alpha_{kl}$ & 0.056 & -0.008 & -0.014 & 0.044 & -0.047 & -0.005 \\ 
   & (0.023) & (0.006) & (0.011) & (0.008) & (0.021) & (0.006) \\ 
  \bottomrule
  \newpage
    \caption*{Productivity Process Estimates} \tabularnewline
  \toprule
 & Germany & Spain & France & Italy & Poland & Romania \\ 
  \midrule
  \textbf{2001 - 2002} & & & & & & \\
  $\delta_0$ & NA & -0.020 & -0.086 & 0.213 & NA & NA \\ 
   & (NA) & (0.011) & (0.021) & (0.037) & (NA) & (NA) \\ 
  $\delta_1$ & NA & 0.719 & 0.747 & 0.717 & NA & NA \\ 
   & (NA) & (0.040) & (0.055) & (0.048) & (NA) & (NA) \\ 
  $\delta_2$ & NA & -0.004 & -0.105 & 0.019 & NA & NA \\ 
   & (NA) & (0.066) & (0.051) & (0.031) & (NA) & (NA) \\ 
  $\delta_3$ & NA & -0.006 & -0.030 & -0.029 & NA & NA \\ 
   & (NA) & (0.024) & (0.011) & (0.009) & (NA) & (NA) \\ 
  \textbf{2003 - 2006} & & & & & & \\
  $\delta_0$ & 0.050 & -0.002 & -0.046 & 0.231 & -0.223 & -0.193 \\ 
   & (0.101) & (0.007) & (0.016) & (0.042) & (0.327) & (0.089) \\ 
  $\delta_1$ & 0.893 & 0.822 & 0.797 & 0.771 & 0.674 & 0.347 \\ 
   & (0.347) & (0.026) & (0.035) & (0.057) & (0.585) & (0.306) \\ 
  $\delta_2$ & 0.096 & -0.026 & -0.138 & -0.031 & -0.122 & -0.368 \\ 
   & (0.478) & (0.065) & (0.036) & (0.018) & (0.483) & (0.304) \\ 
  $\delta_3$ & -0.045 & -0.025 & -0.045 & -0.008 & -0.016 & -0.051 \\ 
   & (0.236) & (0.014) & (0.010) & (0.010) & (0.173) & (0.110) \\ 
  \textbf{2007 - 2009} & & & & & & \\
  $\delta_0$ & 0.066 & -0.014 & -0.058 & 0.220 & -0.288 & -0.179 \\ 
   & (0.128) & (0.011) & (0.015) & (0.036) & (0.320) & (0.041) \\ 
  $\delta_1$ & 0.861 & 0.865 & 0.812 & 0.776 & 0.564 & 0.526 \\ 
   & (0.181) & (0.026) & (0.035) & (0.041) & (0.594) & (0.077) \\ 
  $\delta_2$ & 0.056 & 0.110 & -0.091 & -0.019 & -0.235 & -0.141 \\ 
   & (0.113) & (0.102) & (0.036) & (0.014) & (0.436) & (0.081) \\ 
  $\delta_3$ & -0.039 & 0.015 & -0.017 & -0.019 & -0.063 & -0.033 \\ 
   & (0.040) & (0.035) & (0.017) & (0.004) & (0.121) & (0.011) \\ 
  \textbf{2010 - 2013} & & & & & & \\
  $\delta_0$ & 0.048 & -0.014 & -0.016 & 0.228 & -0.327 & -0.021 \\ 
   & (0.057) & (0.007) & (0.011) & (0.036) & (0.265) & (0.019) \\ 
  $\delta_1$ & 0.949 & 0.852 & 0.854 & 0.812 & 0.283 & 0.756 \\ 
   & (0.092) & (0.016) & (0.027) & (0.040) & (0.390) & (0.051) \\ 
  $\delta_2$ & 0.020 & 0.004 & -0.084 & -0.045 & -0.445 & -0.141 \\ 
   & (0.074) & (0.041) & (0.038) & (0.011) & (0.234) & (0.066) \\ 
  $\delta_3$ & -0.025 & -0.007 & -0.037 & -0.009 & -0.084 & -0.055 \\ 
   & (0.021) & (0.026) & (0.014) & (0.005) & (0.057) & (0.017) \\ 
  \textbf{2014 - 2017} & & & & & & \\
  $\delta_0$ & -0.015 & -0.042 & -0.007 & 0.265 & -0.124 & -0.065 \\ 
   & (0.055) & (0.007) & (0.009) & (0.028) & (0.112) & (0.027) \\ 
  $\delta_1$ & 1.047 & 0.867 & 0.887 & 0.761 & 1.002 & 0.619 \\ 
   & (0.083) & (0.032) & (0.039) & (0.025) & (0.267) & (0.039) \\ 
  $\delta_2$ & -0.023 & -0.023 & -0.139 & -0.047 & 0.126 & -0.067 \\ 
   & (0.060) & (0.088) & (0.036) & (0.007) & (0.191) & (0.066) \\ 
  $\delta_3$ & -0.018 & -0.007 & -0.025 & -0.011 & 0.008 & -0.021 \\ 
   & (0.017) & (0.044) & (0.019) & (0.002) & (0.042) & (0.016) \\ 
  \bottomrule
\end{longtable}
\note{\scriptsize This table reports estimates from the two-stage GNR procedure (Appendix~\ref{Gandhi}). The upper panel presents production function parameters: $\gamma$ coefficients from the first-stage share regression (equation~\ref{eq:share_poly}) and $\alpha$ coefficients from the second-stage GMM (equation~\ref{eq:Y_subst}). The lower panel presents Markov process coefficients $\delta$ for each time period (equation~\ref{eq:omega_markov}), where $\delta_0$ is the drift, $\delta_1$ the AR(1) persistence, and $\delta_2$, $\delta_3$ higher-order terms. Bootstrap standard errors (in parentheses) from 100 non-parametric bootstrap replications. Six countries (Germany, Spain, France, Italy, Poland, Romania), 2001--2017 for Spain, France and Italy and 2005--2017 for Germany, Poland and Romania, whose samples begin in 2004. NA indicates insufficient data for that country-period cell.}

    \newpage
    \fontsize{9pt}{11pt}\selectfont
\setlength{\tabcolsep}{4pt}
\begin{longtable}{lcccccc}
    \caption{Coke, Chemicals, and Pharmaceuticals}
    \label{tab:Coke} \tabularnewline
    \caption*{Production Function Estimates} \tabularnewline
  \toprule
 & Germany & Spain & France & Italy & Poland & Romania \\ 
  \midrule
Observations & 5,653 & 39,794 & 21,373 & 45,809 & 2,916 & 4,797 \\ 
  \midrule
  $\gamma_0$ & 0.019 & 0.596 & 0.549 & 0.020 & 0.539 & 0.654 \\ 
   & (0.022) & (0.030) & (0.025) & (0.004) & (0.071) & (0.068) \\ 
  $\gamma_k$ & -0.084 & -0.089 & -0.042 & -0.096 & -0.146 & -0.081 \\ 
   & (0.026) & (0.007) & (0.004) & (0.011) & (0.017) & (0.014) \\ 
  $\gamma_l$ & 0.091 & -0.060 & -0.123 & 0.186 & -0.054 & -0.082 \\ 
   & (0.026) & (0.012) & (0.008) & (0.022) & (0.038) & (0.026) \\ 
  $\gamma_m$ & 0.070 & 0.134 & 0.167 & 0.018 & 0.190 & 0.150 \\ 
   & (0.025) & (0.007) & (0.007) & (0.002) & (0.021) & (0.017) \\ 
  $\gamma_{kk}$ & -0.003 & -0.003 & -0.001 & 0.005 & -0.014 & -0.001 \\ 
   & (0.001) & (0.001) & (0.000) & (0.001) & (0.002) & (0.001) \\ 
  $\gamma_{ll}$ & -0.016 & -0.002 & 0.007 & -0.040 & -0.001 & 0.000 \\ 
   & (0.005) & (0.002) & (0.001) & (0.005) & (0.005) & (0.002) \\ 
  $\gamma_{mm}$ & 0.008 & 0.008 & 0.012 & 0.004 & 0.014 & 0.009 \\ 
   & (0.003) & (0.000) & (0.001) & (0.000) & (0.002) & (0.001) \\ 
  $\gamma_{kl}$ & 0.016 & 0.014 & 0.006 & 0.021 & 0.026 & 0.010 \\ 
   & (0.005) & (0.002) & (0.001) & (0.002) & (0.004) & (0.002) \\ 
  $\gamma_{km}$ & -0.001 & -0.008 & -0.002 & -0.015 & -0.004 & -0.009 \\ 
   & (0.003) & (0.001) & (0.001) & (0.002) & (0.002) & (0.002) \\ 
  $\gamma_{lm}$ & -0.006 & -0.008 & -0.020 & 0.023 & -0.018 & -0.011 \\ 
   & (0.003) & (0.001) & (0.001) & (0.003) & (0.005) & (0.003) \\ 
  \midrule
  $\alpha_k$ & -0.806 & -0.144 & -0.172 & -0.455 & -0.358 & -0.113 \\ 
   & (0.142) & (0.027) & (0.041) & (0.063) & (0.120) & (0.054) \\ 
  $\alpha_l$ & 0.720 & -0.257 & -0.273 & 0.439 & -0.322 & -0.175 \\ 
   & (0.235) & (0.049) & (0.051) & (0.085) & (0.194) & (0.075) \\ 
  $\alpha_{kk}$ & -0.021 & -0.019 & -0.007 & -0.014 & -0.043 & -0.017 \\ 
   & (0.014) & (0.003) & (0.004) & (0.006) & (0.010) & (0.006) \\ 
  $\alpha_{ll}$ & -0.178 & -0.022 & -0.048 & -0.144 & 0.006 & -0.041 \\ 
   & (0.033) & (0.011) & (0.009) & (0.014) & (0.027) & (0.013) \\ 
  $\alpha_{kl}$ & 0.140 & -0.026 & 0.000 & 0.066 & 0.010 & -0.029 \\ 
   & (0.037) & (0.008) & (0.011) & (0.019) & (0.028) & (0.011) \\ 
  \bottomrule
  \newpage
    \caption*{Productivity Process Estimates} \tabularnewline
  \toprule
 & Germany & Spain & France & Italy & Poland & Romania \\ 
  \midrule
  \textbf{2001 - 2002} & & & & & & \\
  $\delta_0$ & NA & 0.009 & 0.011 & 0.258 & NA & NA \\ 
   & (NA) & (0.006) & (0.013) & (0.069) & (NA) & (NA) \\ 
  $\delta_1$ & NA & 0.920 & 0.877 & 0.799 & NA & NA \\ 
   & (NA) & (0.050) & (0.035) & (0.071) & (NA) & (NA) \\ 
  $\delta_2$ & NA & 0.014 & -0.045 & -0.012 & NA & NA \\ 
   & (NA) & (0.122) & (0.091) & (0.035) & (NA) & (NA) \\ 
  $\delta_3$ & NA & -0.079 & -0.049 & -0.011 & NA & NA \\ 
   & (NA) & (0.051) & (0.038) & (0.010) & (NA) & (NA) \\ 
  \textbf{2003 - 2006} & & & & & & \\
  $\delta_0$ & 1.203 & 0.034 & 0.055 & 0.164 & 0.004 & -0.192 \\ 
   & (0.632) & (0.006) & (0.018) & (0.032) & (0.088) & (0.163) \\ 
  $\delta_1$ & 0.324 & 0.926 & 0.836 & 0.933 & 0.892 & 0.338 \\ 
   & (0.390) & (0.020) & (0.036) & (0.033) & (0.284) & (0.458) \\ 
  $\delta_2$ & 0.032 & 0.035 & -0.128 & 0.007 & -0.074 & -0.597 \\ 
   & (0.202) & (0.057) & (0.068) & (0.016) & (0.446) & (0.438) \\ 
  $\delta_3$ & 0.014 & -0.047 & -0.048 & -0.021 & -0.146 & -0.198 \\ 
   & (0.048) & (0.043) & (0.036) & (0.006) & (0.251) & (0.139) \\ 
  \textbf{2007 - 2009} & & & & & & \\
  $\delta_0$ & 0.504 & 0.018 & 0.016 & 0.157 & 0.015 & -0.229 \\ 
   & (0.618) & (0.006) & (0.011) & (0.051) & (0.047) & (0.120) \\ 
  $\delta_1$ & 0.734 & 0.908 & 0.935 & 0.851 & 0.969 & 0.487 \\ 
   & (0.707) & (0.018) & (0.037) & (0.049) & (0.134) & (0.339) \\ 
  $\delta_2$ & 0.002 & 0.049 & -0.113 & 0.024 & -0.095 & -0.357 \\ 
   & (0.305) & (0.048) & (0.069) & (0.015) & (0.224) & (0.359) \\ 
  $\delta_3$ & 0.006 & -0.008 & -0.080 & -0.008 & -0.273 & -0.150 \\ 
   & (0.045) & (0.056) & (0.042) & (0.005) & (0.165) & (0.132) \\ 
  \textbf{2010 - 2013} & & & & & & \\
  $\delta_0$ & 0.455 & 0.023 & 0.070 & 0.153 & 0.053 & -0.044 \\ 
   & (0.359) & (0.004) & (0.009) & (0.028) & (0.082) & (0.028) \\ 
  $\delta_1$ & 0.753 & 1.002 & 0.933 & 0.902 & 0.810 & 0.882 \\ 
   & (0.179) & (0.022) & (0.038) & (0.028) & (0.174) & (0.083) \\ 
  $\delta_2$ & 0.020 & -0.020 & -0.120 & 0.030 & -0.248 & 0.012 \\ 
   & (0.108) & (0.045) & (0.058) & (0.018) & (0.277) & (0.119) \\ 
  $\delta_3$ & 0.003 & -0.136 & -0.088 & -0.013 & -0.078 & -0.019 \\ 
   & (0.026) & (0.050) & (0.045) & (0.005) & (0.100) & (0.067) \\ 
  \textbf{2014 - 2017} & & & & & & \\
  $\delta_0$ & 0.397 & -0.051 & 0.004 & 0.095 & -0.016 & -0.082 \\ 
   & (0.332) & (0.010) & (0.010) & (0.028) & (0.053) & (0.028) \\ 
  $\delta_1$ & 0.735 & 0.855 & 0.911 & 0.925 & 0.877 & 0.874 \\ 
   & (0.231) & (0.024) & (0.042) & (0.029) & (0.080) & (0.076) \\ 
  $\delta_2$ & 0.037 & -0.115 & -0.061 & 0.029 & -0.035 & -0.060 \\ 
   & (0.120) & (0.038) & (0.065) & (0.019) & (0.095) & (0.152) \\ 
  $\delta_3$ & 0.000 & -0.088 & -0.018 & -0.022 & -0.058 & -0.088 \\ 
   & (0.024) & (0.042) & (0.030) & (0.004) & (0.031) & (0.094) \\ 
  \bottomrule
\end{longtable}
\note{\scriptsize This table reports estimates from the two-stage GNR procedure (Appendix~\ref{Gandhi}). The upper panel presents production function parameters: $\gamma$ coefficients from the first-stage share regression (equation~\ref{eq:share_poly}) and $\alpha$ coefficients from the second-stage GMM (equation~\ref{eq:Y_subst}). The lower panel presents Markov process coefficients $\delta$ for each time period (equation~\ref{eq:omega_markov}), where $\delta_0$ is the drift, $\delta_1$ the AR(1) persistence, and $\delta_2$, $\delta_3$ higher-order terms. Bootstrap standard errors (in parentheses) from 100 non-parametric bootstrap replications. Six countries (Germany, Spain, France, Italy, Poland, Romania), 2001--2017 for Spain, France and Italy and 2005--2017 for Germany, Poland and Romania, whose samples begin in 2004. NA indicates insufficient data for that country-period cell.}

    \newpage
    \fontsize{9pt}{11pt}\selectfont
\setlength{\tabcolsep}{4pt}
\begin{longtable}{lcccccc}
    \caption{Rubber, Plastics, Metallic and Non-Metallic Mineral Products, Fabricated Metal Products}
    \label{tab:Rubber} \tabularnewline
    \caption*{Production Function Estimates} \tabularnewline
  \toprule
 & Germany & Spain & France & Italy & Poland & Romania \\ 
  \midrule
Observations & 25,363 & 333,239 & 175,009 & 431,855 & 18,733 & 41,745 \\ 
  \midrule
  $\gamma_0$ & 0.187 & 0.858 & 0.584 & 0.541 & 0.361 & 0.339 \\ 
   & (0.131) & (0.018) & (0.016) & (0.042) & (0.202) & (0.072) \\ 
  $\gamma_k$ & -0.037 & -0.058 & -0.039 & -0.019 & -0.040 & -0.151 \\ 
   & (0.027) & (0.001) & (0.001) & (0.002) & (0.023) & (0.032) \\ 
  $\gamma_l$ & -0.026 & -0.152 & -0.133 & -0.130 & -0.078 & -0.030 \\ 
   & (0.018) & (0.008) & (0.004) & (0.011) & (0.043) & (0.006) \\ 
  $\gamma_m$ & 0.086 & 0.206 & 0.188 & 0.120 & 0.094 & 0.146 \\ 
   & (0.061) & (0.005) & (0.005) & (0.013) & (0.056) & (0.031) \\ 
  $\gamma_{kk}$ & -0.002 & -0.001 & -0.001 & 0.002 & -0.006 & -0.009 \\ 
   & (0.001) & (0.000) & (0.000) & (0.000) & (0.003) & (0.002) \\ 
  $\gamma_{ll}$ & -0.000 & 0.005 & 0.007 & 0.010 & 0.006 & 0.002 \\ 
   & (0.001) & (0.001) & (0.000) & (0.001) & (0.003) & (0.001) \\ 
  $\gamma_{mm}$ & 0.006 & 0.012 & 0.015 & 0.005 & 0.002 & 0.008 \\ 
   & (0.004) & (0.000) & (0.000) & (0.001) & (0.003) & (0.002) \\ 
  $\gamma_{kl}$ & 0.007 & 0.006 & 0.006 & 0.004 & 0.004 & 0.023 \\ 
   & (0.005) & (0.000) & (0.000) & (0.000) & (0.003) & (0.005) \\ 
  $\gamma_{km}$ & -0.001 & -0.006 & -0.005 & -0.004 & 0.006 & -0.003 \\ 
   & (0.001) & (0.000) & (0.000) & (0.000) & (0.003) & (0.001) \\ 
  $\gamma_{lm}$ & -0.011 & -0.018 & -0.023 & -0.005 & -0.008 & -0.012 \\ 
   & (0.008) & (0.001) & (0.001) & (0.003) & (0.006) & (0.003) \\ 
  \midrule
  $\alpha_k$ & -0.543 & -0.029 & -0.221 & -0.024 & -0.279 & -0.412 \\ 
   & (0.092) & (0.007) & (0.016) & (0.012) & (0.097) & (0.100) \\ 
  $\alpha_l$ & 0.176 & -0.086 & -0.124 & -0.536 & -0.669 & -0.235 \\ 
   & (0.101) & (0.011) & (0.019) & (0.047) & (0.135) & (0.061) \\ 
  $\alpha_{kk}$ & -0.029 & -0.009 & -0.016 & 0.009 & -0.031 & -0.033 \\ 
   & (0.005) & (0.001) & (0.002) & (0.001) & (0.007) & (0.005) \\ 
  $\alpha_{ll}$ & -0.118 & -0.065 & -0.070 & 0.009 & 0.034 & -0.004 \\ 
   & (0.015) & (0.003) & (0.003) & (0.010) & (0.015) & (0.009) \\ 
  $\alpha_{kl}$ & 0.090 & -0.032 & 0.016 & -0.031 & -0.020 & 0.008 \\ 
   & (0.020) & (0.002) & (0.004) & (0.004) & (0.018) & (0.019) \\ 
  \bottomrule
  \newpage
    \caption*{Productivity Process Estimates} \tabularnewline
  \toprule
 & Germany & Spain & France & Italy & Poland & Romania \\ 
  \midrule
  \textbf{2001 - 2002} & & & & & & \\
  $\delta_0$ & NA & 0.011 & -0.007 & -0.193 & NA & NA \\ 
   & (NA) & (0.002) & (0.007) & (0.023) & (NA) & (NA) \\ 
  $\delta_1$ & NA & 0.872 & 0.919 & 0.666 & NA & NA \\ 
   & (NA) & (0.012) & (0.017) & (0.033) & (NA) & (NA) \\ 
  $\delta_2$ & NA & -0.017 & -0.031 & -0.053 & NA & NA \\ 
   & (NA) & (0.034) & (0.061) & (0.025) & (NA) & (NA) \\ 
  $\delta_3$ & NA & -0.076 & -0.014 & -0.012 & NA & NA \\ 
   & (NA) & (0.018) & (0.014) & (0.006) & (NA) & (NA) \\ 
  \textbf{2003 - 2006} & & & & & & \\
  $\delta_0$ & 0.139 & 0.051 & 0.048 & -0.097 & -0.118 & -0.129 \\ 
   & (0.037) & (0.002) & (0.006) & (0.021) & (0.067) & (0.054) \\ 
  $\delta_1$ & 0.865 & 0.928 & 0.903 & 0.683 & 0.985 & 0.640 \\ 
   & (0.132) & (0.008) & (0.017) & (0.027) & (0.162) & (0.126) \\ 
  $\delta_2$ & 0.066 & -0.024 & 0.014 & -0.081 & 0.111 & -0.100 \\ 
   & (0.145) & (0.023) & (0.059) & (0.018) & (0.105) & (0.109) \\ 
  $\delta_3$ & -0.015 & -0.052 & -0.005 & -0.015 & 0.002 & -0.032 \\ 
   & (0.074) & (0.013) & (0.013) & (0.006) & (0.013) & (0.017) \\ 
  \textbf{2007 - 2009} & & & & & & \\
  $\delta_0$ & 0.061 & 0.019 & 0.023 & -0.145 & -0.163 & -0.199 \\ 
   & (0.044) & (0.004) & (0.004) & (0.017) & (0.062) & (0.072) \\ 
  $\delta_1$ & 0.888 & 0.929 & 0.925 & 0.668 & 0.936 & 0.521 \\ 
   & (0.087) & (0.012) & (0.014) & (0.032) & (0.190) & (0.040) \\ 
  $\delta_2$ & 0.012 & -0.002 & -0.058 & -0.071 & 0.055 & -0.066 \\ 
   & (0.071) & (0.039) & (0.039) & (0.019) & (0.116) & (0.028) \\ 
  $\delta_3$ & -0.013 & -0.088 & -0.052 & -0.019 & -0.001 & -0.012 \\ 
   & (0.027) & (0.028) & (0.012) & (0.006) & (0.014) & (0.005) \\ 
  \textbf{2010 - 2013} & & & & & & \\
  $\delta_0$ & 0.127 & -0.020 & 0.054 & -0.124 & -0.073 & -0.066 \\ 
   & (0.054) & (0.003) & (0.006) & (0.021) & (0.044) & (0.048) \\ 
  $\delta_1$ & 0.847 & 0.885 & 0.849 & 0.627 & 0.937 & 0.654 \\ 
   & (0.075) & (0.006) & (0.025) & (0.033) & (0.119) & (0.046) \\ 
  $\delta_2$ & 0.021 & -0.031 & 0.032 & -0.067 & 0.032 & -0.082 \\ 
   & (0.043) & (0.023) & (0.041) & (0.011) & (0.057) & (0.025) \\ 
  $\delta_3$ & -0.004 & -0.040 & 0.005 & -0.015 & -0.002 & -0.012 \\ 
   & (0.020) & (0.021) & (0.025) & (0.005) & (0.006) & (0.005) \\ 
  \textbf{2014 - 2017} & & & & & & \\
  $\delta_0$ & -0.018 & -0.055 & 0.010 & -0.147 & -0.167 & -0.112 \\ 
   & (0.051) & (0.004) & (0.003) & (0.023) & (0.065) & (0.052) \\ 
  $\delta_1$ & 1.052 & 0.709 & 0.990 & 0.617 & 0.872 & 0.629 \\ 
   & (0.132) & (0.013) & (0.024) & (0.034) & (0.135) & (0.044) \\ 
  $\delta_2$ & -0.030 & -0.194 & -0.102 & -0.094 & 0.047 & -0.056 \\ 
   & (0.095) & (0.038) & (0.042) & (0.010) & (0.096) & (0.031) \\ 
  $\delta_3$ & -0.007 & -0.066 & -0.018 & -0.017 & 0.001 & -0.007 \\ 
   & (0.040) & (0.017) & (0.024) & (0.005) & (0.016) & (0.005) \\ 
  \bottomrule
\end{longtable}
\note{\scriptsize This table reports estimates from the two-stage GNR procedure (Appendix~\ref{Gandhi}). The upper panel presents production function parameters: $\gamma$ coefficients from the first-stage share regression (equation~\ref{eq:share_poly}) and $\alpha$ coefficients from the second-stage GMM (equation~\ref{eq:Y_subst}). The lower panel presents Markov process coefficients $\delta$ for each time period (equation~\ref{eq:omega_markov}), where $\delta_0$ is the drift, $\delta_1$ the AR(1) persistence, and $\delta_2$, $\delta_3$ higher-order terms. Bootstrap standard errors (in parentheses) from 100 non-parametric bootstrap replications. Six countries (Germany, Spain, France, Italy, Poland, Romania), 2001--2017 for Spain, France and Italy and 2005--2017 for Germany, Poland and Romania, whose samples begin in 2004. NA indicates insufficient data for that country-period cell.}

    \newpage
    \fontsize{9pt}{11pt}\selectfont
\setlength{\tabcolsep}{4pt}
\begin{longtable}{lcccccc}
    \caption{Electronic, Optical Products and Electrical Equipment}
    \label{tab:Elec} \tabularnewline
    \caption*{Production Function Estimates} \tabularnewline
  \toprule
 & Germany & Spain & France & Italy & Poland & Romania \\ 
  \midrule
Observations & 10,030 & 32,404 & 31,683 & 92,970 & 3,569 & 5,705 \\ 
  \midrule
  $\gamma_0$ & 0.090 & 0.696 & 0.115 & 0.538 & 0.700 & 0.323 \\ 
   & (0.036) & (0.066) & (0.171) & (0.018) & (0.065) & (0.042) \\ 
  $\gamma_k$ & -0.058 & -0.039 & -0.008 & -0.027 & -0.116 & -0.129 \\ 
   & (0.022) & (0.005) & (0.011) & (0.002) & (0.011) & (0.014) \\ 
  $\gamma_l$ & 0.007 & -0.103 & -0.036 & -0.088 & -0.125 & -0.023 \\ 
   & (0.004) & (0.022) & (0.051) & (0.003) & (0.028) & (0.007) \\ 
  $\gamma_m$ & 0.050 & 0.144 & 0.028 & 0.150 & 0.211 & 0.118 \\ 
   & (0.019) & (0.019) & (0.043) & (0.005) & (0.020) & (0.015) \\ 
  $\gamma_{kk}$ & -0.005 & -0.001 & 0.000 & -0.001 & -0.016 & -0.006 \\ 
   & (0.002) & (0.000) & (0.000) & (0.000) & (0.002) & (0.001) \\ 
  $\gamma_{ll}$ & -0.003 & 0.003 & 0.004 & 0.001 & 0.008 & -0.000 \\ 
   & (0.001) & (0.002) & (0.005) & (0.000) & (0.003) & (0.001) \\ 
  $\gamma_{mm}$ & 0.003 & 0.006 & 0.002 & 0.010 & 0.006 & 0.008 \\ 
   & (0.001) & (0.002) & (0.003) & (0.000) & (0.002) & (0.001) \\ 
  $\gamma_{kl}$ & 0.012 & 0.006 & -0.000 & 0.002 & 0.014 & 0.015 \\ 
   & (0.004) & (0.001) & (0.000) & (0.000) & (0.002) & (0.002) \\ 
  $\gamma_{km}$ & -0.000 & -0.001 & -0.002 & -0.002 & 0.011 & -0.008 \\ 
   & (0.000) & (0.001) & (0.003) & (0.000) & (0.003) & (0.001) \\ 
  $\gamma_{lm}$ & -0.005 & -0.008 & -0.003 & -0.013 & -0.017 & -0.006 \\ 
   & (0.002) & (0.004) & (0.005) & (0.000) & (0.004) & (0.001) \\ 
  \midrule
  $\alpha_k$ & -0.503 & -0.061 & -0.329 & -0.039 & -0.190 & -0.486 \\ 
   & (0.150) & (0.022) & (0.063) & (0.017) & (0.088) & (0.091) \\ 
  $\alpha_l$ & -0.537 & -0.147 & -0.670 & -0.292 & -0.519 & -0.275 \\ 
   & (0.261) & (0.039) & (0.080) & (0.028) & (0.145) & (0.113) \\ 
  $\alpha_{kk}$ & -0.056 & -0.004 & -0.021 & -0.001 & -0.030 & -0.046 \\ 
   & (0.011) & (0.002) & (0.005) & (0.002) & (0.011) & (0.009) \\ 
  $\alpha_{ll}$ & -0.018 & -0.057 & 0.006 & -0.042 & 0.027 & -0.004 \\ 
   & (0.029) & (0.008) & (0.013) & (0.005) & (0.019) & (0.014) \\ 
  $\alpha_{kl}$ & 0.077 & -0.004 & -0.001 & -0.031 & -0.022 & -0.001 \\ 
   & (0.034) & (0.008) & (0.015) & (0.005) & (0.020) & (0.018) \\ 
  \bottomrule
  \newpage
    \caption*{Productivity Process Estimates} \tabularnewline
  \toprule
 & Germany & Spain & France & Italy & Poland & Romania \\ 
  \midrule
  \textbf{2001 - 2002} & & & & & & \\
  $\delta_0$ & NA & 0.046 & -0.088 & -0.034 & NA & NA \\ 
   & (NA) & (0.016) & (0.027) & (0.012) & (NA) & (NA) \\ 
  $\delta_1$ & NA & 0.908 & 0.831 & 0.731 & NA & NA \\ 
   & (NA) & (0.036) & (0.045) & (0.028) & (NA) & (NA) \\ 
  $\delta_2$ & NA & -0.131 & -0.122 & -0.147 & NA & NA \\ 
   & (NA) & (0.069) & (0.046) & (0.051) & (NA) & (NA) \\ 
  $\delta_3$ & NA & -0.051 & -0.042 & -0.030 & NA & NA \\ 
   & (NA) & (0.037) & (0.015) & (0.012) & (NA) & (NA) \\ 
  \textbf{2003 - 2006} & & & & & & \\
  $\delta_0$ & -0.017 & 0.030 & -0.037 & -0.005 & -0.167 & -0.026 \\ 
   & (0.166) & (0.008) & (0.022) & (0.010) & (0.189) & (0.181) \\ 
  $\delta_1$ & 0.774 & 0.939 & 0.816 & 0.769 & 0.484 & 0.592 \\ 
   & (0.115) & (0.035) & (0.047) & (0.021) & (0.579) & (0.397) \\ 
  $\delta_2$ & -0.072 & -0.026 & -0.104 & -0.107 & -0.346 & -0.349 \\ 
   & (0.068) & (0.042) & (0.044) & (0.030) & (0.783) & (0.386) \\ 
  $\delta_3$ & -0.013 & -0.039 & -0.029 & -0.030 & -0.074 & -0.142 \\ 
   & (0.021) & (0.064) & (0.009) & (0.006) & (0.471) & (0.129) \\ 
  \textbf{2007 - 2009} & & & & & & \\
  $\delta_0$ & -0.068 & -0.005 & -0.041 & -0.042 & -0.197 & -0.030 \\ 
   & (0.115) & (0.007) & (0.012) & (0.010) & (0.114) & (0.142) \\ 
  $\delta_1$ & 0.841 & 0.929 & 0.934 & 0.782 & 0.774 & 0.474 \\ 
   & (0.086) & (0.030) & (0.039) & (0.018) & (0.286) & (0.097) \\ 
  $\delta_2$ & -0.043 & 0.043 & -0.016 & -0.093 & 0.150 & -0.099 \\ 
   & (0.049) & (0.043) & (0.050) & (0.024) & (0.292) & (0.082) \\ 
  $\delta_3$ & -0.002 & -0.063 & -0.007 & -0.031 & 0.011 & -0.020 \\ 
   & (0.009) & (0.061) & (0.017) & (0.009) & (0.099) & (0.024) \\ 
  \textbf{2010 - 2013} & & & & & & \\
  $\delta_0$ & -0.010 & 0.022 & -0.017 & -0.021 & -0.151 & -0.024 \\ 
   & (0.102) & (0.010) & (0.015) & (0.010) & (0.114) & (0.067) \\ 
  $\delta_1$ & 0.848 & 0.918 & 0.930 & 0.752 & 0.584 & 0.784 \\ 
   & (0.057) & (0.024) & (0.035) & (0.019) & (0.171) & (0.078) \\ 
  $\delta_2$ & -0.019 & -0.124 & -0.006 & -0.138 & -0.212 & -0.115 \\ 
   & (0.035) & (0.079) & (0.068) & (0.025) & (0.167) & (0.050) \\ 
  $\delta_3$ & -0.005 & -0.042 & -0.004 & -0.031 & -0.045 & -0.032 \\ 
   & (0.005) & (0.053) & (0.034) & (0.007) & (0.065) & (0.018) \\ 
  \textbf{2014 - 2017} & & & & & & \\
  $\delta_0$ & 0.007 & -0.121 & -0.016 & -0.028 & -0.187 & -0.029 \\ 
   & (0.044) & (0.023) & (0.015) & (0.011) & (0.141) & (0.095) \\ 
  $\delta_1$ & 0.936 & 0.703 & 0.953 & 0.736 & 0.671 & 0.642 \\ 
   & (0.064) & (0.031) & (0.063) & (0.023) & (0.195) & (0.070) \\ 
  $\delta_2$ & -0.024 & -0.046 & -0.023 & -0.142 & -0.030 & -0.061 \\ 
   & (0.042) & (0.064) & (0.078) & (0.024) & (0.173) & (0.083) \\ 
  $\delta_3$ & -0.002 & -0.002 & -0.012 & -0.028 & -0.008 & 0.000 \\ 
   & (0.005) & (0.039) & (0.025) & (0.006) & (0.069) & (0.017) \\ 
  \bottomrule
\end{longtable}
\note{\scriptsize This table reports estimates from the two-stage GNR procedure (Appendix~\ref{Gandhi}). The upper panel presents production function parameters: $\gamma$ coefficients from the first-stage share regression (equation~\ref{eq:share_poly}) and $\alpha$ coefficients from the second-stage GMM (equation~\ref{eq:Y_subst}). The lower panel presents Markov process coefficients $\delta$ for each time period (equation~\ref{eq:omega_markov}), where $\delta_0$ is the drift, $\delta_1$ the AR(1) persistence, and $\delta_2$, $\delta_3$ higher-order terms. Bootstrap standard errors (in parentheses) from 100 non-parametric bootstrap replications. Six countries (Germany, Spain, France, Italy, Poland, Romania), 2001--2017 for Spain, France and Italy and 2005--2017 for Germany, Poland and Romania, whose samples begin in 2004. NA indicates insufficient data for that country-period cell.}

    \newpage
    \fontsize{9pt}{11pt}\selectfont
\setlength{\tabcolsep}{4pt}
\begin{longtable}{lcccccc}
    \caption{Machinery, Motor Vehicles and Other Transport Equipment}
    \label{tab:Mach} \tabularnewline
    \caption*{Production Function Estimates} \tabularnewline
  \toprule
 & Germany & Spain & France & Italy & Poland & Romania \\ 
  \midrule
Observations & 16,757 & 82,156 & 53,558 & 190,767 & 7,151 & 9,115 \\ 
  \midrule
  $\gamma_0$ & 0.362 & 0.850 & 0.245 & 0.222 & 0.420 & 0.133 \\ 
   & (0.067) & (0.016) & (0.024) & (0.007) & (0.041) & (0.038) \\ 
  $\gamma_k$ & -0.046 & -0.045 & -0.095 & -0.133 & -0.111 & -0.075 \\ 
   & (0.010) & (0.003) & (0.008) & (0.004) & (0.011) & (0.005) \\ 
  $\gamma_l$ & -0.056 & -0.131 & -0.000 & -0.030 & 0.031 & 0.072 \\ 
   & (0.015) & (0.007) & (0.005) & (0.001) & (0.016) & (0.012) \\ 
  $\gamma_m$ & 0.146 & 0.208 & 0.124 & 0.116 & 0.167 & 0.028 \\ 
   & (0.026) & (0.004) & (0.010) & (0.003) & (0.011) & (0.013) \\ 
  $\gamma_{kk}$ & -0.004 & -0.001 & -0.004 & -0.005 & -0.005 & -0.000 \\ 
   & (0.001) & (0.000) & (0.000) & (0.000) & (0.001) & (0.000) \\ 
  $\gamma_{ll}$ & -0.001 & -0.001 & -0.008 & 0.008 & -0.013 & -0.010 \\ 
   & (0.002) & (0.001) & (0.001) & (0.001) & (0.002) & (0.001) \\ 
  $\gamma_{mm}$ & 0.007 & 0.013 & 0.014 & 0.010 & 0.010 & 0.001 \\ 
   & (0.002) & (0.000) & (0.001) & (0.000) & (0.001) & (0.001) \\ 
  $\gamma_{kl}$ & 0.008 & 0.007 & 0.014 & 0.026 & 0.020 & 0.011 \\ 
   & (0.002) & (0.000) & (0.001) & (0.001) & (0.002) & (0.001) \\ 
  $\gamma_{km}$ & 0.004 & -0.005 & -0.011 & -0.006 & -0.004 & -0.008 \\ 
   & (0.002) & (0.000) & (0.001) & (0.000) & (0.001) & (0.001) \\ 
  $\gamma_{lm}$ & -0.019 & -0.018 & -0.006 & -0.007 & -0.012 & 0.007 \\ 
   & (0.003) & (0.001) & (0.001) & (0.000) & (0.002) & (0.002) \\ 
  \midrule
  $\alpha_k$ & -0.251 & -0.028 & -0.398 & -0.425 & -0.272 & -0.741 \\ 
   & (0.096) & (0.017) & (0.051) & (0.030) & (0.090) & (0.102) \\ 
  $\alpha_l$ & -0.018 & -0.021 & -0.251 & -0.455 & 0.271 & 0.280 \\ 
   & (0.174) & (0.021) & (0.063) & (0.036) & (0.118) & (0.116) \\ 
  $\alpha_{kk}$ & -0.020 & -0.005 & -0.036 & -0.021 & -0.018 & -0.053 \\ 
   & (0.008) & (0.002) & (0.005) & (0.003) & (0.008) & (0.008) \\ 
  $\alpha_{ll}$ & -0.097 & -0.081 & -0.036 & 0.058 & -0.087 & -0.075 \\ 
   & (0.026) & (0.003) & (0.009) & (0.006) & (0.016) & (0.013) \\ 
  $\alpha_{kl}$ & 0.040 & -0.017 & 0.013 & 0.045 & 0.030 & 0.091 \\ 
   & (0.026) & (0.004) & (0.012) & (0.009) & (0.022) & (0.020) \\ 
  \bottomrule
  \newpage
    \caption*{Productivity Process Estimates} \tabularnewline
  \toprule
 & Germany & Spain & France & Italy & Poland & Romania \\ 
  \midrule
  \textbf{2001 - 2002} & & & & & & \\
  $\delta_0$ & NA & 0.010 & 0.097 & 0.110 & NA & NA \\ 
   & (NA) & (0.010) & (0.037) & (0.027) & (NA) & (NA) \\ 
  $\delta_1$ & NA & 0.735 & 0.714 & 0.700 & NA & NA \\ 
   & (NA) & (0.051) & (0.051) & (0.032) & (NA) & (NA) \\ 
  $\delta_2$ & NA & 0.017 & -0.081 & -0.017 & NA & NA \\ 
   & (NA) & (0.093) & (0.027) & (0.018) & (NA) & (NA) \\ 
  $\delta_3$ & NA & 0.001 & -0.012 & -0.002 & NA & NA \\ 
   & (NA) & (0.022) & (0.007) & (0.005) & (NA) & (NA) \\ 
  \textbf{2003 - 2006} & & & & & & \\
  $\delta_0$ & 0.070 & 0.034 & 0.089 & 0.131 & 0.182 & 0.143 \\ 
   & (0.039) & (0.008) & (0.027) & (0.017) & (0.082) & (0.043) \\ 
  $\delta_1$ & 0.897 & 0.793 & 0.797 & 0.808 & 0.762 & 0.839 \\ 
   & (0.074) & (0.027) & (0.048) & (0.023) & (0.230) & (0.064) \\ 
  $\delta_2$ & 0.004 & -0.045 & -0.060 & -0.064 & 0.014 & 0.060 \\ 
   & (0.108) & (0.072) & (0.021) & (0.010) & (0.246) & (0.095) \\ 
  $\delta_3$ & -0.002 & -0.025 & -0.014 & -0.009 & -0.051 & -0.010 \\ 
   & (0.052) & (0.021) & (0.008) & (0.003) & (0.125) & (0.081) \\ 
  \textbf{2007 - 2009} & & & & & & \\
  $\delta_0$ & 0.058 & 0.034 & 0.116 & 0.129 & 0.050 & 0.194 \\ 
   & (0.052) & (0.007) & (0.037) & (0.027) & (0.062) & (0.088) \\ 
  $\delta_1$ & 0.870 & 0.840 & 0.729 & 0.757 & 0.840 & 0.690 \\ 
   & (0.080) & (0.023) & (0.052) & (0.034) & (0.218) & (0.049) \\ 
  $\delta_2$ & -0.030 & -0.171 & -0.064 & -0.043 & 0.268 & -0.003 \\ 
   & (0.082) & (0.054) & (0.023) & (0.009) & (0.437) & (0.065) \\ 
  $\delta_3$ & -0.037 & -0.043 & -0.004 & -0.008 & -0.202 & -0.032 \\ 
   & (0.028) & (0.018) & (0.008) & (0.005) & (0.335) & (0.021) \\ 
  \textbf{2010 - 2013} & & & & & & \\
  $\delta_0$ & 0.070 & 0.029 & 0.127 & 0.151 & 0.051 & 0.102 \\ 
   & (0.064) & (0.006) & (0.032) & (0.021) & (0.043) & (0.032) \\ 
  $\delta_1$ & 0.771 & 0.886 & 0.742 & 0.760 & 1.026 & 0.892 \\ 
   & (0.076) & (0.022) & (0.052) & (0.016) & (0.130) & (0.027) \\ 
  $\delta_2$ & 0.091 & -0.141 & -0.052 & -0.056 & -0.032 & 0.035 \\ 
   & (0.061) & (0.061) & (0.020) & (0.007) & (0.143) & (0.026) \\ 
  $\delta_3$ & 0.000 & -0.070 & -0.003 & -0.005 & -0.083 & -0.034 \\ 
   & (0.017) & (0.051) & (0.006) & (0.002) & (0.095) & (0.007) \\ 
  \textbf{2014 - 2017} & & & & & & \\
  $\delta_0$ & 0.009 & -0.037 & 0.118 & 0.136 & 0.099 & 0.110 \\ 
   & (0.039) & (0.013) & (0.038) & (0.019) & (0.050) & (0.042) \\ 
  $\delta_1$ & 0.991 & 0.662 & 0.764 & 0.783 & 1.028 & 0.841 \\ 
   & (0.102) & (0.019) & (0.069) & (0.014) & (0.150) & (0.033) \\ 
  $\delta_2$ & -0.005 & -0.171 & -0.069 & -0.065 & -0.109 & 0.039 \\ 
   & (0.086) & (0.039) & (0.036) & (0.006) & (0.132) & (0.029) \\ 
  $\delta_3$ & -0.023 & -0.048 & -0.008 & -0.006 & -0.098 & -0.029 \\ 
   & (0.024) & (0.021) & (0.012) & (0.002) & (0.061) & (0.008) \\ 
  \bottomrule
\end{longtable}
\note{\scriptsize This table reports estimates from the two-stage GNR procedure (Appendix~\ref{Gandhi}). The upper panel presents production function parameters: $\gamma$ coefficients from the first-stage share regression (equation~\ref{eq:share_poly}) and $\alpha$ coefficients from the second-stage GMM (equation~\ref{eq:Y_subst}). The lower panel presents Markov process coefficients $\delta$ for each time period (equation~\ref{eq:omega_markov}), where $\delta_0$ is the drift, $\delta_1$ the AR(1) persistence, and $\delta_2$, $\delta_3$ higher-order terms. Bootstrap standard errors (in parentheses) from 100 non-parametric bootstrap replications. Six countries (Germany, Spain, France, Italy, Poland, Romania), 2001--2017 for Spain, France and Italy and 2005--2017 for Germany, Poland and Romania, whose samples begin in 2004. NA indicates insufficient data for that country-period cell.}

    \newpage
    \fontsize{9pt}{11pt}\selectfont
\setlength{\tabcolsep}{4pt}
\begin{longtable}{lcccccc}
    \caption{Furniture and Other Manufacturing}
    \label{tab:Furn} \tabularnewline
    \caption*{Production Function Estimates} \tabularnewline
  \toprule
 & Germany & Spain & France & Italy & Poland & Romania \\ 
  \midrule
Observations & 4,906 & 102,639 & 63,528 & 116,022 & 6,959 & 24,629 \\ 
  \midrule
  $\gamma_0$ & 0.395 & 1.177 & 0.596 & 0.570 & 0.339 & 1.018 \\ 
   & (0.076) & (0.016) & (0.024) & (0.018) & (0.029) & (0.032) \\ 
  $\gamma_k$ & -0.061 & 0.021 & -0.048 & -0.059 & -0.157 & -0.137 \\ 
   & (0.012) & (0.006) & (0.002) & (0.003) & (0.009) & (0.012) \\ 
  $\gamma_l$ & -0.071 & -0.268 & -0.130 & -0.139 & -0.034 & -0.175 \\ 
   & (0.022) & (0.006) & (0.005) & (0.005) & (0.010) & (0.010) \\ 
  $\gamma_m$ & 0.147 & 0.231 & 0.176 & 0.170 & 0.154 & 0.289 \\ 
   & (0.025) & (0.004) & (0.007) & (0.006) & (0.010) & (0.011) \\ 
  $\gamma_{kk}$ & -0.004 & 0.003 & -0.006 & -0.001 & -0.013 & -0.006 \\ 
   & (0.001) & (0.000) & (0.000) & (0.000) & (0.001) & (0.001) \\ 
  $\gamma_{ll}$ & 0.002 & 0.013 & 0.006 & 0.012 & 0.001 & 0.006 \\ 
   & (0.002) & (0.001) & (0.000) & (0.001) & (0.001) & (0.001) \\ 
  $\gamma_{mm}$ & 0.011 & 0.012 & 0.012 & 0.013 & 0.014 & 0.015 \\ 
   & (0.002) & (0.000) & (0.000) & (0.000) & (0.001) & (0.001) \\ 
  $\gamma_{kl}$ & 0.009 & -0.007 & 0.004 & 0.003 & 0.019 & 0.026 \\ 
   & (0.002) & (0.001) & (0.000) & (0.000) & (0.001) & (0.003) \\ 
  $\gamma_{km}$ & -0.002 & -0.004 & -0.003 & -0.007 & -0.004 & -0.006 \\ 
   & (0.002) & (0.001) & (0.000) & (0.000) & (0.001) & (0.001) \\ 
  $\gamma_{lm}$ & -0.017 & -0.026 & -0.020 & -0.013 & -0.005 & -0.033 \\ 
   & (0.003) & (0.001) & (0.001) & (0.001) & (0.002) & (0.002) \\ 
  \midrule
  $\alpha_k$ & -0.416 & 0.044 & -0.241 & -0.108 & -0.553 & 0.076 \\ 
   & (0.080) & (0.008) & (0.027) & (0.011) & (0.080) & (0.052) \\ 
  $\alpha_l$ & 0.035 & 0.091 & -0.140 & -0.463 & -0.168 & 0.221 \\ 
   & (0.134) & (0.017) & (0.038) & (0.023) & (0.079) & (0.082) \\ 
  $\alpha_{kk}$ & -0.036 & 0.001 & -0.036 & -0.010 & -0.052 & -0.000 \\ 
   & (0.008) & (0.001) & (0.003) & (0.001) & (0.008) & (0.004) \\ 
  $\alpha_{ll}$ & -0.082 & -0.109 & -0.083 & 0.008 & 0.001 & -0.091 \\ 
   & (0.017) & (0.004) & (0.006) & (0.004) & (0.009) & (0.012) \\ 
  $\alpha_{kl}$ & 0.049 & -0.033 & 0.003 & -0.042 & 0.035 & -0.037 \\ 
   & (0.018) & (0.003) & (0.007) & (0.003) & (0.016) & (0.012) \\ 
  \bottomrule
  \newpage
    \caption*{Productivity Process Estimates} \tabularnewline
  \toprule
 & Germany & Spain & France & Italy & Poland & Romania \\ 
  \midrule
  \textbf{2001 - 2002} & & & & & & \\
  $\delta_0$ & NA & -0.020 & -0.067 & -0.066 & NA & NA \\ 
   & (NA) & (0.006) & (0.027) & (0.012) & (NA) & (NA) \\ 
  $\delta_1$ & NA & 0.768 & 0.648 & 0.675 & NA & NA \\ 
   & (NA) & (0.037) & (0.048) & (0.051) & (NA) & (NA) \\ 
  $\delta_2$ & NA & -0.005 & -0.121 & -0.120 & NA & NA \\ 
   & (NA) & (0.086) & (0.032) & (0.058) & (NA) & (NA) \\ 
  $\delta_3$ & NA & -0.058 & -0.026 & -0.044 & NA & NA \\ 
   & (NA) & (0.043) & (0.009) & (0.017) & (NA) & (NA) \\ 
  \textbf{2003 - 2006} & & & & & & \\
  $\delta_0$ & 0.210 & 0.004 & -0.053 & -0.036 & 0.128 & -0.035 \\ 
   & (0.293) & (0.005) & (0.022) & (0.008) & (0.086) & (0.118) \\ 
  $\delta_1$ & 0.324 & 0.777 & 0.677 & 0.752 & 0.848 & 0.776 \\ 
   & (0.971) & (0.029) & (0.027) & (0.036) & (0.218) & (0.364) \\ 
  $\delta_2$ & 0.811 & -0.036 & -0.133 & -0.055 & -0.035 & -0.100 \\ 
   & (1.238) & (0.068) & (0.022) & (0.044) & (0.248) & (0.418) \\ 
  $\delta_3$ & -0.362 & -0.070 & -0.027 & -0.023 & -0.059 & -0.058 \\ 
   & (0.737) & (0.042) & (0.005) & (0.013) & (0.168) & (0.165) \\ 
  \textbf{2007 - 2009} & & & & & & \\
  $\delta_0$ & 0.091 & 0.013 & -0.027 & -0.040 & 0.303 & -0.005 \\ 
   & (0.046) & (0.005) & (0.016) & (0.009) & (0.110) & (0.068) \\ 
  $\delta_1$ & 0.716 & 0.817 & 0.765 & 0.754 & 0.511 & 0.512 \\ 
   & (0.157) & (0.024) & (0.028) & (0.031) & (0.161) & (0.111) \\ 
  $\delta_2$ & 0.136 & -0.095 & -0.104 & -0.045 & 0.009 & -0.114 \\ 
   & (0.205) & (0.084) & (0.034) & (0.054) & (0.133) & (0.094) \\ 
  $\delta_3$ & 0.036 & -0.071 & -0.027 & -0.027 & 0.005 & -0.037 \\ 
   & (0.169) & (0.038) & (0.013) & (0.016) & (0.061) & (0.037) \\ 
  \textbf{2010 - 2013} & & & & & & \\
  $\delta_0$ & 0.073 & 0.006 & -0.022 & -0.035 & 0.202 & -0.154 \\ 
   & (0.042) & (0.005) & (0.017) & (0.011) & (0.079) & (0.078) \\ 
  $\delta_1$ & 0.989 & 0.793 & 0.752 & 0.683 & 0.820 & 0.385 \\ 
   & (0.099) & (0.021) & (0.031) & (0.023) & (0.105) & (0.097) \\ 
  $\delta_2$ & -0.075 & -0.022 & -0.081 & -0.105 & -0.170 & -0.112 \\ 
   & (0.127) & (0.049) & (0.036) & (0.022) & (0.064) & (0.104) \\ 
  $\delta_3$ & -0.036 & -0.058 & -0.033 & -0.025 & -0.050 & -0.005 \\ 
   & (0.063) & (0.028) & (0.013) & (0.006) & (0.024) & (0.027) \\ 
  \textbf{2014 - 2017} & & & & & & \\
  $\delta_0$ & 0.066 & -0.055 & -0.026 & -0.045 & 0.199 & -0.111 \\ 
   & (0.057) & (0.008) & (0.018) & (0.012) & (0.056) & (0.080) \\ 
  $\delta_1$ & 1.007 & 0.643 & 0.751 & 0.660 & 0.709 & 0.426 \\ 
   & (0.167) & (0.016) & (0.038) & (0.018) & (0.082) & (0.100) \\ 
  $\delta_2$ & -0.081 & -0.155 & -0.101 & -0.117 & -0.052 & -0.090 \\ 
   & (0.194) & (0.036) & (0.020) & (0.017) & (0.046) & (0.088) \\ 
  $\delta_3$ & -0.026 & -0.052 & -0.017 & -0.022 & -0.004 & -0.005 \\ 
   & (0.140) & (0.023) & (0.006) & (0.005) & (0.024) & (0.024) \\ 
  \bottomrule
\end{longtable}
\note{\scriptsize This table reports estimates from the two-stage GNR procedure (Appendix~\ref{Gandhi}). The upper panel presents production function parameters: $\gamma$ coefficients from the first-stage share regression (equation~\ref{eq:share_poly}) and $\alpha$ coefficients from the second-stage GMM (equation~\ref{eq:Y_subst}). The lower panel presents Markov process coefficients $\delta$ for each time period (equation~\ref{eq:omega_markov}), where $\delta_0$ is the drift, $\delta_1$ the AR(1) persistence, and $\delta_2$, $\delta_3$ higher-order terms. Bootstrap standard errors (in parentheses) from 100 non-parametric bootstrap replications. Six countries (Germany, Spain, France, Italy, Poland, Romania), 2001--2017 for Spain, France and Italy and 2005--2017 for Germany, Poland and Romania, whose samples begin in 2004. NA indicates insufficient data for that country-period cell.}
}
\normalsize
\clearpage
\section{Country-Level Dispersion Series}
\label{app:country_disp}

\begin{figure}[p]
    \caption{TFPR Dispersion and Mean Evolution}
    \label{fig:prod}
        \centering
    \begin{subfigure}[h]{0.48\textwidth}
        \centering
        \includegraphics[width=\textwidth]{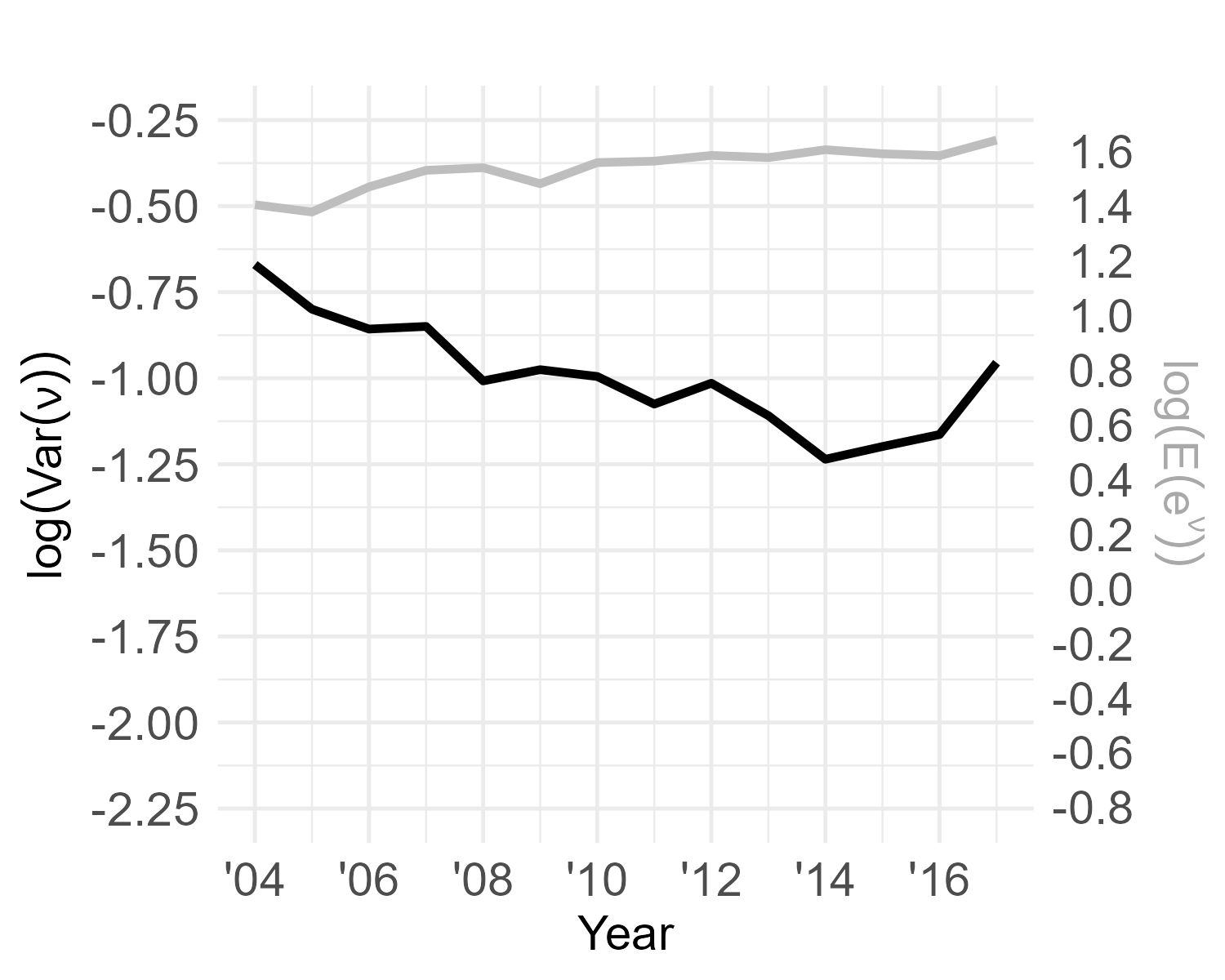}
        \caption{Germany}
        \label{fig:DEU_prod}
    \end{subfigure}
    \hfill
    \begin{subfigure}[h]{0.48\textwidth}
        \centering
        \includegraphics[width=\textwidth]{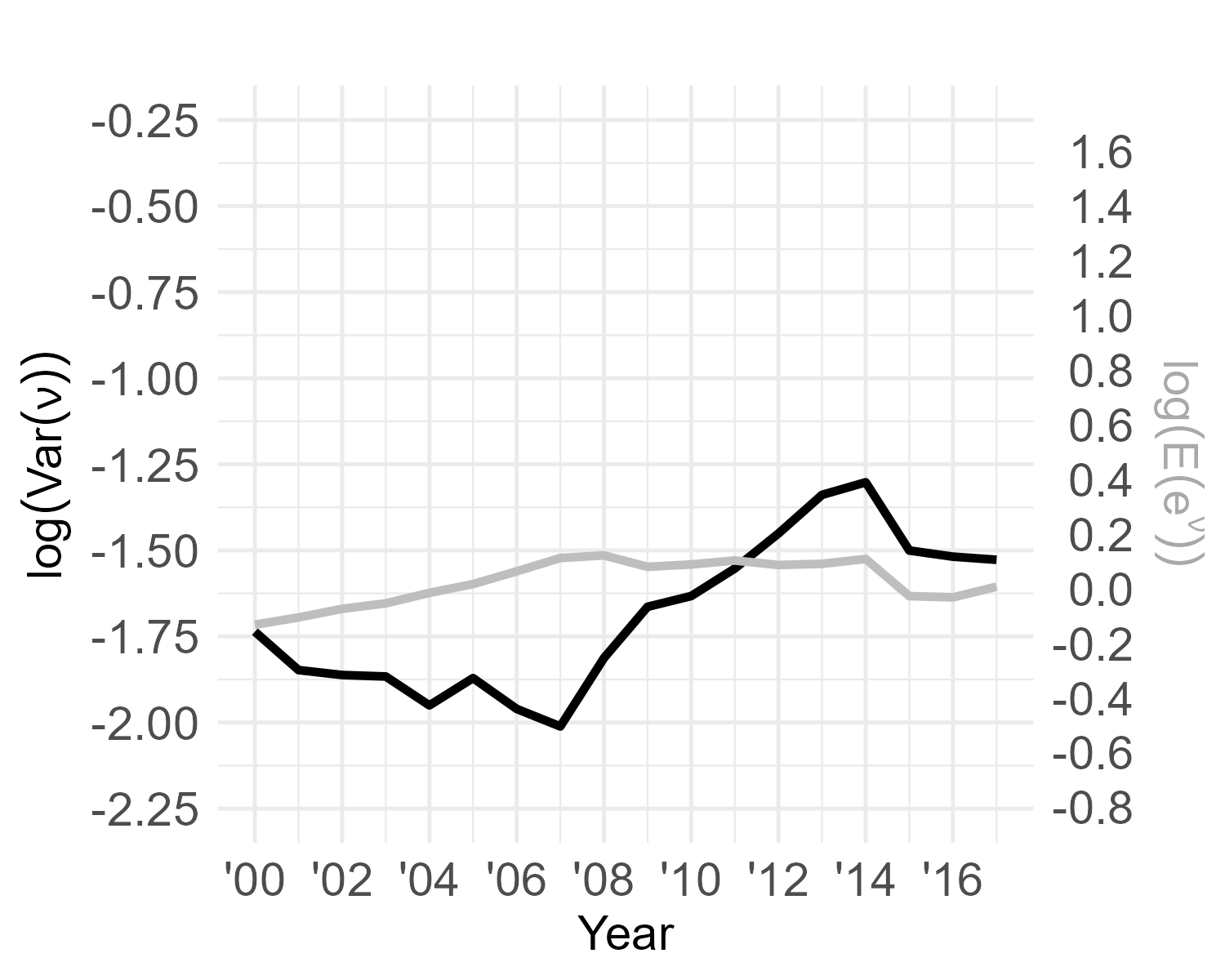}
        \caption{Spain}
        \label{fig:ESP_prod}
    \end{subfigure}

    \vspace{-10pt}
    \begin{subfigure}[h]{0.48\textwidth}
        \centering
        \includegraphics[width=\textwidth]{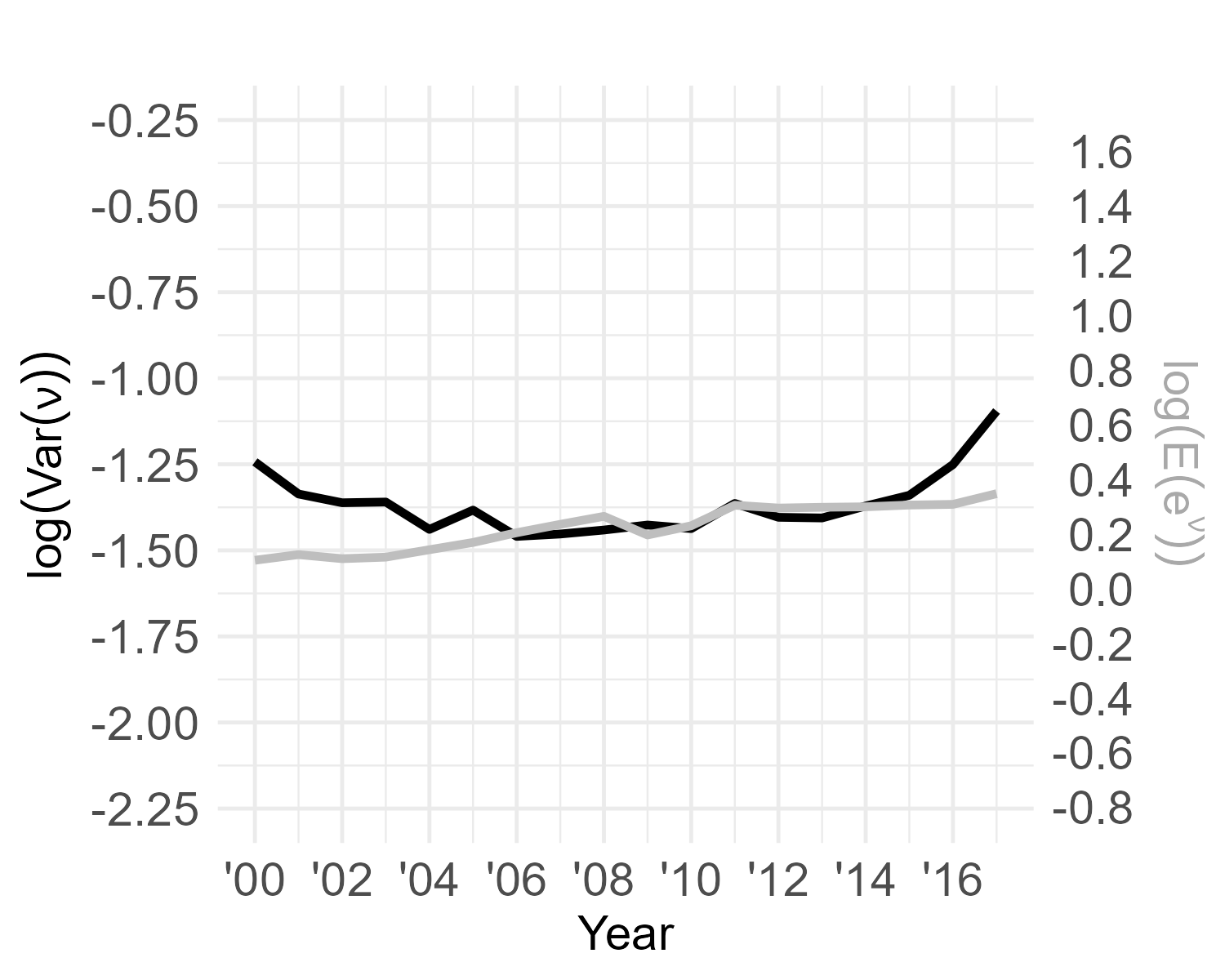}
        \caption{France}
        \label{fig:FRA_prod}
    \end{subfigure}
        \hfill
    \begin{subfigure}[h]{0.48\textwidth}
        \centering
        \includegraphics[width=\textwidth]{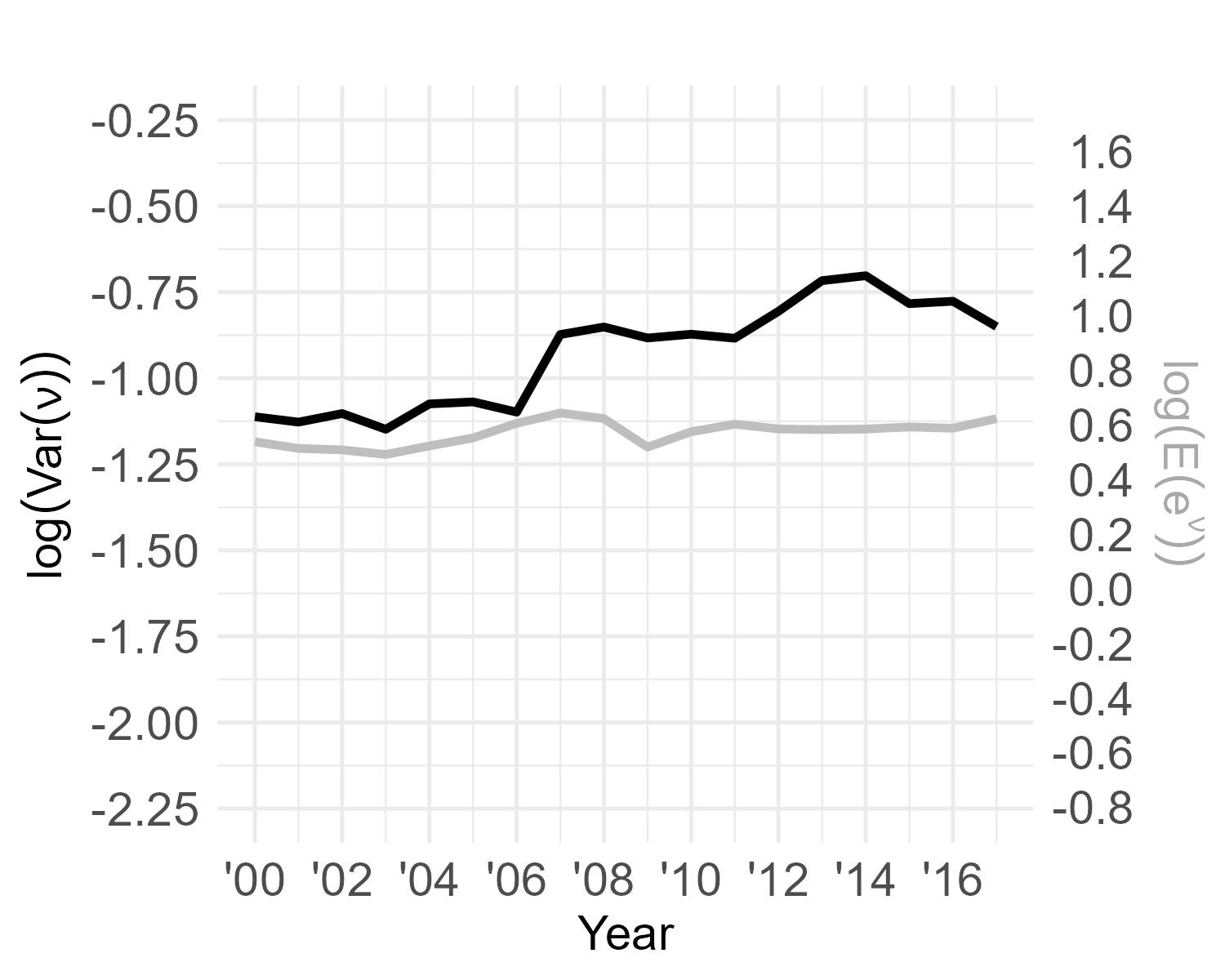}
        \caption{Italy}
        \label{fig:ITA_prod}
    \end{subfigure}

    \vspace{-10pt}
    \begin{subfigure}[h]{0.48\textwidth}
        \centering
        \includegraphics[width=\textwidth]{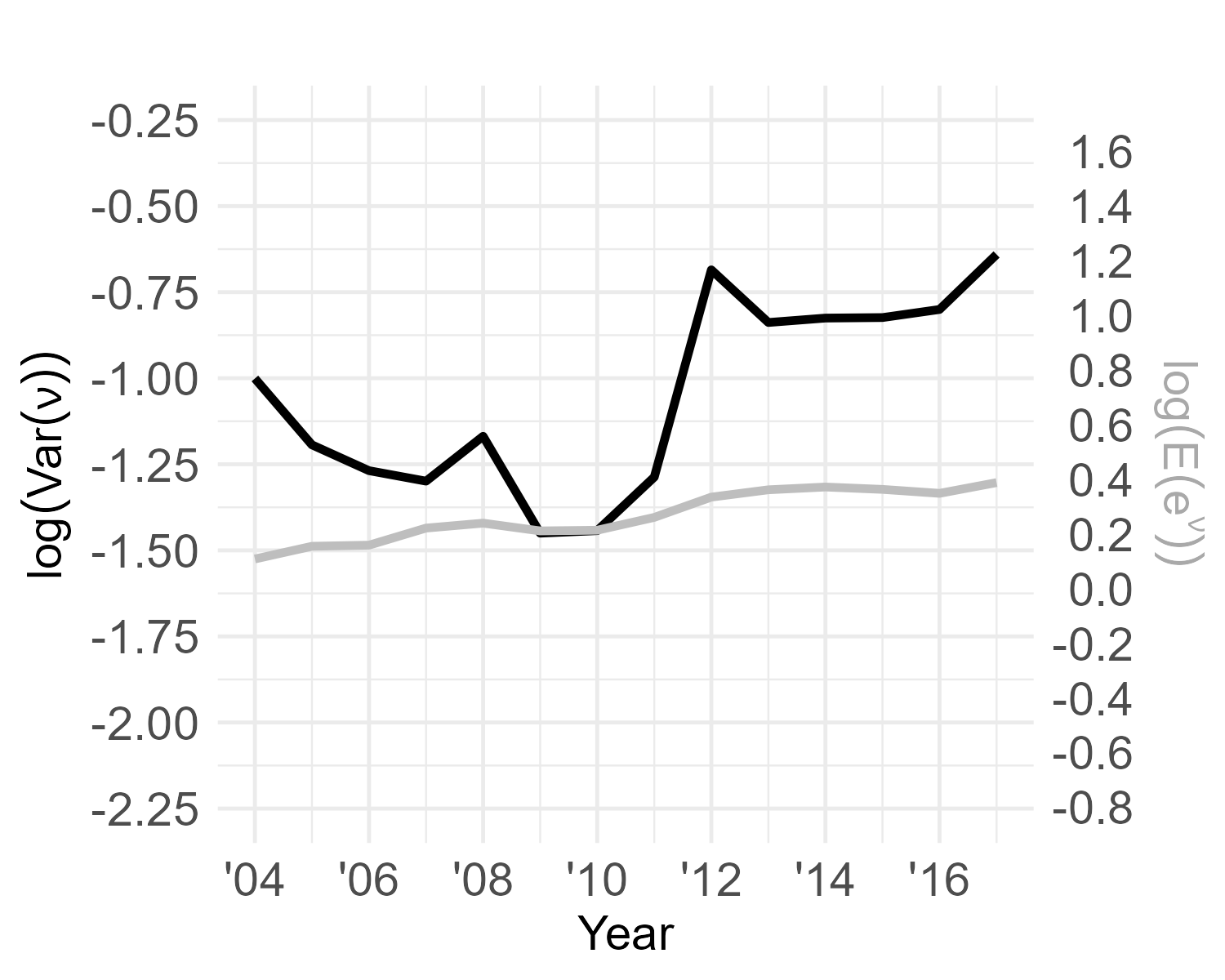}
        \caption{Poland}
        \label{fig:POL_prod}
    \end{subfigure}
    \hfill
    \begin{subfigure}[h]{0.48\textwidth}
        \centering
        \includegraphics[width=\textwidth]{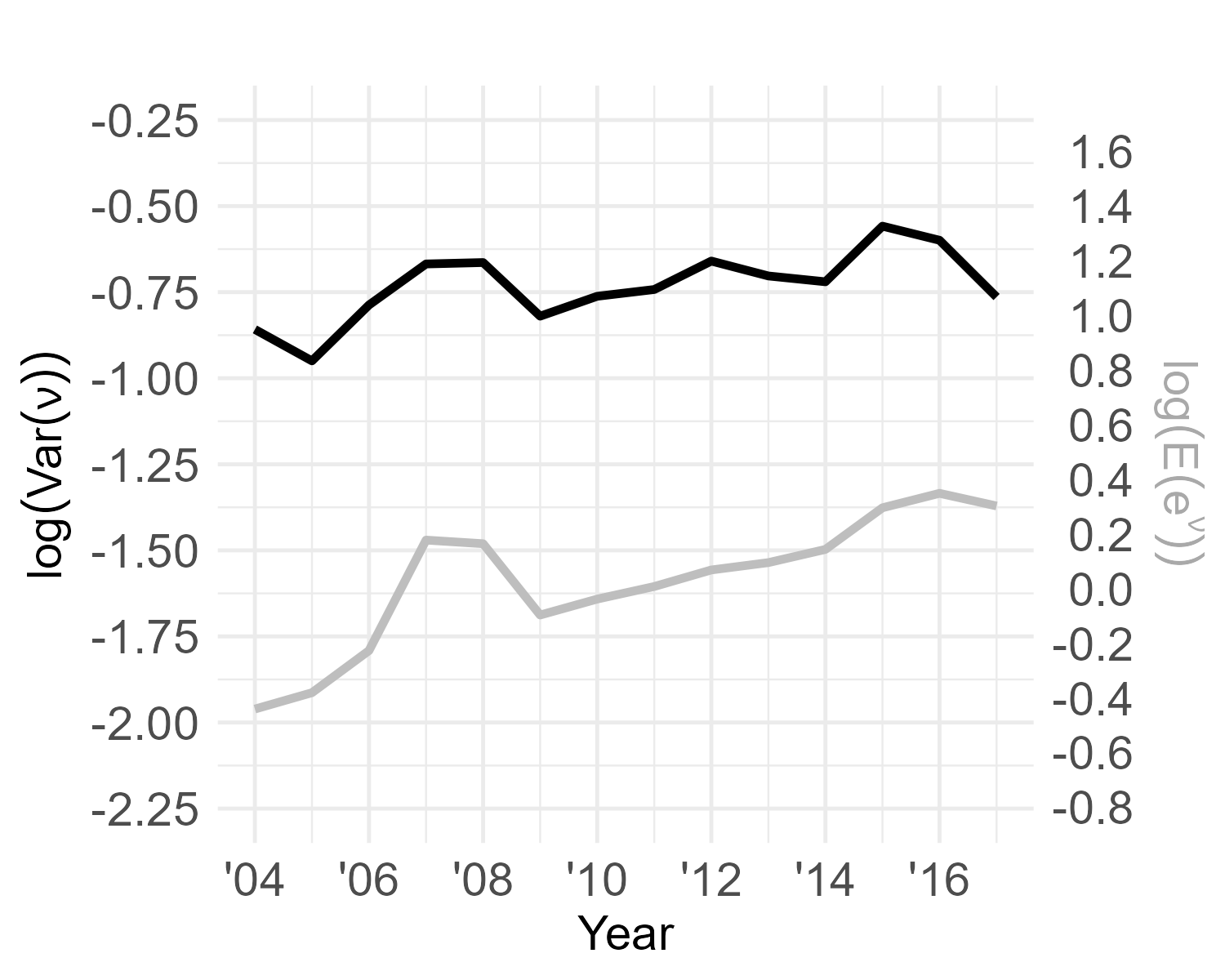}
        \caption{Romania}
        \label{fig:ROM_prod}
    \end{subfigure}
    \note{\scriptsize The solid black line in Figure \ref{fig:prod} displays the evolution of the (log) aggregate variance for TFPR, $\nu$, by country: Germany (Fig. \ref{fig:DEU_prod}), Spain (Fig. \ref{fig:ESP_prod}), France (Fig. \ref{fig:FRA_prod}), Italy (Fig. \ref{fig:ITA_prod}), Poland (Fig. \ref{fig:POL_prod}), and Romania (Fig. \ref{fig:ROM_prod}). The solid gray line shows the (log) average value of TFPR \textit{in levels}, reported on the secondary y-axis. I aggregate variances and means at the country level by taking a weighted average of the industry-specific variances and means, using average total manufacturing revenue shares for each industry as weights. Sample period: 2000--2017 (2004--2017 for Germany, Poland, and Romania).}
    
\end{figure}

\begin{figure}[p]
    \caption{Input Gap Dispersion Evolution}
    \label{fig:MRP}
    \centering
    \begin{subfigure}[h]{0.48\textwidth}
        \centering
        \includegraphics[width=\textwidth]{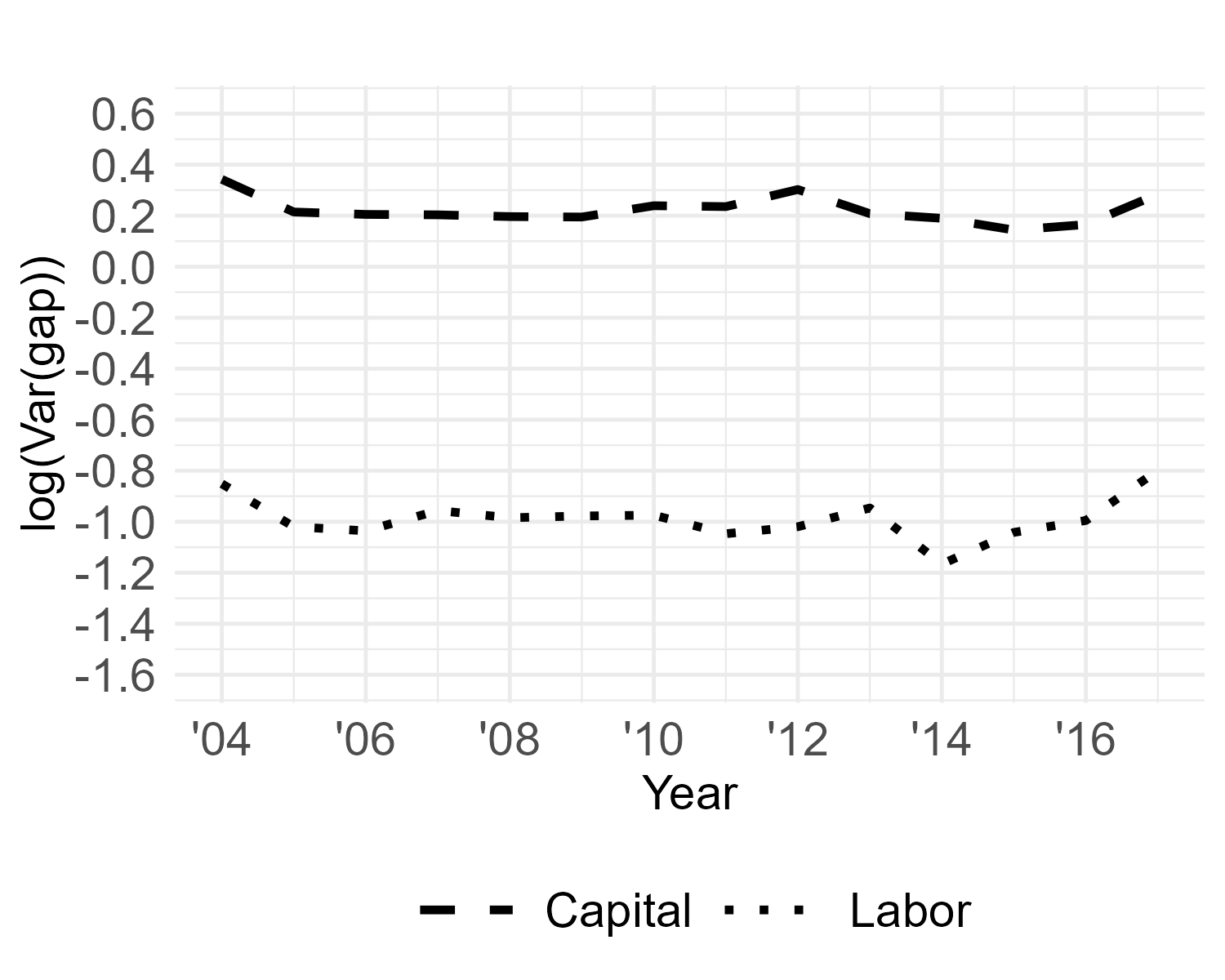}
        \caption{Germany}
        \label{fig:DEU_mrp}
    \end{subfigure}
    \hfill
    \begin{subfigure}[h]{0.48\textwidth}
        \centering
        \includegraphics[width=\textwidth]{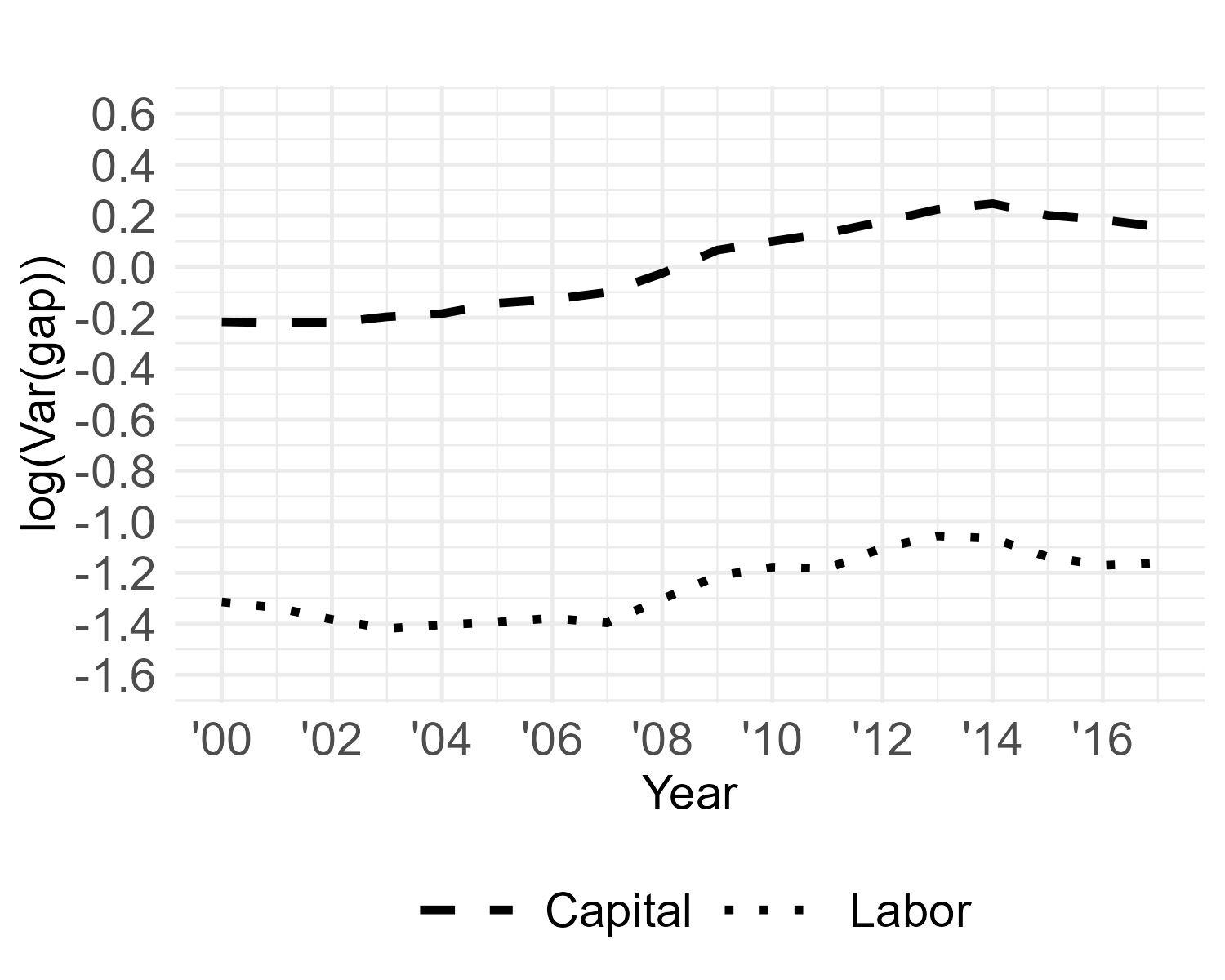}
        \caption{Spain}
        \label{fig:ESP_mrp}
    \end{subfigure}

    \vspace{-10pt}
    \begin{subfigure}[h]{0.48\textwidth}
        \centering
        \includegraphics[width=\textwidth]{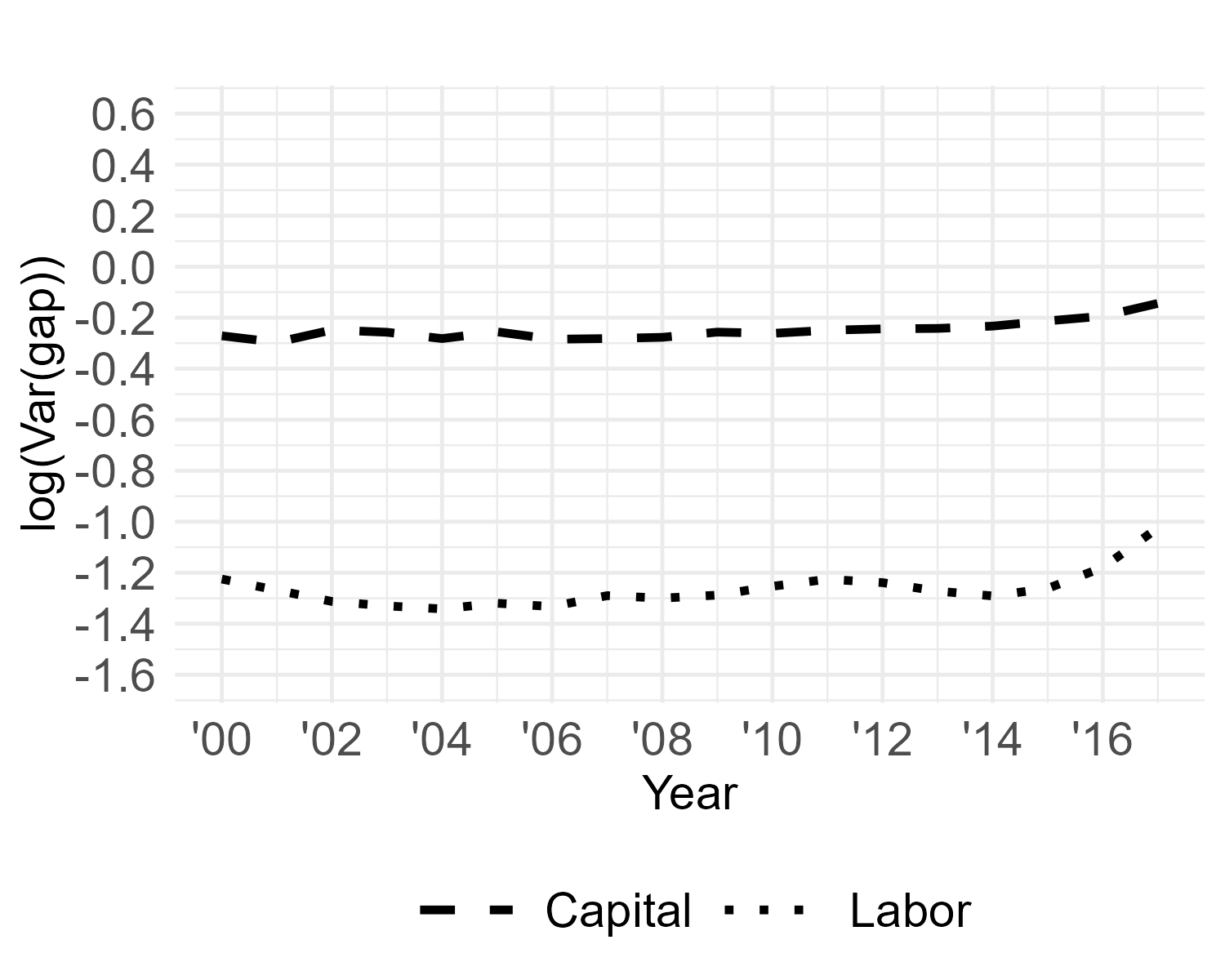}
        \caption{France}
        \label{fig:FRA_mrp}
    \end{subfigure}
        \hfill
    \begin{subfigure}[h]{0.48\textwidth}
        \centering
        \includegraphics[width=\textwidth]{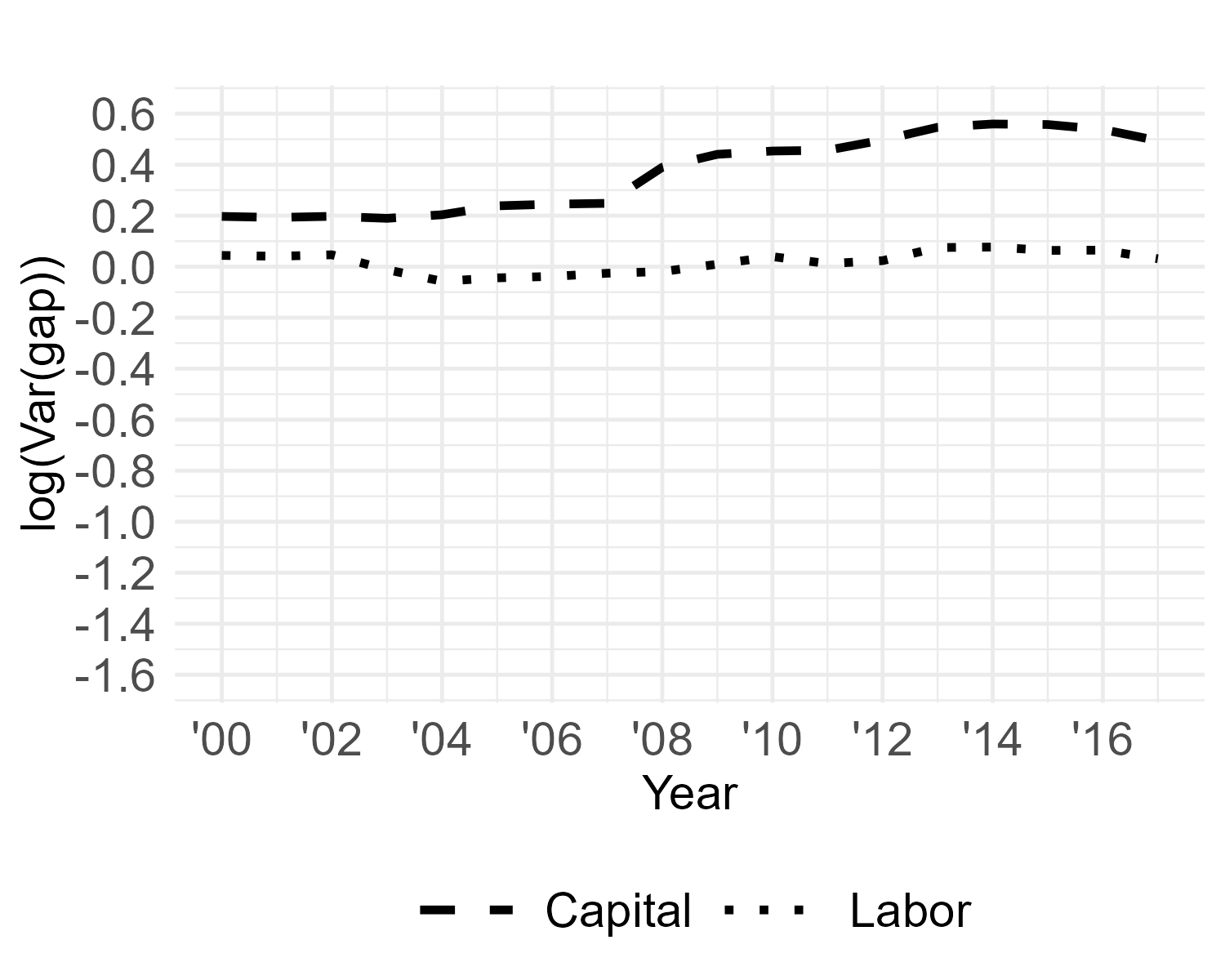}
        \caption{Italy}
        \label{fig:ITA_mrp}
    \end{subfigure}

    \vspace{-10pt}
    \begin{subfigure}[h]{0.48\textwidth}
        \centering
        \includegraphics[width=\textwidth]{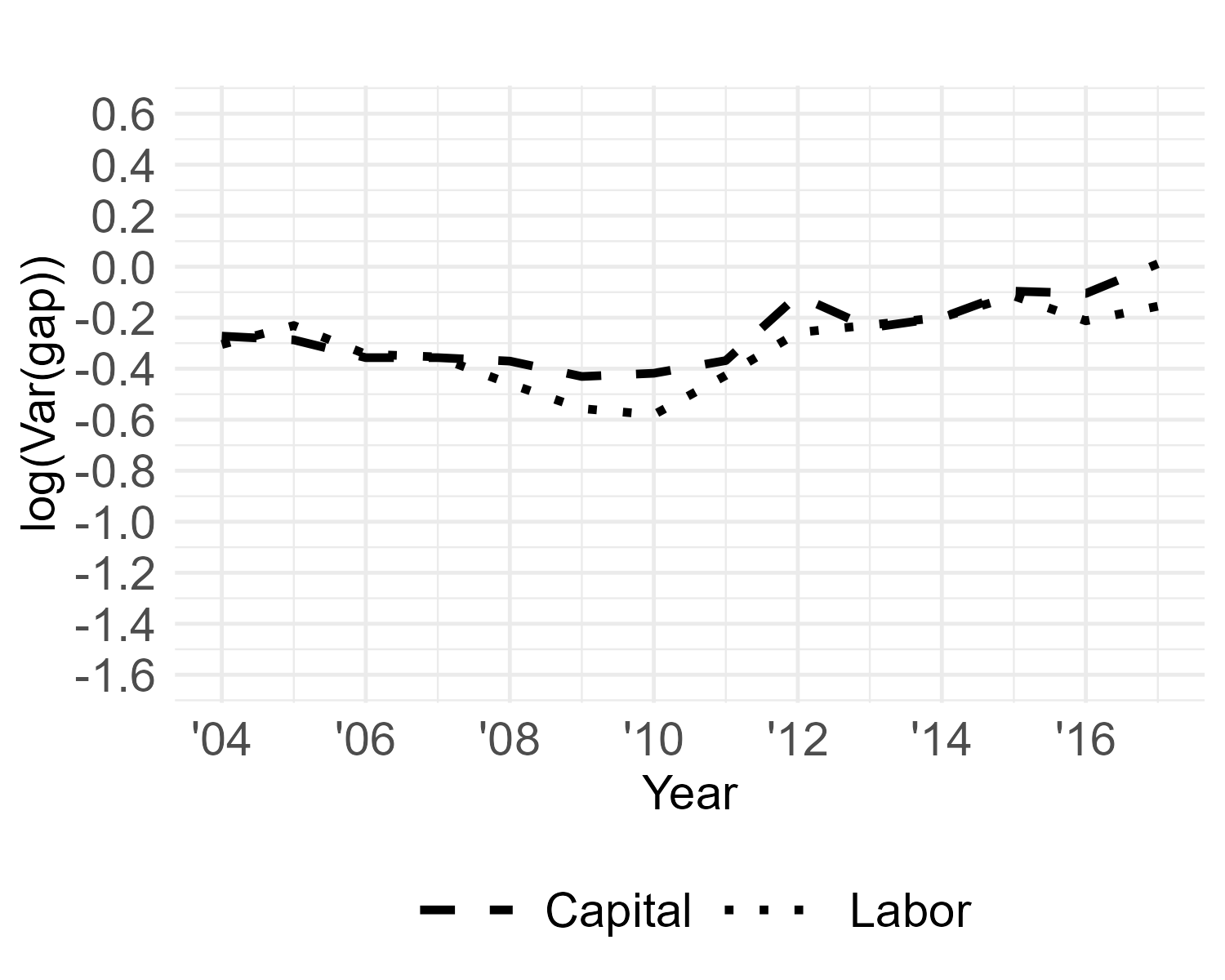}
        \caption{Poland}
        \label{fig:POL_mrp}
    \end{subfigure}
    \hfill
    \begin{subfigure}[h]{0.48\textwidth}
        \centering
        \includegraphics[width=\textwidth]{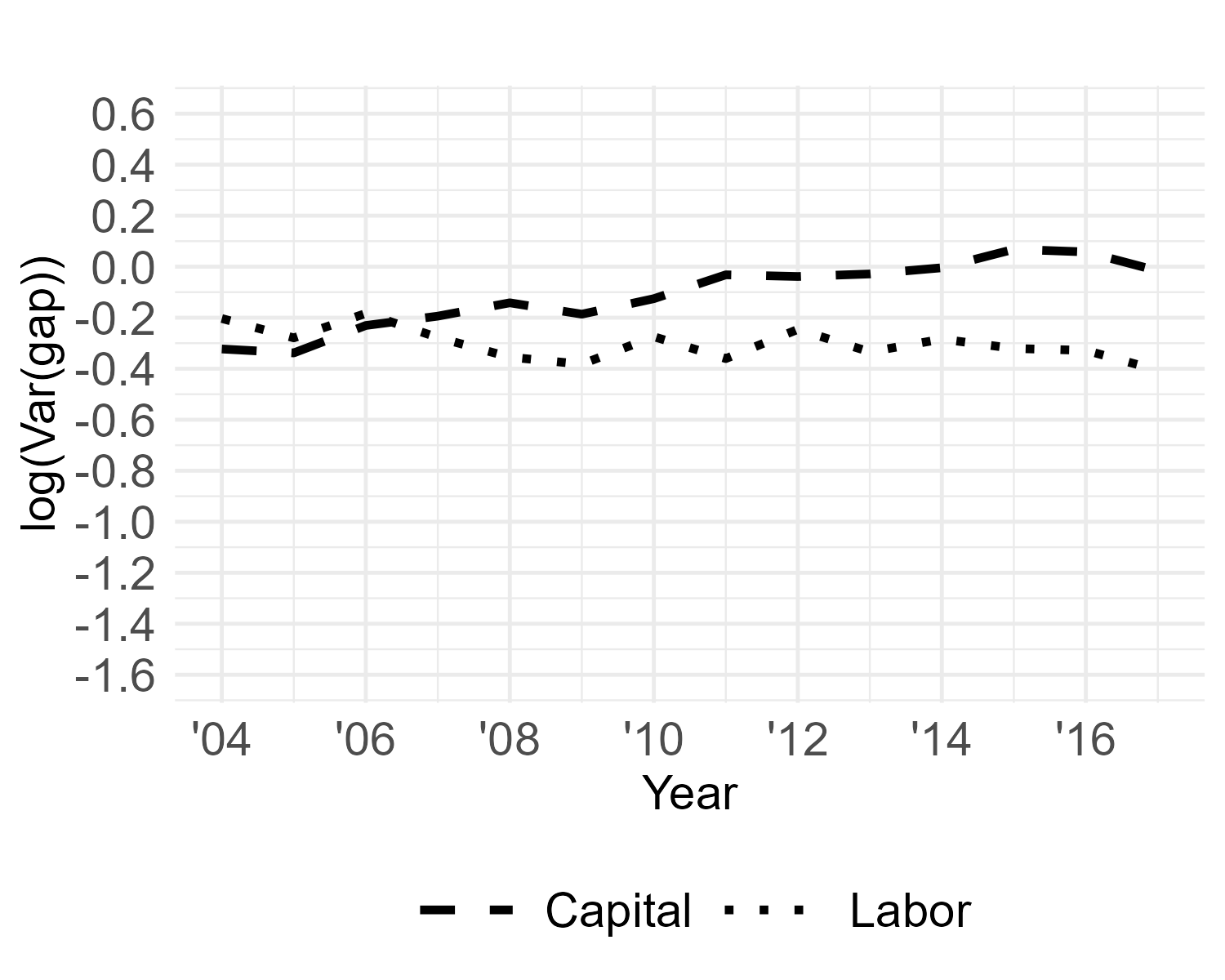}
        \caption{Romania}
        \label{fig:ROM_mrp}
    \end{subfigure}
    \note{\scriptsize Figure \ref{fig:MRP} displays the evolution of the (log) aggregate variance of the input gaps $\mathrm{mrp}^X_{jt} - \log P^X_{jt}$ by country: Germany (Fig. \ref{fig:DEU_mrp}), Spain (Fig. \ref{fig:ESP_mrp}), France (Fig. \ref{fig:FRA_mrp}), Italy (Fig. \ref{fig:ITA_mrp}), Poland (Fig. \ref{fig:POL_mrp}), and Romania (Fig. \ref{fig:ROM_mrp}). The inputs shown are capital (dashed black line) and labor (dotted black line). I aggregate the log variances at the country level by taking a weighted average of the industry-specific variances, using the average shares in total manufacturing revenue for each industry as weights. Sample period: 2000--2017 (2004--2017 for Germany, Poland, and Romania).}
\end{figure}

This appendix reports, country by country, the log aggregate variance series that Figure~\ref{fig:pooled_disp} indexes and pools. Figure~\ref{fig:prod} plots the log aggregate variance of TFPR together with the log of its mean in levels, and Figure~\ref{fig:MRP} plots the log aggregate variance of the capital and labor gaps. Both use the industry weights of Section~\ref{sec:disp_evolution}.

The variance of the labor gap ranges from 26 to 113 percent of the variance of the capital gap across country-years, exceeding it only in Poland and Romania. Within each country, input gap dispersion for capital and labor tends to move concurrently with TFPR dispersion, though with magnitudes that vary across countries and inputs. As detailed in Section~\ref{mechanism}, this co-movement is consistent with the micro-level relationship in which firm-specific TFPR generates variation in the input gaps.

For instance, in Spain and Italy rising TFPR dispersion coincided with the Great Recession;\footnote{Comparable countercyclical patterns are documented for U.S. manufacturing by \cite{kehrig2015cyclical}, who finds that cross-firm TFP variance rises in recessions through both compositional shifts and within-firm productivity changes.} in France the visible TFPR rise is concentrated after 2014, while Romania's TFPR variance fluctuates around a flat trend. Capital gap variance displays an overall increasing trajectory in every country except Germany, with particularly sharp increases in Spain and Italy during the Great Recession.\footnote{This is consistent with \cite{gopinath2017capital}, who document a significant increase in the dispersion of the return to capital among Spanish manufacturing firms between 1999 and 2012, with similar patterns in Italy and Portugal but not in Germany, France, or Norway. They attribute this rise to size-dependent financial frictions through which the post-euro decline in real interest rates channeled capital toward firms with high net worth rather than high productivity. The mechanism documented in this paper may operate alongside financial frictions; the structural model abstracts from financial constraints to isolate the role of productivity timing, and quantifying the two would require a specification that explicitly includes both mechanisms.}

The labor gap variance is less stable: in Spain it rises through the Great Recession and the European debt crisis, in Romania it drifts down over the sample, and in Germany and France it rises sharply in 2017 after a flat stretch. Poland diverges slightly, as its sharp rises in TFPR and input gap dispersion occurred during the European debt crisis rather than the Great Recession. The log of mean TFPR \textit{in levels} falls in 2009 in every country, by 0.03 to 0.10, and by 0.26 in Romania; it recovers, fully or nearly, everywhere, and in Spain falls again after 2014.

\section{The Decomposition by Period}\label{app:shares_period}
Table~\ref{tab:gapshares_period} reports the decomposition of Table~\ref{tab:gapshares} by period of the productivity process, pooled over the six countries. For capital the share of the ex-post shock rises from 11.6 percent in 2001--2002 to 15.6 in the Great Recession and 18.8 from the European debt crisis on, while the share of expected productivity falls from 9.4 to 7.5 percent and that of the ex-ante shock stays at 3.0 to 3.7; the two shocks together account for 14.9 percent of capital gap dispersion in 2001--2002 and 22.1 in 2014--2017. For labor the share of expected productivity falls from 15.3 to 11.6 percent and the two shocks together stay between 4.4 and 5.9 percent.
\begin{table}[!h]
    \centering
    \caption{Share of Input Gap Dispersion Accounted For by Productivity and by Its Components, by Period}
    \label{tab:gapshares_period}
{\setlength{\tabcolsep}{4pt}%
\begin{adjustbox}{max width=\textwidth}
\begin{tabular}{lrccccc}
\toprule
 & Firm-years & TFPR & $g(\omega_{-1})$ & $\eta_{jt}$ & $\varepsilon_{jt}$ & $\eta_{jt} + \varepsilon_{jt}$ \\
\midrule
\multicolumn{7}{l}{\textit{Panel A: capital gap}} \\
  Beginning of the millennium (2001--2002) & 269,678 & 24.1 & 9.4 & 3.3 & 11.6 & 14.9 \\
   & & (0.7) & (0.6) & (0.1) & (0.4) & (0.5) \\
  Pre-crisis (2003--2006) & 596,636 & 25.2 & 9.4 & 3.0 & 12.9 & 15.8 \\
   & & (0.7) & (0.5) & (0.1) & (0.4) & (0.4) \\
  Great Recession (2007--2009) & 489,108 & 27.9 & 8.7 & 3.7 & 15.6 & 19.3 \\
   & & (0.6) & (0.4) & (0.1) & (0.4) & (0.5) \\
  European debt crisis (2010--2013) & 679,386 & 31.1 & 8.9 & 3.4 & 18.8 & 22.2 \\
   & & (0.7) & (0.4) & (0.1) & (0.4) & (0.5) \\
  Post-crisis (2014--2017) & 675,968 & 29.6 & 7.5 & 3.3 & 18.8 & 22.1 \\
   & & (0.7) & (0.3) & (0.1) & (0.4) & (0.5) \\
\midrule
\multicolumn{7}{l}{\textit{Panel B: labor gap}} \\
  Beginning of the millennium (2001--2002) & 269,678 & 14.4 & 15.3 & 4.4 & 1.0 & 5.4 \\
   & & (1.2) & (1.2) & (0.3) & (0.2) & (0.4) \\
  Pre-crisis (2003--2006) & 596,636 & 13.7 & 15.3 & 3.3 & 1.0 & 4.4 \\
   & & (1.1) & (1.2) & (0.2) & (0.2) & (0.3) \\
  Great Recession (2007--2009) & 489,108 & 15.0 & 13.8 & 3.6 & 2.2 & 5.9 \\
   & & (1.1) & (1.1) & (0.2) & (0.3) & (0.4) \\
  European debt crisis (2010--2013) & 679,386 & 14.4 & 13.3 & 2.9 & 2.6 & 5.5 \\
   & & (1.0) & (1.0) & (0.2) & (0.3) & (0.4) \\
  Post-crisis (2014--2017) & 675,968 & 11.7 & 11.6 & 2.3 & 2.1 & 4.4 \\
   & & (0.8) & (0.9) & (0.2) & (0.3) & (0.3) \\
\bottomrule
\end{tabular}
\end{adjustbox}}
\note{\scriptsize Each cell reports the share of within-cell dispersion of the input gap $\mathrm{mrp}^X_{jt} - \log P^X_{jt}$ accounted for by productivity, in percent, on the firm-years of each period of the productivity process, pooled over the six countries: the within-cell $R^2$ of the regression on total TFPR, and the exact regression decomposition of the joint regression on the three components, $\hat{\beta}_k \mathrm{Cov}(x_k, \text{gap}) / \mathrm{Var}(\text{gap})$, as in Table~\ref{tab:gapshares}. The last column sums the shares of the two shocks. Non-parametric bootstrap standard errors in parentheses.}
\end{table}

\clearpage

\section{Pairwise Correlations of the TFPR Components}\label{app:orthogonality}

Table~\ref{tab:orthogonality} reports the weighted average pairwise Pearson correlations between the three TFPR components ($g(\omega_{-1})$, $\eta_{jt}$, and $\varepsilon_{jt}$) across country-industry-year cells. The correlation between expected productivity and the ex-ante shock is economically small (0.012), though statistically significant. The remaining two correlations are negative: $\mathrm{Cor}(g(\omega_{-1}), \varepsilon) = -0.21$ and $\mathrm{Cor}(\eta, \varepsilon) = -0.28$. The two negative correlations hold cell by cell: $\mathrm{Cor}(g(\omega_{-1}), \varepsilon)$ is negative in 90 percent of country-industry-year cells and $\mathrm{Cor}(\eta, \varepsilon)$ in 96 percent\footnote{The two negative correlations have a mechanical candidate: $\varepsilon_{jt}$ is the residual of the share equation and $\omega_{jt}$ is revenue net of the production function and of $\varepsilon_{jt}$, so any error in that residual enters $\omega_{jt}$ and $\eta_{jt}$ with the opposite sign. The polynomial approximation is not the source: with the share equation refitted at degrees 1, 3 and 4 and the second stage held at its estimates, $\mathrm{Cor}(\eta, \varepsilon)$ is $-0.28$ to $-0.32$ for degrees 2 to 4 and $-0.43$ at degree 1, where the fit is worse. Across cells the correlation is more negative where the log materials share is more dispersed and less negative where $\varepsilon_{jt}$ has more variance relative to $\eta_{jt}$.}, whereas $\mathrm{Cor}(g(\omega_{-1}), \eta)$ is positive in 55 percent of cells, consistent with its near-zero average. The decomposition of Table~\ref{tab:gapshares} allocates the covariances these correlations imply to the components through the regression coefficients.

\begin{table}[!h]
\centering
\caption{Pairwise Correlations between the TFPR Components}
\label{tab:orthogonality}
\begin{tabular}{lccc}
\toprule
 & Mean & SE & Frac.\ $> 0$ \\
\midrule
$\mathrm{Cor}(g(\omega_{-1}),\, \eta)$ & 0.012 & (0.002) & 55\% \\
$\mathrm{Cor}(g(\omega_{-1}),\, \varepsilon)$ & -0.206 & (0.009) & 10\% \\
$\mathrm{Cor}(\eta,\, \varepsilon)$ & -0.282 & (0.005) & 4\% \\
\bottomrule
\end{tabular}
\note{\scriptsize Each row reports the weighted average pairwise Pearson correlation between two TFPR components across country-industry-year cells, with weights equal to each industry's share of total manufacturing revenue. Bootstrap standard errors (in parentheses) from 100 non-parametric replications. The last column reports the fraction of country-industry-year cells in which the correlation is positive. Sample period: 2001--2017 for Spain, France and Italy and 2005--2017 for Germany, Poland and Romania; each country loses the first year of its data to the lag.}
\end{table}

\clearpage

\section{Institutional Quality and Productivity Dispersion}\label{app:macro_reg}

This appendix examines whether the dispersion of expected productivity $g(\omega_{-1})$, of the ex-ante shock $\eta_{jt}$, and of the ex-post shock $\varepsilon_{jt}$ varies systematically with the quality of economic institutions, exploiting cross-country variation in institutional indices. I consider nine standardized institutional indices: the six Worldwide Governance Indicators (Government Effectiveness, Regulatory Quality, Rule of Law, Control of Corruption, Voice and Accountability, Political Stability; \citealp{kaufmann2011worldwide}), the World Bank Ease of Doing Business score, the Organisation for Economic Co-operation and Development (OECD) Product Market Regulation indicator, and the World Economic Forum Global Competitiveness Index. 

\begin{figure}[!h]
	\caption{Institutional Quality and Productivity Dispersion: Coefficient Estimates}
	\label{fig:macro_coef}
	\centering
	\begin{subfigure}[h]{0.48\textwidth}
		\centering
		\includegraphics[width=\textwidth]{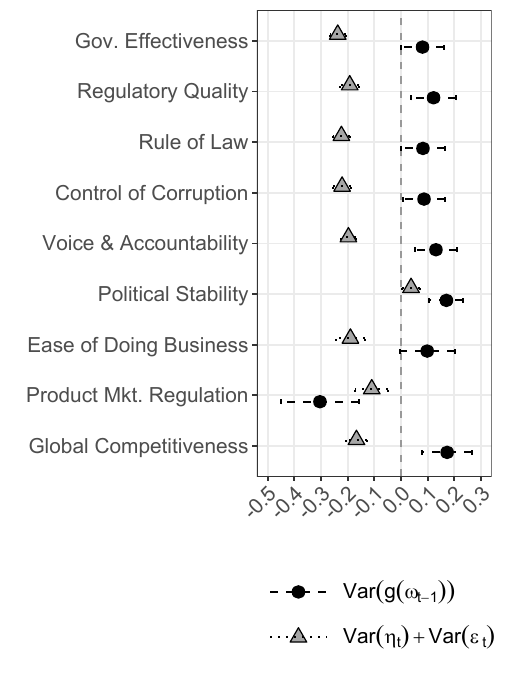}
		\caption{Expected vs.\ shock dispersion}
		\label{fig:macro_coef_2}
	\end{subfigure}
	\hfill
	\begin{subfigure}[h]{0.48\textwidth}
		\centering
		\includegraphics[width=\textwidth]{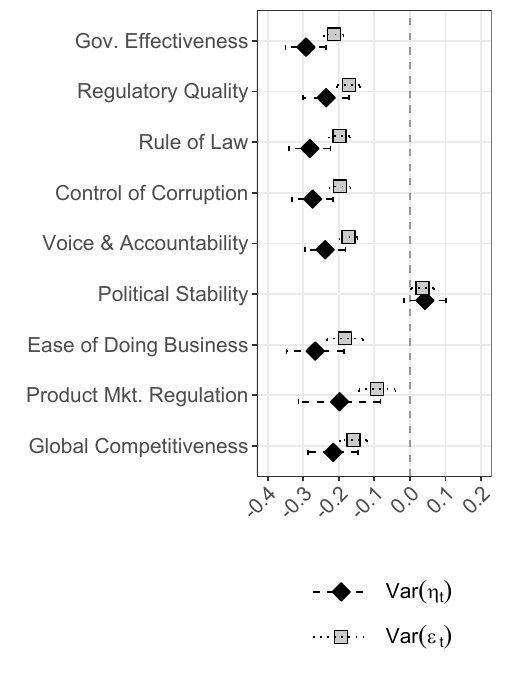}
		\caption{By shock type}
		\label{fig:macro_coef_3}
	\end{subfigure}
	\note{\scriptsize Each point reports the coefficient $\beta$ from a separate regression with dependent variable $\log\mathrm{Var}_{cst}(z)$, where $z$ is a productivity component, on a standardized institutional index, with industry and year fixed effects. Panel (a) contrasts the dispersion of expected productivity $\text{Var}(g(\omega_{-1}))$ with total shock dispersion $\text{Var}(\eta) + \text{Var}(\varepsilon)$. Panel (b) plots coefficients for the ex-ante shock $\eta$ and the ex-post shock $\varepsilon$ separately. Bars denote 95\% confidence intervals from 100 non-parametric bootstrap replications. Six countries: Germany, Spain, France, Italy, Poland, and Romania; country-industry-year cells with at least twenty firms. The Ease of Doing Business score is available from 2010, the Global Competitiveness Index from 2006, and Product Market Regulation through 2013, with a single observation for Romania.}
\end{figure}

For each index, I estimate:
\begin{equation}
\label{model:macro_v2}
\log \mathrm{Var}_{cst}(z) = \iota_s + \iota_t + \beta \, \tilde{I}_{ct} + \xi_{cst}
\end{equation}
where $z \in \{g(\omega_{-1}),\, \eta,\, \varepsilon\}$ is expected productivity, the ex-ante shock, or the ex-post shock, $\mathrm{Var}_{cst}(\cdot)$ is the cross-firm variance in country $c$, industry $s$ at time $t$, $\tilde{I}_{ct}$ is the standardized institutional index for country $c$ at time $t$, $\xi_{cst}$ is an error term, and $\iota_s$ and $\iota_t$ are separate industry and year fixed effects. The regressions use cells with at least twenty firms. Standard errors are obtained from the non-parametric bootstrap of the structural estimation. Figure~\ref{fig:macro_coef} reports the estimated coefficients with 95\% bootstrap confidence intervals for each index-component pair.

Better institutions are associated with lower dispersion of productivity shocks and with higher dispersion of expected productivity, as shown in Panel~(a) of Figure~\ref{fig:macro_coef}. Panel~(b) confirms that this pattern holds for both shock components: a one-standard-deviation improvement in Government Effectiveness is associated with approximately 25\% and 19\% reductions in the variance of $\eta_{jt}$ and $\varepsilon_{jt}$, respectively, with comparable magnitudes for Rule of Law, Control of Corruption, and the Global Competitiveness Index. On the dispersion of expected productivity the coefficients run the other way: a one-standard-deviation improvement in Government Effectiveness is associated with an 8\% higher variance of $g(\omega_{-1})$ ($\beta = 0.081$), and Regulatory Quality, Voice and Accountability, and the Global Competitiveness Index carry larger coefficients ($\beta = 0.122$, $0.131$, and $0.173$). Two indices stand out. Product Market Regulation is the only index with a negative coefficient on the dispersion of expected productivity ($\beta = -0.304$). Political Stability is the only Worldwide Governance Indicator with positive coefficients on shock dispersion, and it carries the largest coefficient on the dispersion of expected productivity among the governance indicators ($\beta = 0.170$).

These patterns are suggestive rather than causal. The analysis exploits variation across only six countries, and the regressions do not control for other country-level determinants (such as financial development or trade openness) that may correlate with both institutional quality and productivity dispersion. With that caveat, the association of stronger governance and higher competitiveness with lower shock dispersion and with higher dispersion of expected productivity is consistent with the interpretation that institutions may operate by reducing the uncertainty firms face while letting productivity differences across firms persist.

\clearpage

\section{Imperfect Competition Robustness}\label{app:ic}

This appendix relaxes Assumption~\ref{assp} (output price-taking behavior) and re-estimates the production function under imperfect competition in the output market, following the procedure in Appendix O6-4 of \cite{gandhi2020identification}. The objective is to assess whether the TFPR and input gap dispersion patterns documented in Section~\ref{s:results} are robust to allowing firms to exercise output market power.

\subsection{Modified Framework}

Under imperfect competition, each firm $j$ faces a downward-sloping demand curve. I parameterize demand using a constant-elasticity structure with a time-varying elasticity parameter $\theta^d_t$:
\begin{equation}\label{eq:ic_demand}
Q_{jt} = D_t \, P_{jt}^{-\theta^d_t}
\end{equation}
where $Q_{jt}$ is physical output, $P_{jt}$ is the firm's output price, $D_t$ is an aggregate demand shifter common to all firms in a given industry-year cell, and $\theta^d_t > 1$ is the price elasticity of demand. The key departure from the baseline model is that firms internalize their effect on price when choosing input allocations. Under perfect competition ($\theta^d_t \to \infty$), the demand curve becomes horizontal and the framework reduces to the baseline in Section~\ref{s:theory}.

Revenue is $R_{jt} = P_{jt} Q_{jt}$. Using the inverse demand function $P_{jt} = (Q_{jt}/D_t)^{-1/\theta^d_t}$, log revenue can be written as
\begin{equation}
y_{jt} = \frac{1}{\theta^d_t} y^D_t + \frac{\theta^d_t - 1}{\theta^d_t} \left[ f(k_{jt}, l_{jt}, m_{jt}) + \omega_{jt} + \varepsilon_{jt} \right]
\end{equation}
where $y^D_t = \log D_t$ is the log aggregate demand shifter. The coefficient $(\theta^d_t - 1)/\theta^d_t < 1$ implies that revenue is a dampened reflection of physical output, since expanding output depresses the firm's price.

The firm's materials allocation problem under IC is
\begin{equation}
\max_{M_{jt}} \left[ E\left( R_{jt} \mid \mathcal{I}_{jt} \right) - \rho_{jt} M_{jt} \right]
\end{equation}
Following the derivation in Appendix~\ref{Gandhi}, but accounting for the demand curvature, the first-order condition delivers a modified log materials cost share equation:
\begin{equation}\label{eq:ic_share}
s_{jt} = \phi_t + \log\left( \frac{\partial f(k_{jt}, l_{jt}, m_{jt})}{\partial m_{jt}} \right) + \log \tilde{\mathcal{E}} - \tilde{\varepsilon}_{jt}
\end{equation}
Compared to the competitive share equation~\eqref{eq:share_reg}, the IC version gains period-specific intercepts $\phi_t = \log((\theta^d_t - 1)/\theta^d_t)$ that absorb the effect of the time-varying demand elasticity on the optimal materials share. The composite shock $\tilde{\varepsilon}_{jt} = \frac{\theta^d_t - 1}{\theta^d_t}\varepsilon_{jt}$ and scalar constant $\tilde{\mathcal{E}} = E[e^{\tilde{\varepsilon}_{jt}}]$ are the IC analogs of $\varepsilon_{jt}$ and $\mathcal{E}$, scaled by the demand curvature. Under perfect competition, $\phi_t = 0$ for all $t$ and equation~\eqref{eq:ic_share} reduces to equation~\eqref{eq:share_reg}.

\subsubsection{Second Stage Under IC}

Since observed revenue reflects both physical output and the endogenous price response, I deflate log revenue by the industry-level output price index $\Pi_t$:
\begin{equation}
r_{jt} = y_{jt} - \log \Pi_t
\end{equation}
The deflated revenue $r_{jt}$ removes common price movements within an industry-year cell. The index is the EUROSTAT producer price index for manufacturing, published by two-digit NACE division; I average the divisions within each of the eight industry groups of Footnote~\ref{industry_list} and merge the result by country, industry group, and year, so that $\Pi_t$ is the group's index in that year.

The analog of the random variable $\mathcal{Y}_{jt}$ in equation~\eqref{eq:Y_def} is constructed using deflated revenue:
\begin{equation}\label{eq:ic_R}
\hat{\mathcal{Y}}^{\mu}_{jt} = r_{jt} - \tilde{\varepsilon}_{jt} - e^{\phi_t} \int \frac{\partial f(k_{jt}, l_{jt}, m_{jt})}{\partial m_{jt}} \, dm_{jt}
\end{equation}
The composite productivity under IC is then
\begin{equation}\label{eq:ic_omega}
\omega^{\mu}_{jt} = \hat{\mathcal{Y}}^{\mu}_{jt} + e^{\phi_t - \mu} \, \mathcal{C}(k_{jt}, l_{jt}) + \left( e^{\phi_t - \mu} - 1 \right) y^{\text{agg}}_t
\end{equation}
where $\mu$ is a constant that pins down the level of markups, $\mathcal{C}(k_{jt}, l_{jt})$ is the constant of integration from equation~\eqref{eq:C_poly} with the same second-order polynomial approximation, and $y^{\text{agg}}_t = \log \sum_j \exp(r_{jt})$ is the log aggregate real revenue within the industry-year cell; $y^{\text{agg}}_t$ proxies for the unobserved aggregate demand shifter $y^D_t$ from equation~\eqref{eq:ic_demand} in all estimation expressions below.

Equations~\eqref{eq:ic_R} and~\eqref{eq:ic_omega} are the IC analogs of equation~\eqref{eq:Y_def}. The $e^{\phi_t}$ factor in $\hat{\mathcal{Y}}^\mu_{jt}$ rescales the materials elasticity integral by the demand curvature wedge, and the $e^{\phi_t - \mu}$ factor in $\omega^\mu_{jt}$ rescales the constant of integration $\mathcal{C}(k_{jt}, l_{jt})$ by the markup, with the residual $(e^{\phi_t - \mu} - 1) y^{\text{agg}}_t$ term capturing the aggregate demand shifter contribution that PPI deflation does not remove. Under perfect competition, $\phi_t \to 0$ and $\mu \to 0$, so both scaling factors equal~$1$, the $y^{\text{agg}}_t$ term vanishes, and the two equations reduce to $\omega_{jt} = \mathcal{Y}_{jt} + \mathcal{C}(k_{jt}, l_{jt})$. See Appendix~O6-4 of \cite{gandhi2020identification} for the full derivation.

The period-specific markup is
\begin{equation}
\text{markup}_t = e^{\phi_t - \mu}
\end{equation}
which measures the ratio of price to marginal cost.\footnote{The markup varies across time periods but not across firms within a given period, since $\phi_t$ and $\mu$ are common parameters. This is a consequence of the CES demand structure, in which all firms within an industry face the same demand elasticity.} Under perfect competition, $\text{markup}_t = 1$ for all $t$.

\subsection{Estimation}\label{app:ic_estimation}

The estimation proceeds in two stages, paralleling the competitive case in Appendix~\ref{Gandhi}.

\paragraph{First stage} The share equation~\eqref{eq:ic_share} is estimated by nonlinear least squares. The materials elasticity $\partial f / \partial m_{jt}$ is approximated by the same second-degree complete polynomial in $(k_{jt}, l_{jt}, m_{jt})$ with ten $\gamma$ parameters used in the baseline. The additional parameters are the $P-1$ period-specific intercepts $\phi_t$, with the first period in the cell's sample as the reference, $\phi_1 = 0$, where $P$ is the number of time periods of Section~\ref{s:empirics} the cell covers: five for Spain and Italy, four for Germany, Poland, and Romania, whose samples begin in 2004, and, for France, four in six industry groups, five in coke, chemicals and pharmaceuticals and three in electronic, optical products and electrical equipment, since its producer price index begins in 2005, 2000 and 2007 respectively. The regression residuals recover $\tilde{\varepsilon}_{jt}$, and the $\tilde{\mathcal{E}}$-correction recovers the $\gamma$ parameters free of the scalar constant, as in the competitive case.

\paragraph{Second stage} The constant of integration $\mathcal{C}(k_{jt}, l_{jt})$ is approximated by a second-order polynomial with five $\alpha$ parameters, as in the baseline. The parameters $\boldsymbol{\alpha}$ and $\mu$ are estimated by just-identified GMM on six moment conditions,
\begin{equation}
\begin{aligned}
&E\left[\eta^{\mu}_{jt} \, k_{jt}^{\tau_k} l_{jt}^{\tau_l}\right] = 0 \quad \forall \, \tau_k, \tau_l, \; 0 < \tau_k + \tau_l \leq 2, \\
&E\left[\eta^{\mu}_{jt} \, y^{\text{agg}}_t\right] = 0,
\end{aligned}
\end{equation}
where $\eta^{\mu}_{jt} = \omega^{\mu}_{jt} - \sum_{a,p} \delta^{\mu}_{ap} (\omega^{\mu}_{jt-1})^a \mathbf{1}(t\in p)$ is the productivity innovation under IC and the period-specific Markov coefficients $\boldsymbol{\delta}^{\mu}_{ap}$, $0 \leq a \leq 3$, are the least-squares coefficients of $\omega^{\mu}_{jt}$ on a cubic in $\omega^{\mu}_{jt-1}$ within each period, as in the competitive case. The first five conditions identify $\boldsymbol{\alpha}$, and the sixth identifies $\mu$ by exploiting the aggregate demand shifter $y^{\text{agg}}_t$ as an instrument: variation in aggregate industry revenue that is orthogonal to firm-level productivity innovations provides the exclusion restriction needed to separate the markup parameter from the production function coefficients.\footnote{Since $y^{\text{agg}}_t = \log\sum_j \exp(r_{jt})$ includes firm $j$'s own deflated revenue, the moment $E[\eta^{\mu}_{jt}\, y^{\text{agg}}_t] = 0$ holds only asymptotically as the number of firms per cell grows and firm $j$'s contribution to the aggregate vanishes. With the cell sizes used here, the implied bias is negligible.}

I estimate the model separately for each of the six countries and eight industry groups (see the industry classification in Footnote~\ref{industry_list}), on the sample of the competitive baseline with two differences. France is covered from 2005 in six industry groups, from 2000 in coke, chemicals and pharmaceuticals and from 2007 in electronic, optical products and electrical equipment, the years in which its producer price index begins; this removes 30 percent of the French firm-years, and no other country loses an observation. One cell, Romania's Electronics \& Electrical Equipment, is excluded because its second stage did not converge.

\subsection{Results}\label{app:ic_results}

\subsubsection{Markups}

Table~\ref{tab:markup_IC} reports summary statistics for the estimated markups, one value per industry-period cell, by country. The revenue-weighted mean ranges from 0.99 (Romania) to 1.16 (Germany), and the median of the cells from 1.01 (Romania) to 1.18 (Germany). The five countries other than Romania show moderate market power, consistent with manufacturing sectors in which firms face differentiated but substitutable products. Across cells the dispersion is modest in those five countries, with standard deviations between 0.07 (Italy) and 0.14 (Spain), and wide in Romania, at 0.22.

\begin{table}[!h]
\centering
\caption{Markup Summary Statistics Under Imperfect Competition} 
\label{tab:markup_IC}
{\setlength{\tabcolsep}{4pt}\begin{adjustbox}{max width=\textwidth}\begin{tabular}{lcccccc}
  \hline
Statistic & Germany & Spain & France & Italy & Poland & Romania \\ 
  \hline
Firm-years & 76,403 & 1,000,134 & 416,847 & 1,261,767 & 55,581 & 162,688 \\ 
  Industry-period cells & 32 & 40 & 31 & 40 & 32 & 28 \\ 
   \hline
Mean & 1.161 & 1.053 & 1.134 & 1.091 & 1.066 & 0.994 \\ 
  Median & 1.176 & 1.091 & 1.112 & 1.091 & 1.090 & 1.008 \\ 
  Std. Dev. & 0.077 & 0.137 & 0.075 & 0.070 & 0.089 & 0.223 \\ 
  P10 & 1.076 & 0.703 & 1.004 & 0.966 & 0.975 & 0.812 \\ 
  P25 & 1.106 & 0.979 & 1.033 & 1.030 & 1.023 & 0.846 \\ 
  P75 & 1.219 & 1.104 & 1.177 & 1.147 & 1.128 & 1.282 \\ 
  P90 & 1.264 & 1.116 & 1.204 & 1.169 & 1.185 & 1.361 \\ 
   \hline
\end{tabular}\end{adjustbox}}
\note{\scriptsize The markup is $\exp(\hat{\phi}_t - \hat{\mu})$, one value per country-industry-period cell. Mean is the average over years of the revenue-weighted mean across industries, each industry weighted by its share of the country's manufacturing revenue in that year; the median, standard deviation, and percentiles are over the industry-period cells. Firm-years are the observations of the imperfect competition estimation. France is covered from 2005 and Romania excludes Electronics \& Electrical Equipment, whose second stage did not converge.}
\end{table}

Table~\ref{tab:markup_industry_IC} disaggregates markups by country and industry. There is considerable heterogeneity across industries within each country: in Germany the mean markup is 1.26 in Food, Beverages \& Tobacco and 1.06 in Coke, Chemicals \& Pharmaceuticals. Markups are at or above one in 38 of the 47 country-industry means; of the nine below one, seven lie between 0.91 and 0.99, and two are lower, Spain's Furniture and Other Manufacturing at 0.70 and Romania's Machinery and Transport Equipment at 0.74. Romania's Electronics \& Electrical Equipment cell is reported as NA because its second stage did not converge (Section~\ref{app:ic_estimation}).

\begin{table}[!h]
\centering
\caption{Mean Markups by Country and Industry} 
\label{tab:markup_industry_IC}
{\setlength{\tabcolsep}{4pt}\begin{adjustbox}{max width=\textwidth}\begin{tabular}{lcccccc}
  \hline
Industry & Germany & Spain & France & Italy & Poland & Romania \\ 
  \hline
1. Food, Beverages and Tobacco & 1.258 & 1.105 & 1.004 & 1.118 & 1.109 & 1.159 \\ 
  2. Textiles, Apparel and Leather & 1.129 & 1.092 & 1.017 & 1.105 & 1.019 & 1.362 \\ 
  3. Wood, Paper and Printing & 1.076 & 1.035 & 1.142 & 0.966 & 1.132 & 0.907 \\ 
  4. Coke, Chemicals and Pharmaceuticals & 1.061 & 1.130 & 1.192 & 0.990 & 1.152 & 1.187 \\ 
  5. Rubber, Plastics, Minerals and Metals & 1.194 & 0.939 & 1.112 & 1.170 & 0.963 & 0.922 \\ 
  6. Electronics and Electrical Equipment & 1.267 & 1.063 & 1.175 & 1.091 & 1.225 & NA \\ 
  7. Machinery and Transport Equipment & 1.171 & 1.107 & 1.198 & 1.043 & 1.024 & 0.736 \\ 
  8. Furniture and Other Manufacturing & 1.170 & 0.700 & 1.051 & 1.149 & 1.069 & 0.993 \\ 
   \hline
\end{tabular}\end{adjustbox}}
\note{\scriptsize Mean of the estimated markup over the industry's time periods, five for Spain, France, and Italy and four for Germany, Poland, and Romania, whose samples begin in 2004. Industries are the groups of Footnote~\ref{industry_list}. NA: Romania's Electronics \& Electrical Equipment, whose second stage did not converge.}
\end{table}

\begin{figure}[ht]
    \caption{Cross-Country Markup Density Under Imperfect Competition}
    \centering
    \includegraphics[width=0.85\textwidth]{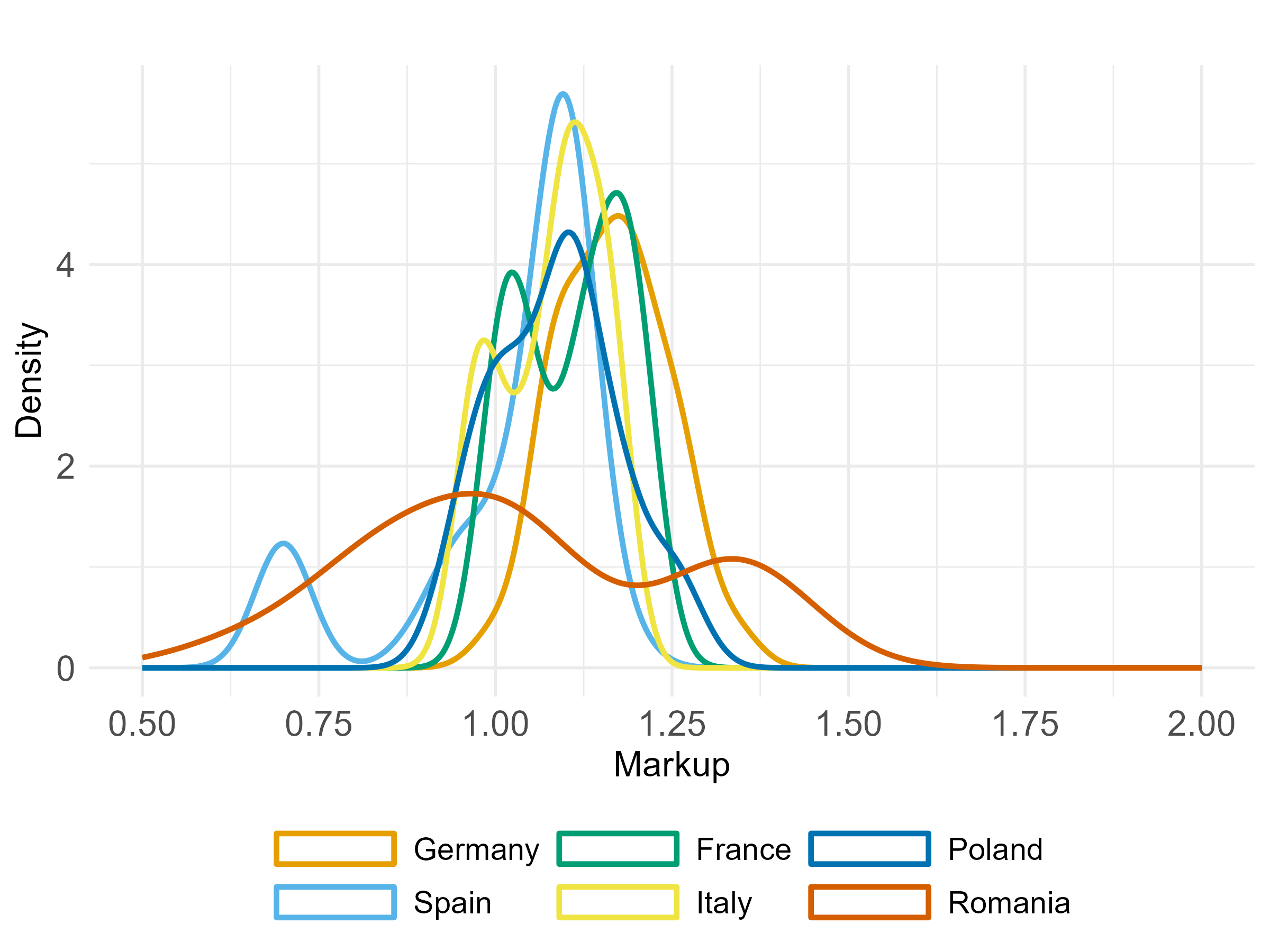}
    \note{\scriptsize Kernel density of the estimated markups, one observation per industry-period cell within each country: 32 cells for Germany and Poland, 40 for Spain and Italy, 31 for France, and 28 for Romania. Markups are defined as $\exp(\hat{\phi}_t - \hat{\mu})$, the ratio of price to marginal cost; a markup of 1 corresponds to perfect competition. Romania excludes Electronics \& Electrical Equipment, whose second stage did not converge; France is covered from 2005. Sample period: 2000--2017 for Spain and Italy, 2004--2017 for Germany, Poland, and Romania, and 2005--2017 for France.}
    \label{fig:markup_density_IC}
\end{figure}

\begin{figure}[ht]
    \caption{Mean Markup Evolution Under Imperfect Competition}
    \centering
    \includegraphics[width=0.85\textwidth]{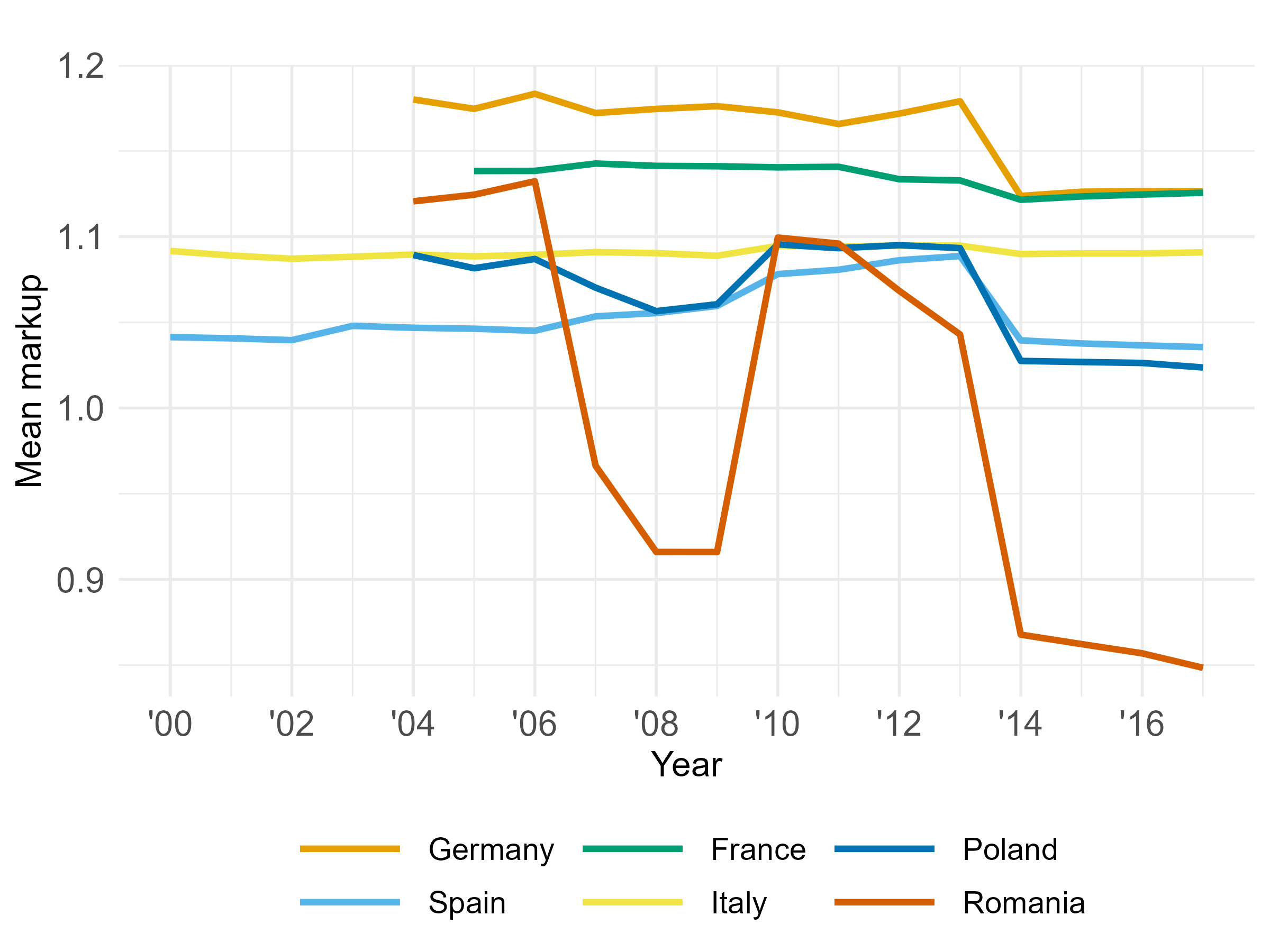}
    \note{\scriptsize Revenue-weighted mean markup by country and year: the markup of each industry-period cell, weighted by the industry's share of the country's manufacturing revenue in that year. The markup is constant within each estimation period, so a series moves at period boundaries and through the weights. Romania excludes Electronics \& Electrical Equipment, whose second stage did not converge. Sample period: 2000--2017 for Spain and Italy, 2004--2017 for Germany, Poland, and Romania, and 2005--2017 for France.}
    \label{fig:markup_time_IC}
\end{figure}

Figure~\ref{fig:markup_density_IC} displays the distribution of the cell markups by country. Five countries concentrate between 0.95 and 1.25, with Germany's cells lying highest, Spain showing a second mode near 0.70, and Romania spreading from 0.62 to 1.36. Figure~\ref{fig:markup_time_IC} traces the revenue-weighted mean markup by year. Since the markup is constant within each estimation period, each series moves at the period boundaries: Italy stays at 1.09 and France between 1.12 and 1.14 throughout; Germany holds 1.17 to 1.18 until 2013 and 1.12 to 1.13 from 2014; Spain and Poland vary within 1.02 to 1.10; and Romania moves the most, from 1.12 in 2004--2006 to 0.92 in 2007--2009, back to 1.04 to 1.10 in 2010--2013, and down to 0.85 in 2014--2017.

\subsubsection{TFPR Dispersion}

\begin{figure}[p]
    \caption{TFPR Dispersion Under Imperfect Competition}
    \label{fig:prod_IC}
    \centering
    \begin{subfigure}[h]{0.43\textwidth}
        \centering
        \includegraphics[width=\textwidth]{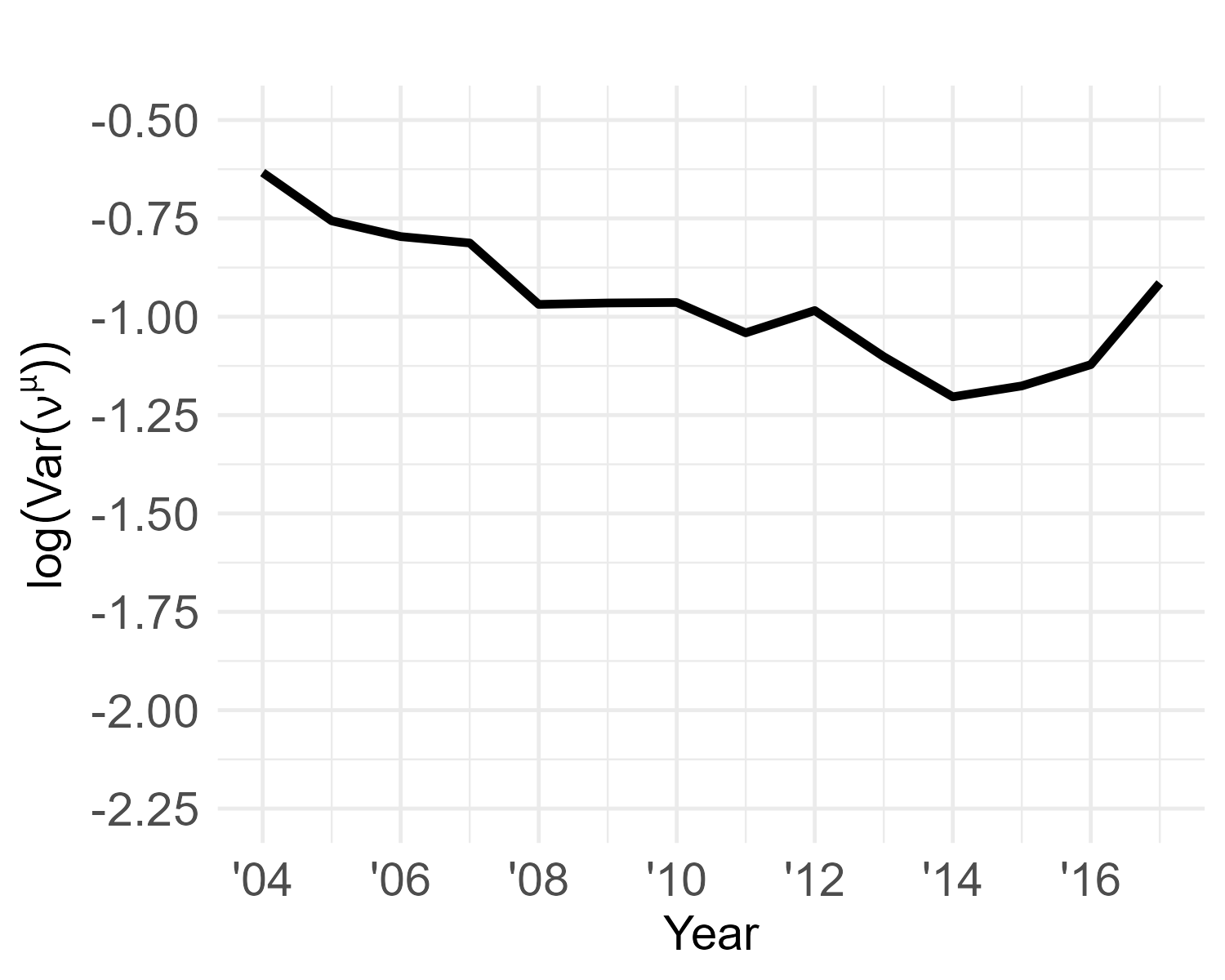}
        \caption{Germany}
        \label{fig:DEU_prod_IC}
    \end{subfigure}
    \hfill
    \begin{subfigure}[h]{0.43\textwidth}
        \centering
        \includegraphics[width=\textwidth]{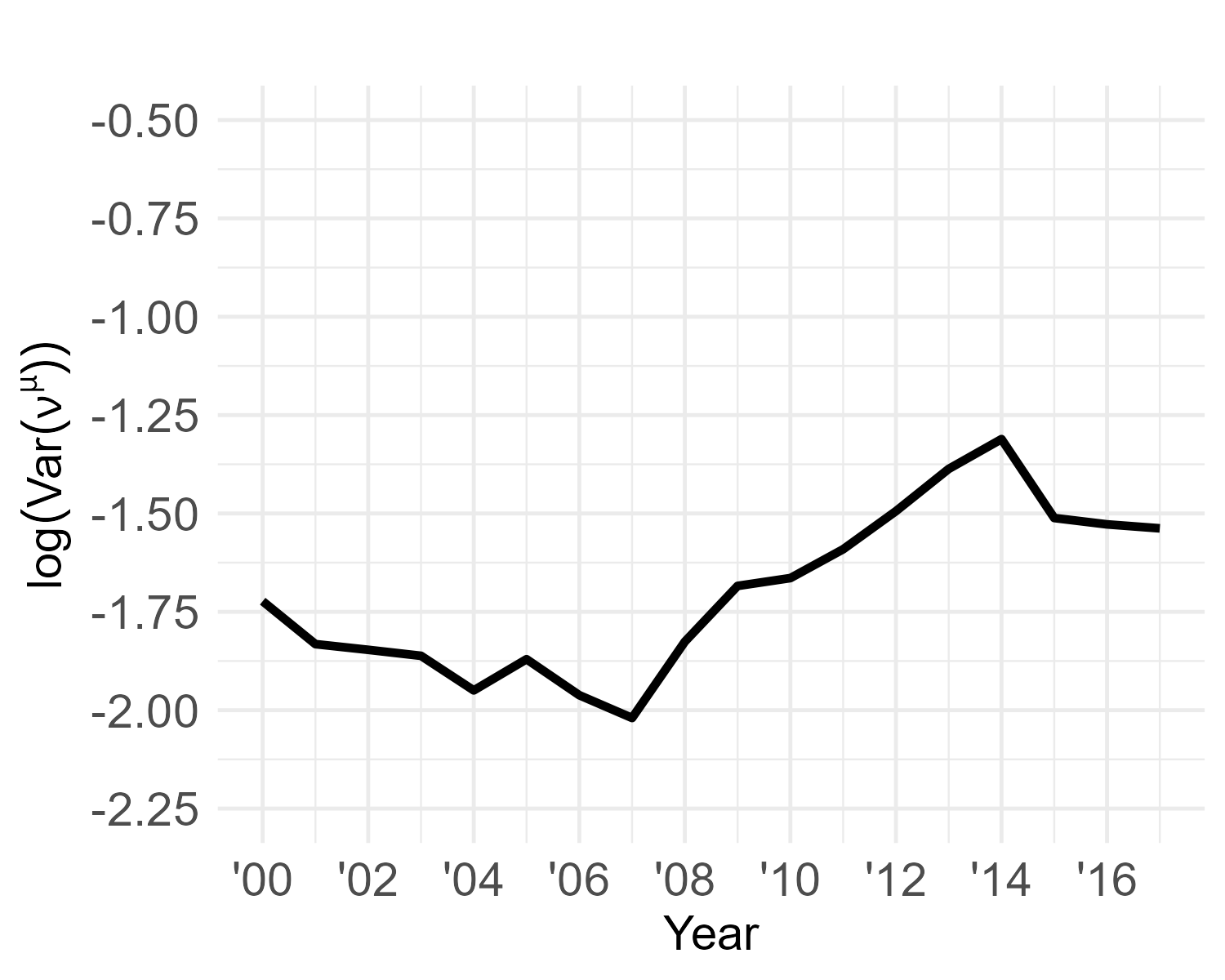}
        \caption{Spain}
        \label{fig:ESP_prod_IC}
    \end{subfigure}

    \begin{subfigure}[h]{0.43\textwidth}
        \centering
        \includegraphics[width=\textwidth]{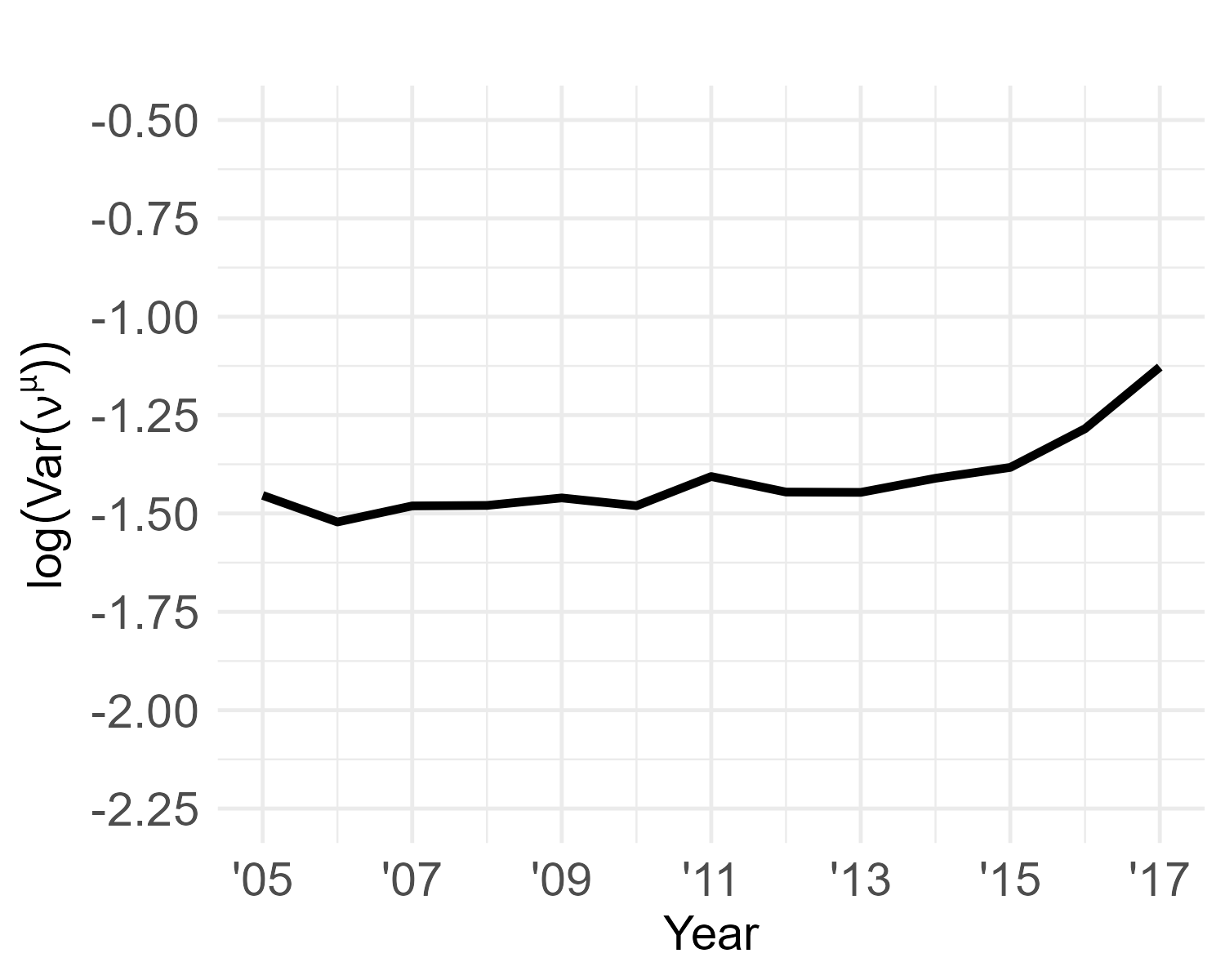}
        \caption{France}
        \label{fig:FRA_prod_IC}
    \end{subfigure}
    \hfill
    \begin{subfigure}[h]{0.43\textwidth}
        \centering
        \includegraphics[width=\textwidth]{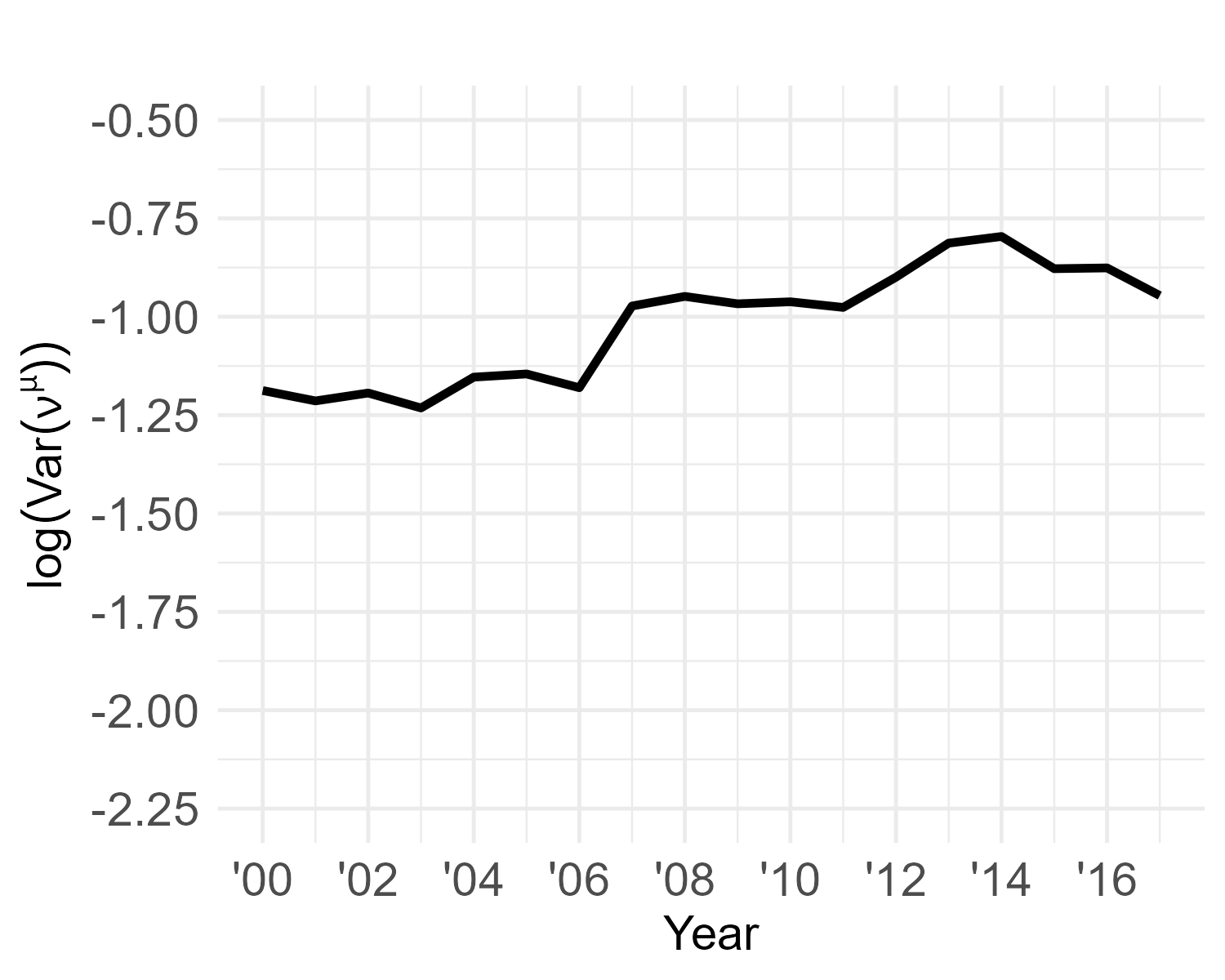}
        \caption{Italy}
        \label{fig:ITA_prod_IC}
    \end{subfigure}

    \begin{subfigure}[h]{0.43\textwidth}
        \centering
        \includegraphics[width=\textwidth]{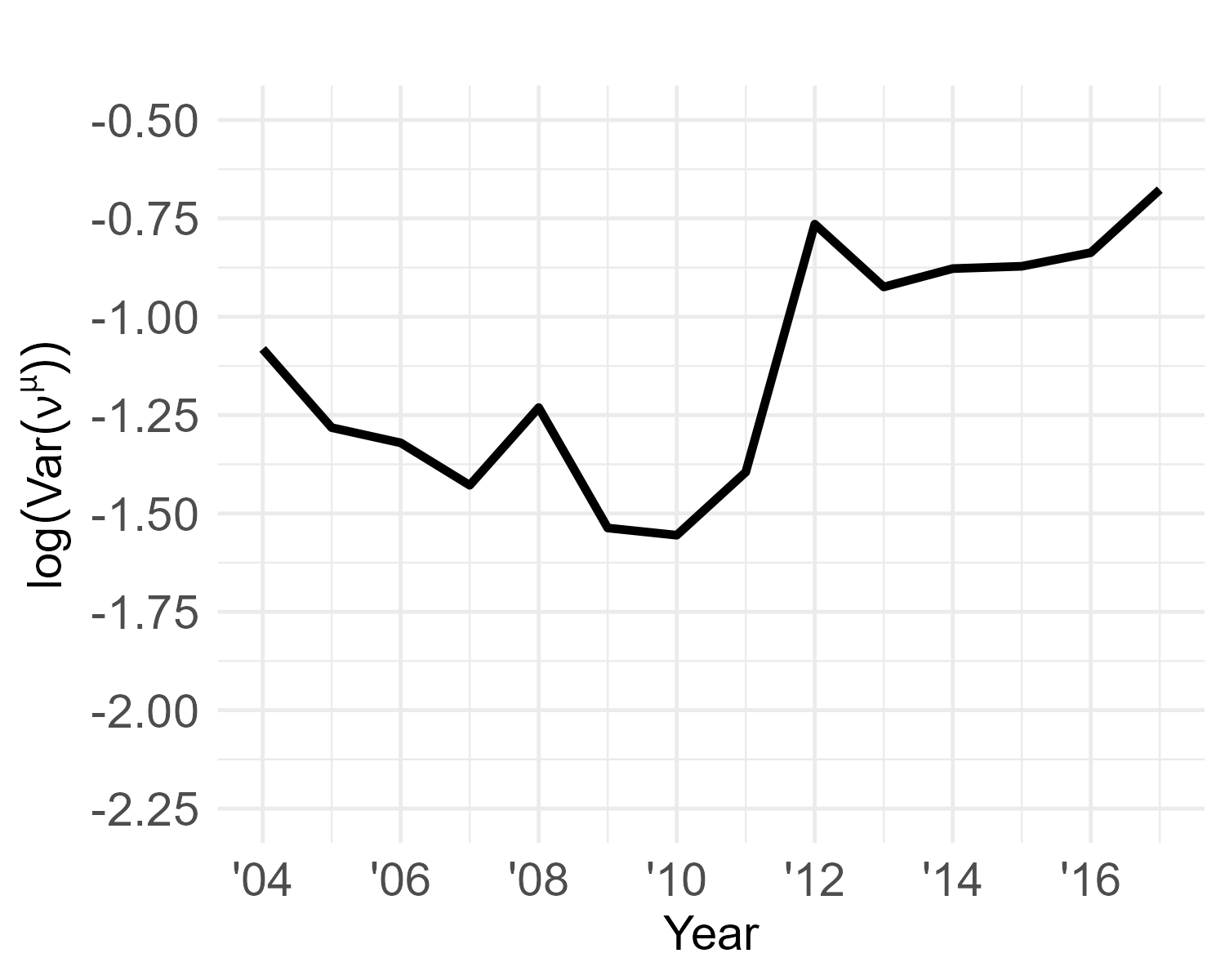}
        \caption{Poland}
        \label{fig:POL_prod_IC}
    \end{subfigure}
    \hfill
    \begin{subfigure}[h]{0.43\textwidth}
        \centering
        \includegraphics[width=\textwidth]{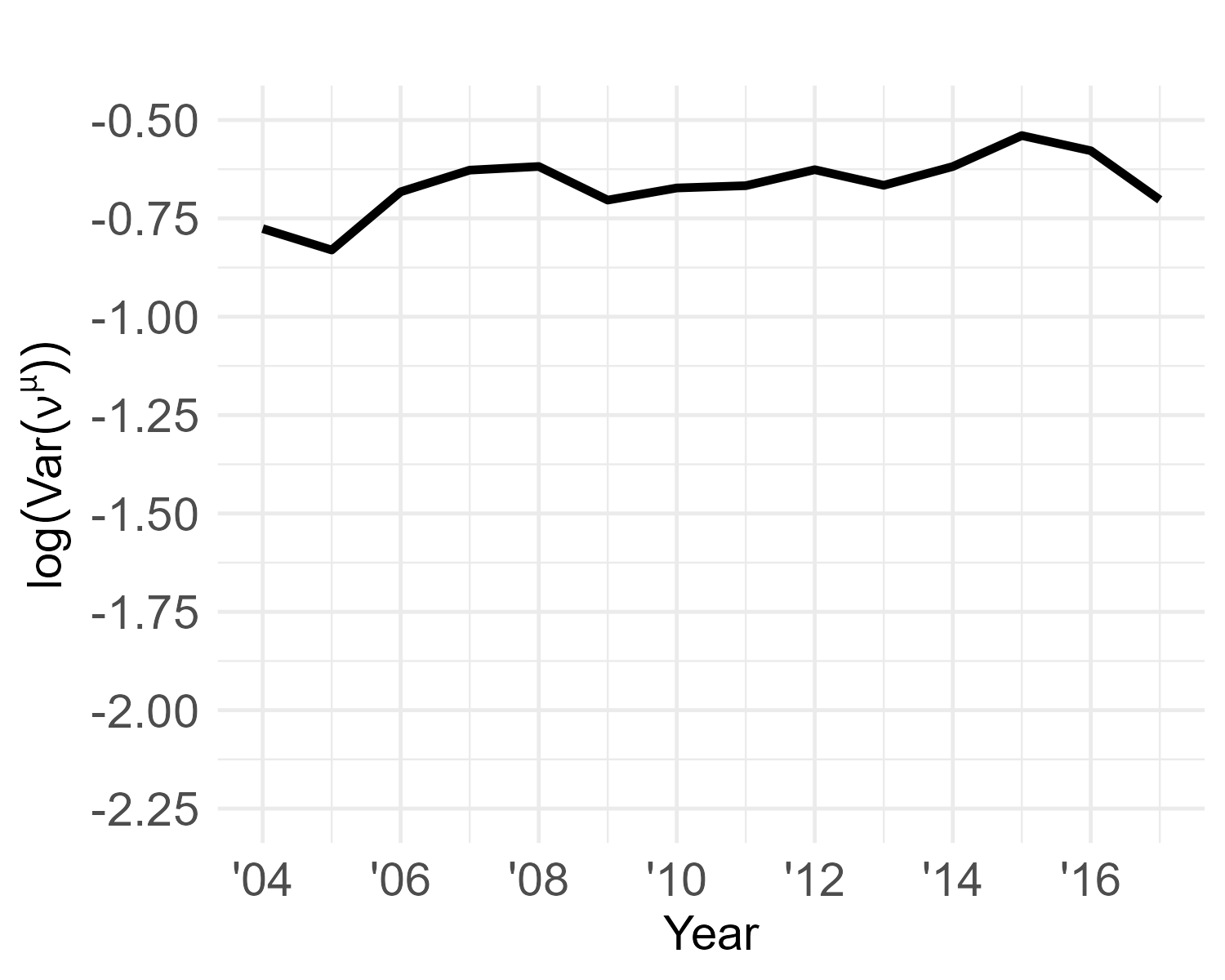}
        \caption{Romania}
        \label{fig:ROM_prod_IC}
    \end{subfigure}
    \note{\scriptsize Each panel displays $\log(\text{Var}_{st}(\nu^{\mu}_{jt}))$, the log variance of TFPR under imperfect competition, by country: Germany (Fig.~\ref{fig:DEU_prod_IC}), Spain (Fig.~\ref{fig:ESP_prod_IC}), France (Fig.~\ref{fig:FRA_prod_IC}), Italy (Fig.~\ref{fig:ITA_prod_IC}), Poland (Fig.~\ref{fig:POL_prod_IC}), and Romania (Fig.~\ref{fig:ROM_prod_IC}). TFPR under IC is $\nu^{\mu}_{jt} = \omega^{\mu}_{jt} + \tilde{\varepsilon}_{jt}$. Variances are computed within each industry-year cell and aggregated at the country level using average total manufacturing revenue shares for each industry as weights, as in Appendix~\ref{app:country_disp}. Romania excludes Electronics \& Electrical Equipment, whose second stage did not converge; France is covered from 2005. Sample period: 2000--2017 for Spain and Italy, 2004--2017 for Germany, Poland, and Romania, and 2005--2017 for France.}
\end{figure}

Figure~\ref{fig:prod_IC} displays the evolution of aggregate TFPR dispersion under imperfect competition, measured as $\log(\text{Var}_{st}(\nu^{\mu}_{jt}))$, for each country. The series track the competitive ones: on the same industry-year cells and with the same aggregation, the IC and competitive series co-move year by year at a correlation of 0.97 or more in every country, and their levels differ by less than 0.1 log points on average. The patterns are those of Appendix~\ref{app:country_disp}: TFPR dispersion rises in Spain and Italy from the Great Recession on and in France after 2014, fluctuates around a flat level in Romania, falls in Germany from 2004 to 2014, and in Poland reaches a trough in 2009--2010 before rising from 2012. Allowing for market power therefore leaves the measured cross-firm variation in productivity, and its evolution, in place.

\subsubsection{Input Gap Dispersion}

\begin{figure}[p]
    \caption{Input Gap Dispersion Under Imperfect Competition}
    \label{fig:mrp_IC}
    \centering
    \begin{subfigure}[h]{0.43\textwidth}
        \centering
        \includegraphics[width=\textwidth]{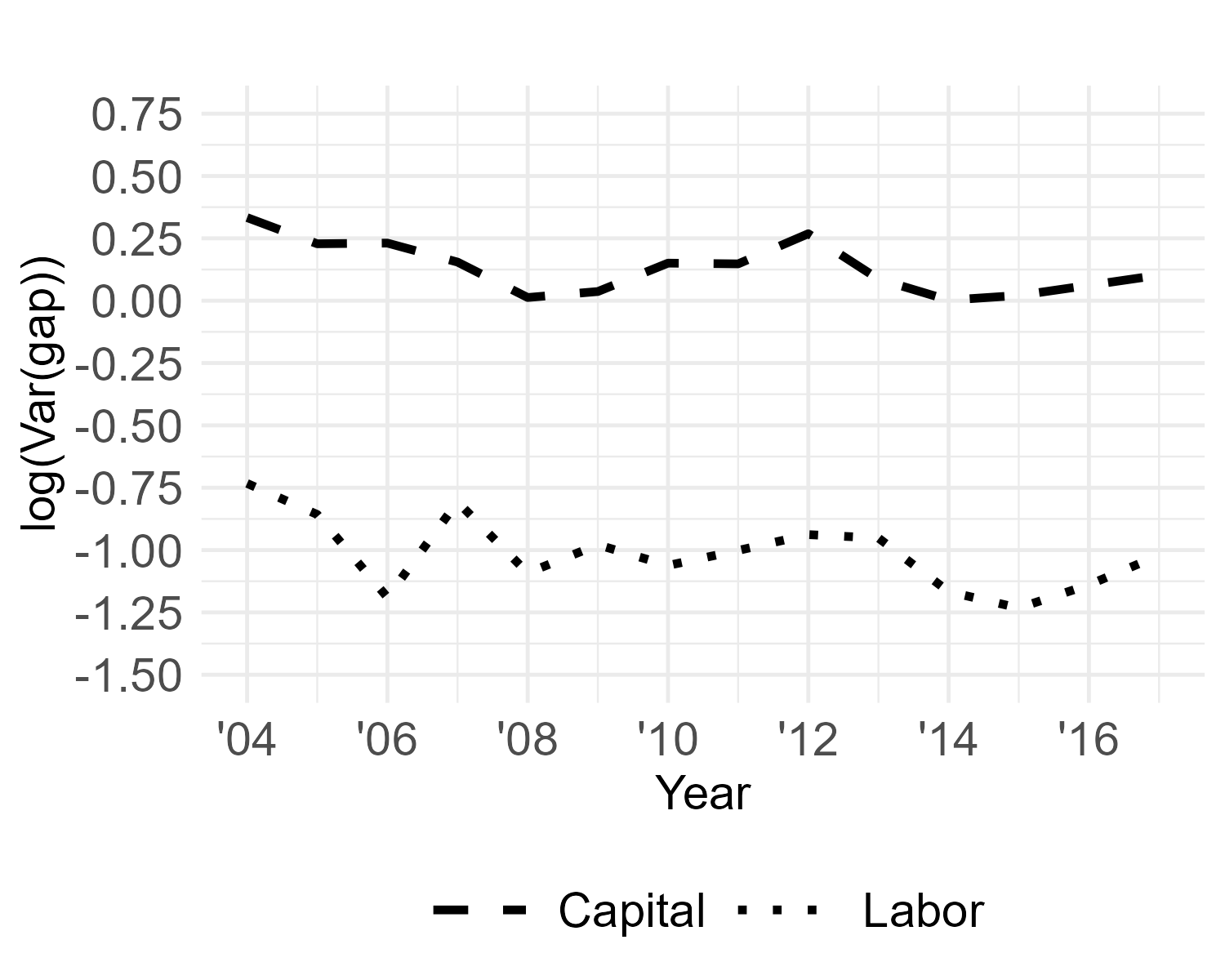}
        \caption{Germany}
        \label{fig:DEU_mrp_IC}
    \end{subfigure}
    \hfill
    \begin{subfigure}[h]{0.43\textwidth}
        \centering
        \includegraphics[width=\textwidth]{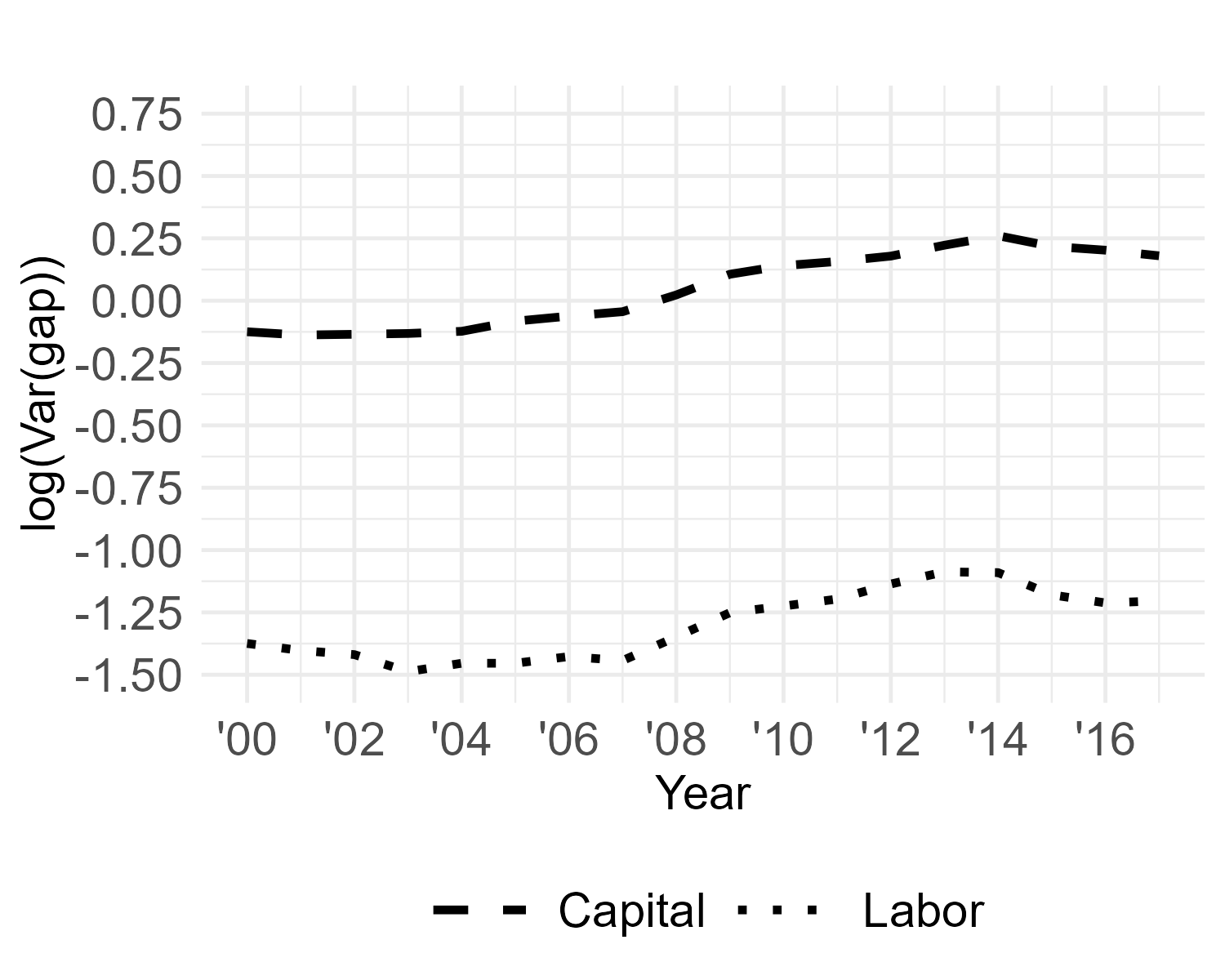}
        \caption{Spain}
        \label{fig:ESP_mrp_IC}
    \end{subfigure}

    \begin{subfigure}[h]{0.43\textwidth}
        \centering
        \includegraphics[width=\textwidth]{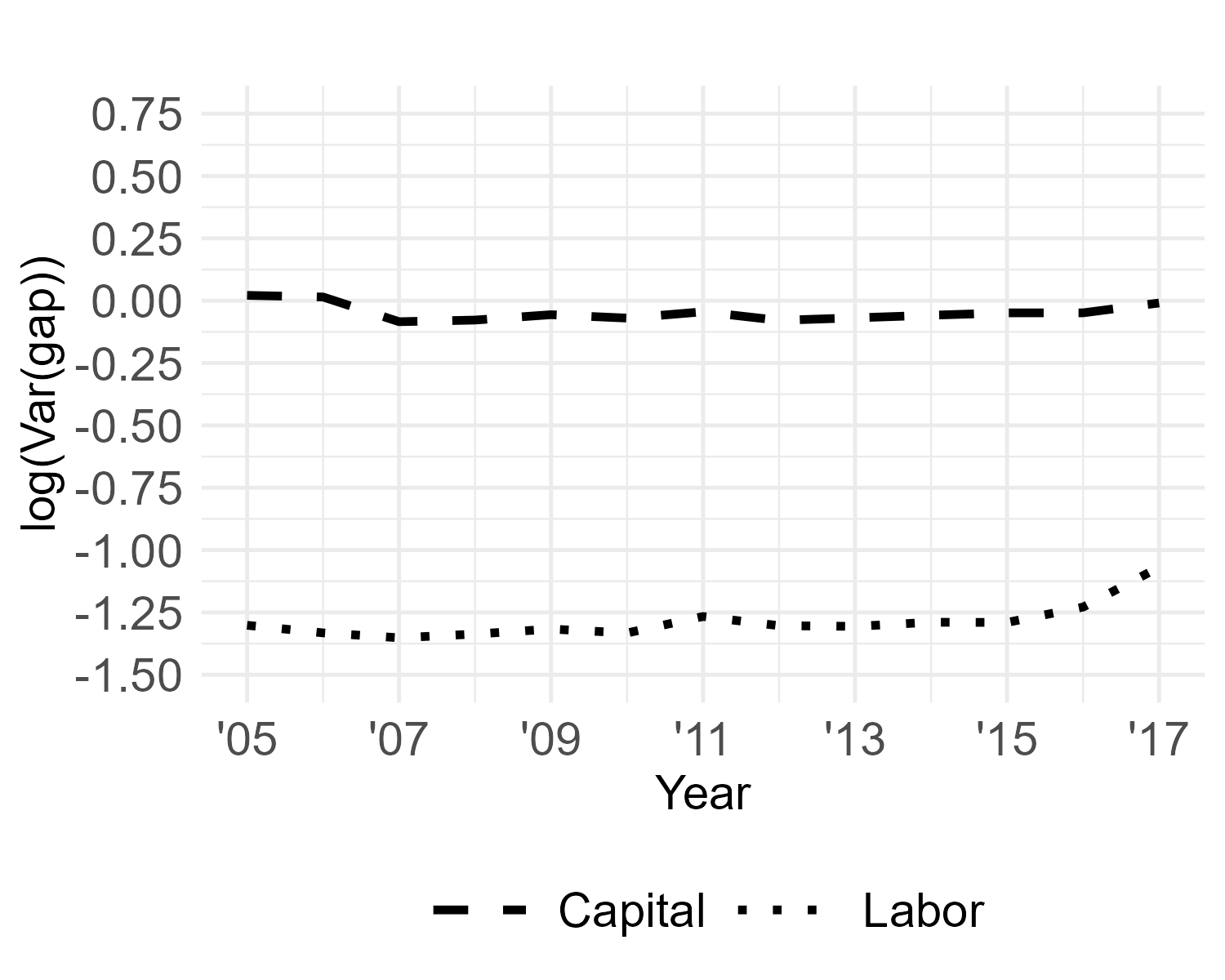}
        \caption{France}
        \label{fig:FRA_mrp_IC}
    \end{subfigure}
    \hfill
    \begin{subfigure}[h]{0.43\textwidth}
        \centering
        \includegraphics[width=\textwidth]{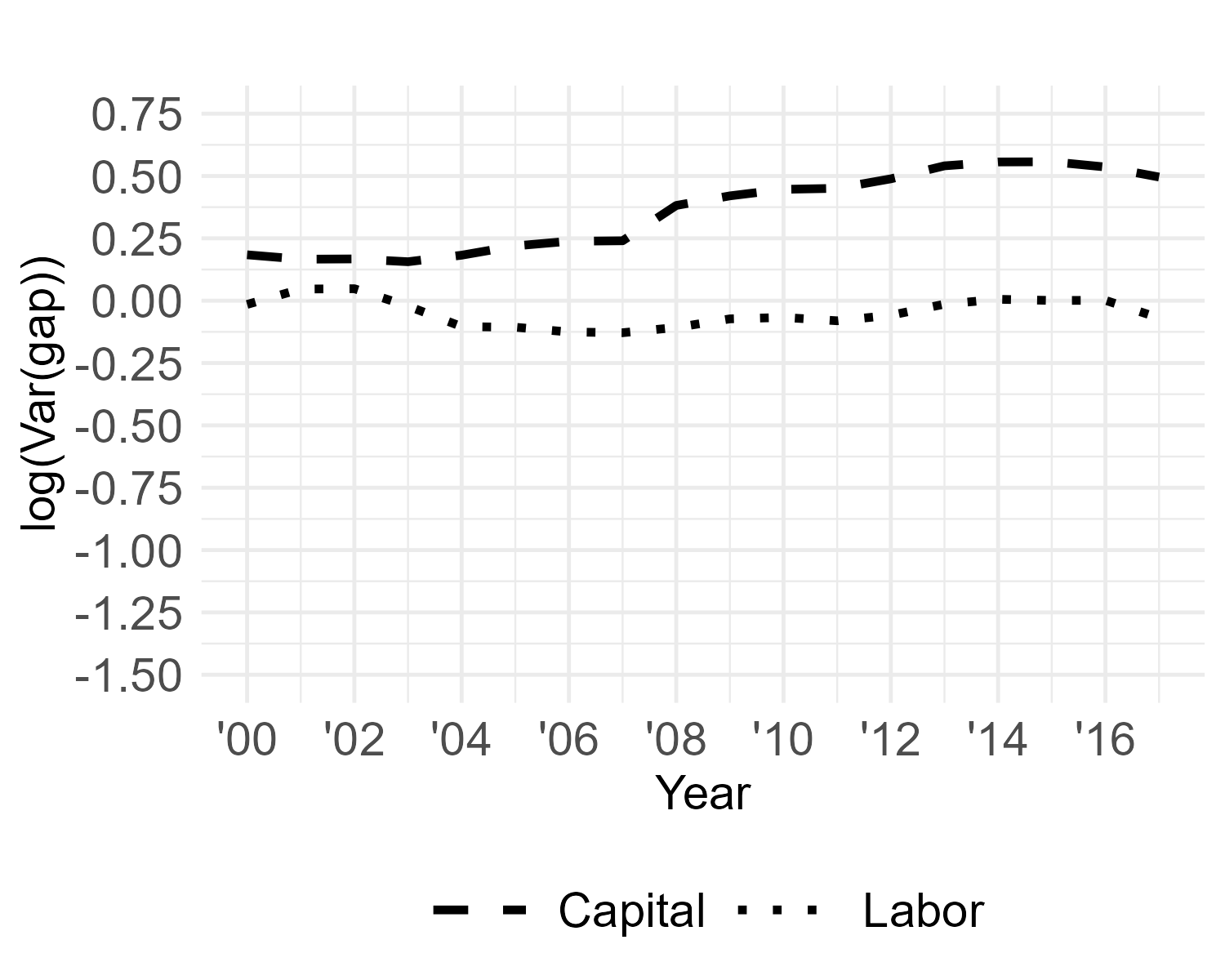}
        \caption{Italy}
        \label{fig:ITA_mrp_IC}
    \end{subfigure}

    \begin{subfigure}[h]{0.43\textwidth}
        \centering
        \includegraphics[width=\textwidth]{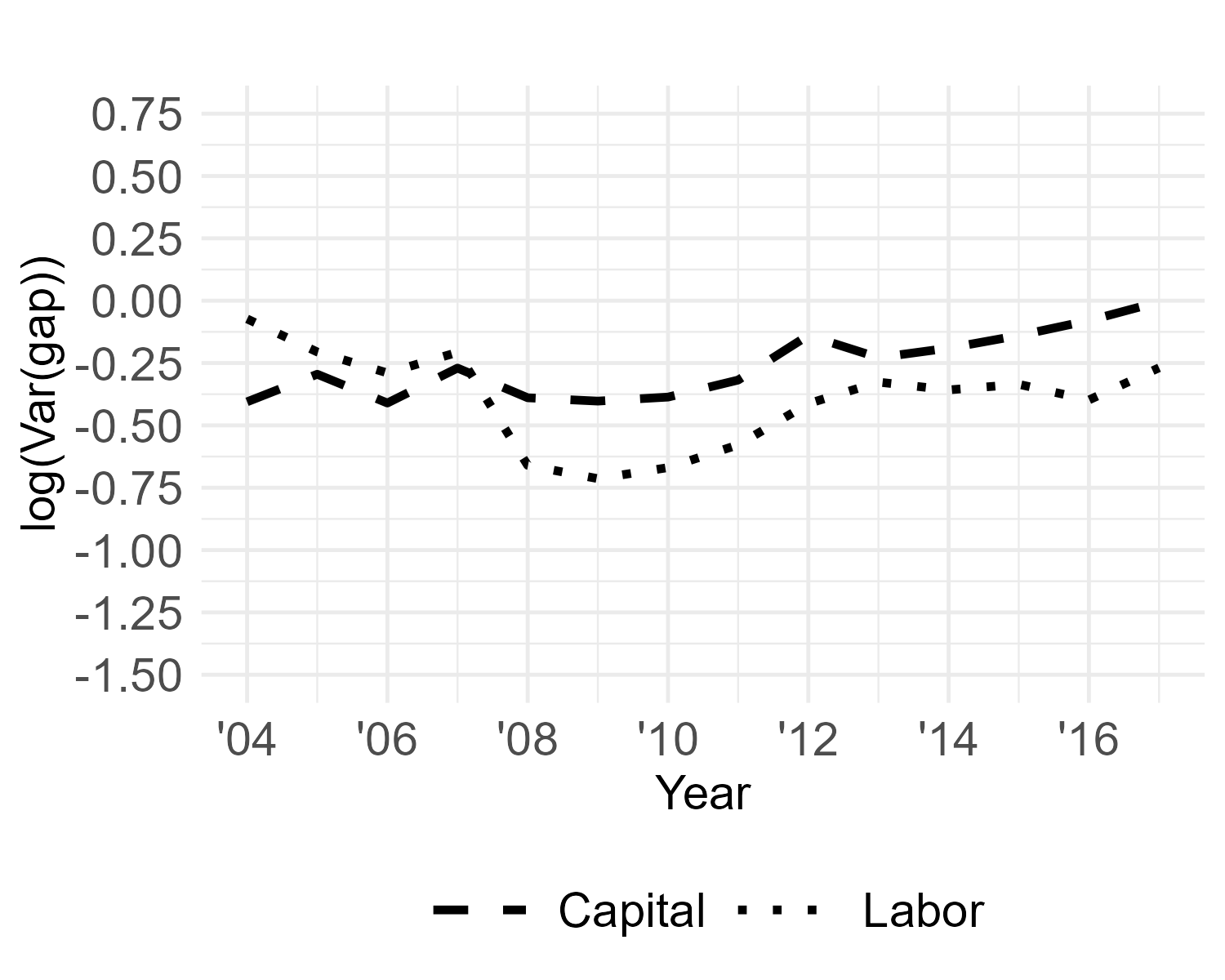}
        \caption{Poland}
        \label{fig:POL_mrp_IC}
    \end{subfigure}
    \hfill
    \begin{subfigure}[h]{0.43\textwidth}
        \centering
        \includegraphics[width=\textwidth]{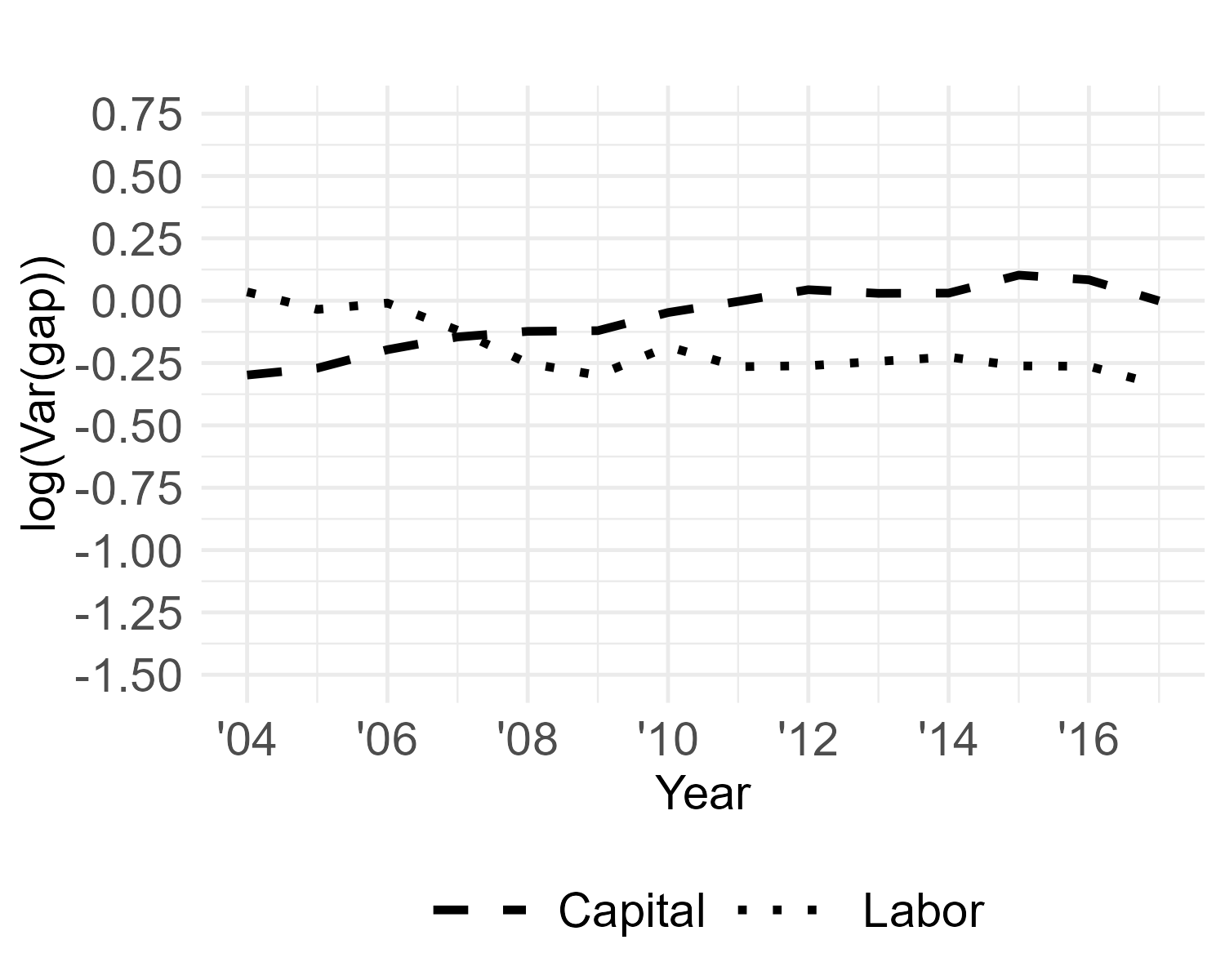}
        \caption{Romania}
        \label{fig:ROM_mrp_IC}
    \end{subfigure}
    \note{\scriptsize Each panel displays $\log(\text{Var}_{st}(\mathrm{mrp}^{X,\mu}_{jt} - \log P^X_{jt}))$, the log variance of the input gap under imperfect competition, for capital (dashed) and labor (dotted), by country: Germany (Fig.~\ref{fig:DEU_mrp_IC}), Spain (Fig.~\ref{fig:ESP_mrp_IC}), France (Fig.~\ref{fig:FRA_mrp_IC}), Italy (Fig.~\ref{fig:ITA_mrp_IC}), Poland (Fig.~\ref{fig:POL_mrp_IC}), and Romania (Fig.~\ref{fig:ROM_mrp_IC}). The gaps use the IC elasticities and the same revenue and input outlays as Appendix~\ref{app:country_disp}; materials is not shown because its gap equals $\tilde{\varepsilon}_{jt}$ up to a cell constant. Variances are computed within each industry-year cell on the firm-years that carry a gap for every input, and aggregated at the country level using each industry's share of the country's manufacturing revenue in that year, as in Appendix~\ref{app:country_disp}. Romania excludes Electronics \& Electrical Equipment, whose second stage did not converge; France is covered from 2005. Sample period: 2000--2017 for Spain and Italy, 2004--2017 for Germany, Poland, and Romania, and 2005--2017 for France.}
\end{figure}

Figure~\ref{fig:mrp_IC} reports the evolution of input gap dispersion for capital (dashed) and labor (dotted) under imperfect competition. The two series keep the baseline ordering: under both models the capital gap is more dispersed than the labor gap in every year in Germany, Spain, France, and Italy, and in every year from 2008 on in Poland and Romania. The levels are close to the competitive series on the same cells: the capital gap series differ by less than 0.1 log points on average in five countries and by 0.22 in France, where the IC series sits at $-0.05$ against $-0.26$; the labor gap series differ by 0.11 in Romania and by less than 0.1 elsewhere. In Germany, the log variance of the capital gap falls from 0.33 in 2004 to 0.00 in 2014 under IC, against 0.40 to 0.16 in the competitive series. Allowing for market power therefore moves the level of input gap dispersion by about a tenth of a log point, 0.22 for France's capital gap, and leaves its evolution in place.

\subsubsection{Firm-Level Pass-Through}

To verify that the pass-through results of Section~\ref{sec:productivity_passthrough} are robust to market power, I re-estimate the regression of Table~\ref{tab:components} on the IC input gaps, built from the IC elasticities, and on the IC productivity components: the Markov conditional mean $g(\omega^{\mu}_{-1})$, the ex-ante shock $\eta^{\mu}_{jt}$, and the ex-post shock $\tilde{\varepsilon}_{jt}$ of equation~(\ref{eq:ic_share}). Table~\ref{tab:components_IC} reports the IC estimates in Panel~A and, in Panel~B, the competitive estimates on the same firm-years.

For capital every pooled coefficient is within 0.1 of its competitive value: total TFPR passes through at 1.196 under IC against 1.220, and the three components at 1.178, 1.240, and 1.197 against 1.217, 1.316, and 1.207. By country the capital coefficients on $g(\omega_{-1})$ and $\eta_{jt}$ move by up to 0.26, in Romania, with no common sign, while the coefficient on $\varepsilon_{jt}$ moves by less than 0.1 everywhere. For labor the coefficient on the ex-post shock is unchanged, 0.337 against 0.345, in every country but Romania, where it falls from 0.598 to 0.378; the coefficients on $g(\omega_{-1})$ and $\eta_{jt}$ are lower under IC in every country, 0.786 against 0.953 and 0.667 against 0.762, so total TFPR passes through at 0.499 against 0.564. The pattern of Section~\ref{sec:productivity_passthrough} holds under both models: the capital coefficients are close to one another, and the labor coefficients fall from $g(\omega_{-1})$ through $\eta_{jt}$ to $\varepsilon_{jt}$.

\begin{table}[!h]
    \centering
    \caption{Pass-Through of TFPR and Its Components into Input Gaps Under Imperfect Competition}
    \label{tab:components_IC}
{\setlength{\tabcolsep}{4pt}%
\begin{adjustbox}{max width=\textwidth, max totalheight=0.78\textheight}
\begin{tabular}{lccccccc}
\toprule
 & Germany & Spain & France & Italy & Poland & Romania & Pooled \\
\midrule
\multicolumn{8}{l}{\textit{Panel A: imperfect competition (IC gaps, IC components)}} \\
  \textbf{Capital}, total TFPR & 1.179 & 1.193 & 1.206 & 1.216 & 1.237 & 1.071 & 1.196 \\
   & (0.043) & (0.017) & (0.021) & (0.016) & (0.028) & (0.013) & (0.010) \\
  \hspace{1em}$g(\omega_{-1})$ & 1.149 & 1.417 & 1.368 & 1.026 & 1.480 & 1.029 & 1.178 \\
   & (0.091) & (0.030) & (0.065) & (0.025) & (0.091) & (0.025) & (0.015) \\
  \hspace{1em}$\eta_{jt}$ & 1.168 & 1.291 & 1.244 & 1.270 & 1.400 & 1.054 & 1.240 \\
   & (0.064) & (0.032) & (0.026) & (0.028) & (0.082) & (0.024) & (0.018) \\
  \hspace{1em}$\varepsilon_{jt}$ & 1.219 & 1.100 & 1.062 & 1.288 & 1.094 & 1.091 & 1.197 \\
   & (0.057) & (0.014) & (0.020) & (0.019) & (0.032) & (0.012) & (0.010) \\
  \textbf{Labor}, total TFPR & 0.672 & 0.432 & 0.533 & 0.511 & 0.405 & 0.515 & 0.499 \\
   & (0.042) & (0.012) & (0.033) & (0.013) & (0.053) & (0.032) & (0.009) \\
  \hspace{1em}$g(\omega_{-1})$ & 0.863 & 0.656 & 0.787 & 0.837 & 0.730 & 0.752 & 0.786 \\
   & (0.036) & (0.018) & (0.028) & (0.027) & (0.088) & (0.051) & (0.016) \\
  \hspace{1em}$\eta_{jt}$ & 0.701 & 0.671 & 0.517 & 0.692 & 0.663 & 0.686 & 0.667 \\
   & (0.045) & (0.014) & (0.019) & (0.021) & (0.063) & (0.035) & (0.013) \\
  \hspace{1em}$\varepsilon_{jt}$ & 0.428 & 0.317 & 0.319 & 0.344 & 0.204 & 0.378 & 0.337 \\
   & (0.036) & (0.009) & (0.013) & (0.016) & (0.035) & (0.030) & (0.009) \\
\midrule
\multicolumn{8}{l}{\textit{Panel B: competitive baseline on the same firm-years}} \\
  \textbf{Capital}, total TFPR & 1.193 & 1.277 & 1.149 & 1.232 & 1.193 & 1.112 & 1.220 \\
   & (0.060) & (0.015) & (0.028) & (0.014) & (0.037) & (0.016) & (0.009) \\
  \hspace{1em}$g(\omega_{-1})$ & 1.154 & 1.445 & 1.243 & 1.109 & 1.326 & 1.285 & 1.217 \\
   & (0.108) & (0.026) & (0.075) & (0.022) & (0.123) & (0.043) & (0.018) \\
  \hspace{1em}$\eta_{jt}$ & 1.169 & 1.402 & 1.218 & 1.325 & 1.285 & 1.206 & 1.316 \\
   & (0.053) & (0.022) & (0.032) & (0.022) & (0.090) & (0.033) & (0.014) \\
  \hspace{1em}$\varepsilon_{jt}$ & 1.252 & 1.190 & 1.050 & 1.270 & 1.097 & 1.053 & 1.207 \\
   & (0.063) & (0.016) & (0.018) & (0.016) & (0.028) & (0.014) & (0.009) \\
  \textbf{Labor}, total TFPR & 0.703 & 0.482 & 0.599 & 0.568 & 0.440 & 0.712 & 0.564 \\
   & (0.029) & (0.013) & (0.040) & (0.020) & (0.072) & (0.019) & (0.013) \\
  \hspace{1em}$g(\omega_{-1})$ & 0.903 & 0.768 & 0.935 & 1.039 & 0.789 & 1.054 & 0.953 \\
   & (0.043) & (0.017) & (0.022) & (0.033) & (0.107) & (0.033) & (0.017) \\
  \hspace{1em}$\eta_{jt}$ & 0.742 & 0.739 & 0.547 & 0.790 & 0.726 & 0.883 & 0.762 \\
   & (0.040) & (0.015) & (0.023) & (0.026) & (0.074) & (0.024) & (0.016) \\
  \hspace{1em}$\varepsilon_{jt}$ & 0.427 & 0.326 & 0.288 & 0.331 & 0.178 & 0.598 & 0.345 \\
   & (0.034) & (0.010) & (0.010) & (0.025) & (0.032) & (0.014) & (0.014) \\
\bottomrule
\end{tabular}
\end{adjustbox}}
\note{\scriptsize Each cell reports $\beta$ from a firm-level regression of the input gap $\mathrm{mrp}^X_{jt} - \log P^X_{jt}$ on productivity, absorbing country-industry-year fixed effects as in Table~\ref{tab:components}: on total TFPR, $\omega_{jt} + \varepsilon_{jt}$, or on its three components, $g(\omega_{-1})$, $\eta_{jt}$, and $\varepsilon_{jt}$, entered jointly. Panel A uses the IC gaps and the IC components $g(\omega^{\mu}_{-1})$, $\eta^{\mu}_{jt}$, and $\tilde{\varepsilon}_{jt}$; Panel B uses the competitive gaps and components on the same 2,511,131 firm-years. The sample excludes France before 2005 in six industry groups and before 2007 in Electronics \& Electrical Equipment (Coke, Chemicals \& Pharmaceuticals is covered from 2000) and Romania's Electronics \& Electrical Equipment, which the IC estimation lacks. Standard errors in parentheses are non-parametric bootstrap standard errors across 100 replications. Materials is omitted because its gap equals the ex-post shock up to a cell constant under both models.}
\end{table}

\subsubsection{Discussion}

The IC robustness exercise produces four main findings. First, estimated markups are moderate and economically plausible, centered around 1.0--1.2 across countries and industries. This is consistent with the manufacturing sector, in which firms possess some degree of product differentiation but face substantial competitive pressure.

Second, TFPR dispersion under IC replicates the competitive baseline in level and in evolution: the two series co-move at correlations of 0.97 or more and differ by less than 0.1 log points on average in every country. Because the IC extension jointly re-estimates the production function coefficients ($\gamma$, $\alpha$, $\delta$) together with the markup parameters, TFPR under IC, $\nu^{\mu}_{jt} = \omega^{\mu}_{jt} + \tilde{\varepsilon}_{jt}$, is redistributed across firms differently from the competitive baseline; cross-firm variation is not mechanically preserved, and the close empirical match between the IC and competitive dispersion patterns is therefore non-trivial.

Third, the input gap dispersion results are unchanged. The capital gap remains more dispersed than the labor gap in every year in four countries and from 2008 on in Poland and Romania, and the levels of both series stay within about a tenth of a log point of the competitive ones, except France's capital gap at 0.22. The IC framework accounts for endogenous pricing through the period-specific intercepts $\phi_t$ in the first-stage share equation and the deflation of revenue by the producer price index in the second stage. Since input gap dispersion is preserved after these adjustments, the patterns documented in Section~\ref{s:results} are not attributable to industry-period-level deviations from price-taking.

Fourth, the firm-level pass-through re-estimated on the IC gaps and the IC components preserves the results of Section~\ref{sec:productivity_passthrough}: every pooled capital coefficient is within 0.1 of its competitive value on the same firm-years, the pooled labor coefficient on the ex-post shock is unchanged, and the pooled labor coefficients on $g(\omega_{-1})$ and $\eta_{jt}$ are lower under IC by 0.17 and 0.10; by country the two fall by 0.03 to 0.30, most in Romania, where the coefficient on the ex-post shock also falls, from 0.60 to 0.38.
\clearpage

\section{Comparison of Production Function Estimators}\label{app:mc}
\normalsize
In this appendix I evaluate the finite-sample performance of the \cite{gandhi2020identification} estimator (henceforth GNR) against two widely used alternatives: the factor shares approach (\citealp{de2021industrial}) and the proxy-variable method of \cite{ackerberg2015identification} (henceforth ACF), extended to gross output as in Section III of GNR. I conduct Monte Carlo experiments under two three-input data-generating processes (DGPs), a translog and a Cobb-Douglas specification. The experiments assess each estimator's ability to measure the dispersion of input elasticities, marginal revenue products, and TFPR, and to recover the pass-through of TFPR and of each of its components into the input gaps.

\subsection{Data-Generating Processes}\label{app:mc_dgp}

Both DGPs feature three inputs (log capital, $k$, log labor, $l$, and log materials, $m$) and share the productivity structure, the input timing, and the wage response described in Section \ref{s:theory}. I simulate 500 firms over 200 periods, discarding the first 170 as burn-in. I conduct 100 independent replications.

\subsubsection{Production Technology}

\paragraph{Translog} The log revenue function is
\begin{equation}\label{eq:mc_translog}
y_{jt} = \gamma_k k_{jt} + \gamma_l l_{jt} + \gamma_m m_{jt} + \gamma_{kk} k_{jt}^2 + \gamma_{ll} l_{jt}^2 + \gamma_{mm} m_{jt}^2 + \gamma_{kl} k_{jt} l_{jt} + \gamma_{km} k_{jt} m_{jt} + \gamma_{lm} l_{jt} m_{jt} + \omega_{jt} + \varepsilon_{jt}
\end{equation}
with first-order parameters $(\gamma_k, \gamma_l, \gamma_m) = (0.12, 0.23, 0.55)$, own second-order parameters $\gamma_{kk} = \gamma_{ll} = \gamma_{mm} = 0.015$, and interactions $(\gamma_{kl}, \gamma_{km}, \gamma_{lm}) = (-0.010, -0.032, -0.020)$.\footnote{These $\gamma$ symbols denote production function coefficients, following \cite{gandhi2020identification}'s notation. They are distinct from the $\gamma$ coefficients in Appendix~\ref{Gandhi}, which parameterize the polynomial approximation of $\partial f / \partial m$.} The implied returns to scale, evaluated at sample means, are 0.880.

\paragraph{Cobb-Douglas (CD)} The log revenue function is
\begin{equation}\label{eq:mc_cd}
y_{jt} = \alpha_k^{\scriptscriptstyle\text{CD}} k_{jt} + \alpha_l^{\scriptscriptstyle\text{CD}} l_{jt} + \alpha_m^{\scriptscriptstyle\text{CD}} m_{jt} + \omega_{jt} + \varepsilon_{jt}
\end{equation}
with $(\alpha_k^{\scriptscriptstyle\text{CD}}, \alpha_l^{\scriptscriptstyle\text{CD}}, \alpha_m^{\scriptscriptstyle\text{CD}}) = (0.12, 0.23, 0.55)$ and returns to scale 0.90. Because the Cobb-Douglas elasticities are constant, any dispersion in the estimated elasticities under this DGP is spurious, which makes it a direct test of whether an estimator introduces spurious heterogeneity.

\subsubsection{Productivity}

In both DGPs, the persistent component of TFPR follows an AR(1) process as in Assumption \ref{asspr}:
\begin{equation}\label{eq:mc_omega}
\omega_{jt} = d_0 + d_1 \, \omega_{jt-1} + \eta_{jt}, \qquad \eta_{jt} \sim N(0, \sigma^2_\eta)
\end{equation}
with $d_1 = 0.80$ and $\sigma^2_\eta = 0.04$. The translog DGP sets $d_0 = 0$, while the Cobb-Douglas DGP sets $d_0 = 0.20$ (implying a stationary mean of 1.0). The ex-post productivity shock is $\varepsilon_{jt} \sim N(0, \sigma^2_\varepsilon)$, with $\sigma^2_\varepsilon = 0.07$ (translog) and $\sigma^2_\varepsilon = 0.04$ (Cobb-Douglas).

\subsubsection{Input Timing, Capital, and Labor}

Capital is predetermined, consistent with Assumption \ref{assa}. The law of motion is $k_{jt} = \log(\exp(k_{jt-1})(1-\delta_j) + I_{jt-1})$, where investment follows a reduced-form rule
\begin{equation}\label{eq:mc_inv}
I_{jt} = \exp(\varphi_0 + \varphi_k \, k_{jt} + \varphi_\omega \, \omega_{jt} + \zeta_{jt})
\end{equation}
with $(\varphi_0, \varphi_k, \varphi_\omega) = (-1.70, 0.60, 0.50)$ and $\zeta_{jt} \sim N(0, 0.0025)$. Firm-specific depreciation rates $\delta_j \in \{0.05, 0.075, 0.10, 0.125, 0.15\}$, assigned uniformly across firms, generate cross-sectional capital heterogeneity.

Labor is likewise chosen one period ahead, consistent with Assumption \ref{assa}, through a log-linear partial-adjustment rule
\begin{equation}\label{eq:mc_labor}
l_{jt+1} = \lambda_{0j} + \lambda_l \, l_{jt} + \lambda_\omega \, \omega_{jt} + \xi_{jt}
\end{equation}
with $(\lambda_l, \lambda_\omega) = (0.60, 0.50)$ and $\xi_{jt} \sim N(0, 0.0025)$. The firm-specific intercept $\lambda_{0j}$ is drawn uniformly from five values that place the steady-state log labor input, $\lambda_{0j}/(1 - \lambda_l)$, on the grid $\{0.5, 1.2, 1.9, 2.6, 3.2\}$, the range spanned by the capital steady states. Both predetermined inputs therefore respond to last period's productivity and are fixed when the period's two shocks realize.

Materials are a flexible input allocated after $\omega_{jt}$ is realized, consistent with Assumption \ref{assa}. The firm chooses $m_{jt}$ to solve the static first-order condition equating the expected marginal revenue product of materials to the input price $\rho_t$, where the expectation is over the ex-post shock, so that $\mathcal{E} = E[e^{\varepsilon_{jt}}] = e^{\sigma^2_\varepsilon/2}$ enters the condition and the log materials share is
\begin{equation}\label{eq:mc_share}
s_{jt} = \log \mathrm{elas}^M_{jt} + \log \mathcal{E} - \varepsilon_{jt}.
\end{equation}
In the translog DGP, this FOC yields an implicit equation for $m_{jt}$ solved via damped fixed-point iteration. In the Cobb-Douglas DGP, the FOC admits a closed-form solution: $m_{jt} = (\log \alpha_m^{\scriptscriptstyle\text{CD}} + \log \mathcal{E} + \alpha_k^{\scriptscriptstyle\text{CD}} k_{jt} + \alpha_l^{\scriptscriptstyle\text{CD}} l_{jt} + \omega_{jt} - \log \rho_t) / (1 - \alpha_m^{\scriptscriptstyle\text{CD}})$.

\subsubsection{Input Prices}

To provide an exclusion restriction for the proxy-variable estimator, I introduce time-varying aggregate materials prices following an AR(1) process:
\begin{equation}\label{eq:mc_prices}
\log \rho_t = \phi_\rho \, \log \rho_{t-1} + \eta_{\rho,t}, \qquad \eta_{\rho,t} \sim N(0, \sigma^2_\rho)
\end{equation}
with $\phi_\rho = 0.6$ and $\sigma^2_\rho = 0.001$. The proxy-variable estimator needs this variation, since $\rho_t$ shifts the demand for materials without entering the production function, and $\sigma^2_\rho$ is set so that the shifts are sizable.

The base wage $\bar w_t$ follows an AR(1) process with the same persistence and innovation variance, around a level that sets the mean wage-bill share equal to the labor elasticity. The realized wage responds to the ex-post shock as in Assumption~\ref{assp}, $w_{jt} = \bar w_t \, e^{\theta \varepsilon_{jt}}$, with $\theta = 0.41$, the value implied by the wage split of Section~\ref{sec:passthrough}. The log materials share is $s_{jt} = \log \rho_t + m_{jt} - y_{jt}$ and the log wage-bill share is $s^L_{jt} = \log w_{jt} + l_{jt} - y_{jt}$.

\subsection{Estimators}\label{app:mc_estimators}

I compare three estimators that differ in their identification strategies, functional form assumptions, and ability to separate the components of TFPR. Each estimator delivers firm-level input elasticities, from which I construct marginal revenue products and TFPR.

\subsubsection{GNR Estimator}

The GNR estimator follows the two-stage procedure detailed in Appendix~\ref{Gandhi} and Section~\ref{s:empirics}. In the first stage, the log materials share equation~\eqref{eq:share_reg} is estimated by nonlinear least squares using a second-degree complete polynomial in $(k_{jt}, l_{jt}, m_{jt})$ with ten $\gamma$ parameters, the specification of the main analysis. The regression residuals recover the ex-post shock $\varepsilon_{jt}$, and the $\mathcal{E}$-correction recovers the production function parameters free of the scalar constant.

In the second stage, the constant of integration $\mathcal{C}(k_{jt}, l_{jt})$ is approximated by a second-order polynomial in capital and labor with five $\alpha$ parameters. The productivity law of motion~\eqref{eq:mc_omega} is approximated by a cubic polynomial in $\omega_{jt-1}$ with four $\delta$ parameters. The $\alpha$ and $\delta$ parameters are estimated jointly via GMM using nine moment conditions: $E[\eta_{jt}\, k_{jt}^{\tau_k} l_{jt}^{\tau_l}] = 0$ for $0 < \tau_k + \tau_l \leq 2$ (identifying $\boldsymbol{\alpha}$) and $E[\eta_{jt}\, \hat{\mathcal{Y}}^a_{jt-1}] = 0$ for $0 \leq a \leq 3$ (identifying $\boldsymbol{\delta}$), yielding a just-identified system.

The GNR estimator delivers three distinct objects: firm-level input elasticities from the estimated $\gamma$ and $\alpha$ parameters via equation~\eqref{eq:elas}; the persistent productivity component $\omega_{jt}$; and the ex-post shock $\varepsilon_{jt}$. This separation is central to the TFPR decomposition in the main analysis.

\subsubsection{Factor Shares Estimator}

The factor shares (FS) approach equates the materials revenue elasticity to the observed expenditure share:
\begin{equation}\label{eq:mc_fs_elas}
\widehat{\mathrm{elas}}^{M,\mathrm{FS}}_{jt} = \exp(s_{jt}) = \frac{\rho_t M_{jt}}{Y_{jt}}
\end{equation}
and the labor revenue elasticity to the observed wage-bill share:
\begin{equation}\label{eq:mc_fs_elasl}
\widehat{\mathrm{elas}}^{L,\mathrm{FS}}_{jt} = \exp(s^L_{jt}) = \frac{w_{jt} L_{jt}}{Y_{jt}}
\end{equation}
Under constant returns to scale (CRS), the capital elasticity is the residual:
\begin{equation}\label{eq:mc_fs_elask}
\widehat{\mathrm{elas}}^{K,\mathrm{FS}}_{jt} = 1 - \widehat{\mathrm{elas}}^{M,\mathrm{FS}}_{jt} - \widehat{\mathrm{elas}}^{L,\mathrm{FS}}_{jt}
\end{equation}
No estimation is required; elasticities are read directly from the data. TFPR is then computed as a Solow residual:
\begin{equation}\label{eq:mc_fs_tfpr}
\hat{\nu}^{\mathrm{FS}}_{jt} = y_{jt} - \widehat{\mathrm{elas}}^{K,\mathrm{FS}}_{jt} \, k_{jt} - \widehat{\mathrm{elas}}^{L,\mathrm{FS}}_{jt} \, l_{jt} - \widehat{\mathrm{elas}}^{M,\mathrm{FS}}_{jt} \, m_{jt}
\end{equation}

This approach has three limitations in the present context. First, equations~\eqref{eq:mc_fs_elas} and~\eqref{eq:mc_fs_elasl} hold only in expectation conditional on $\mathcal{I}_{jt}$: the realized materials share contains $-\varepsilon_{jt}$, and the realized wage-bill share contains $(\theta - 1)\varepsilon_{jt}$ once the wage responds, so the elasticities recovered from the shares carry the ex-post shock. Second, imposing CRS biases the capital elasticity whenever the true technology exhibits non-constant returns, since $\widehat{\mathrm{elas}}^{K,\mathrm{FS}}_{jt}$ absorbs the full difference between unity and the two observed shares. Third, the factor shares approach cannot separate $\omega_{jt}$ from $\varepsilon_{jt}$ within the Solow residual $\hat{\nu}^{\mathrm{FS}}_{jt}$. The only TFPR variance decomposition available is therefore:
\begin{equation}\label{eq:mc_fs_decomp}
\text{Var}(\hat{\nu}^{\mathrm{FS}}_{jt}) \approx \text{Var}(\hat{\nu}^{\mathrm{FS}}_{jt-1}) + \text{Var}(\hat{\nu}^{\mathrm{FS}}_{jt} - \hat{\nu}^{\mathrm{FS}}_{jt-1})
\end{equation}
where the first term captures cross-firm productivity levels and the second captures period-to-period changes.

\subsubsection{Proxy-Variable Estimator}

The proxy-variable method uses a control function approach to address the endogeneity of materials allocation with respect to productivity realizations. ACF propose the procedure for value-added production functions and advise against applying it to gross output functions in which materials enter; the version here is the extension GNR analyze in their Section III, in which time-series variation in the materials price is the only source of identification for the materials terms. In the first stage, revenue is projected onto a flexible polynomial in state variables to partial out the ex-post shock:
\begin{equation}\label{eq:mc_proxy_s1}
y_{jt} = \phi(k_{jt}, l_{jt}, m_{jt}, \log \rho_t) + \varepsilon_{jt}
\end{equation}
where $\phi(\cdot)$ is approximated by a third-degree polynomial in $(k_{jt}, l_{jt}, m_{jt})$ with $\log \rho_t$ entering linearly. The fitted values $\hat{\phi}_{jt}$ capture $f(k_{jt}, l_{jt}, m_{jt}) + \omega_{jt}$.

In the second stage, the production function parameters are identified by exploiting the assumption that $\omega_{jt}$ follows a first-order Markov process. For a candidate parameter vector $\boldsymbol\theta$, productivity is constructed as
\begin{equation}\label{eq:mc_proxy_omega}
\omega_{jt}(\boldsymbol\theta) = \hat{\phi}_{jt} - f(k_{jt}, l_{jt}, m_{jt}; \boldsymbol\theta)
\end{equation}
The conditional expectation $E[\omega_{jt} \mid \omega_{jt-1}]$ is approximated by a cubic polynomial in $\omega_{jt-1}(\boldsymbol\theta)$, and the innovation $\eta_{jt}(\boldsymbol\theta) = \omega_{jt}(\boldsymbol\theta) - E[\omega_{jt} \mid \omega_{jt-1}(\boldsymbol\theta)]$ is the residual. Since $\eta_{jt}$ is realized after capital and labor are chosen but before materials, the moment conditions are:
\begin{equation}\label{eq:mc_proxy_moments}
E[\eta_{jt}(\boldsymbol\theta) \cdot Z_{jt}] = 0
\end{equation}
where the predetermined inputs $k_{jt}$ and $l_{jt}$ and the lagged materials $m_{jt-1}$ serve as instruments.

Under the translog specification, $\boldsymbol\theta = (\gamma_k, \gamma_l, \gamma_m, \gamma_{kk}, \gamma_{ll}, \gamma_{mm}, \gamma_{kl}, \gamma_{km}, \gamma_{lm})$ and $Z_{jt} = (k_{jt}, l_{jt}, k_{jt}^2, l_{jt}^2, k_{jt} l_{jt}, \allowbreak \, m_{jt-1}, m_{jt-1}^2, k_{jt} m_{jt-1}, l_{jt} m_{jt-1})$, yielding a just-identified system with nine moments for nine parameters. Under the Cobb-Douglas specification, $\boldsymbol\theta = (\alpha_k^{\scriptscriptstyle\text{CD}}, \alpha_l^{\scriptscriptstyle\text{CD}}, \alpha_m^{\scriptscriptstyle\text{CD}})$ and $Z_{jt} = (k_{jt}, l_{jt}, m_{jt-1})$, again just-identified with three moments for three parameters.

The proxy-variable estimator can, in principle, separate $\omega_{jt}$ from $\varepsilon_{jt}$ and recover the full TFPR decomposition. However, identification of $\boldsymbol\theta$ relies on the excluded instrument $\log \rho_t$ generating sufficient independent variation in $m_{jt}$ conditional on $(k_{jt}, l_{jt}, \omega_{jt})$. When materials price variation is limited, the GMM objective surface becomes nearly flat, leading to weak identification and imprecise parameter estimates.

\subsection{Monte Carlo Results}\label{app:mc_results}

I assess each estimator along four dimensions: input elasticities, marginal revenue products, TFPR dispersion and persistence, and the TFPR variance decomposition. Results are reported for both the translog and Cobb-Douglas DGPs.

\subsubsection{Input Elasticities}

\begin{figure}[ht]
\caption{Input elasticity distributions --- Translog DGP}
    \centering
    \begin{subfigure}[h]{0.32\textwidth}
        \centering
        \includegraphics[width=\textwidth]{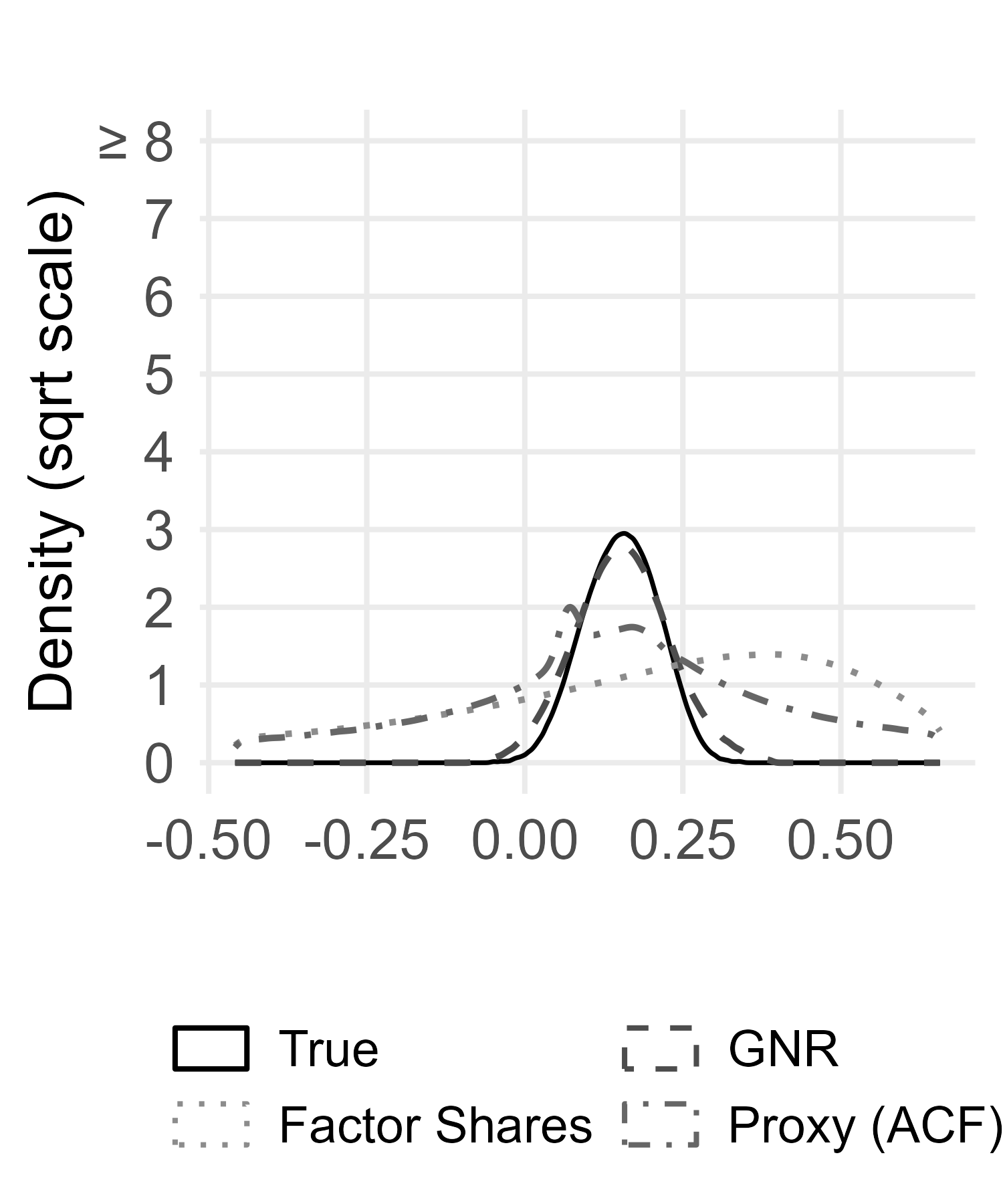}
        \caption{Capital elasticity}
        \label{fig:mc_elas_K_tl}
    \end{subfigure}
    \hfill
    \begin{subfigure}[h]{0.32\textwidth}
        \centering
        \includegraphics[width=\textwidth]{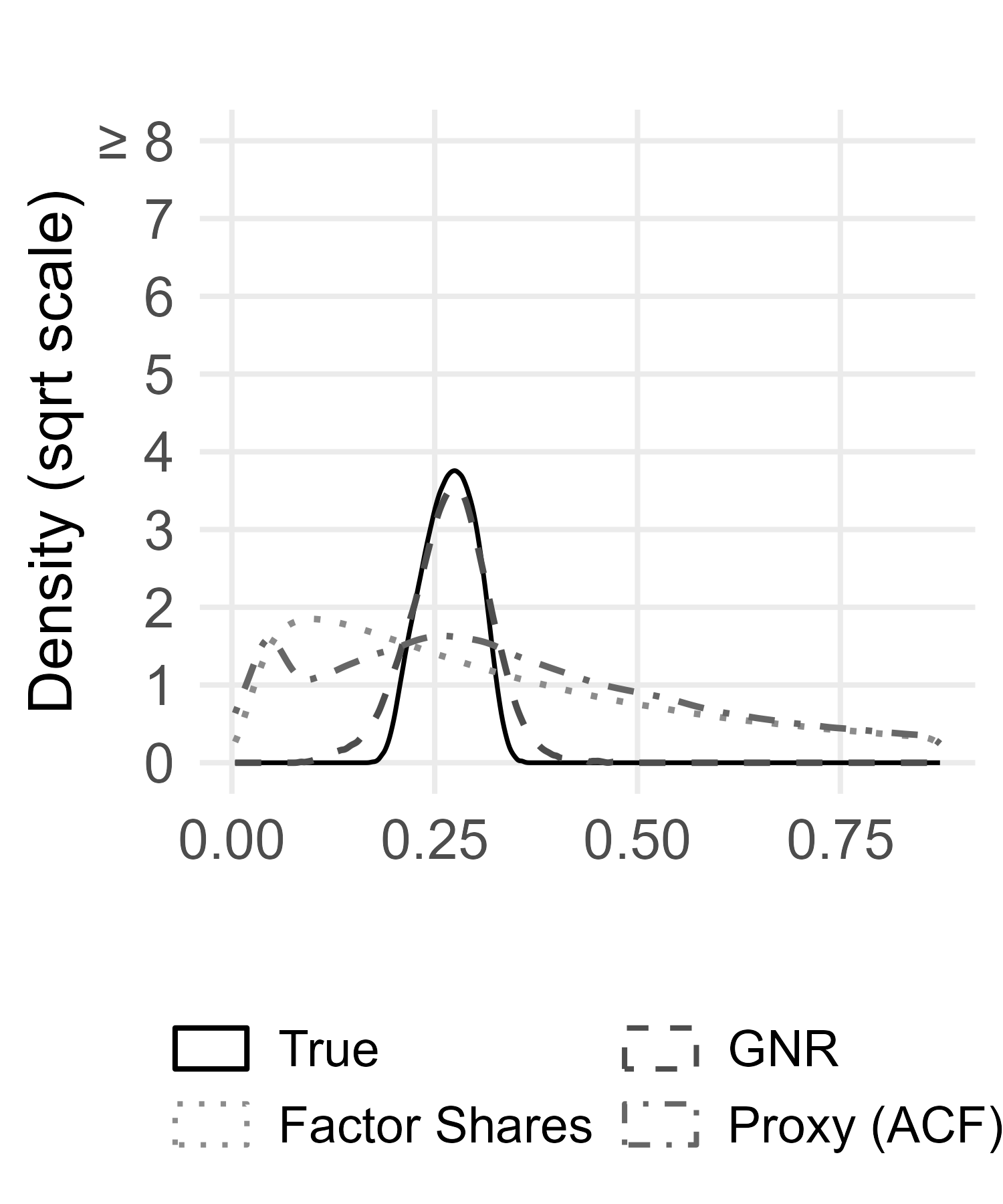}
        \caption{Labor elasticity}
        \label{fig:mc_elas_L_tl}
    \end{subfigure}
    \hfill
    \begin{subfigure}[h]{0.32\textwidth}
        \centering
        \includegraphics[width=\textwidth]{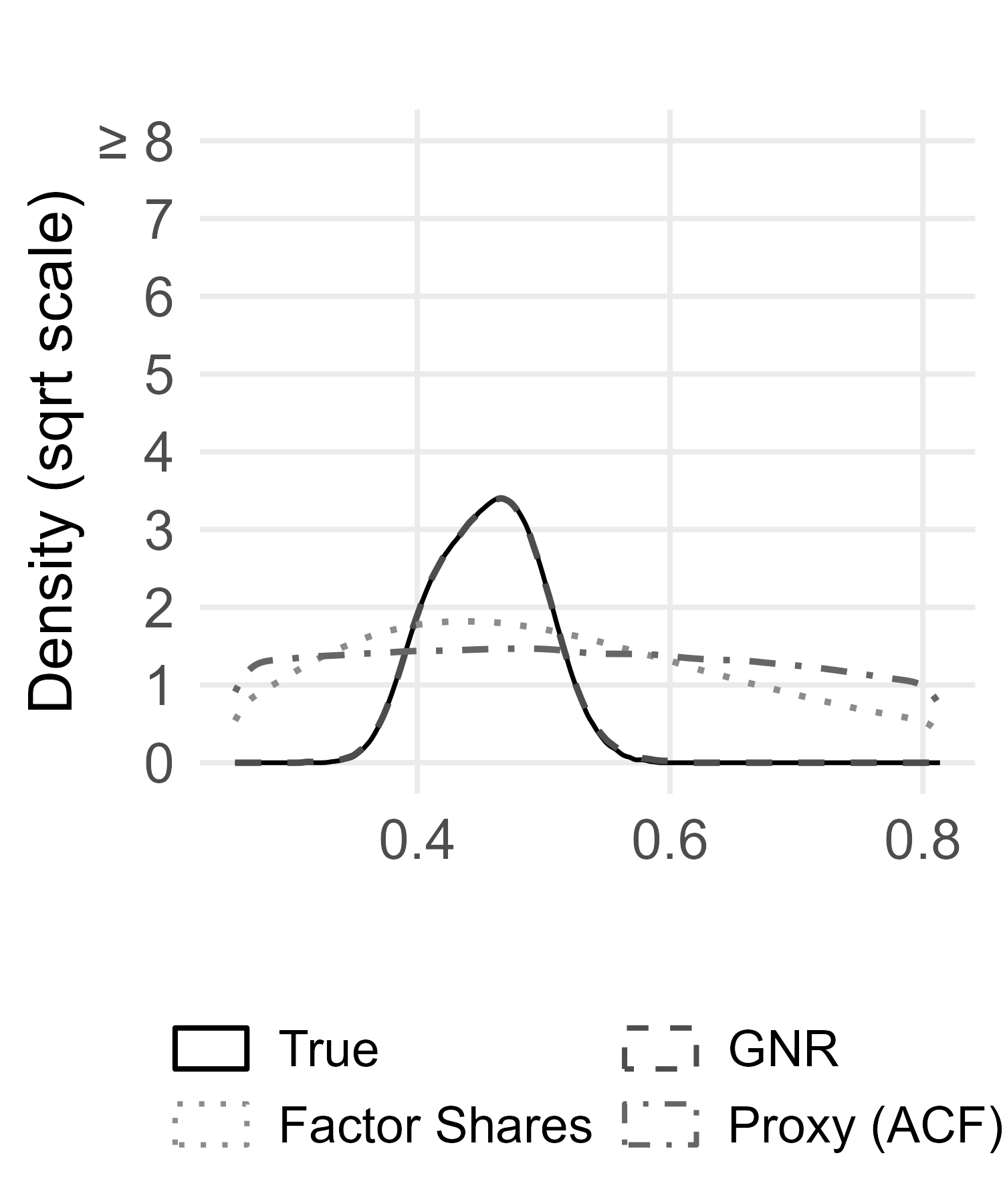}
        \caption{Materials elasticity}
        \label{fig:mc_elas_M_tl}
    \end{subfigure}
     \note{\scriptsize Each curve pools firm-level estimates across 100 Monte Carlo replications (500 firms $\times$ 30 periods each). ``True'' denotes the elasticity implied by the known DGP parameters. The $y$-axis density values are reported on a square root scale to improve visual resolution in the tails while preserving the relative ordering of peaks.}
    \label{fig:mc_elas_translog}
\end{figure}

Figure~\ref{fig:mc_elas_translog} displays the pooled density of firm-level revenue elasticity estimates across all 100 replications under the translog DGP. The GNR estimator closely tracks the true elasticity distributions for capital, labor, and materials: the means across firms are 0.153, 0.271, and 0.456 against true values of 0.154, 0.270, and 0.456, and the within-replication standard deviations are within 0.007 of the true ones. The factor shares estimator produces materials and labor elasticities centered near the truth but with inflated dispersion, as the ex-post shock $\varepsilon_{jt}$ contaminates both expenditure shares: the within-replication standard deviation is 0.137 for materials against a true 0.032, and 0.230 for labor against 0.026, the wage response adding to the noise in the wage-bill share. The CRS assumption forces the capital elasticity upward (the factor shares mean of 0.243 overstates the true mean of 0.154 by 58\%) because capital must absorb the full difference between unity and the two observed shares, and 15\% of firm-years receive a non-positive capital elasticity. The proxy-variable estimator exhibits near-zero mean bias but substantial imprecision: the cross-replication standard deviation of the mean materials elasticity is 0.315, roughly 175 times that of GNR (0.002), and that of the mean labor elasticity is 0.150 against 0.009, rendering firm-level elasticity estimates unreliable for distributional analysis.

\begin{figure}[ht]
\caption{Input elasticity distributions --- Cobb-Douglas DGP}
    \centering
    \begin{subfigure}[h]{0.32\textwidth}
        \centering
        \includegraphics[width=\textwidth]{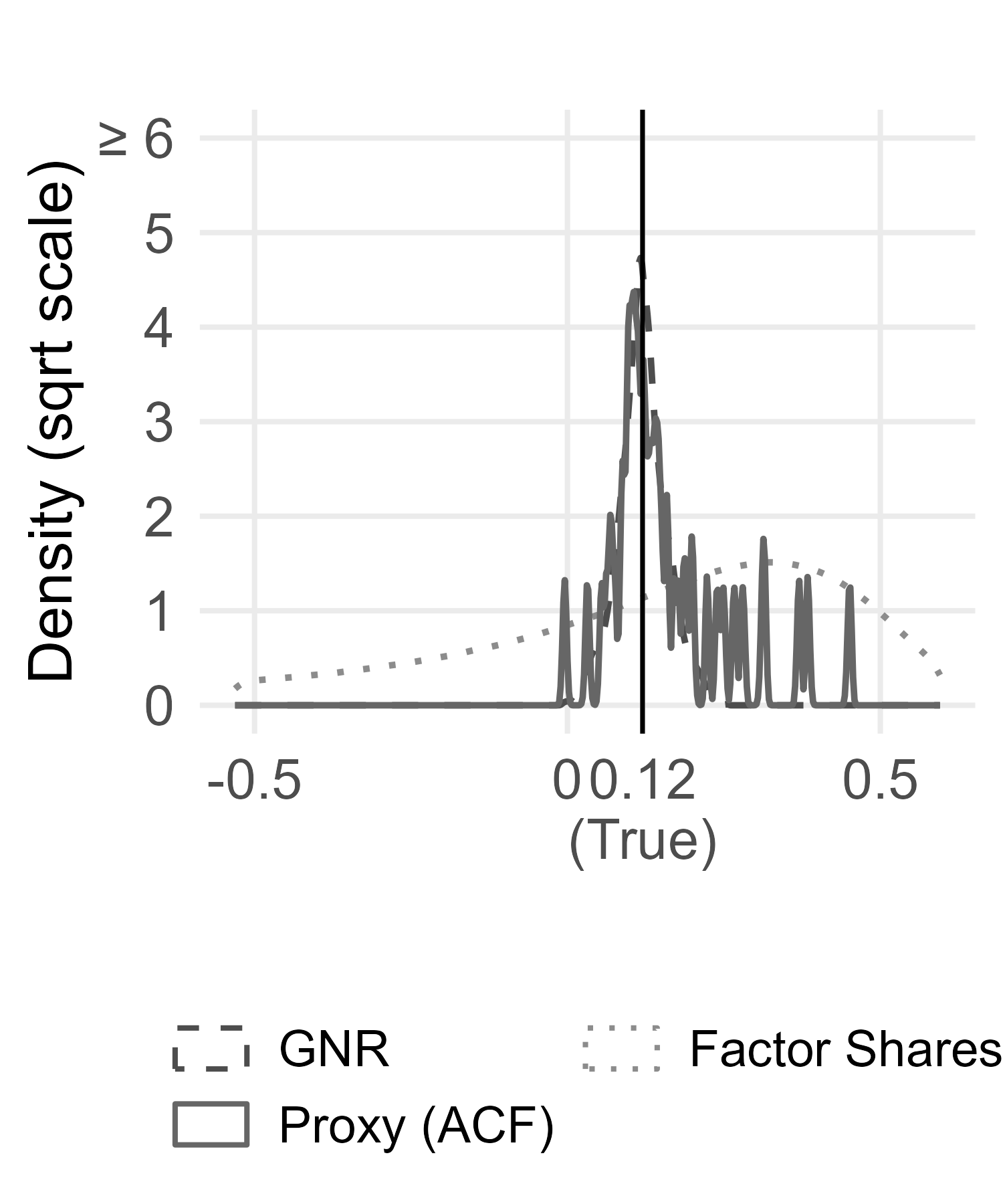}
        \caption{Capital elasticity}
        \label{fig:mc_elas_K_cd}
    \end{subfigure}
    \hfill
    \begin{subfigure}[h]{0.32\textwidth}
        \centering
        \includegraphics[width=\textwidth]{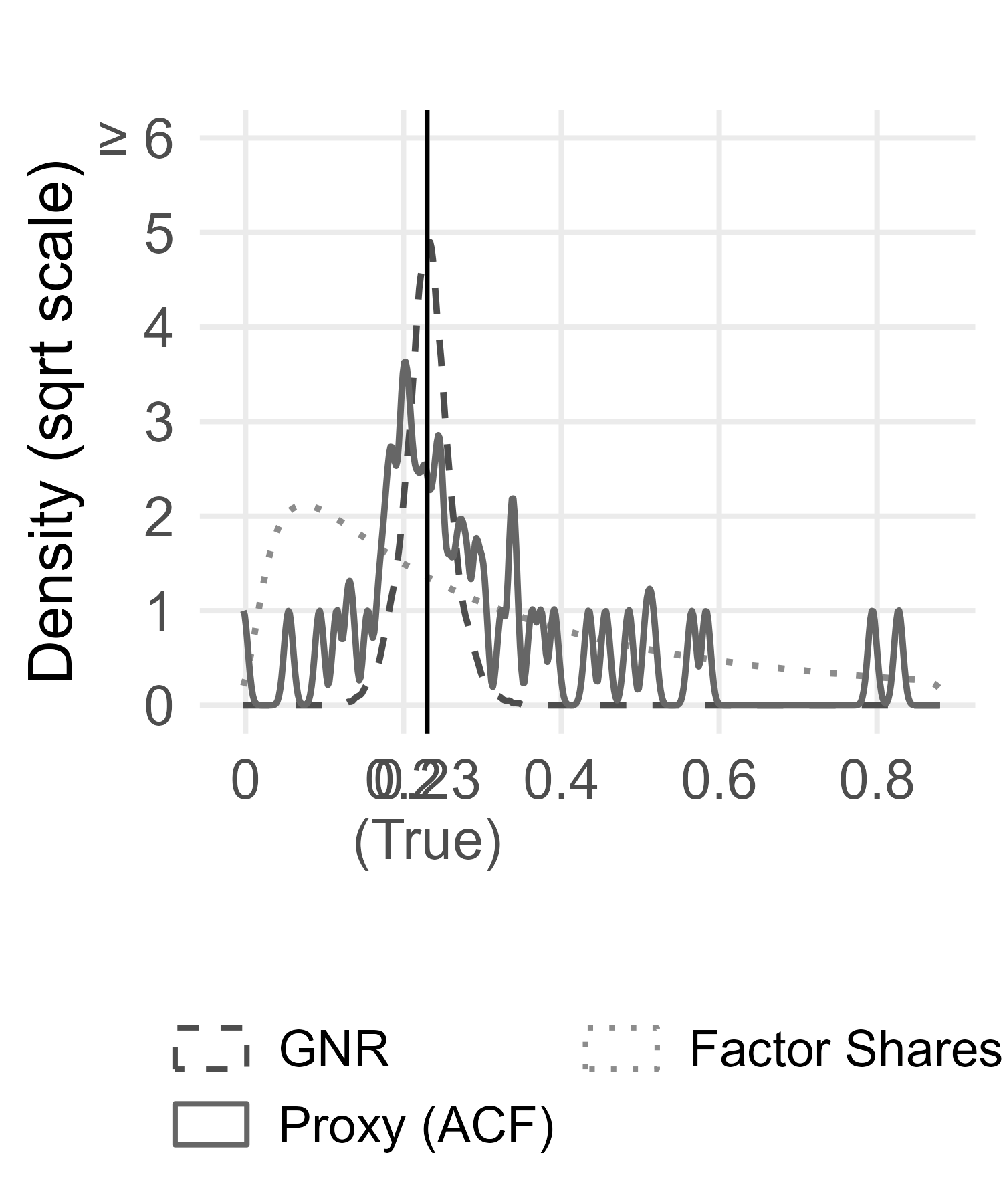}
        \caption{Labor elasticity}
        \label{fig:mc_elas_L_cd}
    \end{subfigure}
    \hfill
    \begin{subfigure}[h]{0.32\textwidth}
        \centering
        \includegraphics[width=\textwidth]{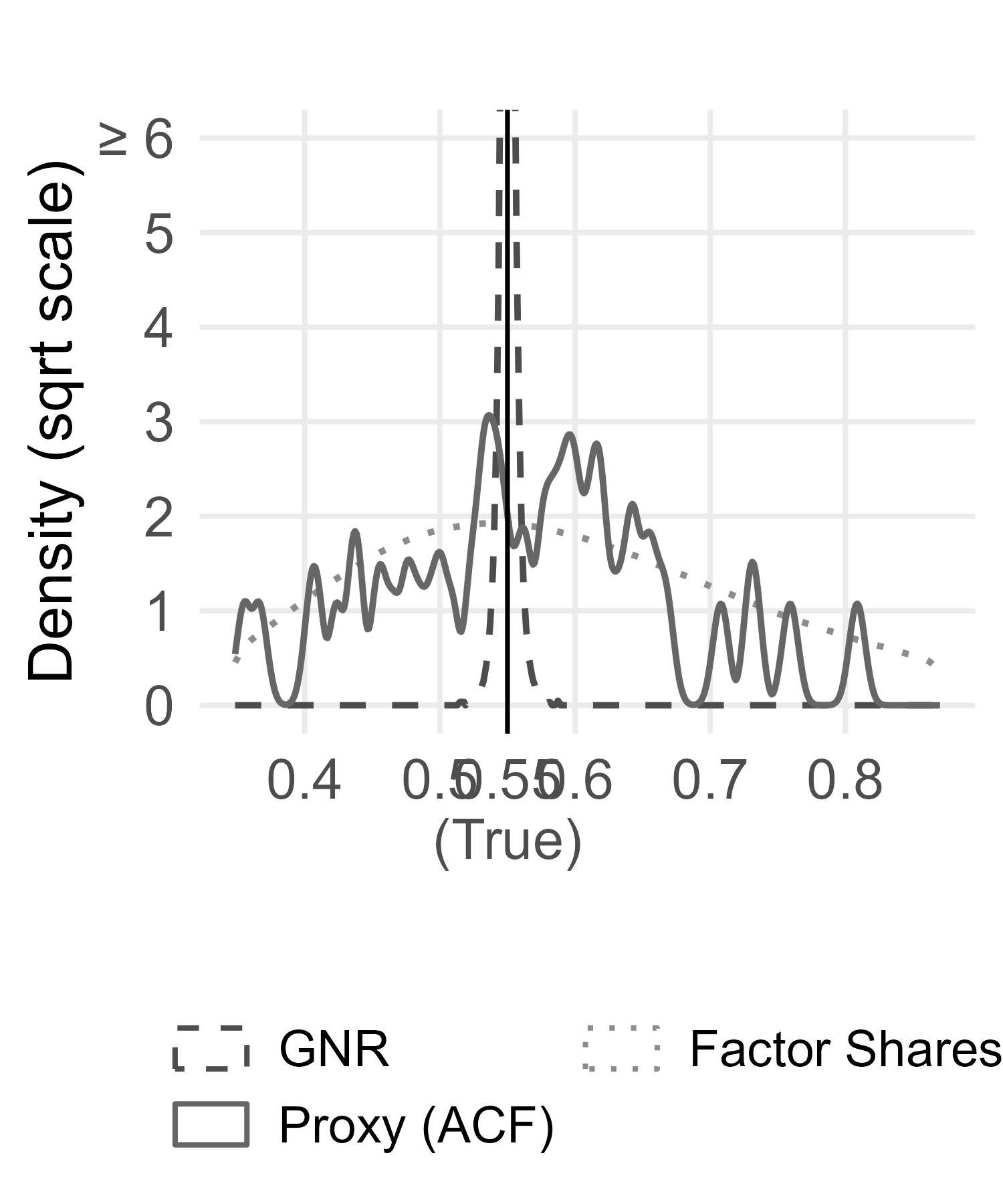}
        \caption{Materials elasticity}
        \label{fig:mc_elas_M_cd}
    \end{subfigure}
    \note{\scriptsize Each curve pools firm-level estimates across 100 Monte Carlo replications (500 firms $\times$ 30 periods each). Vertical lines indicate the true constant elasticities. The $y$-axis density values are reported on a square root scale. The GNR density peaks are truncated to preserve visual comparability across estimators.}
    \label{fig:mc_elas_CD}
\end{figure}

Figure~\ref{fig:mc_elas_CD} repeats the exercise under the Cobb-Douglas DGP, where true elasticities are constants ($\alpha_k^{\scriptscriptstyle\text{CD}} = 0.12$, $\alpha_l^{\scriptscriptstyle\text{CD}} = 0.23$, $\alpha_m^{\scriptscriptstyle\text{CD}} = 0.55$). The true elasticities appear as vertical lines, so any dispersion in the estimated densities is entirely spurious. The GNR estimator, despite using a flexible translog polynomial, recovers the constant elasticities with near-zero dispersion (within-replication standard deviations of 0.018, 0.015, and 0.003 for capital, labor, and materials). The factor shares estimator generates substantial within-replication dispersion (standard deviations of 0.116 for materials and 0.210 for labor) that is entirely spurious, driven by the ex-post shock passing through the expenditure shares. The CRS bias on capital is even more pronounced than under translog: the factor shares mean of 0.211 overstates the true 0.12 by 76\%. The proxy-variable estimator again shows wide cross-replication variance.

\subsubsection{Marginal Revenue Products}

\begin{figure}[ht]
\caption{Marginal revenue product distributions --- Translog DGP}
    \centering
    \begin{subfigure}[h]{0.32\textwidth}
        \centering
        \includegraphics[width=\textwidth]{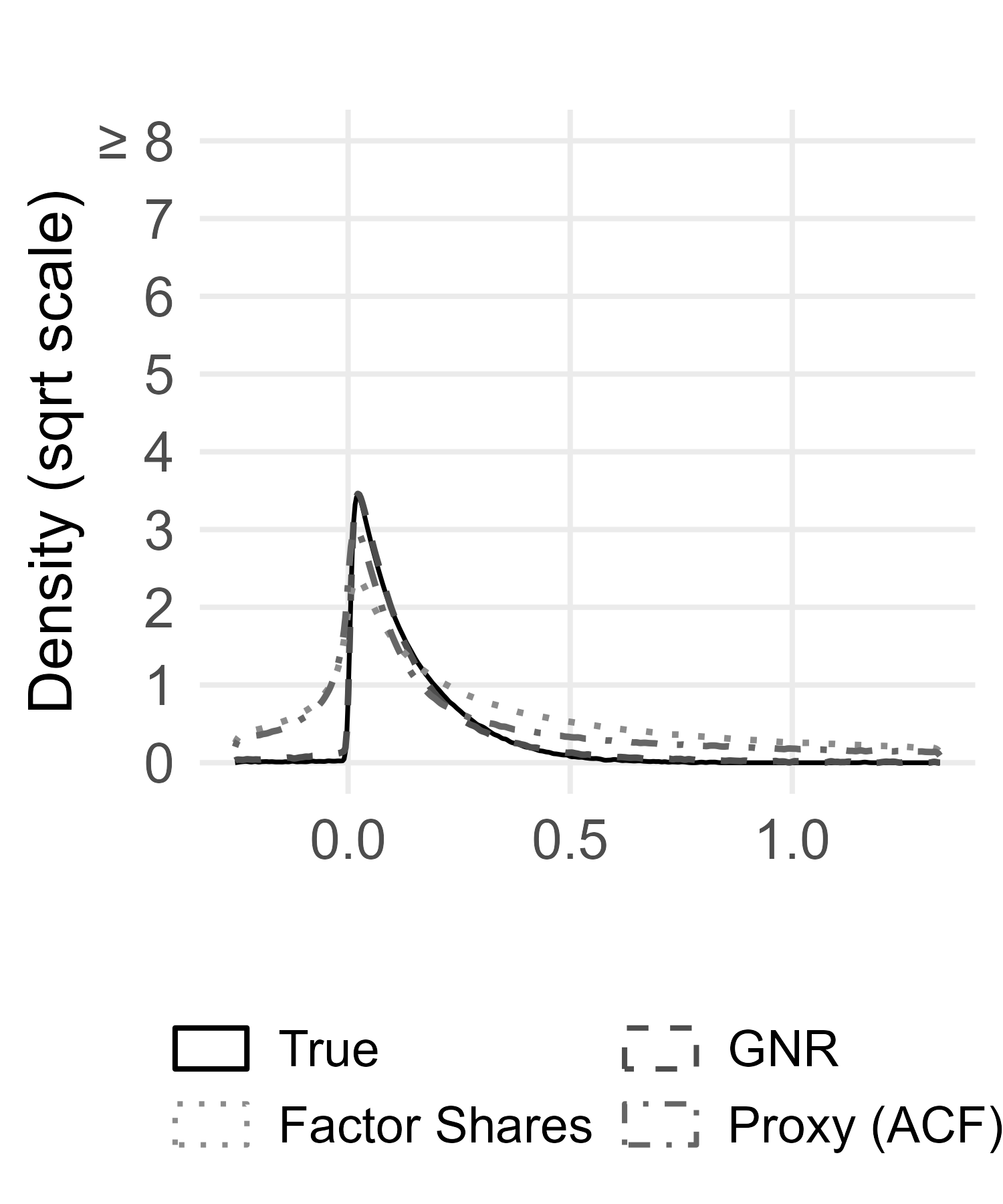}
        \caption{Capital MRP}
        \label{fig:mc_mrp_K_tl}
    \end{subfigure}
    \hfill
    \begin{subfigure}[h]{0.32\textwidth}
        \centering
        \includegraphics[width=\textwidth]{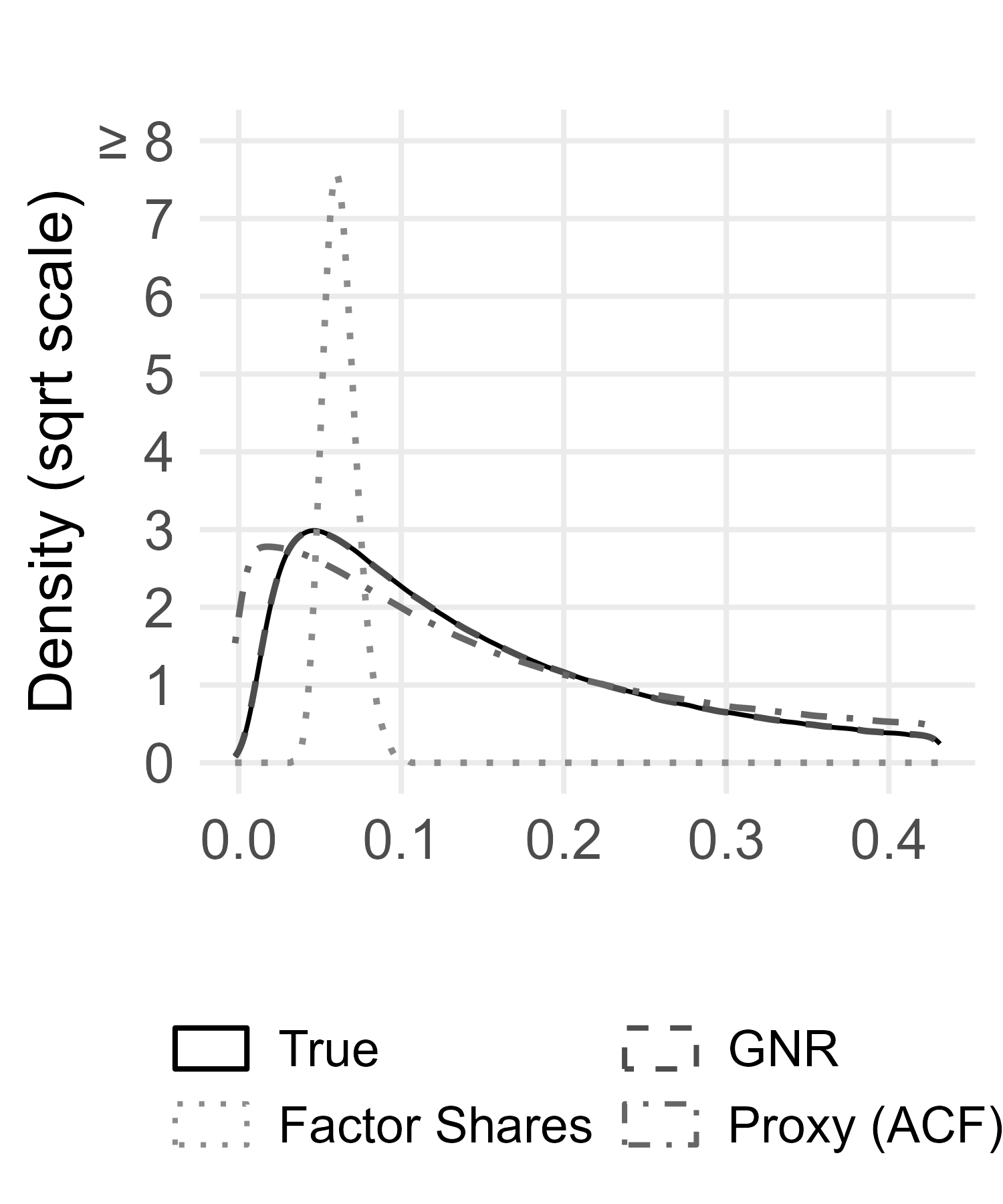}
        \caption{Labor MRP}
        \label{fig:mc_mrp_L_tl}
    \end{subfigure}
    \hfill
    \begin{subfigure}[h]{0.32\textwidth}
        \centering
        \includegraphics[width=\textwidth]{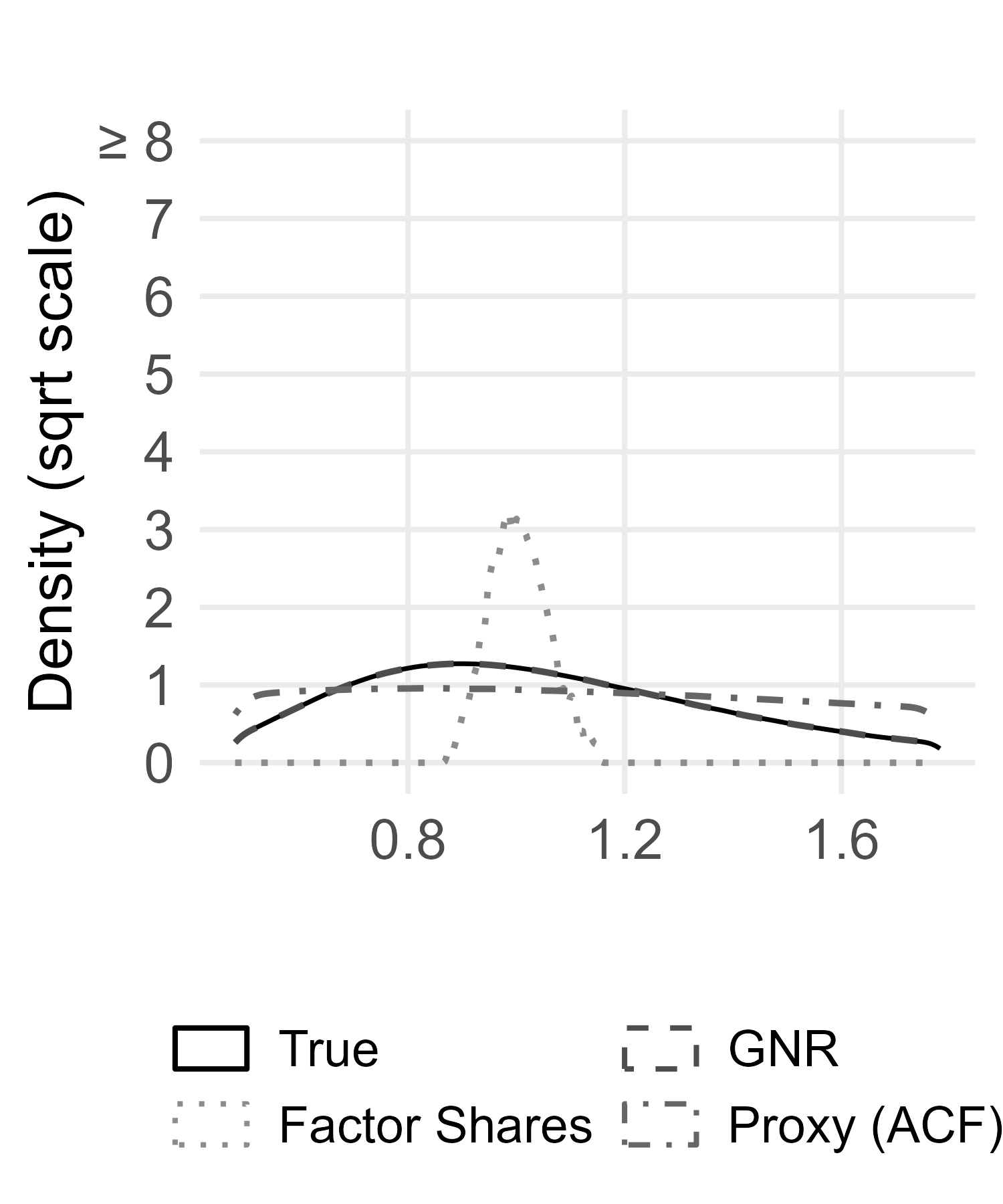}
        \caption{Materials MRP}
        \label{fig:mc_mrp_M_tl}
    \end{subfigure}
    \note{\scriptsize Each curve pools firm-level estimates across 100 Monte Carlo replications (500 firms $\times$ 30 periods each). ``True'' denotes the MRP computed from the true DGP parameters. The $y$-axis density values are reported on a square root scale to improve visual resolution in the tails while preserving the relative ordering of peaks.}
    \label{fig:mc_mrp_translog}
\end{figure}

\begin{figure}[ht]
    \caption{Marginal revenue product distributions --- Cobb-Douglas DGP}
    \centering
    \begin{subfigure}[h]{0.32\textwidth}
        \centering
        \includegraphics[width=\textwidth]{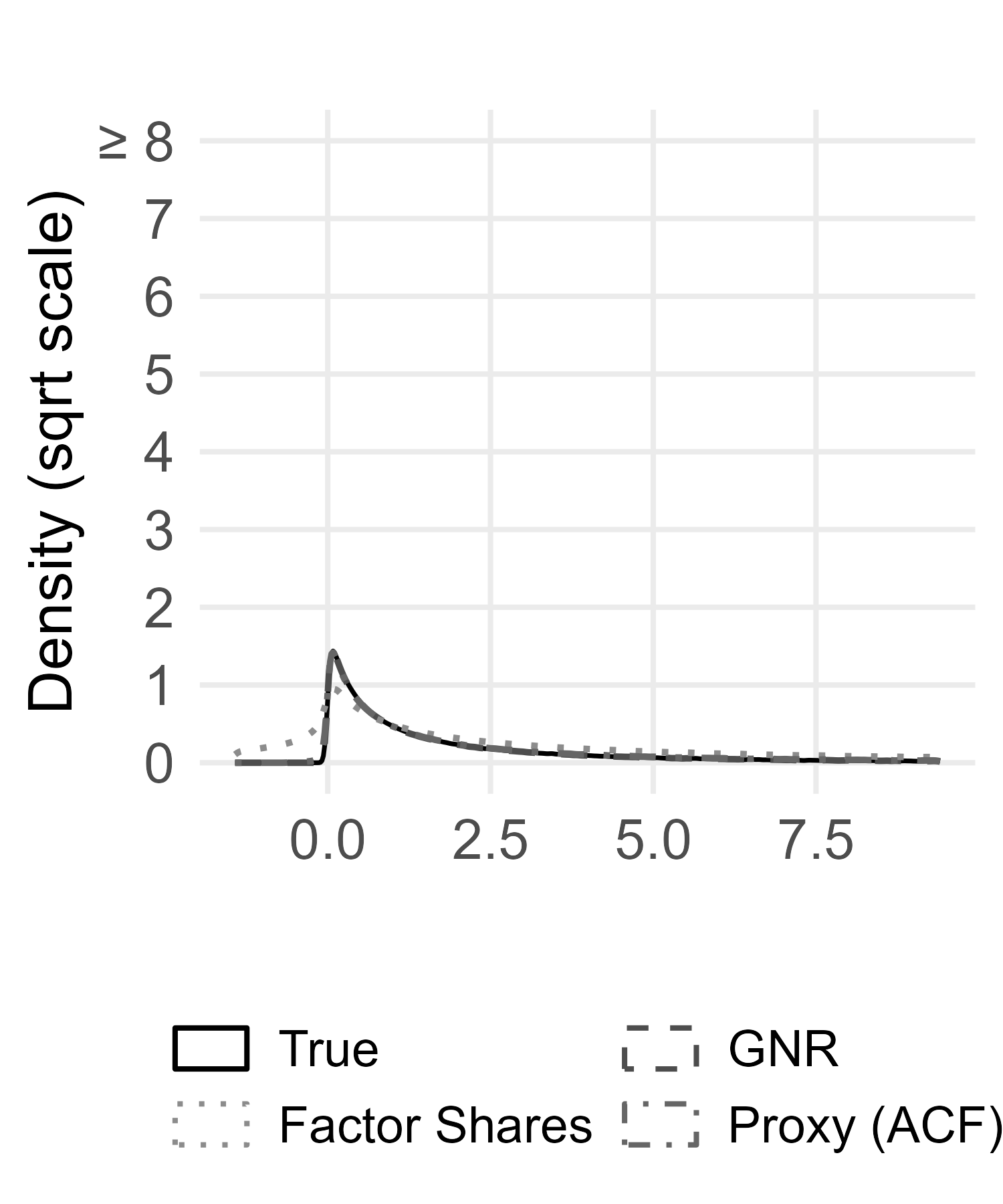}
        \caption{Capital MRP}
        \label{fig:mc_mrp_K_cd}
    \end{subfigure}
    \hfill
    \begin{subfigure}[h]{0.32\textwidth}
        \centering
        \includegraphics[width=\textwidth]{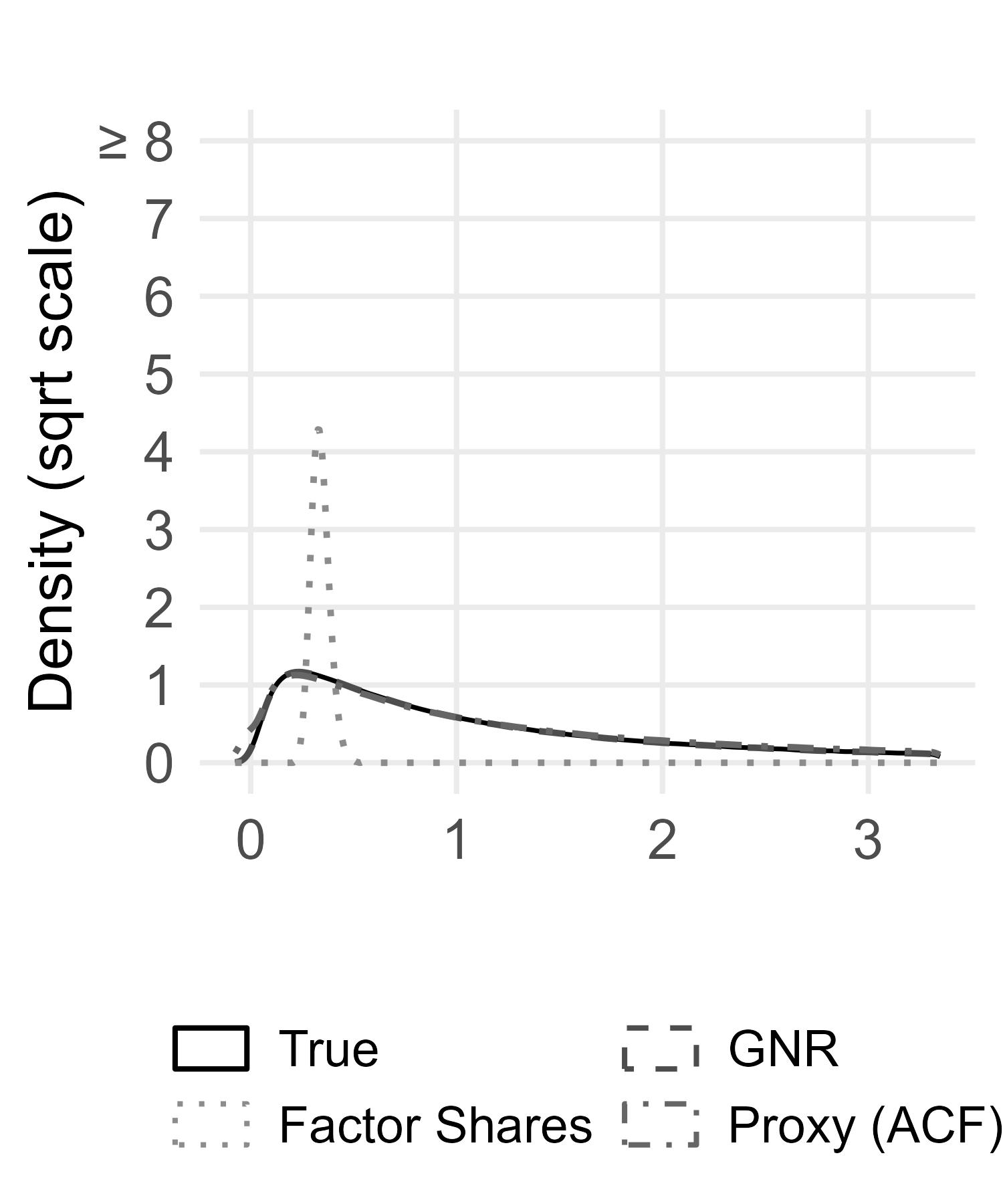}
        \caption{Labor MRP}
        \label{fig:mc_mrp_L_cd}
    \end{subfigure}
    \hfill
    \begin{subfigure}[h]{0.32\textwidth}
        \centering
        \includegraphics[width=\textwidth]{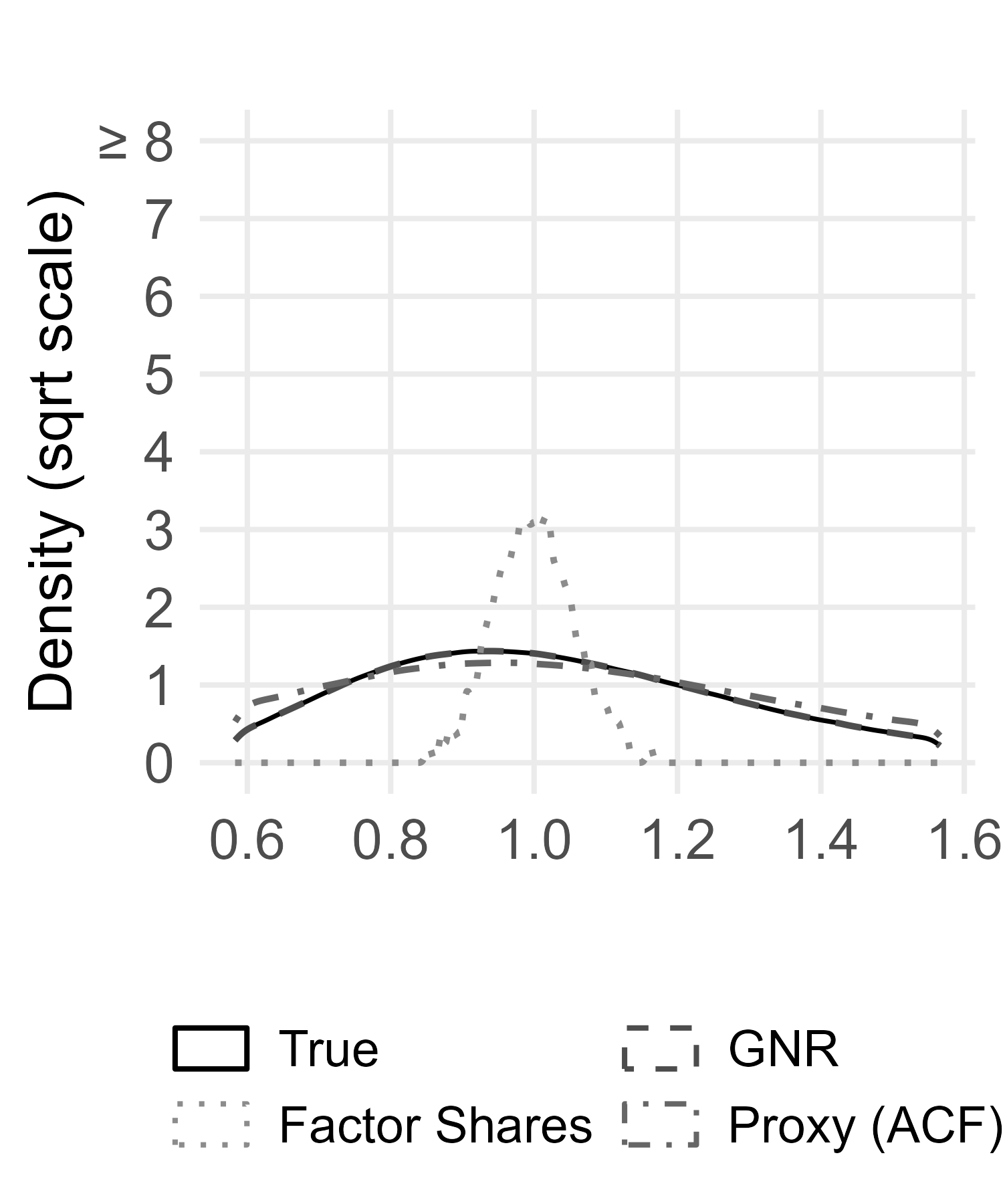}
        \caption{Materials MRP}
        \label{fig:mc_mrp_M_cd}
    \end{subfigure}
    \note{\scriptsize Each curve pools firm-level estimates across 100 Monte Carlo replications (500 firms $\times$ 30 periods each). ``True'' denotes the MRP computed from the true DGP parameters. The $y$-axis density values are reported on a square root scale. Factor shares density peaks for the labor and materials MRP are truncated to preserve visual comparability across estimators.}
    \label{fig:mc_mrp_CD}
\end{figure}

The elasticity differences across estimators carry over to marginal revenue products via equation~\eqref{eq:MP}. Figures~\ref{fig:mc_mrp_translog} and~\ref{fig:mc_mrp_CD} display the MRP distributions under the translog and Cobb-Douglas DGPs, respectively. Under both DGPs, the GNR estimator closely replicates the true MRP distributions for capital, labor, and materials. The factor shares estimator produces materials and labor MRP distributions that collapse to the input prices, since $\mathrm{mrp}^{M,\mathrm{FS}} = y - m + s_{jt} = \log\rho_t$ and $\mathrm{mrp}^{L,\mathrm{FS}} = y - l + s^L_{jt} = \log w_{jt}$ by construction, so that all remaining variation comes from input prices, and for labor from the wage response to the ex-post shock, rather than from productivity heterogeneity. The proxy-variable estimator produces MRP distributions with heavier tails, reflecting the imprecision of the underlying elasticity estimates.

\subsubsection{TFPR Dispersion and Persistence}

\begin{figure}[ht]
    \caption{TFPR distributions by estimator}
    \centering
    \begin{subfigure}[h]{0.48\textwidth}
        \centering
        \includegraphics[width=\textwidth]{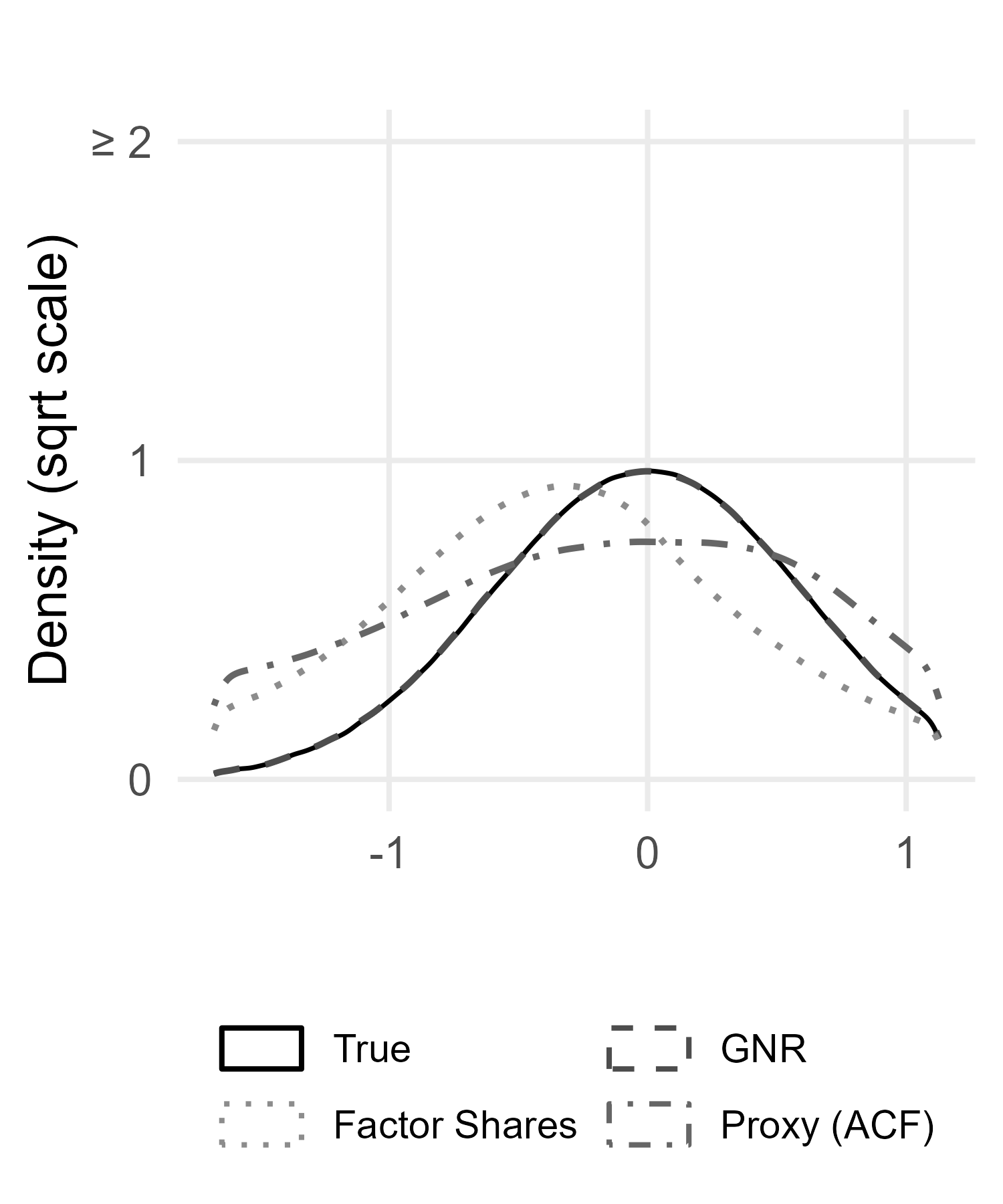}
        \caption{Translog DGP}
        \label{fig:mc_tfpr_tl}
    \end{subfigure}
    \hfill
    \begin{subfigure}[h]{0.48\textwidth}
        \centering
        \includegraphics[width=\textwidth]{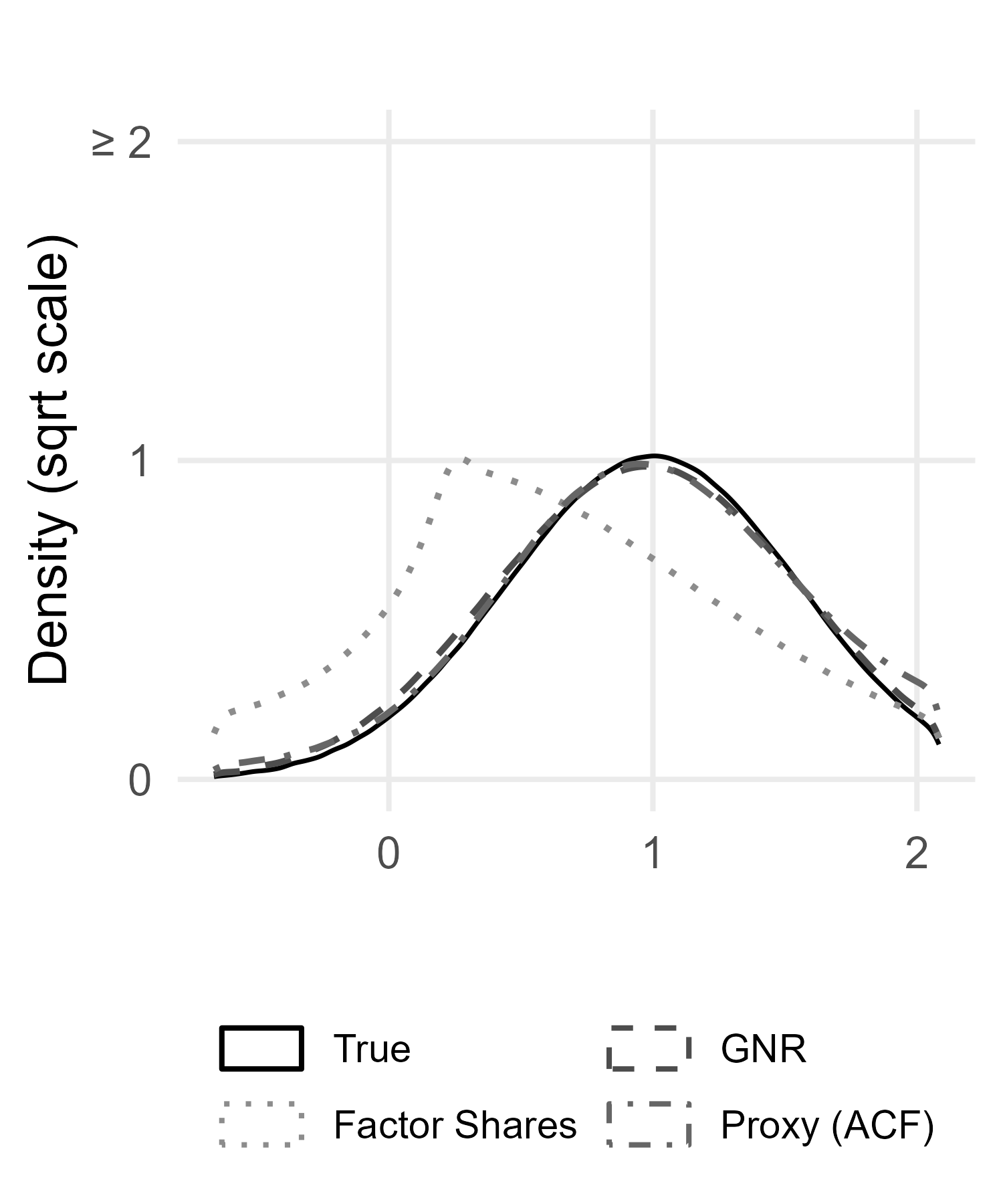}
        \caption{Cobb-Douglas DGP}
        \label{fig:mc_tfpr_cd}
    \end{subfigure}
        \note{\scriptsize Each curve pools firm-level estimates across 100 Monte Carlo replications (500 firms $\times$ 30 periods each). ``True'' denotes the TFPR computed from the true DGP parameters. The $y$-axis density values are reported on a square root scale to improve visual resolution in the tails while preserving the relative ordering of peaks.}
    \label{fig:mc_tfpr}
\end{figure}

Figure~\ref{fig:mc_tfpr} compares the TFPR distributions. Under both DGPs the factor shares estimator produces a TFPR distribution more dispersed than the truth, with a variance 2.0 times the true one under translog and 2.4 times under Cobb-Douglas, because the Solow residual inherits the noise of the two contaminated shares through the elasticities that multiply $k$, $l$, and $m$. The proxy-variable estimator generates a wider distribution with heavier tails, 2.8 and 1.5 times the true variance. The GNR estimator tracks the true TFPR distribution closely under both DGPs, with a variance ratio of 1.00.

The estimators differ in their ability to recover productivity persistence. The true AR(1) coefficient is $d_1 = 0.80$. GNR recovers it in both DGPs (0.801 translog, 0.789 Cobb-Douglas). The factor shares estimator underestimates it substantially: 0.680 under translog (15\% below the truth) and 0.566 under Cobb-Douglas (29\% below). The proxy-variable estimator recovers it on average (0.775 under translog and 0.798 under Cobb-Douglas) but with a cross-replication standard deviation of 0.090 under translog, ten times GNR's. The factor shares bias is larger under Cobb-Douglas than under translog.

\subsubsection{TFPR Variance Decomposition}

\begin{figure}[ht]
\caption{TFPR variance decomposition by estimator}
    \centering
    \begin{subfigure}[h]{0.48\textwidth}
        \centering
        \includegraphics[width=\textwidth]{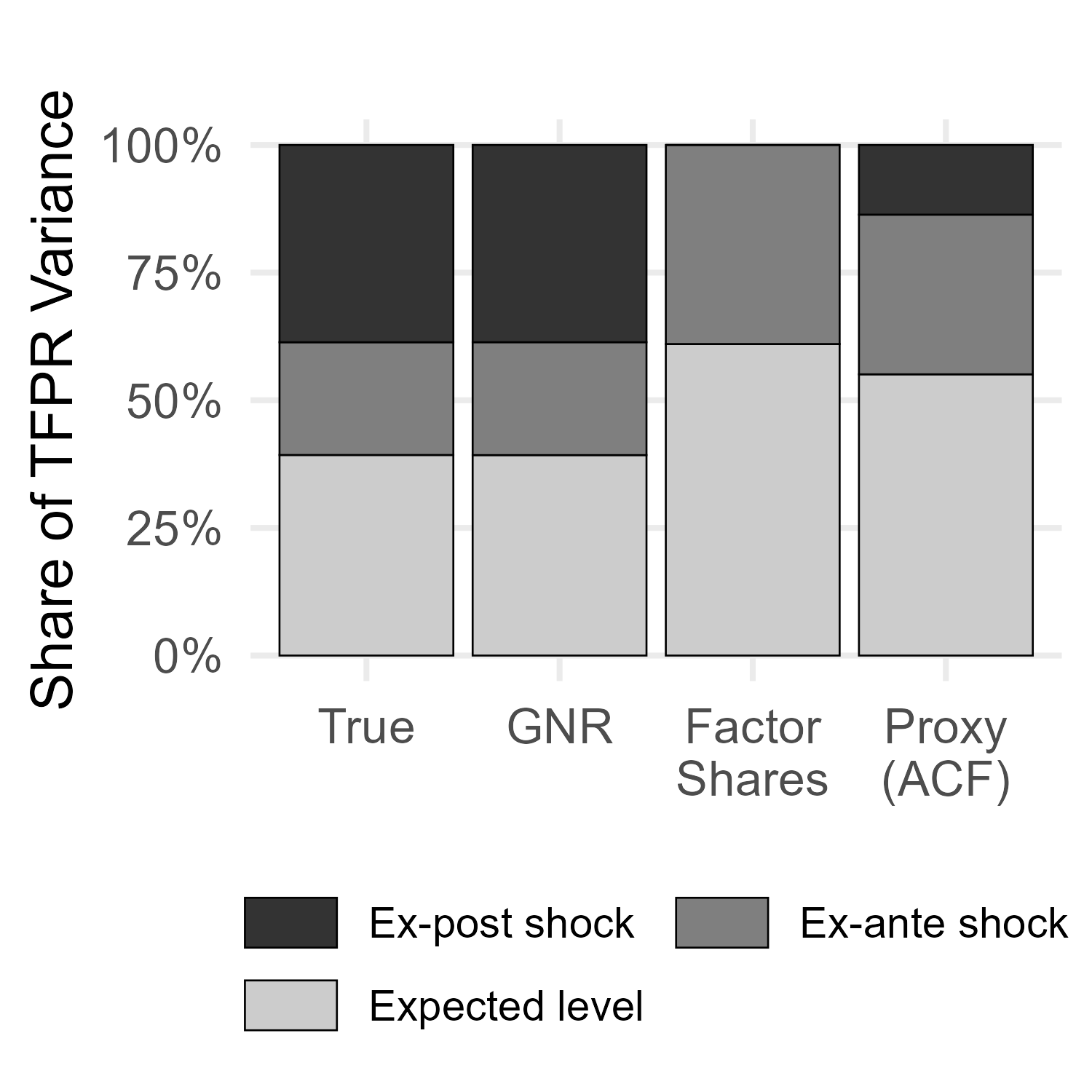}
        \caption{Translog DGP}
        \label{fig:mc_decomp_tl}
    \end{subfigure}
    \hfill
    \begin{subfigure}[h]{0.48\textwidth}
        \centering
        \includegraphics[width=\textwidth]{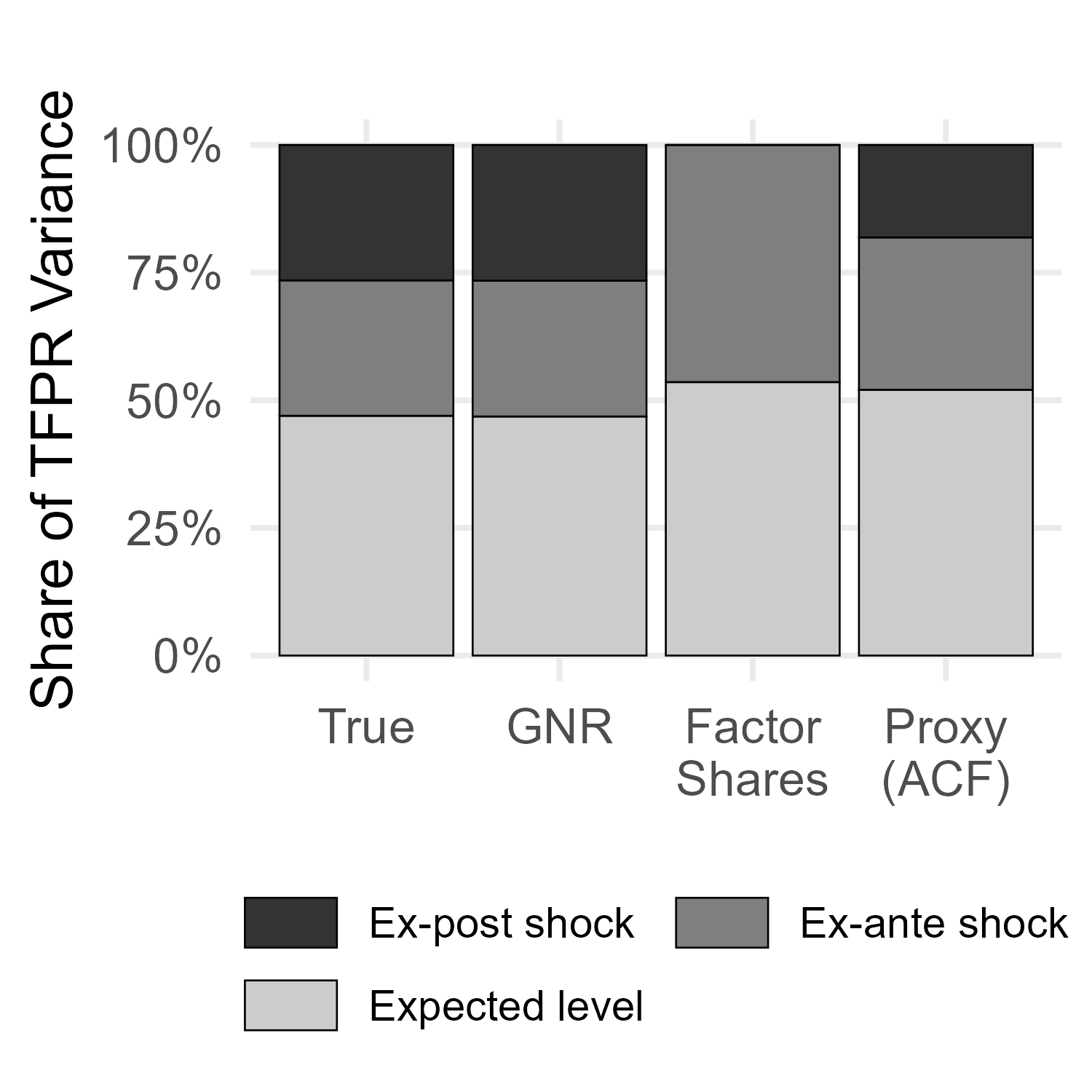}
        \caption{Cobb-Douglas DGP}
        \label{fig:mc_decomp_cd}
    \end{subfigure}
    \note{\scriptsize Bars show the share of cross-firm TFPR variance attributed to each component, averaged across 100 replications. For GNR and Proxy (ACF), the three components are the expected productivity level $\text{Var}(g(\omega_{-1}))$, the ex-ante shock $\text{Var}(\eta)$, and the ex-post shock $\text{Var}(\varepsilon)$. For Factor Shares, which cannot separate $\omega_{jt}$ from $\varepsilon_{jt}$, the decomposition is $\text{Var}(\hat{\nu}^{\mathrm{FS}}_{jt-1})$ and $\text{Var}(\hat{\nu}^{\mathrm{FS}}_{jt} - \hat{\nu}^{\mathrm{FS}}_{jt-1})$.}
    \label{fig:mc_decomp}
\end{figure}

Figure~\ref{fig:mc_decomp} decomposes the cross-firm TFPR variance into its components, expressed as shares of total variance. For the GNR and proxy-variable estimators, which can separate $\omega_{jt}$ from $\varepsilon_{jt}$, the decomposition is
\begin{equation*}
\text{Var}(\nu_{jt}) \approx \text{Var}\big(g_t(\omega_{jt-1})\big) + \text{Var}(\eta_{jt}) + \text{Var}(\varepsilon_{jt})
\end{equation*}
corresponding to the expected productivity level, the ex-ante shock, and the ex-post shock. For the factor shares estimator, which cannot separate these components, the decomposition is limited to
\begin{equation*}
\text{Var}(\hat\nu^{\mathrm{FS}}_{jt}) \approx \text{Var}(\hat\nu^{\mathrm{FS}}_{jt-1}) + \text{Var}(\hat\nu^{\mathrm{FS}}_{jt} - \hat\nu^{\mathrm{FS}}_{jt-1})
\end{equation*}

The GNR estimator replicates the true variance shares in both DGPs to within 0.14 percentage points. Under the true decomposition, the ex-post shock accounts for 39\% of TFPR variance, the ex-ante shock for 22\%, and the expected level for 39\% under translog (27\%, 27\%, and 47\% under Cobb-Douglas), so the two shocks jointly account for the majority of TFPR variance. The factor shares decomposition replaces the three-component split with the two-component approximation above. Because $\varepsilon_{jt}$ contaminates the Solow residual, the expected level component $\text{Var}(\hat{\nu}^{\mathrm{FS}}_{jt-1})$ is substantially inflated (it captures $\text{Var}(\omega_{jt-1}) + \text{Var}(\varepsilon_{jt-1})$ rather than $\text{Var}(\omega_{jt-1})$ alone), while the first-difference component absorbs both the ex-ante shock $\eta_{jt}$ and the change in the ex-post shock $\varepsilon_{jt} - \varepsilon_{jt-1}$; the level component takes 61\% under translog and 54\% under Cobb-Douglas. The proxy-variable estimator recovers the qualitative pattern of the true decomposition but under-attributes variance to the ex-post shock component, 14\% against the true 39\% under translog and 18\% against 27\% under Cobb-Douglas.

These Monte Carlo results highlight three limitations of standard approaches that are directly relevant to the analysis in this paper. First, the factor shares approach reads the labor and materials elasticities off shares the ex-post shock contaminates, and the CRS assumption then biases the capital elasticity and distorts the capital MRP distribution. Second, the factor shares inability to separate $\omega_{jt}$ from $\varepsilon_{jt}$ attenuates estimated productivity persistence and inflates the expected level component of the TFPR variance decomposition, obscuring the role of productivity changes across firms. Third, the proxy-variable estimator, even with generous price variation for identification, delivers estimates too imprecise for reliable inference on elasticity heterogeneity and MRP dispersion. The GNR estimator avoids all three issues, motivating its use in the main analysis.

\subsubsection{Firm-Level Benchmarks}\label{app:mc_reg}

The simulated panels also provide a model-implied benchmark for the firm-level regression of Section~\ref{sec:passthrough}. Table~\ref{tab:MC_passthrough} regresses each input gap on productivity within replication-period cells, for the true components of the data-generating process and for the components each estimator recovers, in the two specifications of Table~\ref{tab:components}: total TFPR, and the three components entered jointly.

Under the timing of the data-generating process, capital and labor are fixed when the ex-post shock lands and materials is chosen before it, so the capital and materials gaps respond to the true $\varepsilon$ one for one, while the labor gap at the realized wage responds at $1 - \theta = 0.59$ because the wage absorbs the remainder, as in Assumption~\ref{assp}. On the true components, the capital gap returns 1.001 on $\varepsilon$, and the labor gap 0.589 at the realized wage and 0.999 at the base wage, all within one Monte Carlo standard deviation of the benchmark. The materials gap returns exactly one, as the identity of Section~\ref{materials_identity} requires. The coefficients on total TFPR are 1.369 for capital and 1.146 for labor, and the capital gap returns 1.676 on $g(\omega_{-1})$ and 1.467 on $\eta$.

Of the three estimators, only GNR reproduces every benchmark, at 1.000 for capital and 0.590 for labor. Its coefficients on $g(\omega_{-1})$ and $\eta$, 1.629 and 1.432 for the capital gap, sit within one Monte Carlo standard deviation of the values for the true components. The proxy-variable estimator reproduces the coefficients on $\varepsilon$ (0.999 for capital, 0.591 for labor) but not those on the other two components. For the capital gap, its coefficients on $g(\omega_{-1})$ and $\eta$ are 0.829 and 0.722, with standard deviations across replications of 4.52 and 3.63. Factor shares recovers no ex-post shock and sets the labor and materials gaps to zero by construction, so it can be benchmarked on the capital gap and total TFPR only, where it returns 2.310 against 1.369 for the true components.

\begin{table}[!h]
    \centering
    \caption{Firm-Level Pass-Through on Monte Carlo Data}
    \label{tab:MC_passthrough}
{\footnotesize\setlength{\tabcolsep}{6pt}%
\begin{adjustbox}{max width=\textwidth}
\begin{tabular}{lcccc}
\toprule
 & & \multicolumn{3}{c}{Components, entered jointly} \\
\cmidrule(lr){3-5}
 & TFPR & $g(\omega_{-1})$ & $\eta$ & $\varepsilon$ \\
\midrule
\multicolumn{5}{l}{\textit{Panel A: capital gap}} \\
  Model-implied value & & & & 1 \\
  True DGP & 1.369 & 1.676 & 1.467 & 1.001 \\
   & (0.028) & (0.054) & (0.026) & (0.020) \\
  GNR & 1.342 & 1.629 & 1.432 & 1.000 \\
   & (0.058) & (0.121) & (0.066) & (0.005) \\
  Proxy variable & 0.831 & 0.829 & 0.722 & 0.999 \\
   & (0.918) & (4.516) & (3.626) & (0.025) \\
  Factor shares & 2.310 &  &  &  \\
   & (0.029) &  &  &  \\
\midrule
\multicolumn{5}{l}{\textit{Panel B: labor gap at the realized wage, $\theta = 0.41$}} \\
  Model-implied value & & & & $1 - \theta = 0.59$ \\
  True DGP & 1.146 & 1.325 & 1.804 & 0.589 \\
   & (0.020) & (0.038) & (0.020) & (0.015) \\
  GNR & 1.145 & 1.321 & 1.803 & 0.590 \\
   & (0.032) & (0.078) & (0.011) & (0.003) \\
  Proxy variable & 0.595 & 0.501 & 0.768 & 0.591 \\
   & (0.400) & (2.267) & (2.947) & (0.017) \\
\multicolumn{5}{l}{\textit{Labor gap at the base wage, $\theta = 0$}} \\
  Model-implied value & & & & 1 \\
  True DGP & 1.305 & 1.325 & 1.804 & 0.999 \\
   & (0.019) & (0.038) & (0.020) & (0.015) \\
  GNR & 1.304 & 1.321 & 1.804 & 1.000 \\
   & (0.032) & (0.078) & (0.011) & (0.003) \\
  Proxy variable & 0.659 & 0.501 & 0.768 & 1.001 \\
   & (0.431) & (2.289) & (2.953) & (0.017) \\
\midrule
\multicolumn{5}{l}{\textit{Panel C: materials gap (identity)}} \\
  Model-implied value & & & & 1 \\
  True DGP & 0.386 & 0.000 & 0.000 & 1.000 \\
   & (0.006) & (0.000) & (0.000) & (0.000) \\
  GNR & 0.386 & 0.000 & 0.000 & 1.000 \\
   & (0.007) & (0.000) & (0.000) & (0.000) \\
  Proxy variable & 0.203 & -0.073 & -0.042 & 1.001 \\
   & (0.486) & (0.791) & (0.649) & (0.022) \\
\bottomrule
\end{tabular}
\end{adjustbox}%
}
\note{\scriptsize Each cell reports the coefficient from a regression of the input gap, $\log \mathrm{elas}^X_{jt} + y_{jt} - x_{jt} - \log P^X_{jt}$, on the productivity measure in the column header, on the simulated panel of the three-input translog data-generating process of Appendix~\ref{app:mc_dgp} (100 replications of 500 firms over 30 periods), pooled across replications with replication-period cell fixed effects; the TFPR column regresses the gap on total TFPR, and the three component columns enter $g(\omega_{-1})$, $\eta$, and $\varepsilon$ together and report the coefficient on each. Parentheses give the standard deviation of the coefficient across the 100 replications. Rows differ in the productivity measure: the true components of the data-generating process, and the components recovered by each estimator of Appendix~\ref{app:mc_estimators}. Factor shares recovers no ex-post shock and sets the labor and materials elasticities equal to the observed shares, so it appears only in the TFPR column of the capital panel. Under the timing of the data-generating process capital and labor are fixed when the ex-post shock lands and materials is chosen before it; the wage is settled after it and responds with elasticity $\theta$, so the model-implied coefficient on $\varepsilon$ is 1 for capital, $1 - \theta$ for labor at the realized wage, 1 for labor at the base wage, and 1 for materials, shown in the first row of each panel; the model implies no value for the coefficients on total TFPR, $g(\omega_{-1})$, or $\eta$.}
\end{table}

\subsection{Empirical Results}\label{app:mc_empirical}

I now apply the factor shares estimator to the same firm-level data used in the main analysis and compare the resulting TFPR and input gap dispersion estimates to those obtained from the GNR estimator in Section~\ref{s:results}. The Monte Carlo experiments above show that the factor shares approach introduces biases under controlled conditions; this section compares the two estimators on the data.

\subsubsection{Implementation}

I compute the factor shares elasticities of equations~\eqref{eq:mc_fs_elas}--\eqref{eq:mc_fs_elask} in the \textit{sectoral} implementation, the form in which the factor shares approach is normally applied: every firm in a country-industry-year cell receives the same expenditure shares, computed from the cell's aggregate expenditure and revenue. The materials and labor elasticities are the cell's materials expenditure and wage bill, each divided by the cell's revenue, and the capital elasticity is the CRS residual $1 - \widehat{\mathrm{elas}}^{M,\mathrm{FS}}_{jt} - \widehat{\mathrm{elas}}^{\,L,\mathrm{FS}}_{jt}$. TFPR is the Solow residual of equation~\eqref{eq:mc_fs_tfpr}. The excess TFPR dispersion and the attenuated persistence of the Monte Carlo arise under each firm's own shares, which the ex-post shock enters in full; a cell's share is the revenue-weighted mean of its firms' shares, so the shock enters it only through its mean across the cell's firms, which removes the excess dispersion of the factor shares TFPR relative to the true one and reduces the attenuation of its persistence.

\subsubsection{GNR against Sectoral Factor Shares}

Table~\ref{tab:fs_comparison} compares GNR with the sectoral factor shares benchmark on identical samples, so the two differ only in how the elasticities are obtained. GNR delivers less input gap dispersion than the sectoral factor shares benchmark. The within-cell standard deviation of the capital gap is 1.122 under GNR against 1.308 under factor shares, and that of the materials gap 0.450 against 0.748, a ratio of 0.60. Labor is the one input for which the two are close, at 0.759 against 0.733. The sectoral benchmark holds the elasticity fixed within a cell, so its within-cell gap dispersion is the dispersion of the log ratio of revenue to outlay on the input. GNR lets the elasticity vary across firms and, for capital and materials, attributes to technology part of the cross-firm variation in shares that the sectoral factor shares benchmark reads as input gaps.

\begin{table}[!h]
    \centering
    \caption{Dispersion of Input Gaps: GNR against Factor Shares}
    \label{tab:fs_comparison}
\begin{tabular}{lccc}
\toprule
 & Capital & Labor & Materials \\
\midrule
  GNR & 1.122 & 0.759 & 0.450 \\
  Factor shares, sectoral & 1.308 & 0.733 & 0.748 \\
   \bottomrule
\end{tabular}
\note{\scriptsize Each entry is the within-cell standard deviation of the log gap $\mathrm{mrp}^X_{jt} - \log P^X_{jt}$, pooled across country-industry-year cells, computed on identical samples and differing only in how the input elasticities are obtained. The factor shares row gives every firm in a country-industry-year cell the cell's expenditure shares, computed from aggregate expenditure and revenue. Pooled AR(1) persistence of the recovered productivity process is 0.818 under GNR against 0.793 under factor shares.}
\end{table}

\subsubsection{TFPR Dispersion}

\begin{figure}[ht]
    \caption{TFPR dispersion and mean evolution}
    \centering
    \begin{subfigure}[h]{0.48\textwidth}
        \centering
        \includegraphics[width=\textwidth]{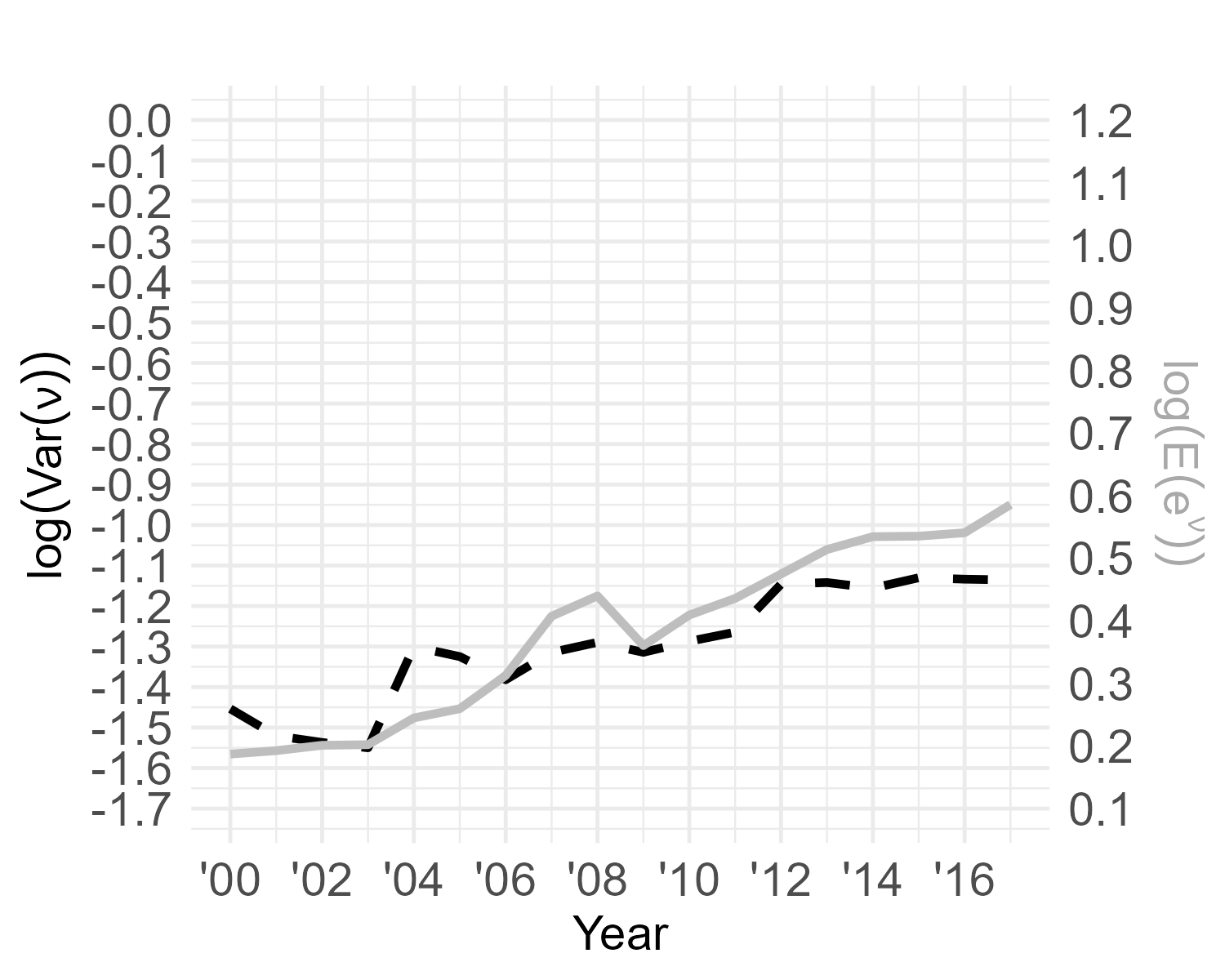}
        \caption{Baseline --- GNR}
        \label{fig:emp_tfpr_gnr}
    \end{subfigure}
    \hfill
    \begin{subfigure}[h]{0.48\textwidth}
        \centering
        \includegraphics[width=\textwidth]{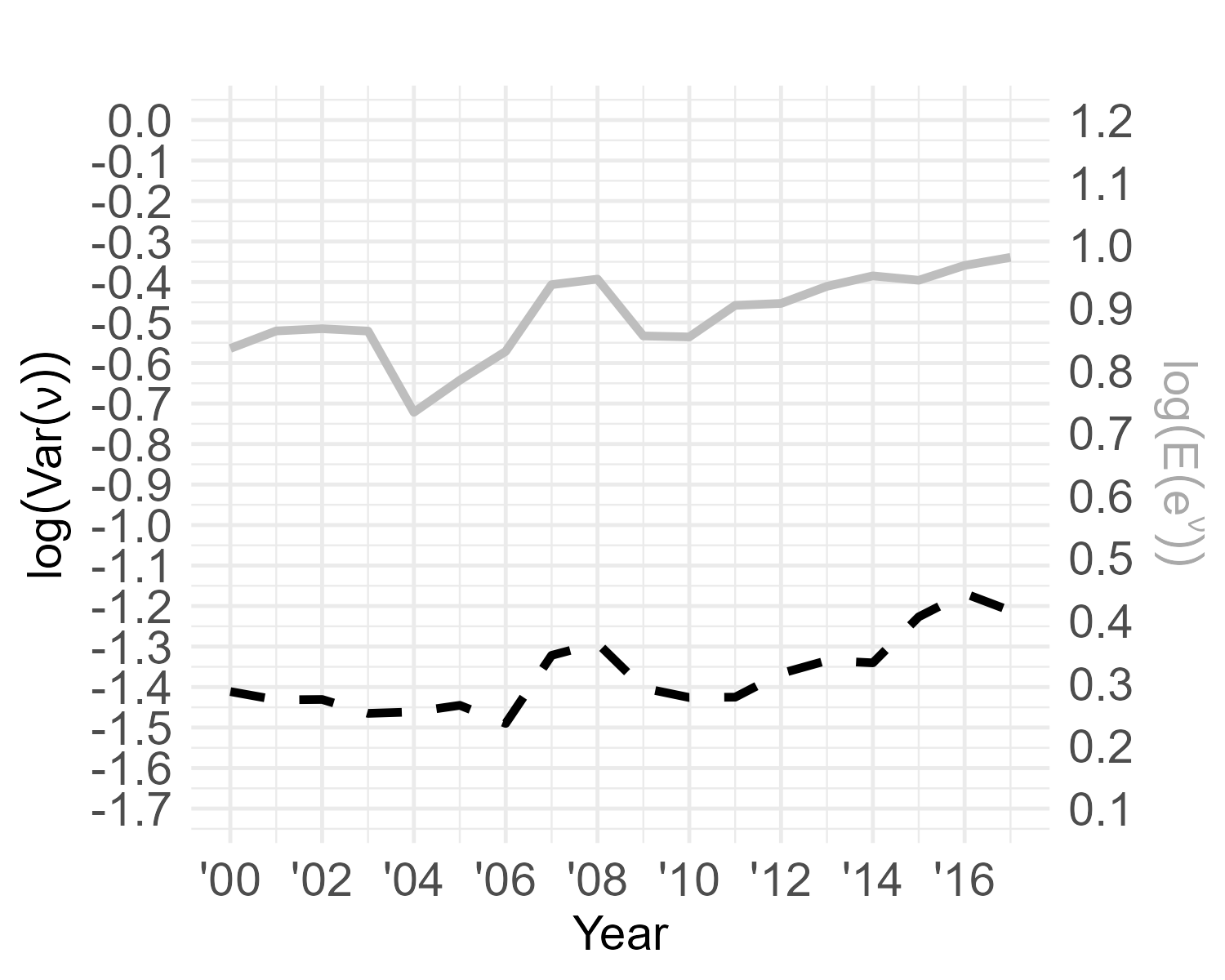}
        \caption{Factor Shares}
        \label{fig:emp_tfpr_fs}
    \end{subfigure}
    \note{\scriptsize The dashed black line shows $\log(\text{Var}_{st}(\nu_{jt}))$ (primary axis); the solid grey line shows $\log(E_{st}[\exp(\nu_{jt})])$ (secondary axis). In the left panel, $\nu_{jt}$ is the GNR-estimated TFPR from Section~\ref{s:results}; in the right panel, $\nu_{jt}$ is the factor shares Solow residual with the cell's expenditure shares. Variances and means are computed within each country-industry-year cell across six countries (France, Spain, Italy, Germany, Poland, Romania) and eight NACE industry groups over 2000--2017 (2004--2017 for Germany, Poland, and Romania), pooled using industry revenue shares as weights.}
    \label{fig:emp_tfpr}
\end{figure}

Figure~\ref{fig:emp_tfpr} displays the evolution of aggregate TFPR dispersion, measured as $\log\text{Var}_{st}(\nu)$ (black dashed line, primary axis), and mean TFPR level, measured as $\log E_{st}[\exp(\nu)]$ (grey solid line, secondary axis), under the GNR estimator and the factor shares approach.

The two estimators deliver similar TFPR dispersion: the log variance ranges between $-1.5$ and $-1.2$ under factor shares and between $-1.6$ and $-1.1$ under GNR. Factor shares is above GNR by 0.04 to 0.11 log points in 2000 to 2003 and below it by up to 0.22 from 2004 on, and both series trend upward over the period. The Monte Carlo experiments predict higher dispersion under factor shares because the ex-post shock $\varepsilon_{jt}$ enters each firm's materials and wage-bill shares, and the Solow residual carries that noise through the elasticities that multiply $k_{jt}$, $l_{jt}$ and $m_{jt}$. A cell's share is the revenue-weighted mean of its firms' shares, in which the shock averages out. The two residuals then differ in the technology: one elasticity per cell for each input under factor shares, with the capital elasticity set to one minus the two shares, against the firm-specific translog elasticities of GNR, which need not sum to one. The factor shares mean TFPR level is higher than GNR's, 0.73 to 0.98 against 0.19 to 0.59, and rises less over the period. On persistence the data go the same way as the simulation, with a smaller gap: on the same firm-year pairs, with the lag taken at the firm's previous observation, the pooled AR(1) coefficient is 0.818 for $\omega_{jt}$ under GNR and 0.793 for the factor shares Solow residual, against 0.80 and 0.68 in Appendix~\ref{app:mc_results}.

\subsubsection{Input Gap Dispersion}

\begin{figure}[ht]
    \caption{Input gap dispersion evolution}
    \centering
    \begin{subfigure}[h]{0.48\textwidth}
        \centering
        \includegraphics[width=\textwidth]{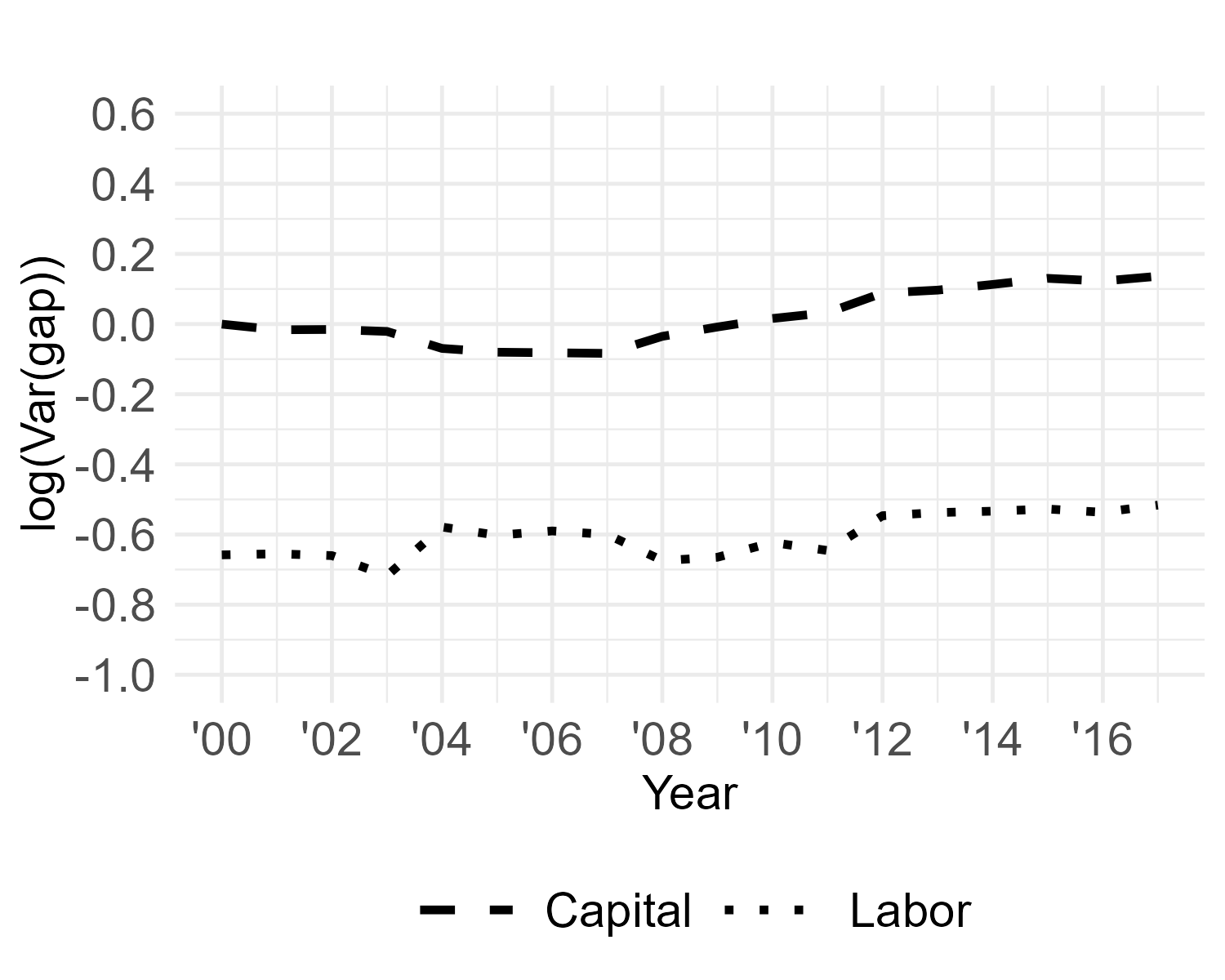}
        \caption{Baseline --- GNR}
        \label{fig:emp_gap_gnr}
    \end{subfigure}
    \hfill
    \begin{subfigure}[h]{0.48\textwidth}
        \centering
        \includegraphics[width=\textwidth]{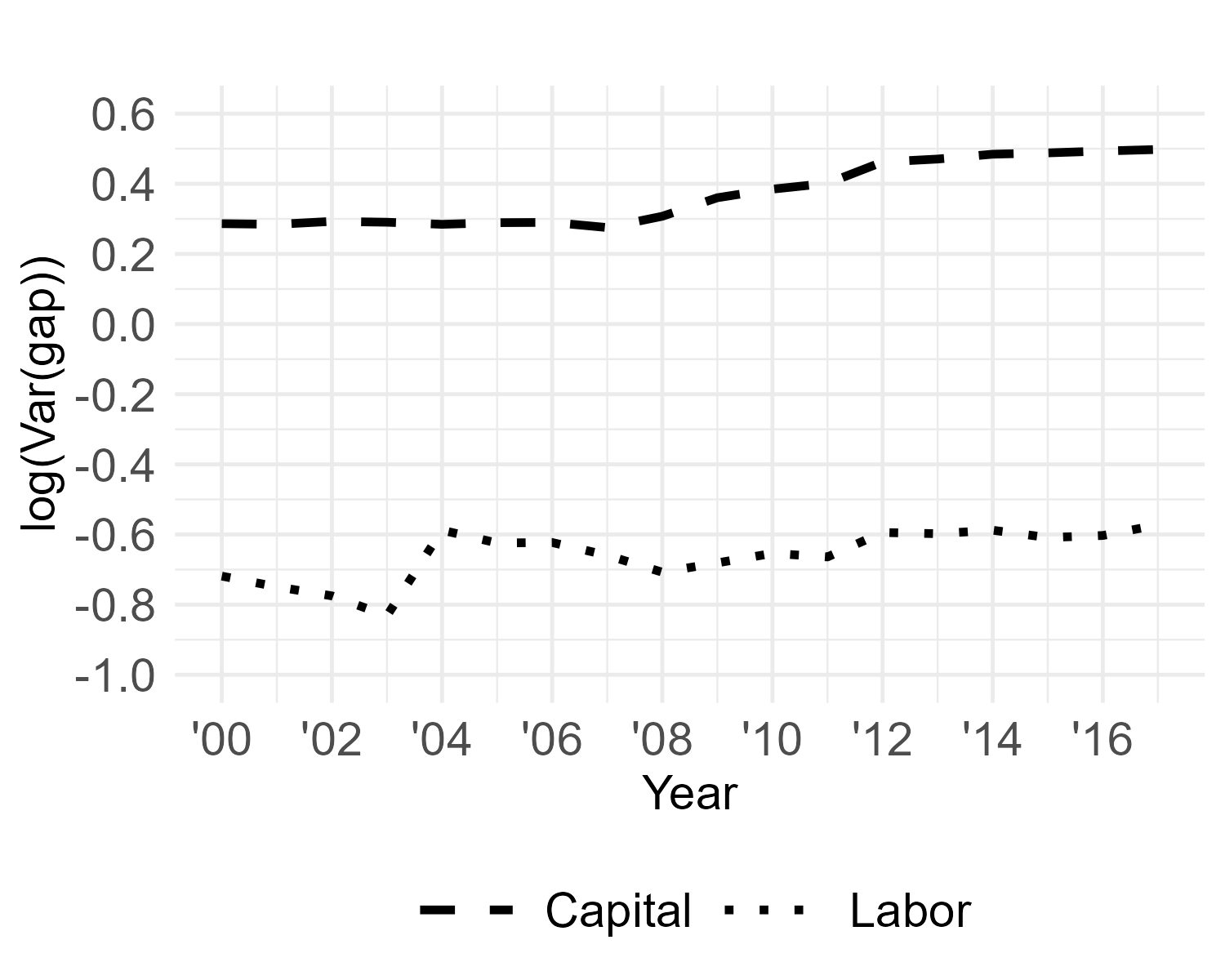}
        \caption{Factor Shares}
        \label{fig:emp_gap_fs}
    \end{subfigure}
    \note{\scriptsize Each line shows $\log(\text{Var}_{st}(\mathrm{mrp}^X_{jt} - \log P^X_{jt}))$, the within-cell variance of the input gap of equation~(\ref{eq:gap}), for capital (dashed) and labor (dotted). In the left panel, elasticities are from the GNR estimator; in the right panel, from the factor shares approach with the cell's expenditure shares. Variances are computed within each country-industry-year cell across six countries and eight NACE industry groups over 2000--2017 (2004--2017 for Germany, Poland, and Romania), pooled using industry revenue shares as weights.}
    \label{fig:emp_gap}
\end{figure}

Input gap dispersion is measured as the cross-sectional variance of the input gap $\mathrm{mrp}^X_{jt} - \log P^X_{jt}$ of equation~(\ref{eq:gap}) within each country-industry-year cell. Figure~\ref{fig:emp_gap} compares its evolution for capital and labor under the two estimators. Under both, the capital gap (dashed) is more dispersed than the labor gap (dotted), and all four series rise over the period. The figure is the year-by-year counterpart of Table~\ref{tab:fs_comparison}: the sectoral benchmark delivers more dispersion of the capital gap than GNR in every year, by 0.29 to 0.37 log points, and slightly less dispersion of the labor gap, by 0.01 to 0.11 log points.

\clearpage

\end{document}